\documentclass[12pt]{article}
\usepackage{amsmath}
\usepackage{amsfonts}
\usepackage{graphicx,psfrag,epsf}
\usepackage[usenames, dvipsnames]{color}
\usepackage{url} 
\usepackage[shortlabels]{enumitem} 
\usepackage{mdframed}
\usepackage{caption}
\usepackage{subcaption}
\usepackage{siunitx}
\usepackage{float}
\usepackage[normalem]{ulem}
\newcommand{\stkout}[1]{\ifmmode\text{\sout{\ensuremath{#1}}}\else\sout{#1}\fi}
\usepackage{mathtools}
\usepackage{soul}
\usepackage{bm}
\usepackage{booktabs}
\usepackage{dsfont}
\usepackage{multirow}

\graphicspath{{./}}

\usepackage[dvipsnames]{xcolor} %
\usepackage{diagbox}
\usepackage{bbm}
\usepackage{algorithm}
\usepackage[noend]{algpseudocode}
\algrenewcommand\algorithmicrequire{\textbf{Input:}}

\algrenewcommand\algorithmicensure{\textbf{Output:}}

\usepackage[utf8]{inputenc}
\usepackage[english]{babel}
\usepackage{amsthm}
\usepackage{mathtools}

\usepackage{hyperref}
\hypersetup{
    colorlinks=true,
    linkcolor=blue,
    citecolor=blue,
    filecolor=magenta,      
    urlcolor=blue,
}
\usepackage{lineno}
\usepackage{setspace}
\usepackage{placeins}  

\newcommand{\bs}{\bm{s}}

\makeatletter
\renewcommand*\env@matrix[1][\arraystretch]{%
  \edef\arraystretch{#1}%
  \hskip -\arraycolsep
  \let\@ifnextchar\new@ifnextchar
  \array{*\c@MaxMatrixCols c}}
\makeatother

\usepackage[natbibapa]{apacite}
\begin{document}
	
	\setstretch{1.5}

\title{
Spatial Extremes on Domains with Physical Barriers}
\author{Chien-Chung Wang and Benjamin A. Shaby \\ Department of Statistics, Colorado State University}
\date{\today}
\maketitle


\begin{abstract}
		We propose a model for spatial extremes what realistically accommodates nonconvex domains.  Most existing models assume that tail dependence between locations depends only on their separation.  Even if extended to be anisotropic or nonstationary, they still cannot accommodate features between points that may mediate the dependence. This assumption limits their applicability in settings where physical barriers like mountains, coastlines, or islands may attenuate or even block dependence in extreme events.
		
		In this work, we use the idea of a visibility graph to modify an existing spatial random scale model.  This results in a tail dependence model that naturally accommodates nonconvex spatial domains, where the nonconvex features are interpreted as physical barriers.  Furthermore, the model inherits key properties from the scale mixture like the ability to represent asymptotic independence at long ranges and either asymptotic dependence or asymptotic independence at short ranges.

        We apply our model to two applications in which physical barriers feature prominently in determining the spatial structure of extreme events.  The first is precipitation in the South Island of New Zealand, where the Southern Alps shelter the eastern region from the effects of atmospheric rivers which arrive from the west. The second is significant wave heights in the North Atlantic, where the Faroe Islands tend to block the effects of extratropical cyclones traveling into the North Sea.
		
\end{abstract}

\noindent%

\section{Introduction} \label{sec:intro}
    
    Extreme environmental events are shaped not only by the distance between locations but also by the physical features separating them. Mountain ranges can disrupt atmospheric systems and create sharp contrasts in precipitation extremes, while islands can shield coastal waters from severe ocean waves. Yet most spatial extreme value models characterize dependence primarily through Euclidean distance, potentially overlooking these directional effects. In this paper, we develop a spatial extremes model that explicitly accounts for physical barriers through a visibility graph. By incorporating geometric connectivity into a flexible random scale mixture framework, our approach accommodates both asymptotic dependence and independence while allowing extremal dependence to vary according to the physical structure of a nonconvex domain. We illustrate the approach through analyses of extreme precipitation over New Zealand's South Island and significant wave heights around the Faroe Islands. 
	
	A central challenge in spatial extreme value analysis is characterizing the dependence between extreme events at different locations. We measure pairwise extremal dependence using
	\[
	\chi_u(\bm{s}, \bm{s}') = \Pr\left( F_{\bm{s}}(X(\bm{s})) > u \mid F_{\bm{s}'}(X(\bm{s}')) > u \right),
	\]
	where $F_{\bm{s}}$ denotes the marginal distribution function at location $\bm{s}$, and $u$ is a high quantile threshold. A pair of locations are asymptotically dependent (AD) if $\chi = \lim_{u \to 1}\chi_u > 0$, indicating persistent dependence at increasingly extreme levels, and asymptotically independent (AI) if $\chi = 0$. Since the limiting dependent cannot be estimated directly from finite data, we assess extremal dependence through $\chi_u$ at high, finite thresholds. 
	
	Existing modeling approaches often represent two contrasting regimes. Gaussian processes are asymptotically independent and can therefore underestimate the occurrence of joint extremes when substantial tail dependence is present. Conversely, max-stable and generalized Pareto process models are asymptotically dependent and may impose stronger tail dependence than is supported by environmental data. Random scale mixture models offer a flexible alternative by accommodating both dependence regimes \citep{Huser2019}. In particular, we consider the spatially varying formulation developed by \citet{Shi2026, Shi2026log}, 
	\begin{equation}
		X(\bm{s}) = R(\bm{s})^{\delta(\bm{s})} \cdot g(Z(\bm{s})),
		\label{eq:random_scale_model}
	\end{equation}
	where $Z(\bm{s})$ is an asymptotically independent process, $R(\bm{s})$ is a heavy-tailed random scale, $\delta(\bm{s})$ is a spatially varying tail index, and $g(\cdot)$ is a link function. The relative tail behavior of the Gaussian process and random scale components governs the extremal dependence of $X(\bm{s})$. This construction provides a flexible transition between asymptotic independence and dependence, while allowing the strength of extremal dependence to vary spatially \citep{Engelke2019, Shi2026}.
	
	Despite this flexibility, existing random scale mixture models generally characterize spatial dependence through Euclidean separation. Consequently, two pairs of locations at the same distance are treated similarly, even when one pair has an unobstructed path between its locations and the other is separated by a physical barrier. This assumption can be restrictive on domains containing mountain ranges, islands, or other obstacles that impede the physical processes generating extreme events. 
	
\begin{figure}[ht]
		\centering
		\includegraphics[width=\textwidth]{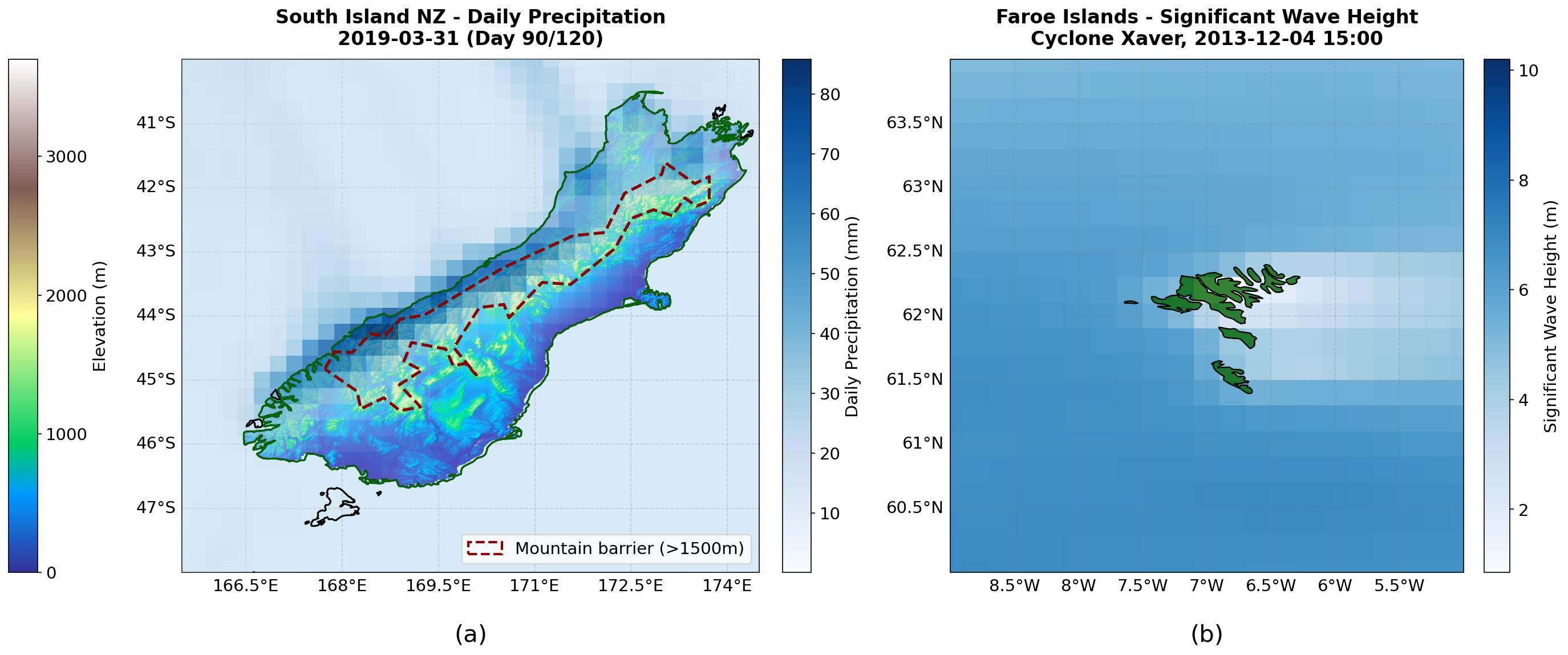} 
		\caption{(a) Daily precipitation (mm) over the South Island of New Zealand during an atmospheric river event: the Southern Alps (brown band) block the system from the leeward (eastern) side. (b) Significant wave height (m) near the Faroe Islands during Cyclone Xaver: waves in the islands' lee (east) are noticeably smaller.}
		\label{fig:reanalysis_data}
	\end{figure}

\begin{figure}[ht]
		\centering
		\begin{tabular}{cc}
			\includegraphics[width=0.3\textwidth, clip=true, trim=0 0 0 40]
			{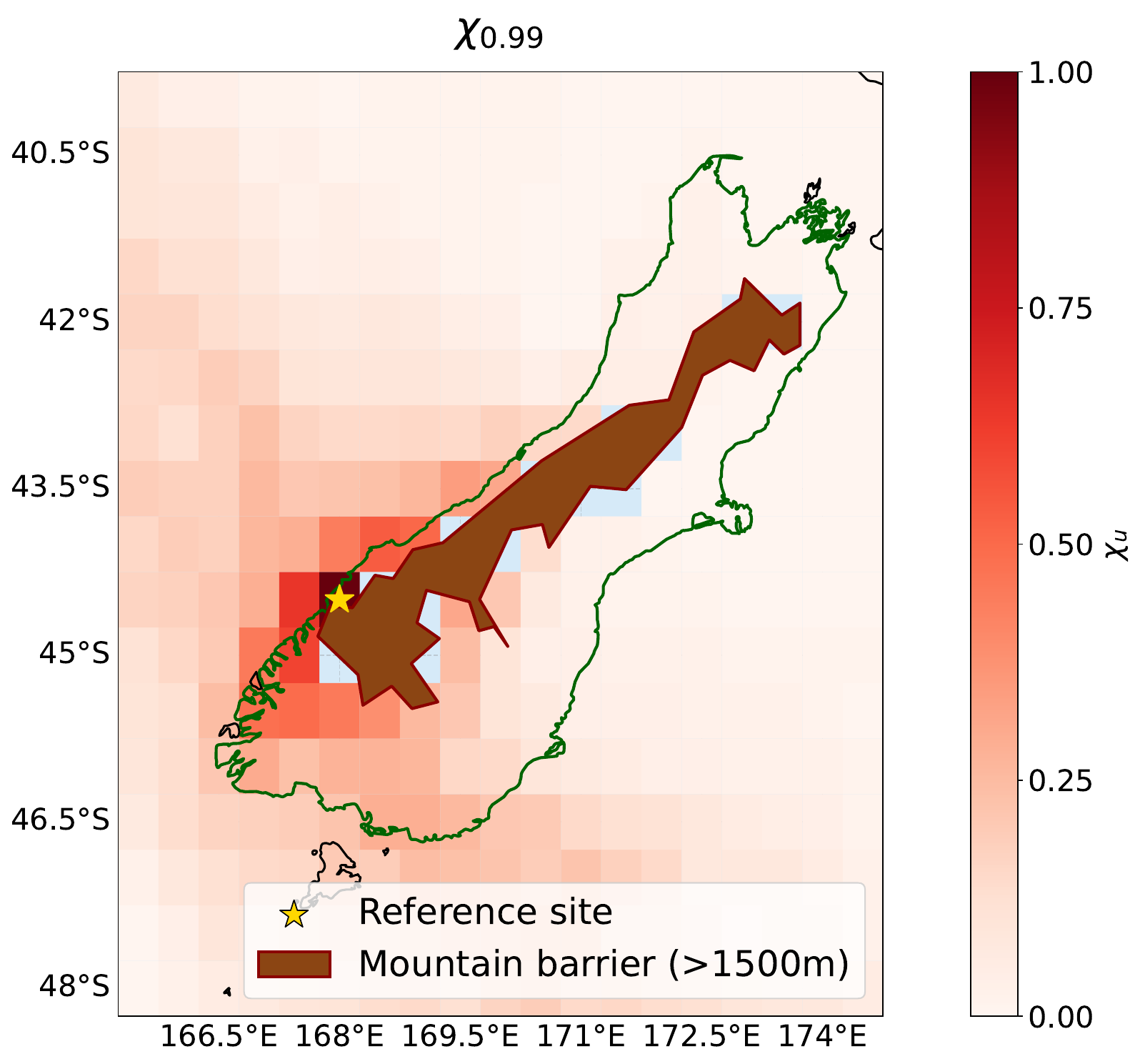} & 
			\includegraphics[width=0.3\textwidth, clip=true, trim=0 0 0 40]
			{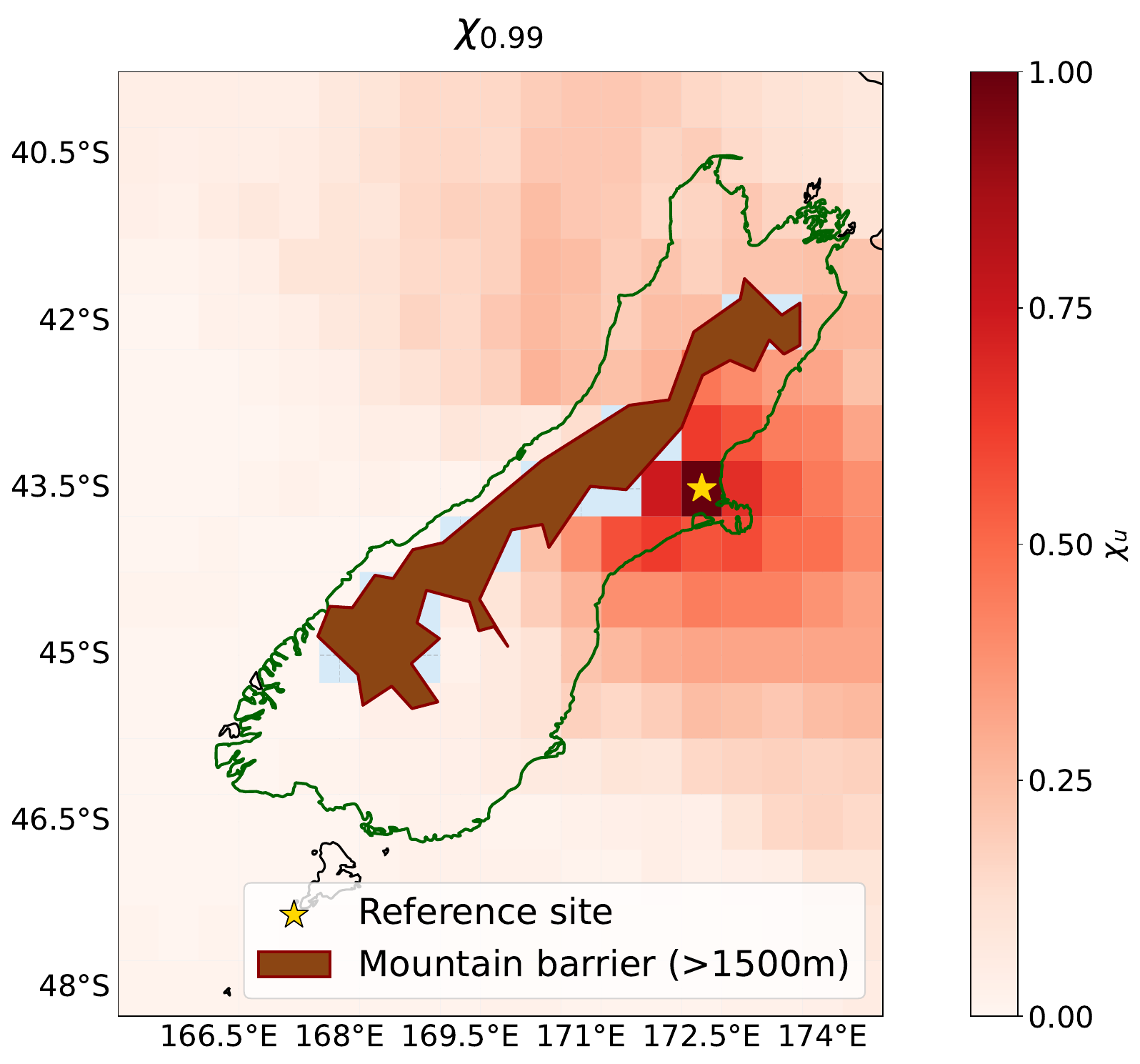}
			\\
			\includegraphics[width=0.3\textwidth, clip=true, trim=0 0 0 40]
			{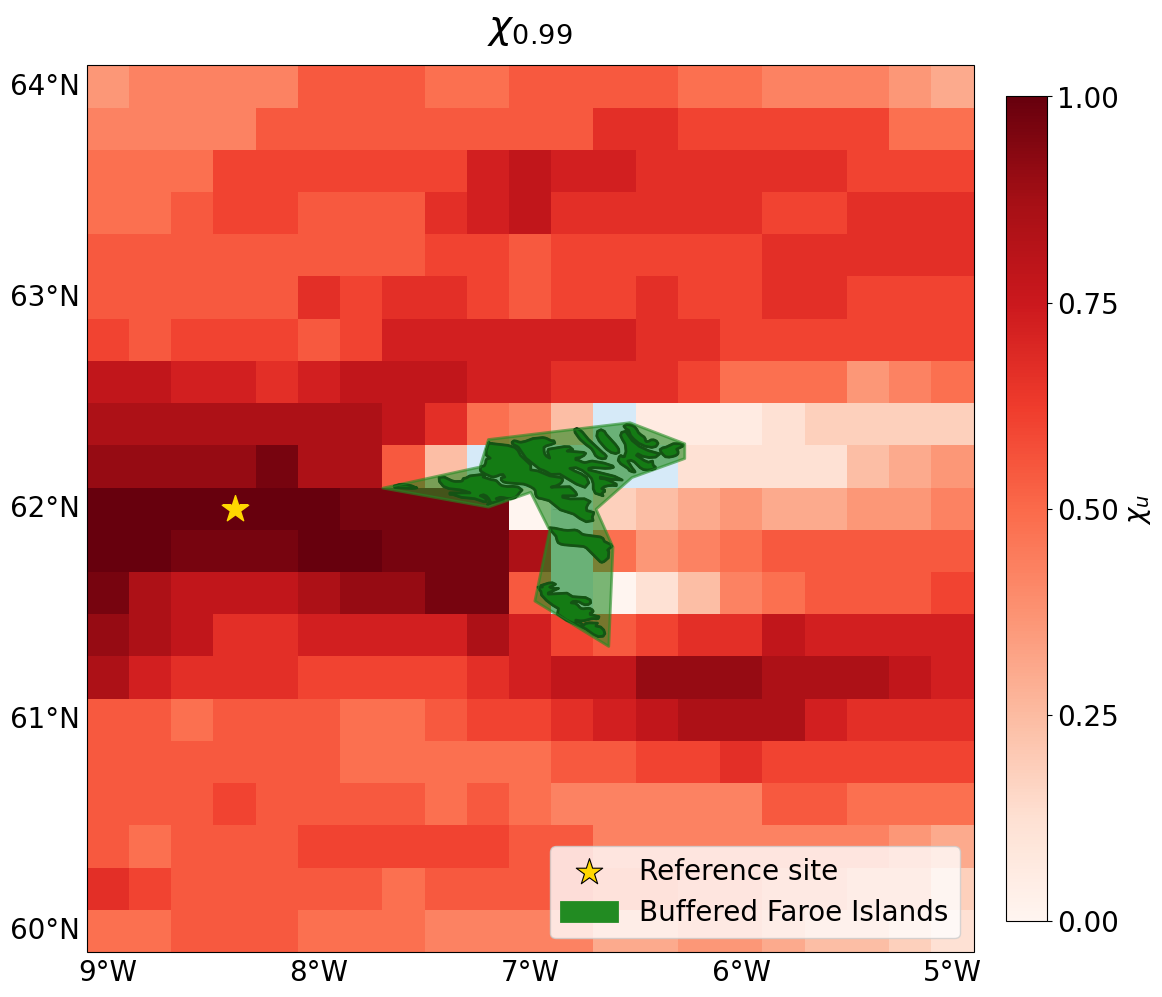} & 
			\includegraphics[width=0.3\textwidth, clip=true, trim=0 0 0 40]
			{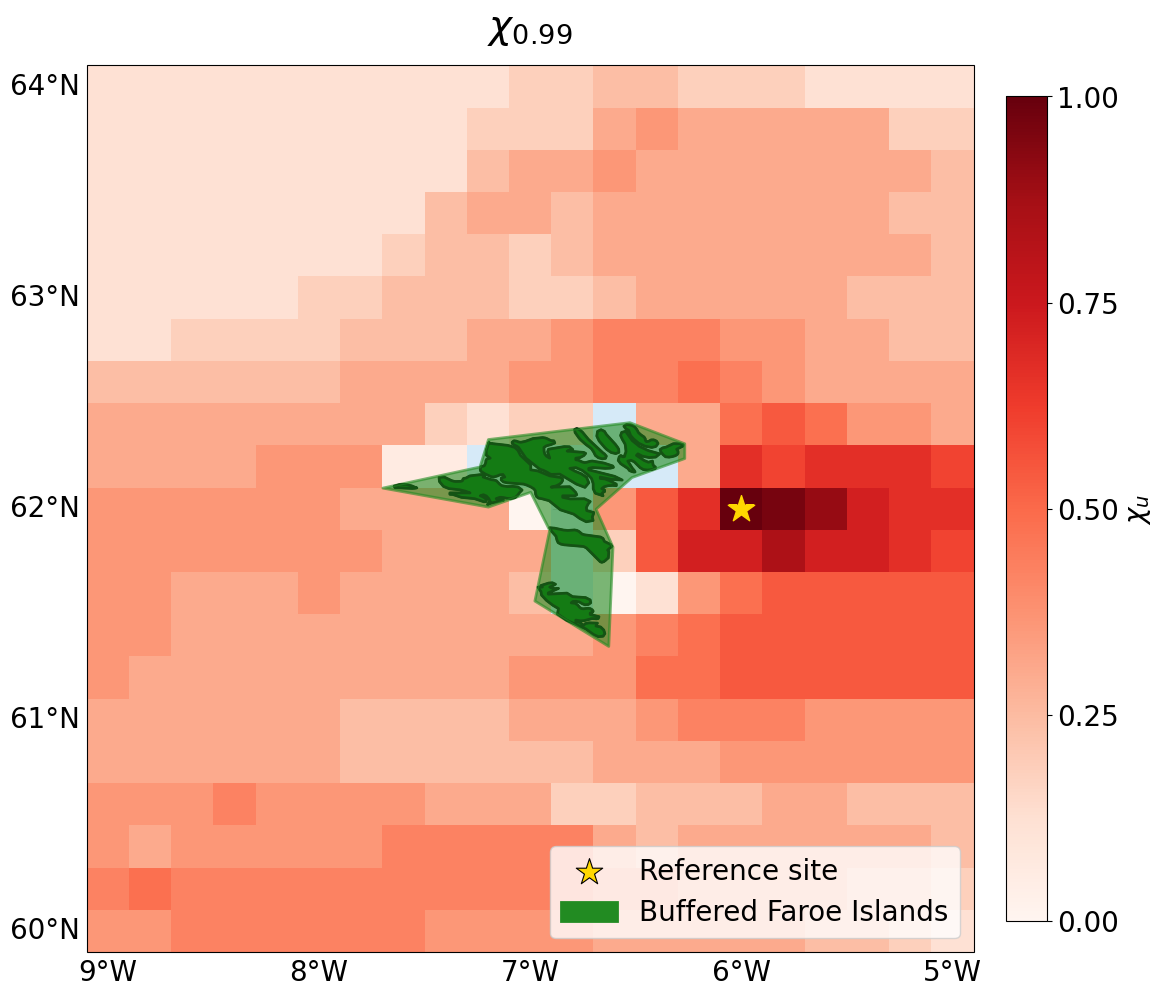}
		\end{tabular}
		\caption{Pairwise $\chi_{0.99}$ relative to a reference site (gold star), for two South Island, New Zealand sites (top; daily precipitation throughout the atmospheric river season, January--April, 1950--2026) and two Faroe Islands sites (bottom; 3-hourly significant wave height throughout the winter storm season, October--March, 1980--2026).}
		\label{fig:empirical_chiu_99}
	\end{figure}

	Our two applications illustrate this limitation. Over New Zealand's South Island, the Southern Alps separate the moisture-rich western side from the substantially drier eastern side (Figure~\ref{fig:reanalysis_data}(a)). Around the Faroe Islands, the islands shield waters on their eastern side from incoming ocean waves (Figure~\ref{fig:reanalysis_data}(b)). Empirical estimates of $\chi_{0.99}$ (Figure~\ref{fig:empirical_chiu_99}) reveal directional patterns in both settings. For a fixed reference location, extremal dependence is generally stronger toward locations with an unobstructed path and weaker toward locations shielded by the barrier, even at comparable Euclidean distances. These patterns motivate a dependence structure that accounts for physical connectivity rather than relying on distance alone. 
	
	From a geometric perspective, physical barriers create nonconvex spatial domains in which Euclidean distance may not adequately describe the relationship between locations. Replacing Euclidean distance with an alternative measure, such as geodesic distance, in a conventional covariance function does not necessarily preserve positive definiteness. Several remedies have been developed for spatial modeling on nonconvex domains, including multidimensional scaling \citep{Cox2008}, geodesic covariance approximations \citep{Davis2019}, graph-based Gaussian processes \citep{Dunson2022}, stochastic differential equation and partial differential equation approaches \citep{Niu2019, Bakka2019, Borovitskiy2020}, and barrier-aware nearest-neighbor Gaussian processes \citep{Jin2024}. However, incorporating physical barriers into the flexible tail dependence structures required for spatial extremes remains challenging. 
	
	We address this challenge by combining the visibility graph approach of \citet{Gilbert2024} with the random scale mixture framework of \citet{Shi2026, Shi2026log}. The visibility graph encodes whether pairs of locations are connected by an unobstructed line segment within the spatial domain. In the Gaussian process construction of \citet{Gilbert2024}, this graph determines conditional independence relationships. Locations without direct visibility are conditionally independent given the remaining field. Building on this idea, we incorporate visibility constraints into both the Gaussian process and random scale components of the spatial extremes model. The resulting framework allows physical barriers to influence extremal dependence while retaining the flexibility to represent both asymptotic dependence and independence and spatially varying tail behavior. We demonstrate its practical utility through simulation and the two environmental applications described above. 
	
	The remainder of the paper is organized as follows. Section~\ref{sec:model} introduces the proposed visibility model, and Section~\ref{sec::estimation} describes the inferential procedure. Section~\ref{sec:simulation} presents a simulation study. Section~\ref{sec:precip_application} and~\ref{sec:wave_application} present the precipitation and wave height applications, respectively, and Section~\ref{sec:discussion} concludes the paper.

\section{Spatial Extremes Model}\label{sec:model}

	
	
    Our starting point is the model from \citet{Shi2026}.  This model is a ``scale aware'' version of \eqref{eq:random_scale_model} in that it is asymptotically independent at long spatial lags and either asymptotically dependent or asymptotically independent at short spatial lags, according to the value of $\delta(\bm{s})$.  It achieves this property by constructing $R(\bs)$ from compact kernels, as 
    \begin{equation}
		R(\bm{s}) = \sum_{k=1}^{K} w_k(\bm{s}, r_k) S_k,
		\label{eq:random_scale-shi}
	\end{equation}
	where $K$ is the number of kernels placed in the domain, $S_k$ has a L{\'e}vy distribution with location parameter zero and scale parameter $\gamma$, and $w_k(\bm{s}, r_k)$ is a compact kernel centered at knot $k$ with support radius $r_k$, for $k = 1, \ldots, K$. 
    In addition, $\sum_{k=1}^{K} w_k(\bm{s}, r_k, \lambda, \alpha; \mathcal{G}) = 1$ for all $\bm{s} \in \mathcal{V}$.  \cite{Shi2026} complete the construction of \eqref{eq:random_scale_model} by specifying $\delta(\bm{s})$ as a linear combination of Gaussian kernels and $Z(\bm{s})$ as a Gaussian process, and the link function $g(\cdot) = \Phi(\cdot) / (1 - \Phi(\cdot))$.

    We extend the \citet{Shi2026} transformed random scale version of \eqref{eq:random_scale_model} by incorporating a visibility graph structure from \citet{Gilbert2024} to better capture extremal dependence in complex nonconvex domains.
    
	\subsection{Visibility Graph Model}\label{subsec:graphical}
 
	Let $\mathcal{D}$ be a nonconvex spatial domain on $\mathbb{R}^2$, and let $\mathcal{V} = \{\bm{s}_1, \ldots, \bm{s}_d\}$ denote a finite set of locations within $\mathcal{D}$. We define the adjacency matrix $A$ associated with $\mathcal{V}$ as 
	\begin{align}
		(A)_{ij} = 
		\begin{cases}
			1, & \mathrm{if}\; (i, j) \in \mathcal{E};
			\\
			0, & \mathrm{if}\; (i, j) \notin \mathcal{E},
		\end{cases}
		\label{adj_matrix}
	\end{align}
	where $\mathcal{E}$ represents the set of unordered pairs of locations $(\bm{s}_i, \bm{s}_j)$ that are visibly connected within $\mathcal{D}$. Two locations are considered visibly connected if the straight line segment between them lies entirely inside $\mathcal{D}$, without intersecting any barrier or excluded region. If $\bm{s}_i$ and $\bm{s}_j$ are not visibly connected, a barrier exists that prevents direct line-of-sight between the two points (see Figure \ref{fig:sim_domain_centers}(a) for illustration). The pair $(\mathcal{V}, \mathcal{E})$ defines an undirected graph $\mathcal{G}$, which is commonly referred to as a visibility graph in computational geometry and artificial intelligence literature.

    \subsection{The AI Component}
    
	Given any positive definite matrix $K$ and a graph $\mathcal{G}$ whose nodes correspond to the indices of $K$, there exists a unique positive definite matrix $L$ that satisfies the following properties:
	\begin{align}
		\begin{split}
			L_{ii} =&\; K_{ii} \; \mathrm{for\; all} \; i;
			\\
			L_{ij} =&\; K_{ij} \; \mathrm{for\; all} \; i \neq j \; \mathrm{and} \; (i, j) \in \mathcal{E};
			\\
			(L^{-1})_{ij} =&\; 0 \; \mathrm{for\; all} \; i \neq j \; \mathrm{and} \; (i, j) \notin \mathcal{E}
		\end{split}
		\label{eq:nonconvex_matrix_properties}
	\end{align}
	\citep{Dempster1972}. In our context, $K$ can be chosen as any valid covariance matrix, for example $K = C(\mathcal{V}, \mathcal{V})$ for a specified covariance function $C(\cdot, \cdot)$. 
	
	The first property preserves the marginal variance at each site, ensuring marginal stationary over $\mathcal{V}$. The second property enforces that for pairs of locations connected in $\mathcal{G}$, the covariance entries agree with those of the underlying Euclidean covariance function $C$. The third property encodes conditional independence: if two sites are not connected in $\mathcal{G}$, their corresponding entries in the precision matrix $L^{\; -1}$ are zeros, meaning they are conditionally independent given all other sites. We denote this construction as 
	\begin{align}
		L = \mathrm{CovSel}(C(\mathcal{V}, \mathcal{V}), \mathcal{G}),
		\label{eq:nonconvex_cov}
	\end{align}
	where $\mathrm{CovSel}(\cdot, \cdot)$ represents the covariance selection in \cite{Dempster1972}. This matrix $L$ is unique and satisfies all properties in (\ref{eq:nonconvex_matrix_properties}). Following \citet{Gilbert2024}, we can then specify the spatial process $\{ Z(\bm{s}) : \bm{s} \in \mathcal{V} \}$ as
	\begin{align}
		Z(\mathcal{V}) \sim N(0, L), \quad L = \mathrm{CovSel}(C(\mathcal{V}, \mathcal{V}), \mathcal{G}),
		\label{eq:nonconvex_GP}
	\end{align}
	which defines a Gaussian process over the possibly nonconvex domain $\mathcal{D}$ that explicitly accounts for line-of-sight connectivity.	

    We insert $Z(\bm{s}) \equiv Z(\mathcal{V})$ into \eqref{eq:random_scale_model}, resulting in a model for $X(\bm{s})$ that respects the nonconvex geometry of domains with barriers through the visibility graph.  However, this only applies to the asymptotically independent regime of $X(\bm{s})$ (corresponding to small $\delta(\bm{s})$), which, again, inherits its tail dependence structure from  $Z(\bm{s})$.  For the asymptotically dependent regime to also respect the nonconvex geometry, we must modify $R(\bm{s})$ as well.

\subsection{The AD Component}

    To incorporate the nonconvex geometry into  $R(\bm{s})$, we simply modify the compact kernels in \eqref{eq:random_scale-shi} such that a knot contributes to a site $\bm{s}$ only if the segment connecting the knot to the site does not pass through a barrier (see Figure \ref{fig:anisotropy_wendland}).  In this way, these visibility-clipped kernels also depend on the visibility graph $\mathcal{G}$.
    
	In addition, for applications like ours which feature strongly directional dynamics, it is necessary to accommodate anisotropy.  To flexibly capture anisotropy, we adopt an anisotropic Wendland kernel function. Specifically, for a knot location $\bm{s}_k \in \mathcal{D}$, the basis function is defined as:
	\begin{equation}
		w_k(\bm{s}, r_k, \lambda, \alpha; \mathcal{G}) \propto a\left( 1 - \frac{|| (\bm{s} - \bm{s}_k)^T B B^T (\bm{s} - \bm{s}_k) ||^2}{r_k} \right)_{+}^{\lfloor \frac{\bm{s}}{2} \rfloor + 2k + 1},
		\label{eq:anisotropy_wendland}
	\end{equation}
	where the order of covariance function is fixed at $k = 2$, the kernel radius is denoted by $r_k$, and $a$ ensures that only line-of-sight connectivity is respected, i.e., $a = 1$ if the line segment between $\bm{s}$ and $\bm{s}_k$ lies entirely within the domain, and $a = 0$ otherwise. The geometric anisotropy is governed through the linear transformation
	\begin{align*}
		B^T = 
		\begin{pmatrix}
			1 & 0 \\ 0 & \lambda^{-1}
		\end{pmatrix}
		\begin{pmatrix}
			\cos \alpha & \sin \alpha \\ - \sin \alpha & \cos \alpha
		\end{pmatrix},
	\end{align*}
	where $\lambda$ represents the anisotropy ratio and $\alpha$ represents the rotation angle. This construction allows the effective range of dependence to differ across direction: when $\lambda > 1$, dependence decays more slowly along the primary axis of anisotropy, whereas $\lambda < 1$ indicates faster decay. The rotation parameter $\alpha$ aligns the anisotropy with the prevailing spatial orientation, enabling the kernel to adapt to directional features in the data. Importantly, the line-of-sight restriction embedded in $a$ ensures that dependence cannot propagate across physical barriers, so that anisotropy is expressed only where physical connectivity is preserved. This formulation thus combines geometric flexibility with physically realistic restrictions, making it well suited for modeling spatial tail dependence in nonconvex domains.
	
	The random scale factors, which now depend on the visibility graph $\mathcal{G}$, are constructed as
	\begin{equation}
		R(\bm{s}, \mathcal{G}) = \sum_{k=1}^{K} w_k(\bm{s}, r_k, \lambda, \alpha; \mathcal{G}) S_k,
		\label{eq:random_scale}
	\end{equation}
	where $K$ and $S_k$ are as in \eqref{eq:random_scale-shi}, and $w_k(\bm{s}, r_k, \lambda, \alpha; \mathcal{G})$ are defined in \eqref{eq:anisotropy_wendland}, for $k=1, \ldots, K$. In addition, while \cite{Shi2026} only place knots within the domain, we additionally allow knots outside the boundary to mitigate edge effects.
	

	This spatial extreme model generalizes \cite{Shi2026} by embedding visibility-induced constraints, anisotropy, and improved treatment of edge effects, leading to more realistic representations of extremal dependence in complex spatial domains.
	
	\subsection{Hierarchical Model}
	Following \cite{Huser2019}, we adopt a copula-based approach to construct the joint distribution of $X(\bm{s})$. Since the copula is invariant to monotone marginal transformations, it is convenient to work with the logarithmic transformation:
	\begin{equation}
		\tilde{X}(\bm{s}) := \delta(\bm{s}) \tilde{R}(\bm{s}, \mathcal{G}) + \tilde{W}(\bm{s}),
		\label{eq:log_vis_model}
	\end{equation}
	where $\tilde{R}(\bm{s}, \mathcal{G}) := \log(R(\bm{s}, \mathcal{G}))$ is independent of $\tilde{W}(\bm{s}) := \log(g(Z(\bm{s})))$. This additive form facilitates analytical tractability. The resulting marginal distribution of $\tilde{X}(\bm{s})$ admits explicit forms for the cdf and pdf, given in (\ref{eq:cdf}) and (\ref{eq:pdf}), respectively, as
	\begin{align}
		F_{\tilde{X}_j}(\tilde{x}) =&\; 1 - \sqrt{\frac{\bar{\gamma}_j}{2 \pi}} \int_{0}^{\infty} r^{\delta - 3/2} (e^{\tilde{x}} + r^\delta)^{-1} \exp\left\{ \frac{-\bar{\gamma}_j}{2r} \right\} dr;
		\label{eq:cdf}
	\end{align}
	\begin{align}
		f_{\tilde{X}_j}(\tilde{x}) =&\; \sqrt{\frac{\bar{\gamma}_j}{2\pi}} e^{\tilde{x}} \int_{0}^{\infty} r^{\delta - 3/2} (r^\delta + e^{\tilde{x}})^{-2} \exp\left\{ \frac{-\bar{\gamma}_j}{2r} \right\} dr,
		\label{eq:pdf}
	\end{align}
	where $\bar{\gamma}_j = \left[ \sum_{k=1}^{K} (w_k(\bm{s}, r_k, \lambda, \alpha; \mathcal{G}) \gamma_j)^{1/2} \right]^2$.

	Let $\{Y(\bm{s}): \bm{s} \in \mathcal{S}\}$ denote the process of marginal extreme observations with distribution function $F_{Y \mid \bm{\theta}_Y}(y)$ and density $f_{Y \mid \bm{\theta}_Y}$, where $\bm{\theta}_Y$ denotes the corresponding marginal parameters. The marginal distribution may be specified as generalized extreme value (GEV) distribution for block maxima or as a generalized Pareto distribution (GPD) for threshold exceedances. In the applications presented in Section \ref{sec:precip_application} and \ref{sec:wave_application}, we use GEV margins for precipitation data over the South Island of New Zealand and GPD margins for significant wave height data over the region surrounding the Faroe Islands, respectively.
	
	To couple the marginal distribution with the dependence structure described above, we adopt a hierarchical framework based on probability integral transforms. Specifically, 
	\[
		F_{Y \mid \bm{\theta}_Y}(Y) = F_{X \mid \bm{\phi}, h, \bm{\theta}_R, \bm{\theta}_C}(X) = F_{\tilde{X} \mid \bm{\phi}, h, \bm{\theta}_R, \bm{\theta}_C}(\tilde{X}),
	\]
	where $\bm{\theta}_R = (r, \lambda, \alpha, \bm{S})$, $\bm{S} = (S_1, \ldots, S_K)$, and $\bm{\theta}_C$ are covariance parameters. Consequently, the latent process is obtained from the marginal observations through
	\[
		X(\bm{s}) = F_{\tilde{X} \mid \bm{\phi}, h, \bm{\theta}_R, \bm{\theta}_C}^{\; -1} \circ F_{Y \mid \bm{\theta}_Y}(Y(\bm{s})).
	\]
	This construction separates the specification of the marginal distribution from the spatial dependence model while preserving their compatibility through the probability integral trasform.
	
	Conditioning on the latent scaling variables $\bm{S}_t$, the conditional likelihood of the observations $\bm{Y}_t = (Y_{t, 1}, \ldots, Y_{t, d})^T$ is then obtained through the change-of-variables formula and can be written as 
	\begin{equation}
		f_{\bm{Y}_t}(\bm{y}_t) = f_{\bm{Z}_t}\left( \Phi^{-1}\left( [ 1 + e^{-\bm{\tilde{W}}_t}]^{-1} \right) \right) \left\vert \det \left( \frac{\partial \bm{Z}_t}{\partial \bm{Y}_t} \right) \right\vert
		\label{eq:liklihood_t}
	\end{equation}
	The first term in \eqref{eq:liklihood_t} arise from the Gaussian copula representation of the transformed process, whereas the second term accounts for the marginal transformation from $Y$ to $\tilde{X}$. 
	
	Assuming independence across temporal replicates, the likelihood for $T$ replicates is given by 
	\[
		L(\bm{\theta}) = \prod_{t=1}^{T} f_{\bm{Y}_t}(\bm{y}_t),
	\]
	where $\bm{\theta}$ collects the marginal, tail dependence, covariance, and latent scaling parameters. The resulting formulation permits the same spatial dependence framework to be used with different marginal models, including GEV and GPD distributions, without altering the underlying visibility-based dependence structure.  When the response is GPD, full likelihood-based inference under \eqref{eq:random_scale_model} is computationally prohibitive when the number of sites is large, which can be addressed by introducing a nugget effect that makes inference tractable at the scale of a realistic spatial network while preserving the key tail-dependence properties of the model \citep{Zhang2022,Shi2026log}.

\section{Parameter Estimation}\label{sec::estimation}
	
	\subsection{Computation of Nonconvex Covariance Function}\label{subsec::nonconvex_covariance}
	
	To construct a valid covariance matrix consistent with a visibility graph, one natural approach is the iterative proportional scaling (IPS) algorithm \citep{Speed1986}. 
	However, IPS requires is computationally intensive even for moderate-sized spatial datasets and impractical at large scales.  To overcome this limitation, we adopt the graph-based framework of \cite{Gilbert2024}, which constructs nonconvex covariance matrices efficiently using the cliques and separators of the visibility graph. Their approach requires the visibility graph to be chordal (decomposable). When the graph is not decomposable, as is often the case for nonconvex spatial domains, additional edges can be introduced to obtain a chordal extension. Importantly, \cite{Gilbert2024} show that this augmentation typically introduces minimal distortion. Figure \ref{fig:sim_domain_centers}(a) illustrates a representative example of a T-shaped barrier domain, where line-of-sight restrictions yield a non-decomposable visibility graph that is chordalized before covariance computation.
	
	
	\subsection{Bayesian Inference}
	
	We use Markov chain Monte Carlo (MCMC) to estimate the posterior distribution of the model parameters. Since conjugate priors are not available, all parameters are updated using random walk Metropolis steps, with adaptive tuning \citep{Shaby2010}.
	
	To complete the model, we specify the following prior distributions for the model parameters: $\bm{\phi} \sim \text{Unif}(0, 3)$, $h \sim \text{Half \text{-} Normal}(0, 1)$, $\bm{\theta}_{Y} \sim N(0, 1)$, $r \sim \text{Half \text{-} Normal}(0, 4)$, $\lambda' \sim \text{Beta}(2, 2)$, $\alpha \sim \text{Unif}(0, \pi)$, $\rho \sim \text{Half \text{-} Cauchy}(1)$, $\nu \sim \text{Unif}(0, 2)$, and $S_k \sim \text{L{\'e}vy}(0, 1)$ for $k = 1, \ldots, K$. Since we assume conditional temporal independence, the latent scaling variables $\{ \bm{S}_t \}_{t = 1}^T$ can be updated in parallel across replicates $t$, thereby, improving computational efficiency.
	

\section{Simulation Study}\label{sec:simulation}
	
	We conducted a simulation study to evaluate the proposed hierarchical spatial extremes model and assess the reliability of posterior inference under a known data-generating process. Specifically, we examine the empirical coverage of posterior credible intervals when the marginal responses follow GEV distributions.
	
\begin{figure}[ht]
		\centering
		\includegraphics[width=0.7\textwidth]{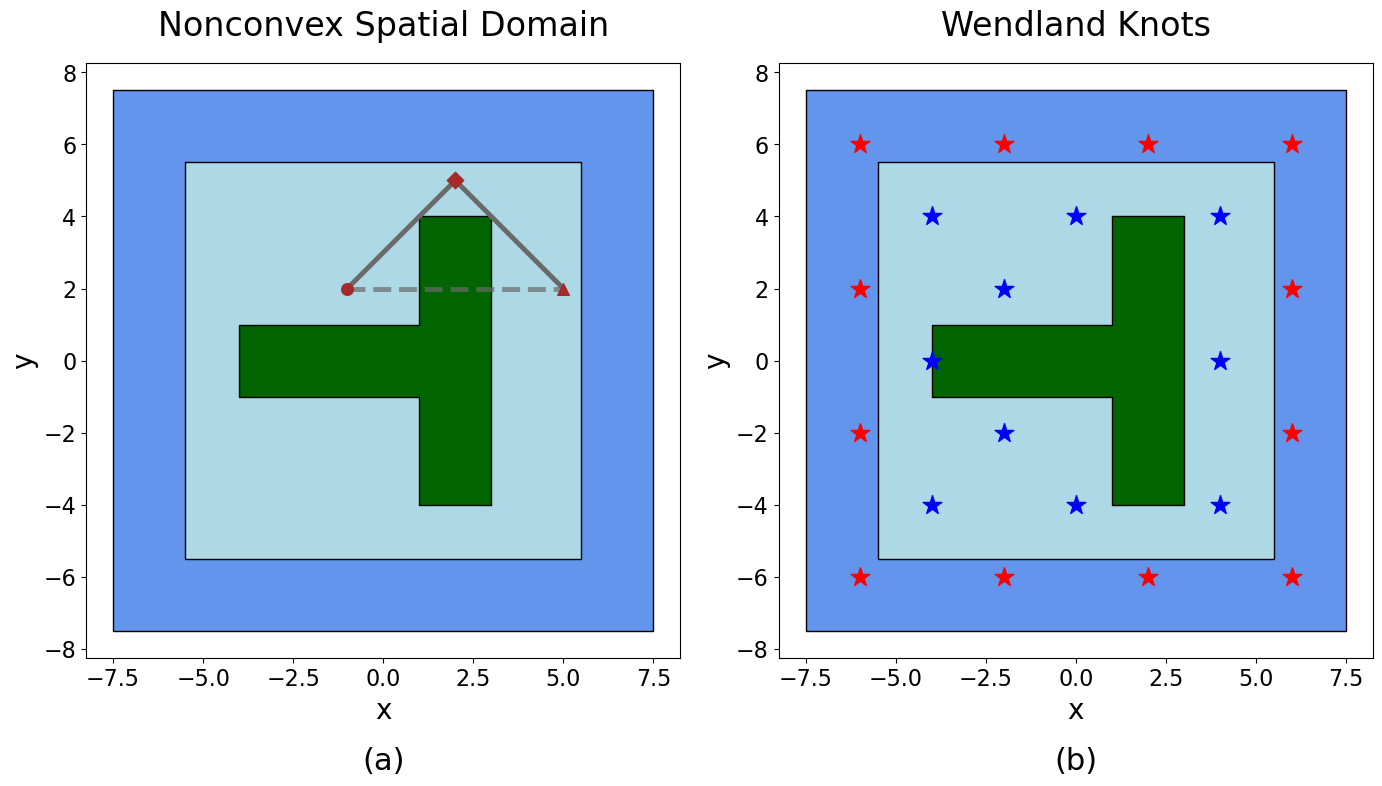}
		\caption{(a) Example of a nonconvex spatial domain containing a T-shape barrier (green region) that restricts line-of-sight connectivity. Solid lines indicate pairs of sites that are visibly connected, while the dashed line indicates a pair of sites that are not visibly connected due to obstruction by the barrier. (b) Uniformly distributed knot locations $K = 22$ on the spatial domain. Blue stars represent the knots placed inside the domain, and red stars represent the knots placed outside the domain.}
		\label{fig:sim_domain_centers}
	\end{figure}

\begin{figure}[ht]
		\centering
		\begin{tabular}{cccc}
			\includegraphics[scale=0.25]{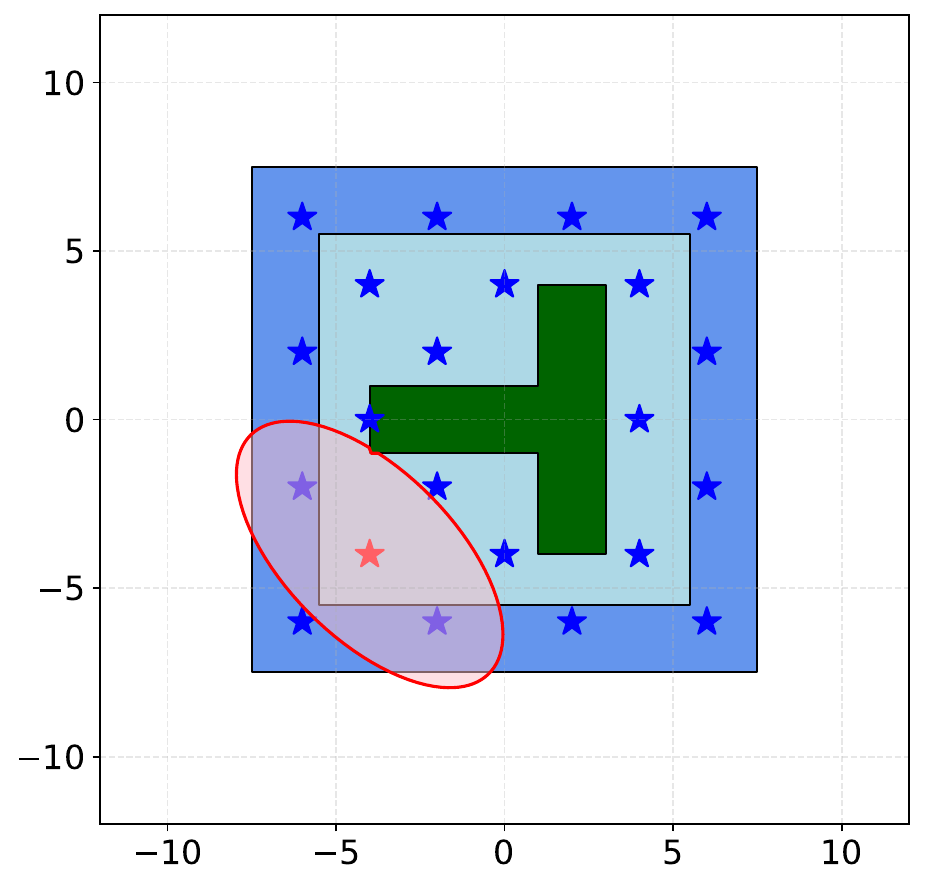} & 
			\includegraphics[scale=0.25]{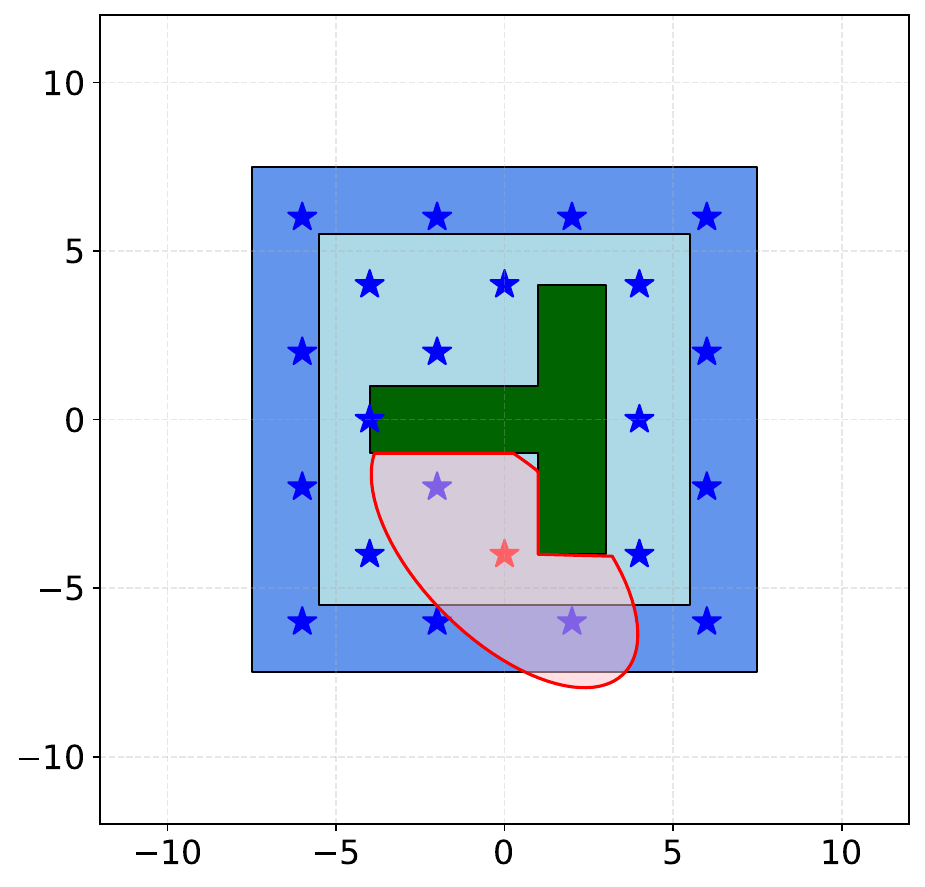} & 
			\includegraphics[scale=0.25]{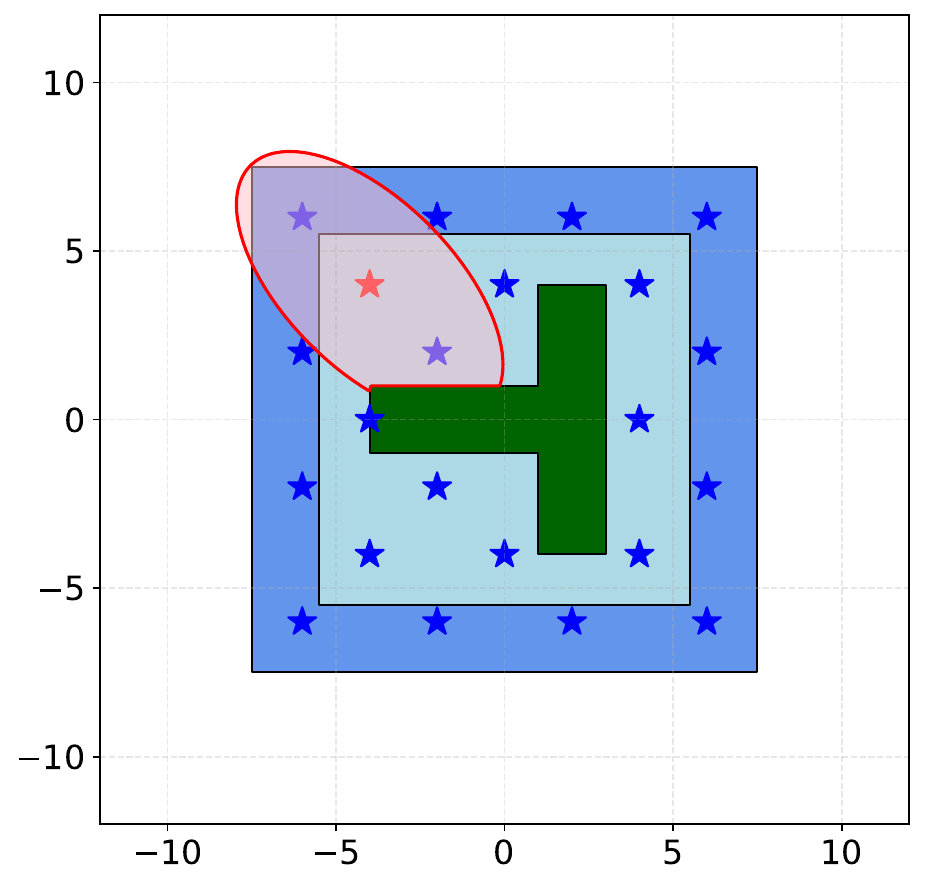} &
			\includegraphics[scale=0.25]{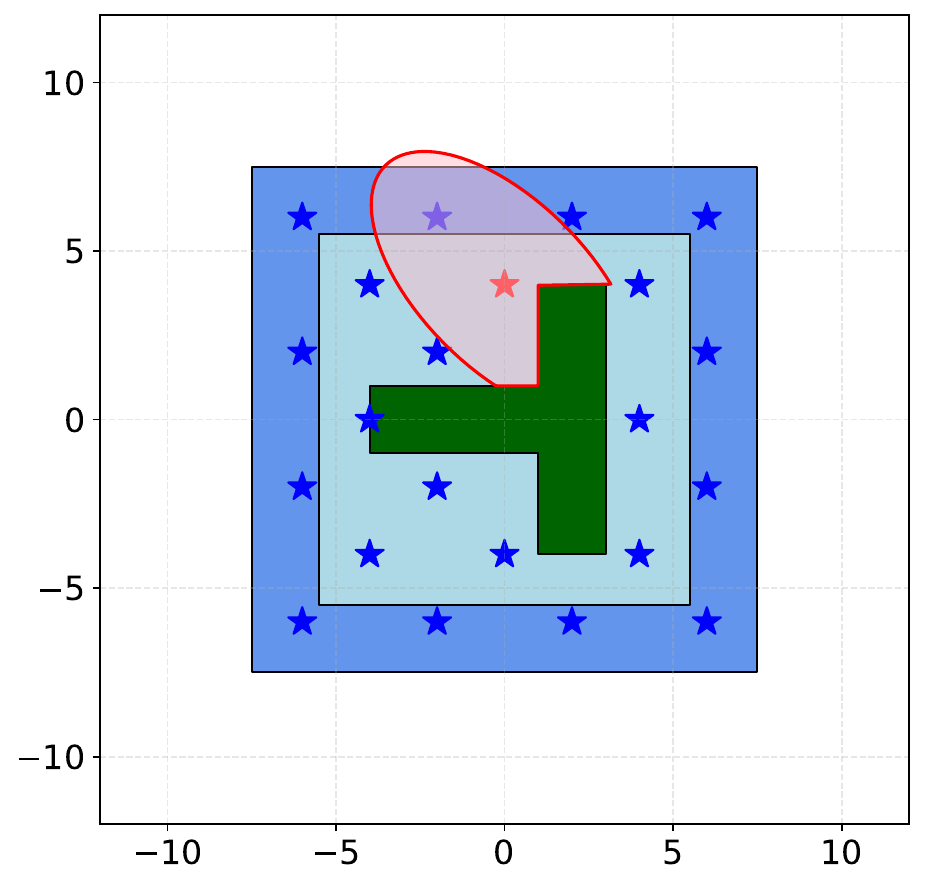}
		\end{tabular}
		\caption{Examples of anisotropic Wendland kernel  support for selected knot locations.}
		\label{fig:anisotropy_wendland}
	\end{figure}

	We consider 79 sites over the nonconvex domain $\mathcal{D} = [-5, 5] \times [-5, 5]$, which contains a T-shaped barrier (Figure \ref{fig:sim_domain_centers}(a)), with 64 time replicates. The latent Gaussian process $Z_t(\bm{s})$ follows Equation (\ref{eq:nonconvex_GP}) with an isotropic stationary Mat{\'e}rn covariance function having range $\rho = 20$ and smoothness $\nu = 1$. For each replicate, $K = 22$ knot locations $\{ \bm{s}_k \}_{k=1}^{K}$ are uniformly distributed across $\mathcal{D}$ (Figure \ref{fig:sim_domain_centers}(b)). Independent L{\'e}vy random variables with scale parameter $\gamma = 1$ are generated at these knot locations and interpolated to all sites via anisotropic Wendland kernel functions with radius $r = 5$, anisotropy ratio $\lambda = 0.5$, and rotation angle $\alpha = 135^\circ$ (Figure \ref{fig:anisotropy_wendland}).
	
	For numerical stability and prior specification, we reparameterize the anisotropy ratio as $\lambda' = \lambda / (1 + \lambda) \in (0, 1)$. The tail dependence parameter is fixed at $\delta(\bm{s}) = 0.6$, while the marginal GEV parameters are spatially and temporally constant at $(\mu(\bm{s}), \sigma(\bm{s}), \xi(\bm{s})) = (0, 1, 0.5)$. Although these are treated as constants in the data generation, they are still estimated in the MCMC inference to assess identifiability and estimation performance.
	
\begin{figure}[ht]
		\centering
		\includegraphics[width=0.8\textwidth]{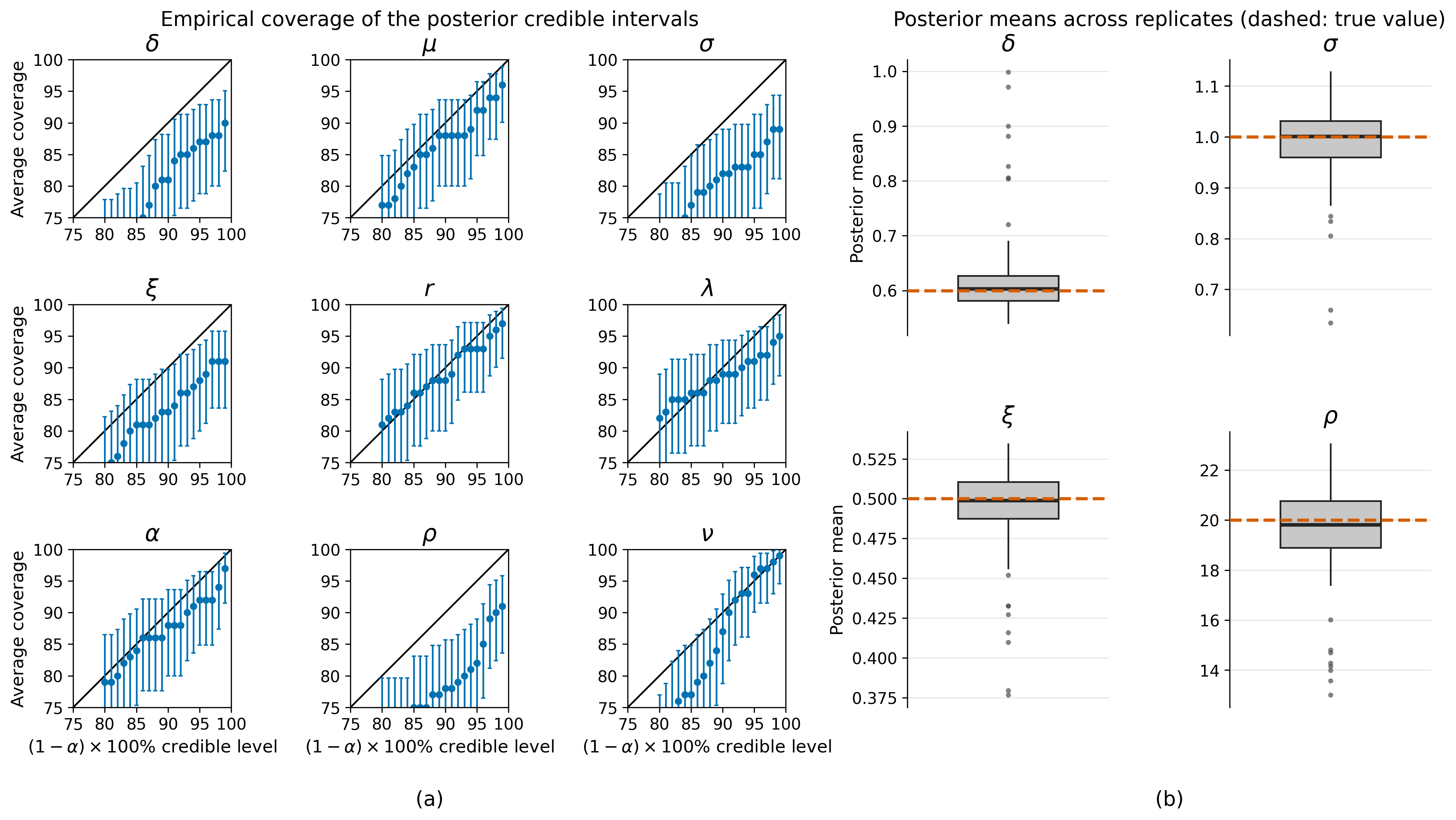}
		\caption{(a) Empirical coverage of posterior credible intervals from 100 simulated datasets. (b) Posterior means of parameters $\delta$, $\sigma$, $\xi$, and $\rho$ across 100 simulated datasets. Red dashed lines indicate true parameter values.}
		\label{fig:sim_coverage_analysis}
	\end{figure}

	We generate 100 independent datasets and evaluate the empirical coverage of nominal $100 \times (1 - \alpha)\%$ posterior credible intervals for all model parameters, including the tail dependence parameter $\delta$, Wendland parameters $(r, \lambda, \alpha)$, Mat{\'e}rn parameters $(\rho, \nu)$, and GEV parameters $(\mu, \sigma, \xi)$. Figure \ref{fig:sim_coverage_analysis}(a) summarizes the results with the diagonal line indicating nominal coverage and blue points with binomial confidence intervals showing empirical coverage. Overall, coverage is satisfactory, although mild under-coverage is observed for $\delta$, $\sigma$, $\xi$, and $\rho$.
	
	To further examine these parameters, Figure~\ref{fig:sim_coverage_analysis}(b) presents their posterior means across the 100 simulations. The boxplots indicate that the posterior estimates are approximately unbiased with medians close to their true values (denoted by red dashed lines). Nonetheless, a small subset of replicates shows higher variability, likely due to incomplete convergence in some MCMC chains. This suggests that while the estimation procedure is generally robust, longer MCMC runs or additional adaptation steps in LAP may improve convergence and posterior accuracy for these parameters.

\section{New Zealand Precipitation}  \label{sec:precip_application}
	
	We first consider daily total precipitation over the South Island of New Zealand.  We consider the period from 1950 through 2026 and restrict the analysis to the atmospheric river (AR) season, defined as January through April. This seasonal restriction is motivated by the important role of landfalling atmospheric rivers in producing extreme precipitation over the South Island. During this period, atmospheric rivers transport large volumes of moisture toward New Zealand from the north-west and contribute substantially to extreme rainfall events \citep{Reid2021extreme}. Focusing on the AR season allows us to characterize extreme precipitation within a period in which this physical mechanism is particularly relevant, while reducing potential heterogeneity associated with different seasonal precipitation regimes. 
    
    The data we analyzed is from the ERA5 reanalysis, produced by the European Center for Medium-Range Weather Forecasts \citep{Hersbach2023}. The ERA5 product provides global atmospheric reanalysis data to an original spatial resolution of $0.25^\circ \times 0.25^\circ$ on a regular latitude-longitude grid and at an hourly temporal resolution. We consider the spatial domain $[165.5^\circ \text{E}, 174.5^\circ \text{E}] \times [48^\circ \text{S}, 40^\circ \text{S}]$ which encompasses the South Island and its surrounding region. The Southern Alps form a prominent mountain range extending approximately along the southwest-northeast axis of the island. As illustrated in Figure~\ref{fig:reanalysis_data}(a), this mountain range can affect the spatial propagation of atmospheric river precipitation and produce a pronounced rain-shadow effect. We therefore use the Southern Alps as the physical barrier in the visibility-based analysis, defined as the region with elevation exceeding 1500m.

	We aggregated the hourly total precipitation observations to daily totals (in millimeters). Let $P_{t,m}(\bm{s}_i)$ denote the daily total precipitation at location $\bm{s}_i$, day $m = 1, \ldots, M_t$, and year $t = 1950, \ldots, 2026$, where $M_t \in \{199, 120\}$ is the number of days in the January-April AR season, with the difference accounting for leap years. For each location, we extract the seasonal maxima
	\begin{align*}
		Y_t(\bm{s}_i) = \max_{m=1,\ldots,M_t} \left\{ P_{t,m}(\bm{s}_i) \right\},
	\end{align*}
	yielding $T = 77$ annual maxima, or time replicates, at each spatial location. These seasonal block maxima form the response data for the subsequent spatial extremes analysis.
	
	For the spatial analysis, we subsample the original $0.25^\circ \times 0.25^\circ$ grid by retaining every other grid location in each spatial direction, resulting in a regular $0.5^\circ \times 0.5^\circ$ grid. We then exclude grid locations with elevation exceeding 1500m, corresponding to locations within the defined mountain barrier. This preprocessing results in $d = 309$ spatial locations for analysis.
	
	\subsection{Marginal GEV Model} \label{subsec:gev_margin}
	
\begin{figure}[htbp]
		\centering
		\includegraphics[width=0.8\textwidth]{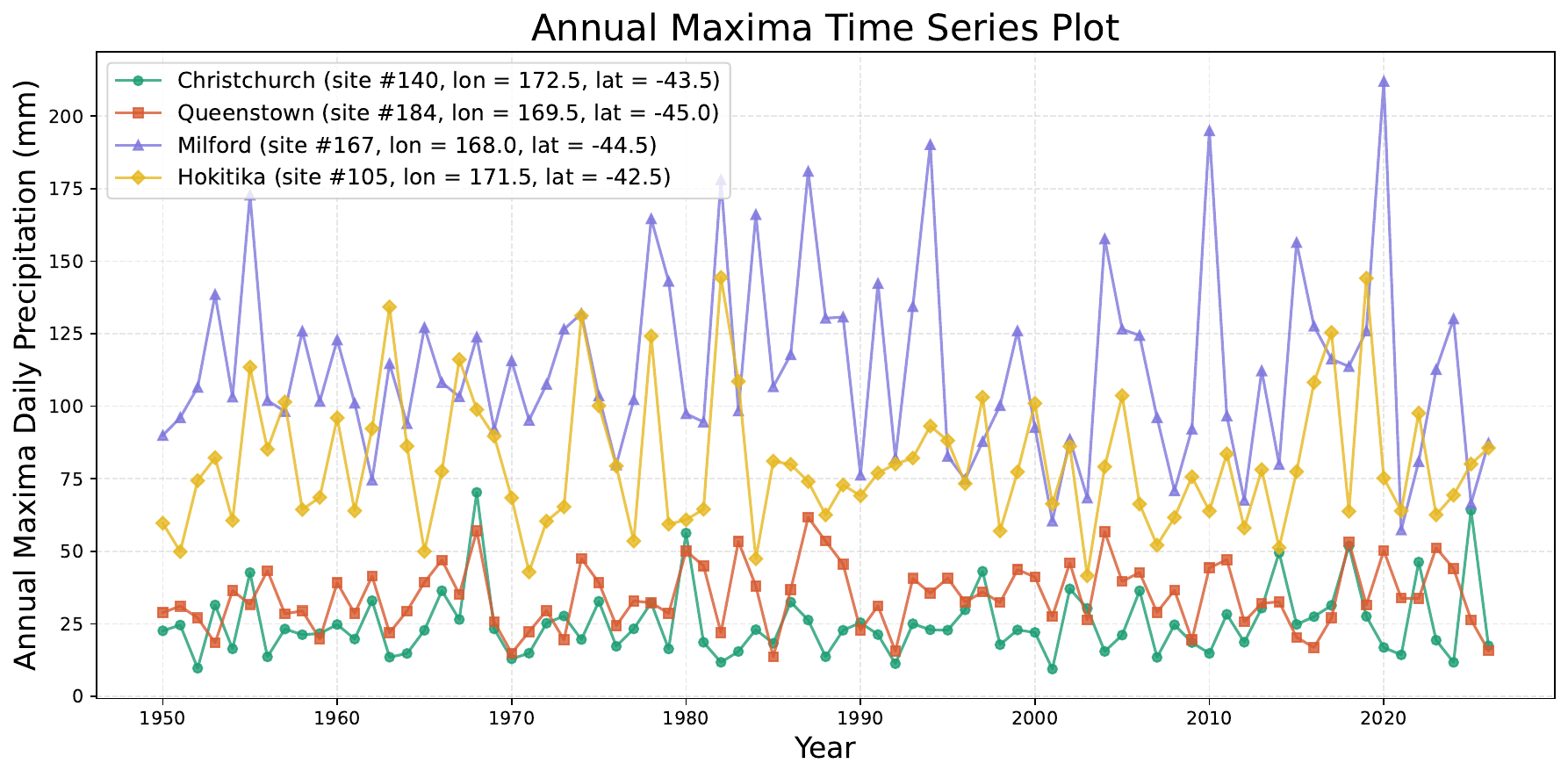}
		\caption{Time series of annual maxima (AR season, January-April) at four representative sites, two leeward (Christchurch, Queenstown) and two windward (Milford Sound, Hokitika), from 1950--2026. No systematic trend is apparent at any site, supporting temporally constant GEV margins.}
		\label{fig:annual_max_ts}
	\end{figure}

	We model the seasonal maxima $Y_t(\bm{s})$ using a generalized extreme value (GEV) distribution, 
	\begin{align*}
		Y_t(\bm{s}) \sim \operatorname{GEV} \left( \mu(\bm{s}), \sigma(\bm{s}), \xi(\bm{s}) \right)
	\end{align*}
	where $\mu(\bm{s})$, $\sigma(\bm{s}) > 0$, and $\xi(\bm{s})$ denote the spatially varying location, scale, and shape parameters, respectively. Before specifying the marginal model, we examine temporal variation in the annual maxima. Figure~\ref{fig:annual_max_ts} shows the seasonal maxima from 1950-2026 at four representative sites, including leeward sites near Christchurch and Queenstown and windward sites near Milford Sound and Hokitika. No clear systematic temporal trend is apparent at these locations. We therefore treat the GEV parameters as stationary in time, i.e., with marginal parameters constant at each location across the 77-year record, while allowing them to vary spatially. 
	
	To account for spatial variation associated with the physical barrier, we modeled the three GEV parameters using spatial covariates. Specifically, we consider an indicator for whether a site lies on the windward side of the barrier (including near-coastline sites), the distance from the site to the centroid of the barrier, the sine and cosine of the bearing angle from the barrier centroid to the site, and the interaction between the windward indicator and the distance. Let $\mathds{1}_{\text{windward}}(\bm{s})$ denote the windward-side indicator, $\operatorname{dist}(\bm{s})$ the distance from $\bm{s}$ to the barrier centroid, and $\theta(\bm{s})$ the corresponding bearing angle. The location parameter is specified as 
	\begin{align*}
		\mu(\bm{s}) =\beta_0 +\beta_1 \mathds{1}_{\mathrm{w}}(\bm{s}) +\beta_2\operatorname{dist}(\bm{s}) +\beta_3\sin\{\theta(\bm{s})\} +\beta_4\cos\{\theta(\bm{s})\} +\beta_5 \mathds{1}_{\mathrm{w}}(\bm{s})\operatorname{dist}(\bm{s}).
	\end{align*}
	The scale parameter is modeled on the log scale and the shape parameter is modeled using a rescaled-logit transformation that constrains $\xi(\bm{s}) \in (-0.5, 1)$. Both transformed parameters use the same linear covariate structure as $\mu(\bm{s})$. The selection and construction of the spatial covariates, together with additional candidate covariates and alternative specifications considered in the analysis, are described in Section~\ref{supp:precip_gev_covariates} of the Supplementary Material.
	
	\subsection{Results} \label{subsec:precip_result}
	
	We fit the visibility model jointly to the $d = 309$ spatial locations. For the dependence component, anisotropic Wendland and isotropic Gaussian kernels are constructed using 24 knots placed on a regular grid over the study domain, excluding knots that fall within the Southern Alps barrier. Posterior inference is performed using a joint MCMC scheme for all marginal and dependence parameters. We run the chain for 60{,}000 iterations, discard the first 30{,}000 iterations as burn-in, and retain every third iteration thereafter, yielding $B = 10{,}000$ posterior samples. Posterior means and 95\% equal-tailed credible intervals for the model parameters, together with trace plots and additional model diagnostics, are provided in the Supplementary Material Section~\ref{supp:precip_fit}. 
	
\begin{figure}[htbp]
		\centering
		\begin{tabular}{cc cc}
			\includegraphics[width=0.2\textwidth, clip=true, trim=0 0 30 40]{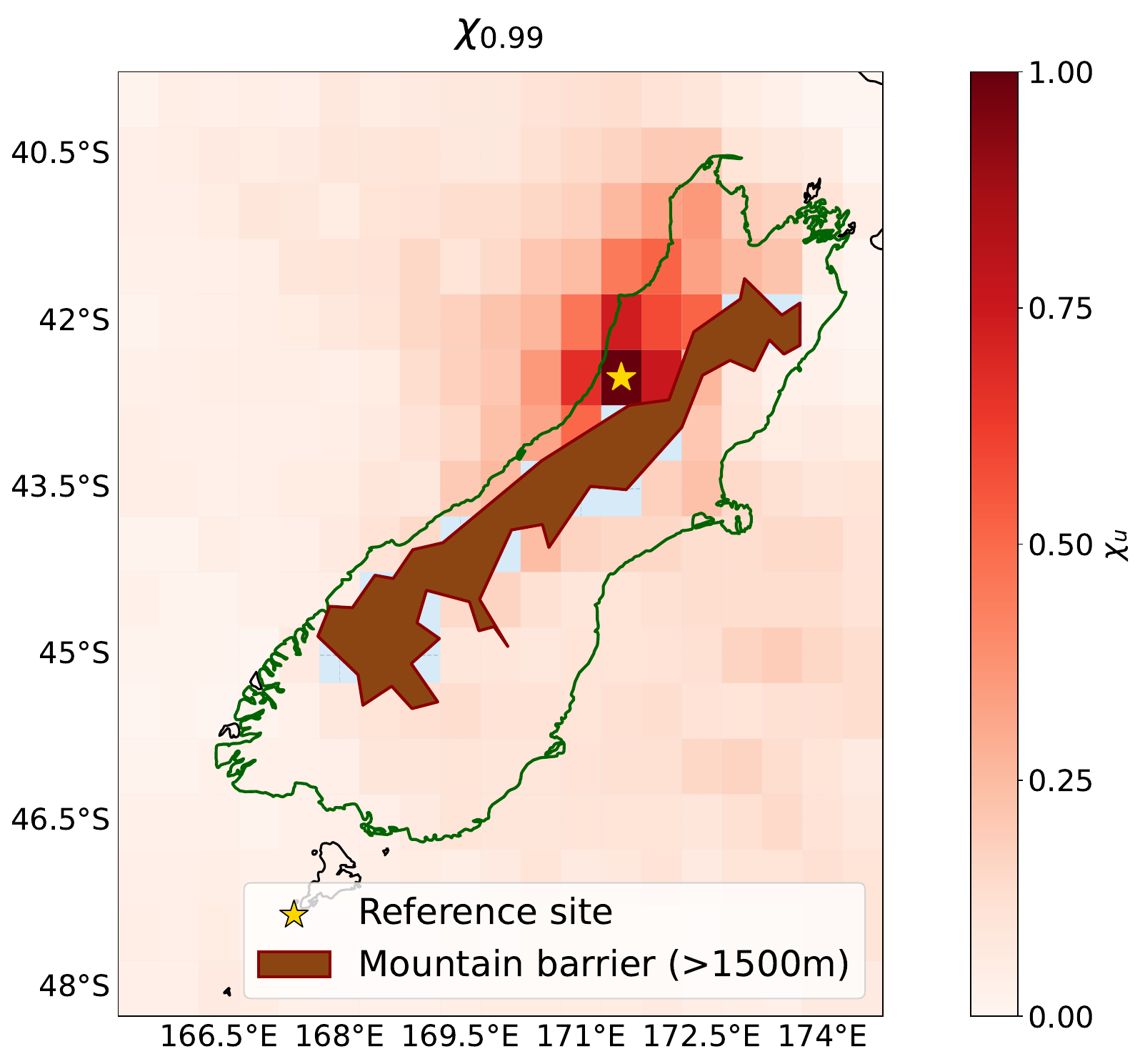} & 
			\includegraphics[width=0.2\textwidth, clip=true, trim=0 0 30 40]{precip_emp_chiu_inland_center_167_99.pdf} & 
			\includegraphics[width=0.2\textwidth, clip=true, trim=0 0 30 40]{precip_emp_chiu_inland_center_140_99.pdf} & 
			\includegraphics[width=0.2\textwidth, clip=true, trim=0 0 30 40]{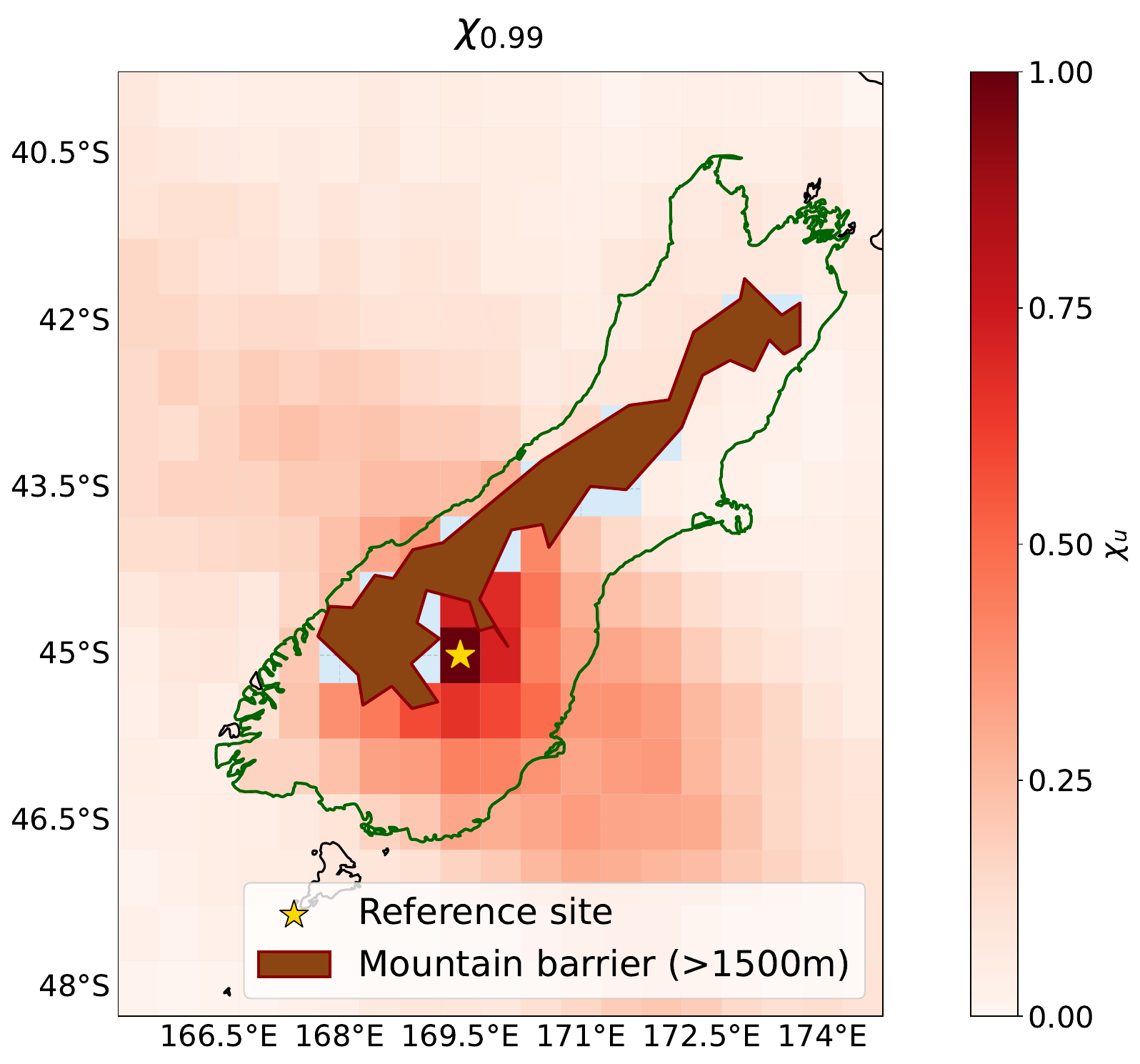}
			\\
			\includegraphics[width=0.2\textwidth, clip=true, trim=0 0 30 40]{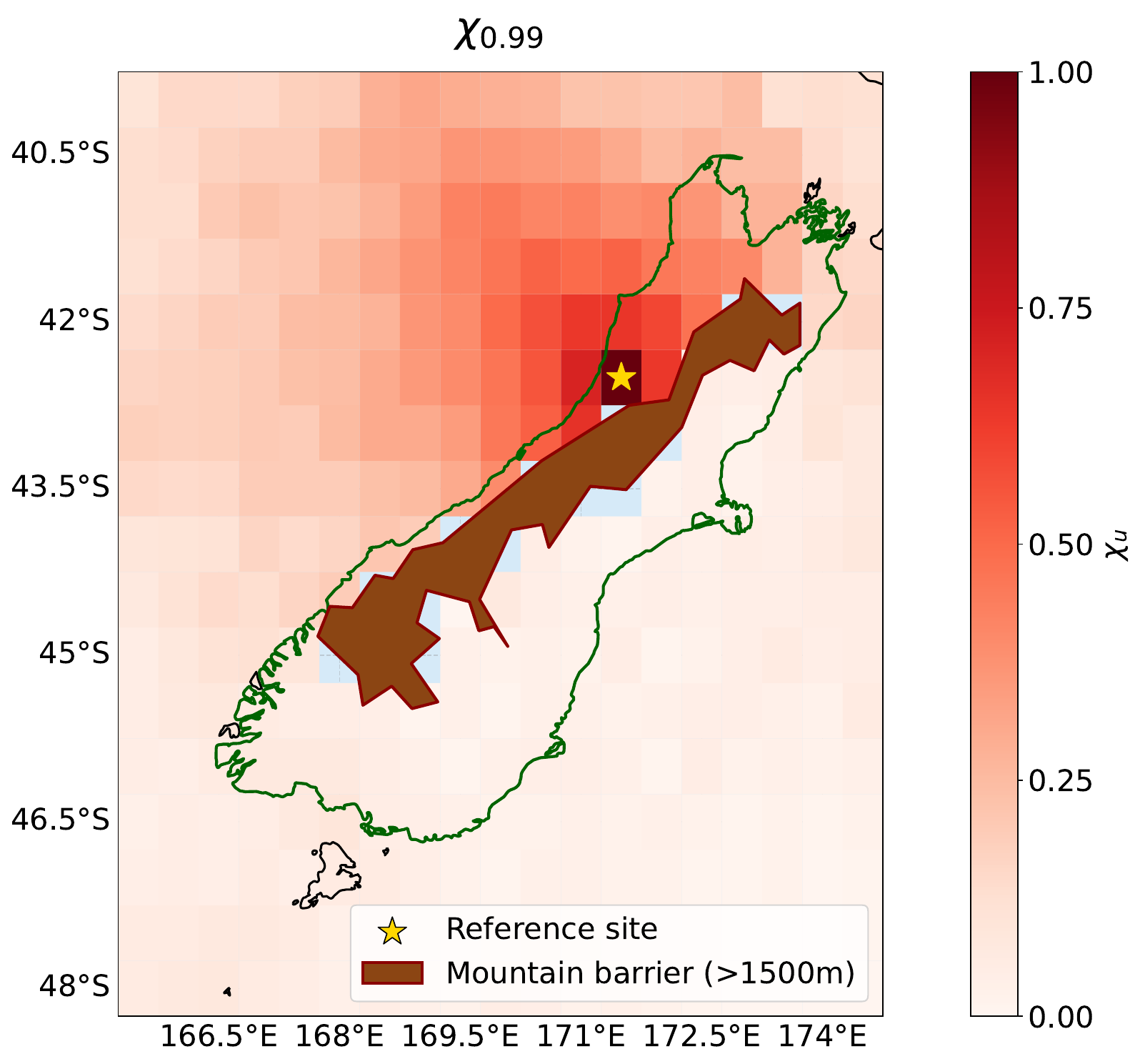} & 
			\includegraphics[width=0.2\textwidth, clip=true, trim=0 0 30 40]{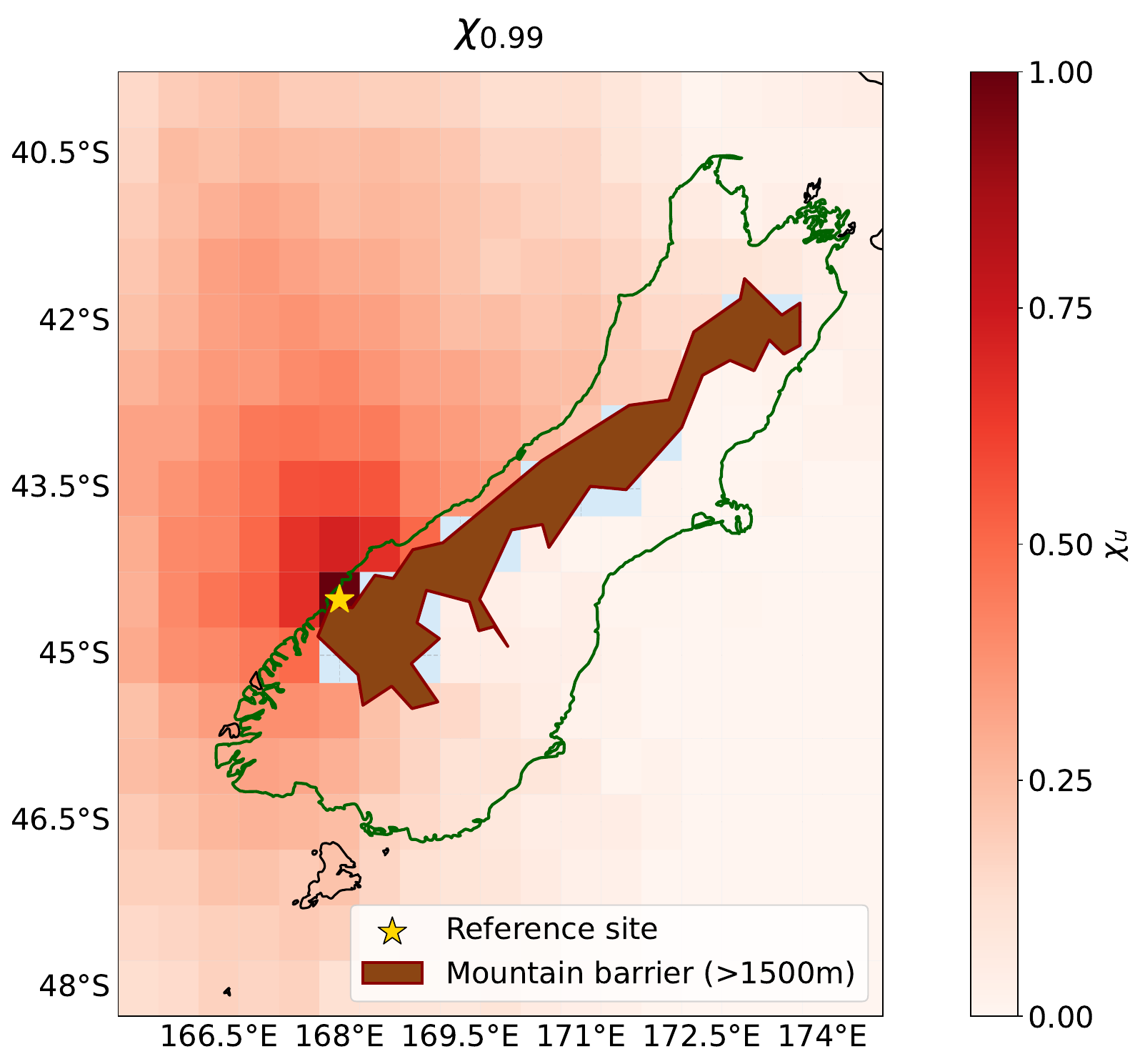} & 
			\includegraphics[width=0.2\textwidth, clip=true, trim=0 0 30 40]{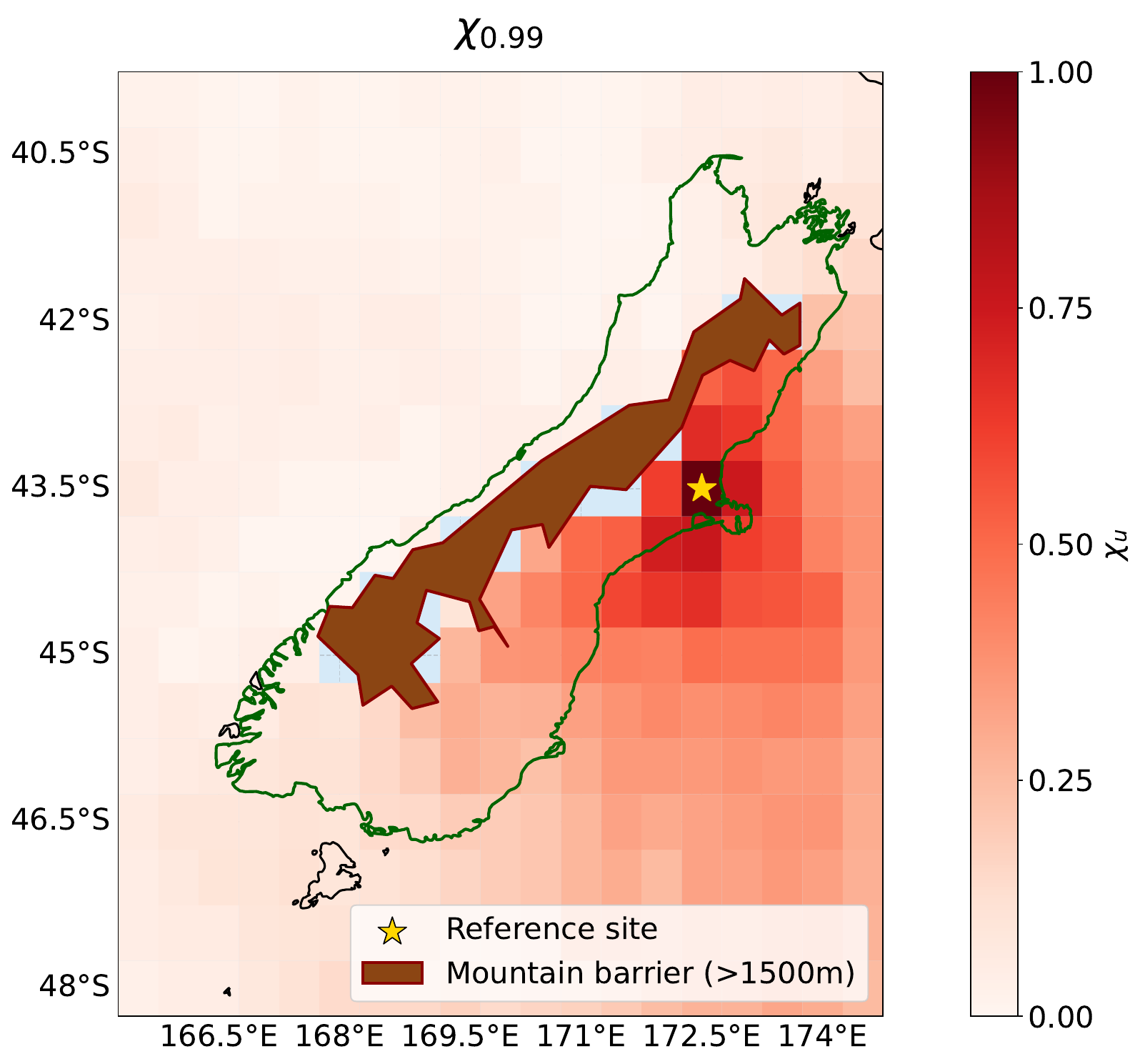} & 
			\includegraphics[width=0.2\textwidth, clip=true, trim=0 0 30 40]{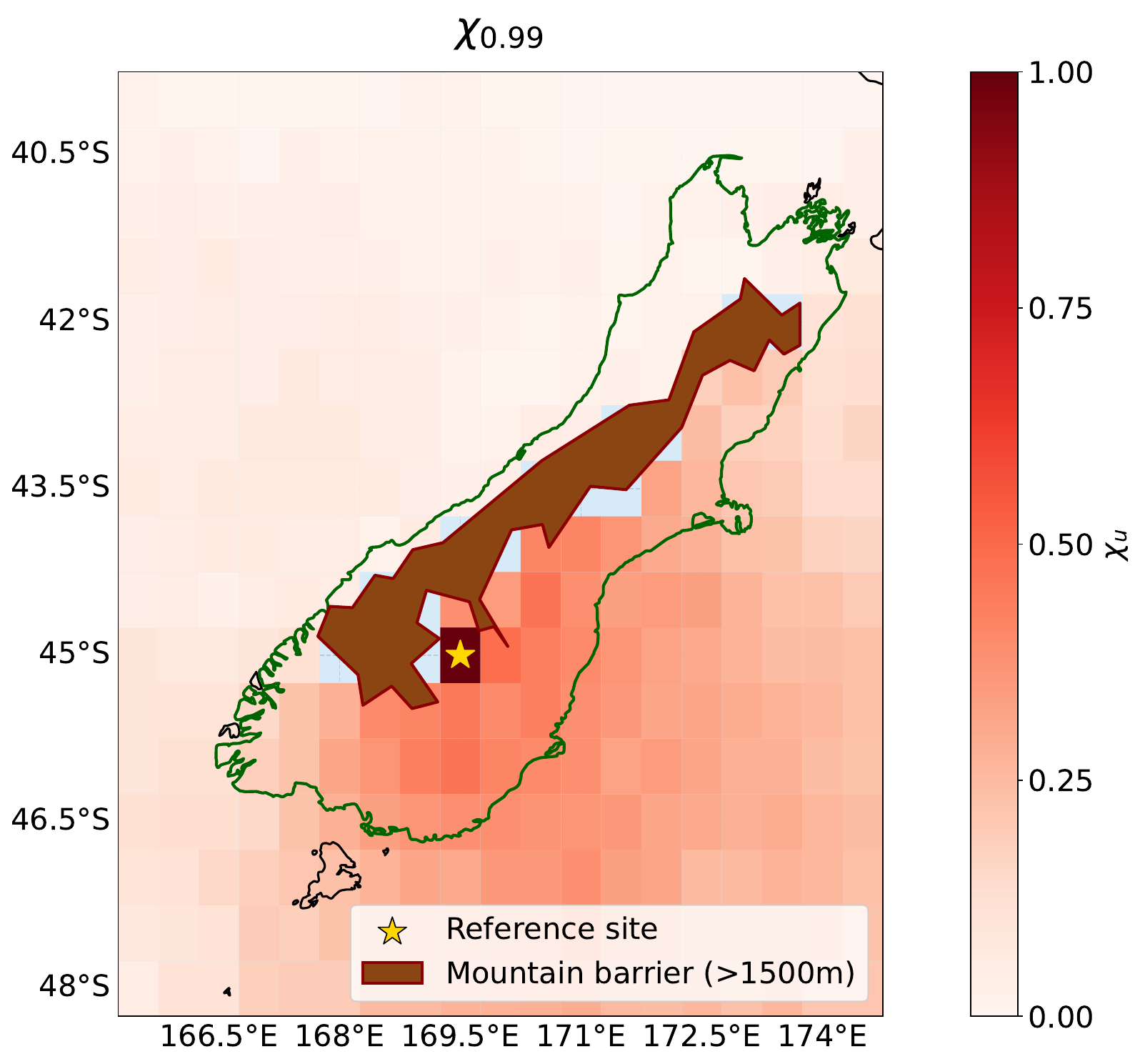}
		\end{tabular}
		\caption{Pairwise empirical (top) and visibility-model-based (bottom) $\chi_{0.99}$. Columns (left to right) relative to reference sites near Hokitika, Milford Sound, Christchurch, and Queenstown.}
		\label{fig:precip_chiu_emp_est}
	\end{figure}

	We first examine whether the fitted model reproduces the spatial structure of extremal dependence and, in particular, the attenuation induced by the Southern Alps. Figure~\ref{fig:precip_chiu_emp_est} compares pairwise empirical and model-based $\chi_u$ at $u = 0.99$ relative to four reference sites near Hokitika and Milford Sound on the windward side and near Christchurch and Queenstown on the leeward side of the barrier. The empirical estimates are computed from the daily precipitation totals over the AR season, while the model-based estimates are obtained from posterior predictive draws of the fitted visibility model. Additional $\chi_u$ fitness assessment at $u = 0.90$ and $0.95$ are reported in the Supplementary Material Section~\ref{supp:precip_chiu}. 
	
	For the Christchurch and Queenstown reference sites, the empirical $\chi_u$ surface exhibits a clear directional pattern, with stronger extremal dependence on the same side of the Southern Alps and weaker dependence across the mountain barrier. The model-based surface reproduces this directional attenuation, rather than producing dependence that is determined primarily by Euclidean distance. The contrast becomes particularly apparent at the higher threshold $u = 0.99$, where the region across the barrier exhibits substantially weaker extremal dependence than nearby locations on the same side. The model also captures the broad anisotropic structure of the empirical surface, although the magnitude of dependence is not reproduced uniformly across the domain. 
	
	We observe a similar pattern for the Hokitika and Milford Sound reference sites, but with the roles of the two sites reversed. The empirical surface shows relatively strong dependence toward locations on the windward side and attenuated dependence across the Southern Alps. The visibility model reproduces this barrier-induced separation and the associated directional pattern. Comparing the two reference sites is particularly informative; the fitted dependence structure responds to the side of the barrier on which the reference site is located, demonstrating that the visibility representation allows the physical barrier to influence extremal dependence beyond what can be explained by Euclidean separation alone. 
	
	The model-based and empirical surfaces are not expected to coincide exactly. The empirical $\chi_u$ estimates are subjected to substantial sampling variability, especially at $u = 0.99$, whereas the model-based surfaces represent dependence implied by the fitted process. Overall, the visibility model captures the principal spatial feature of the data, extremal dependence is systematically attenuated for pairs separated by the Southern Alps, while retaining the anisotropic spatial structure observed in the empirical estimates.
	
	\subsection{Model Comparison} \label{subsec:precip_comparison}
	
\begin{figure}[htbp]
		\centering
		\includegraphics[width=0.8\textwidth]{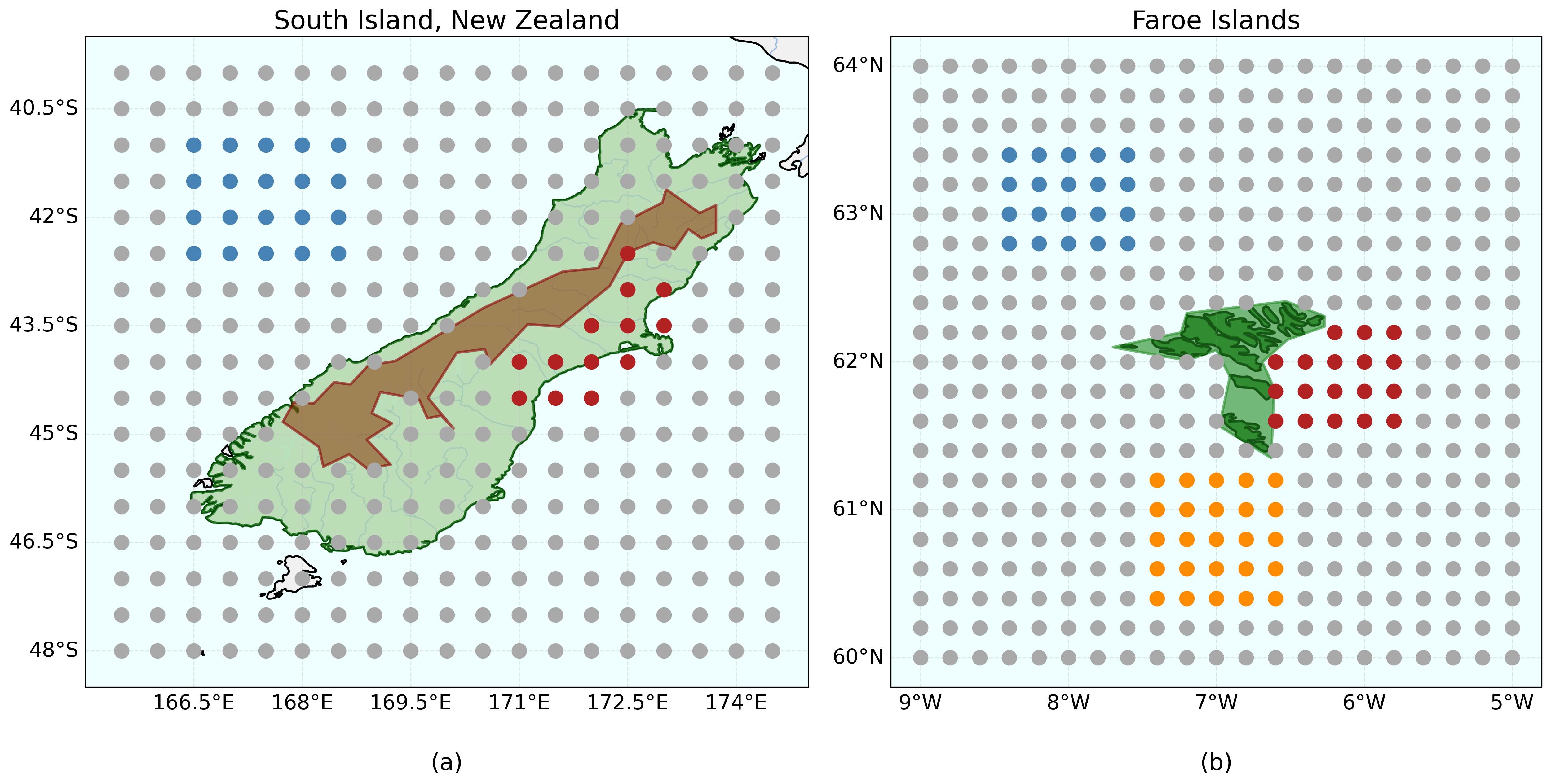}
		\caption{Spatial configurations of the precipitation and wave height data heldout analyses for (a) the South Island, New Zealand. Gray points indicate the 276 training sites, and blue and red points indicate the the two holdout blocks with 33 holdout sites in total. (b) The Faroe Islands. Gray points indicate the 369 training sites, and blue, red, and orange points indicate the three holdout blocks, with 63 holdout sites in total.}
		\label{fig:train_test_split}
	\end{figure}

	To assess out-of-sample performance, we divide the $d = 309$ spatial locations into a training set of 276 sites and a holdout set of 33 sites. Rather than selecting the holdout locations at random, we use spatial block sampling, with one contiguous block on the windward side and one on the leeward side of the Southern Alps (Figure~\ref{fig:train_test_split}(a)). Under spatial dependence, a randomly selected holdout site is typically surrounded by nearby training sites, so prediction at the withheld location can largely reduce to short-range spatial interpolation. In contrast, withholding contiguous spatial blocks requires the model to predict an entire region using information from locations outside the block, providing a more stringent assessment of spatial extrapolation. Using blocks on both sides of the barrier further allows us to assess predictive performance across the two sides of the Southern Alps. 
	
	We compare the visibility model with \citet{Shi2026} and \citet{Huser2019}. For the Shi et al. model, we use an isotropic stationary Mat{\'e}rn covariance function together with 24 isotropic Wendland and Gaussian kernel knots. Unlike the visibility model, this specification does not incorporate the visibility graph or the Southern Alps as a physical barrier. The Huser-Wadsworth model likewise uses an isotropic stationary Mat{\'e}rn covariance structure, but its tail dependence parameter is constant over space, whereas both the visibility and Shi et al.\ models allow the tail dependence structure to vary spatially. The detailed model specifications, posterior diagnostics, and additional parameter estimates for all three models are provided in the Supplementary Material Section~\ref{supp:precip_model_comparison}.
	
	We evaluate the models using the widely applicable information criterion (WAIC, \citet{watanabe2010, gelman2014}) on the training data and two predictive scores on the holdout data, the predictive log-likelihood and a threshold-weighted continuous ranked probability score (twCRPS, \citet{gneiting2011, lerch2017}). Definitions and computational details for all three criteria are given in Section~\ref{supp:scoring_criteria} of the Supplementary Material. For each criterion, model differences are computed relative to the visibility model, with uncertainty assessed using years as exchangeable resampling unit. 
	
	Table~\ref{tab:precip_comparison_score} summarizes the resulting scores. The visibility model does not provide a uniform improvement in aggregate predictive performance. The Shi et al.\ model has a distinguishably higher holdout predictive log-likelihood, while its WAIC is essentially indistinguishable from that of the visibility model. The Huser-Wadsworth model has a distinguishably worse WAIC but a distinguishably higher holdout predictive log-likelihood. For twCRPS-smooth, neither competing model differs significantly from the visibility model. Thus, based on aggregate scoring criteria alone, there is no clear overall predictive advantage for the barrier-aware model. 
	
	\begin{table}[ht]
		\centering
		\caption{Held-out predictive performance at the 33 test sites (pooled over all two holdout blocks), also expressed as differences from the visibility model (Diff = model $-$ visibility model). All model parameters are jointly estimated. For WAIC and twCRPS-smooth (lower is better), a negative difference favors the row model over the visibility model; for predictive log-likelihood (higher is better), a positive difference favors the row model. \textbf{Bold} entries represent the best score and \textit{italic} entries are distinguishable from the visibility model at the 2$\times$SE level.}
		\label{tab:precip_comparison_score}
			\begin{tabular}{lrrr}
				\toprule
				& Train WAIC & Holdout PredLL & twCRPS-smooth \\
				\midrule
				Visibility  & \textbf{140{,}730.6} & $-$8130.8 & 0.879 \\
				Shi et al. & 140{,}739.7 & \textbf{$-$8025.1} & \textbf{0.867} \\
				Huser--Wadsworth & 141{,}352.9 & $-$8067.2 & 0.879 \\
				\midrule
				Shi $-$ Visibility  & $+9.1\pm98.4$ & \textit{+105.6 $\pm$ 21.1} & $-0.012\pm0.014$ \\
				HW $-$ Visibility & \textit{+622.3 $\pm$ 89.4} & \textit{+63.5 $\pm$ 21.6} & $-0.001\pm0.017$ \\
				\bottomrule
			\end{tabular}%
	\end{table}
	
	Because these aggregate scores combine the marginal and dependence components of the model, they do not directly reveal whether differences in predictive performance arise from the marginal GEV specification or from the spatial dependence structure. We therefore decompose the holdout log-likelihood into a copula component and marginal component in (\refeq{eq:copula_margin}). The resulting decomposition (Table~\ref{tab:copula_margin}) shows a consistent pattern across the full holdout set and separately for the windward and leeward blocks. The visibility model has a higher copula score than both competing models, while both competing models have higher marginal scores. Thus, although the visibility model does not achieve the best overall predictive score, its dependence component provides a better description of the spatial dependence in the holdout data; its aggregate predictive disadvantage is instead associated with the marginal GEV fit. 
	\begin{align}
		\log p(\bm{y}_t) = \underbrace{\tfrac{1}{2} \log |\bm{Q}_{\text{cond}}| - \tfrac{1}{2} (\bm{z} - \bm{\mu}_{\text{cond}})^\top \bm{Q}_{\text{cond}} (\bm{z} - \bm{\mu}_{\text{cond}})}_{\text{copula score}}
		+ \underbrace{\log\vert \det ( \tfrac{\partial \bm{Z}_t}{\partial \bm{Y}_t} ) \vert}_{\text{margin score}}, 
		\label{eq:copula_margin}
	\end{align}
	
	\begin{table}[ht]
		\centering
		\caption{Copula/margin decomposition at the 33 test sites, also expressed as differences from the visibility model (Diff = model $-$ visibility model). All model parameters are jointly estimated. For the copula and margin score (higher is better), a positive difference favors the row model over the visibility model. All the differences are distinguishable.}
		\label{tab:copula_margin}
		\begin{tabular}{llrr}
			\toprule
			Subset & & Copula score & Margin score \\
			\midrule
			\multirow{3}{*}{Full} & Visibility & 2631.1 & $-$10{,}568.3 \\
			& Shi $-$ Visibility & $-1040.6 \pm 23.3$ & $+1086.6 \pm 7.5$ \\
			& HW $-$ Visibility & $-452.4 \pm 22.0$ & $+488.6 \pm 8.3$ \\
			\midrule
			\multirow{3}{*}{Windward} & Visibility & 1794.2 & $-$6555.5 \\
			& Shi $-$ Visibility & $-693.7 \pm 11.1$ & $+694.1 \pm 5.5$ \\
			& HW $-$ Visibility & $-308.1 \pm 7.7$ & $+303.8 \pm 4.3$ \\
			\midrule
			\multirow{3}{*}{Leeward} & Visibility & 807.2 & $-$4044.0 \\
			& Shi $-$ Visibility & $-340.8 \pm 18.4$ & $+402.4 \pm 5.4$ \\
			& HW $-$ Visibility & $-147.4 \pm 19.8$ & $+175.8 \pm 6.6$ \\
			\bottomrule
		\end{tabular}
	\end{table}
	
\begin{figure}[ht]
		\centering
		\includegraphics[width=0.9\textwidth]{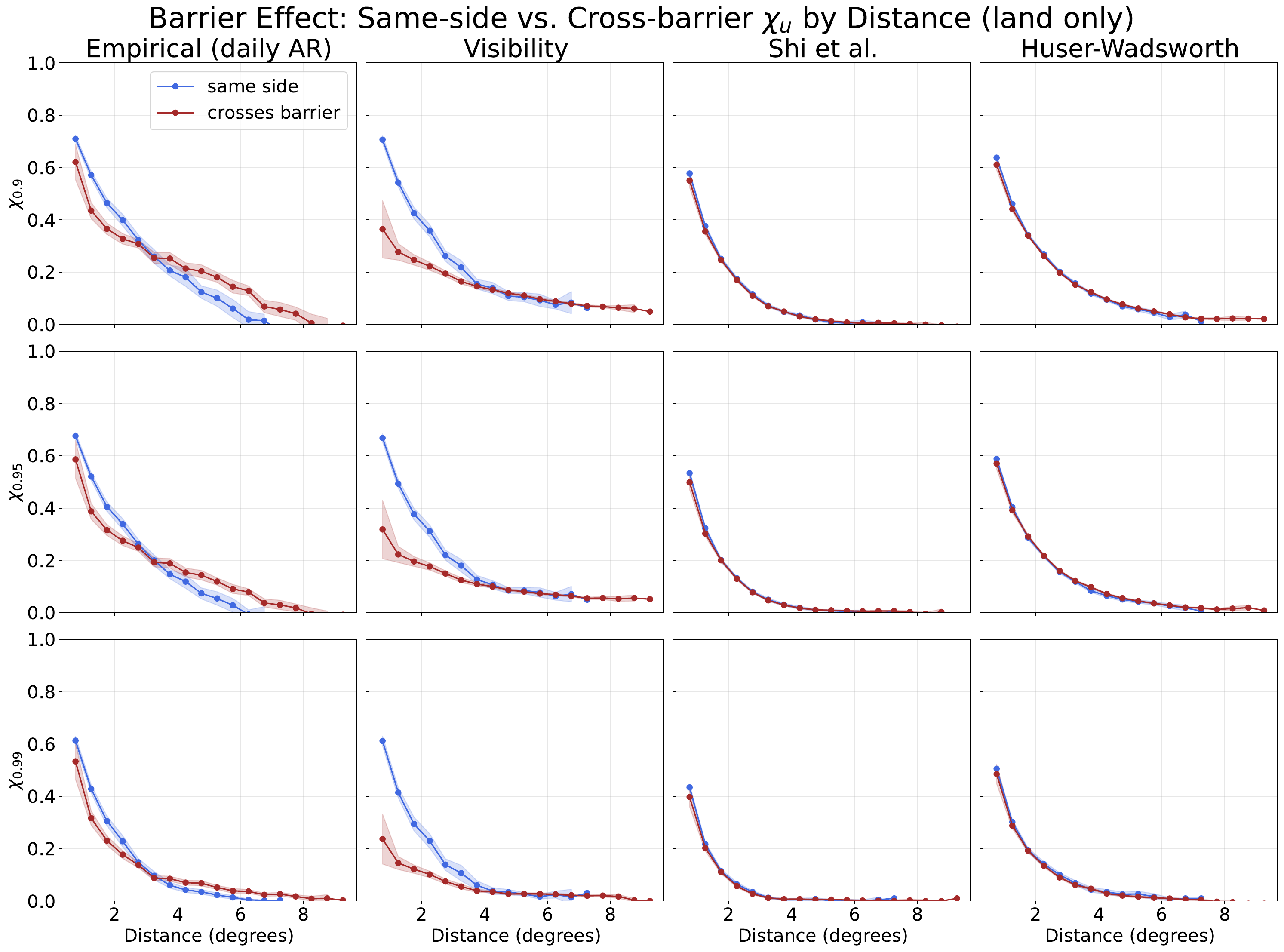}
		\caption{$\chi_u$ vs.\ distance, same-side vs.\ cross-barrier, by model and threshold $u\in\{0.9,0.95,0.99\}$.}
		\label{fig:barrier_dist}
	\end{figure}

	We next examine directly whether the competing models reproduce the barrier-induced extremal dependence observed in the data. Figure~\ref{fig:barrier_dist} shows $\chi_u$ as a function of Euclidean distance for pairs of sites on the same side of the Southern Alps and pairs whose connecting segment crosses the barrier, at $u = \{0.9, 0.95, 0.99\}$. This comparison is useful because it controls for the primary distance-based mechanism of spatial dependence; if Euclidean distance alone adequately explained the observed extremal dependence, the two curves should be approximately coincident at comparable distances. 
	
	The empirical estimates instead show a clear separation between same-side and cross-barrier pairs, particularly at short and intermediate distances. At a given Euclidean separation, pairs on the same side of the Southern Alps tend to exhibit stronger extremal dependence than pairs separated by the barrier. The separation becomes especially informative at the higher thresholds, where the extremal dependence is weaker overall but the difference between the two classes remains evident. The visibility model reproduces this qualitative separation across all three thresholds, its same-side dependence is stronger than its cross-barrier dependence at comparable distances, with the two curves gradually converging as distance increases. 
	
	In contrast, the curves from the Shi et al.\ and Huser-Wadsworth models are nearly indistinguishable for same-side and cross-barrier pairs at a given distance. Although these models allow dependence to decay with Euclidean distance, neither explicitly represents the physical barrier, so pairs at similar separation receive essentially the same dependence treatment regardless of whether the Southern Alps lie between them. The Shi et al.\ model allows the tail dependence structure to vary spatially, but without barrier information this variation does not reproduce the systematic same-side versus cross-barrier separation observed in the data. 
	
\begin{figure}[ht]
		\centering
		\begin{tabular}{cccc}
			Empirical & Visibility & Shi et al. & Huser--Wadsworth \\
			\includegraphics[width=0.2\textwidth, clip=true, trim=0 0 30 40]{precip_emp_chiu_inland_center_105_99.pdf} &
			\includegraphics[width=0.2\textwidth, clip=true, trim=0 0 30 40]{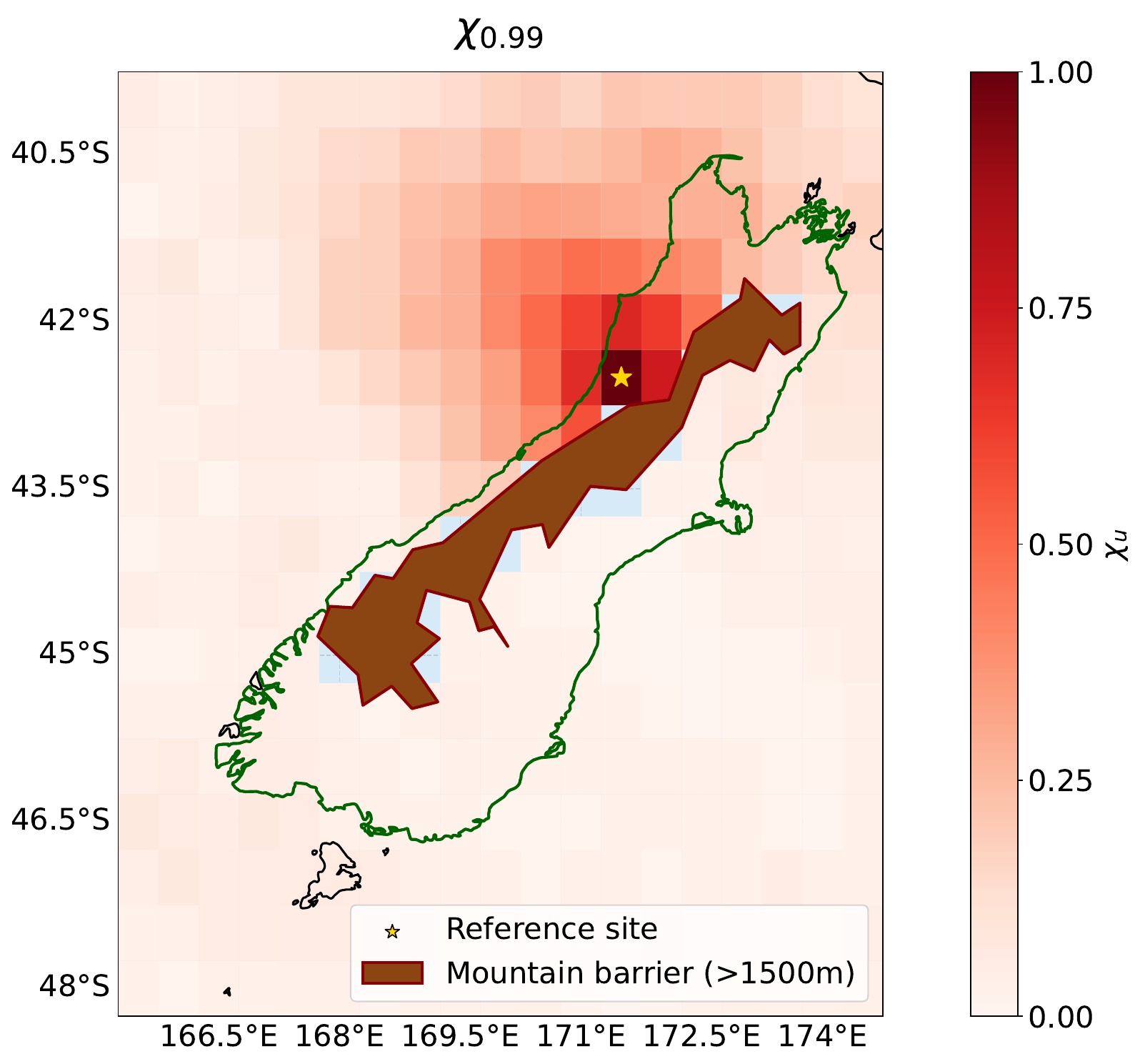} &
			\includegraphics[width=0.2\textwidth, clip=true, trim=0 0 30 40]{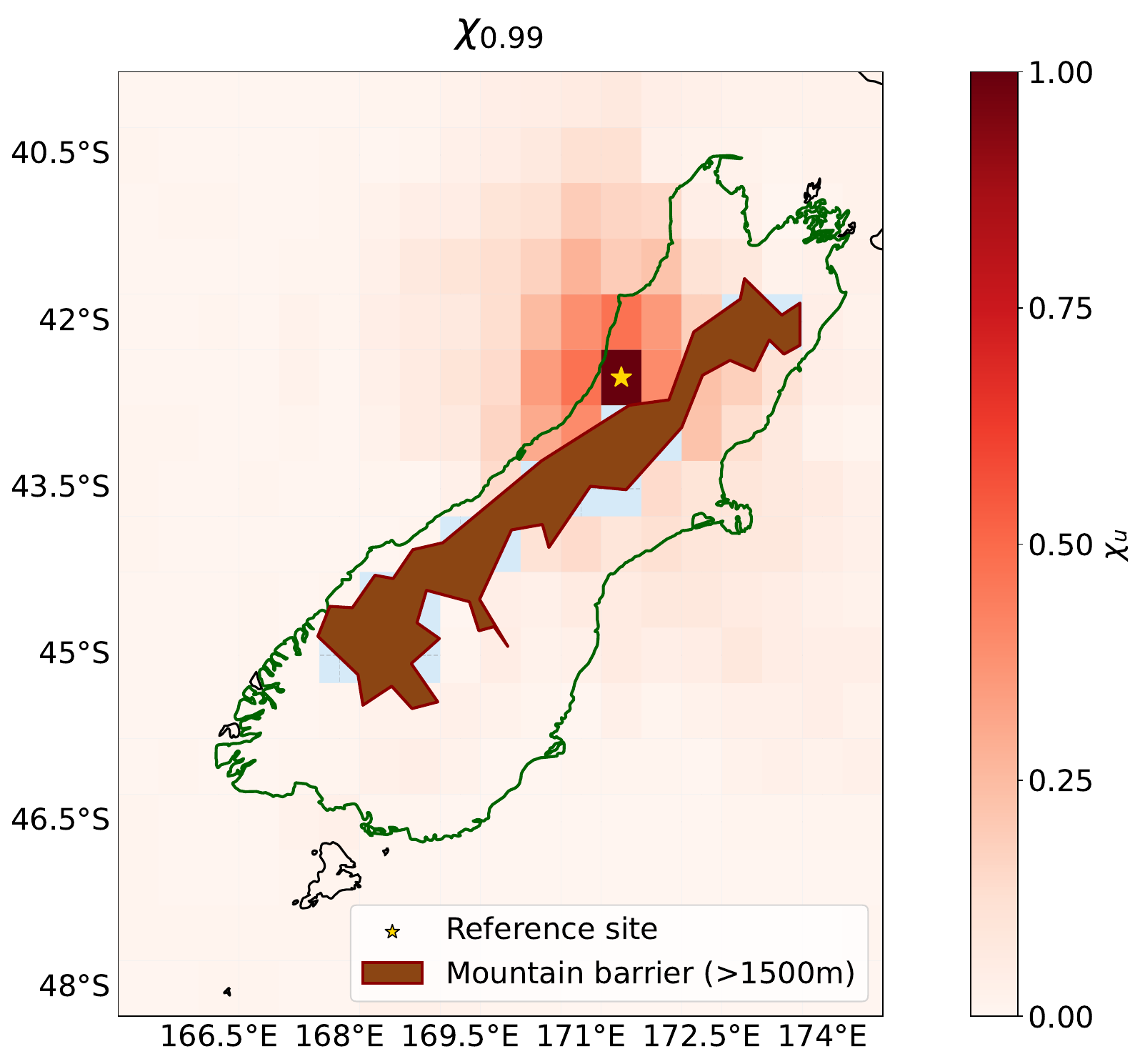} &
			\includegraphics[width=0.2\textwidth, clip=true, trim=0 0 30 40]{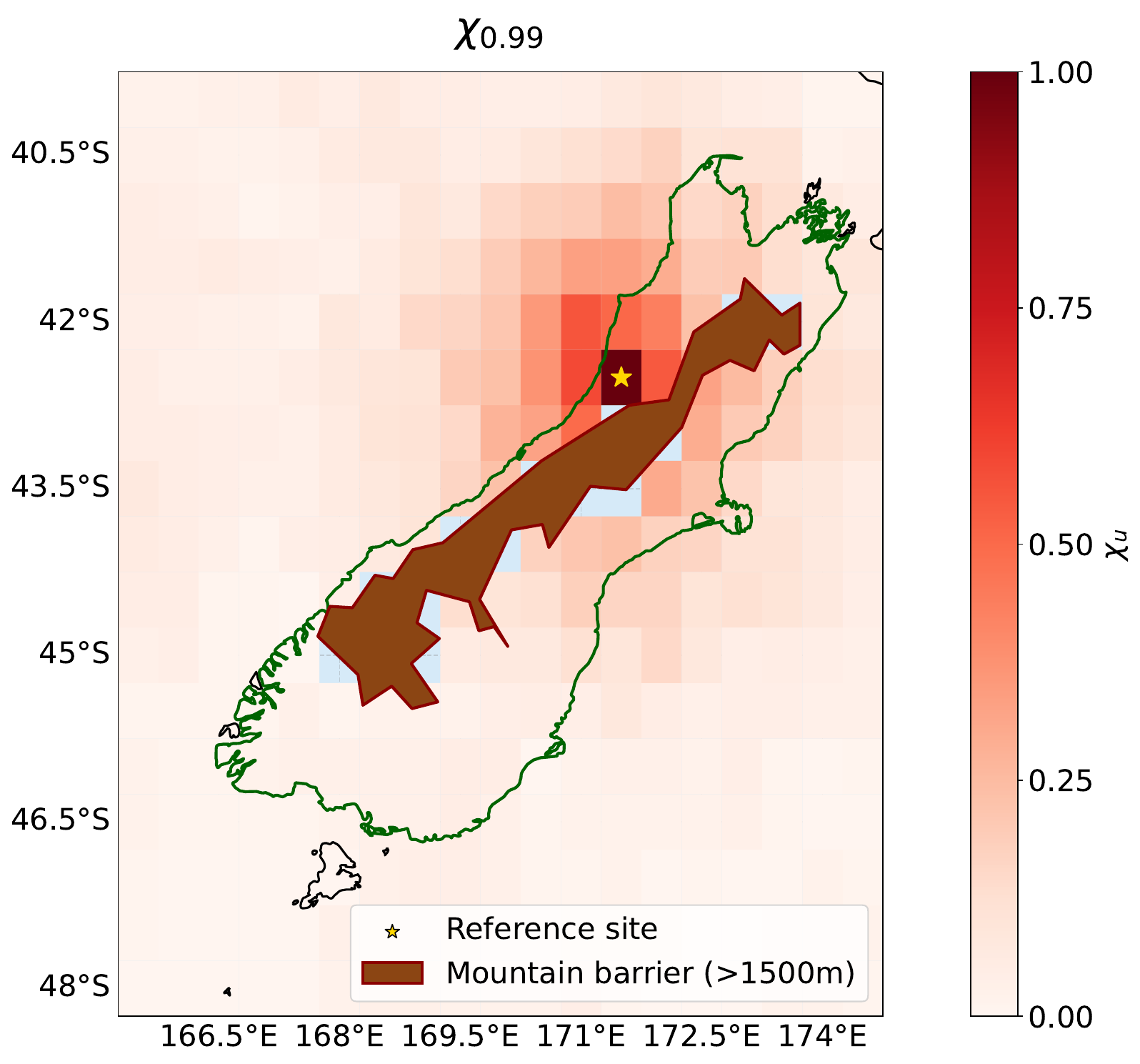}
			\\
			\includegraphics[width=0.2\textwidth, clip=true, trim=0 0 30 40]{precip_emp_chiu_inland_center_167_99.pdf} &
			\includegraphics[width=0.2\textwidth, clip=true, trim=0 0 30 40]{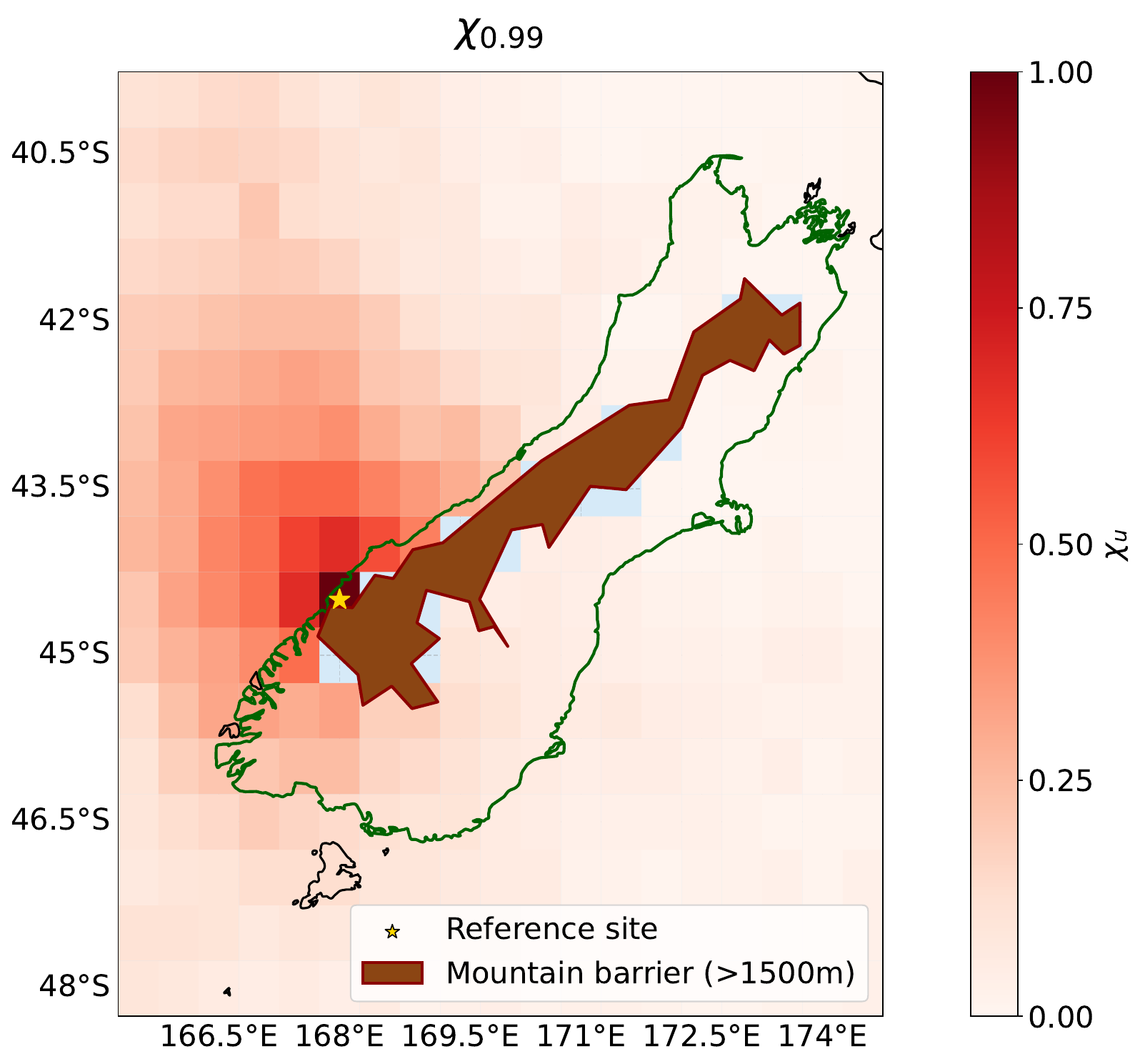} &
			\includegraphics[width=0.2\textwidth, clip=true, trim=0 0 30 40]{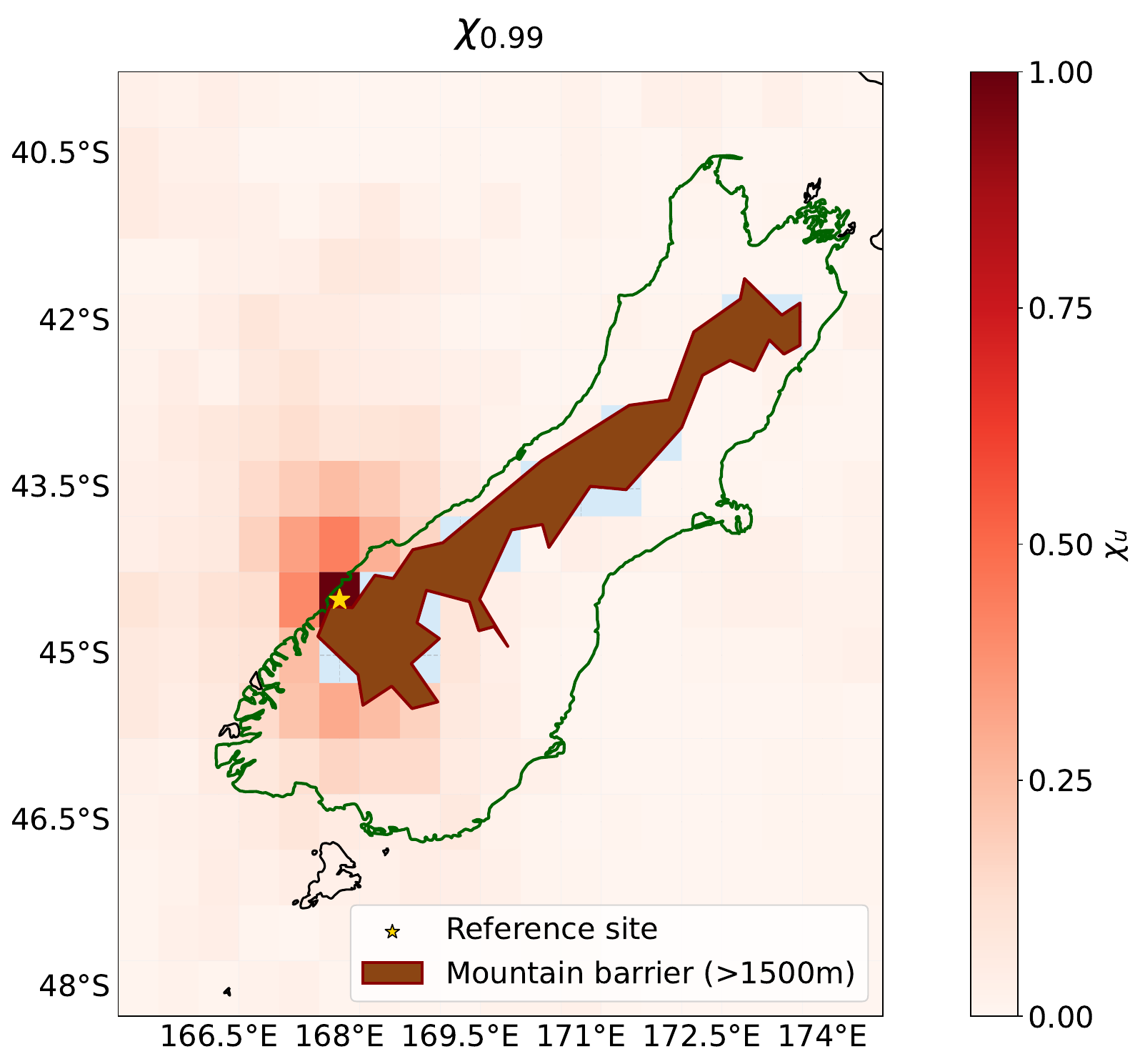} &
			\includegraphics[width=0.2\textwidth, clip=true, trim=0 0 30 40]{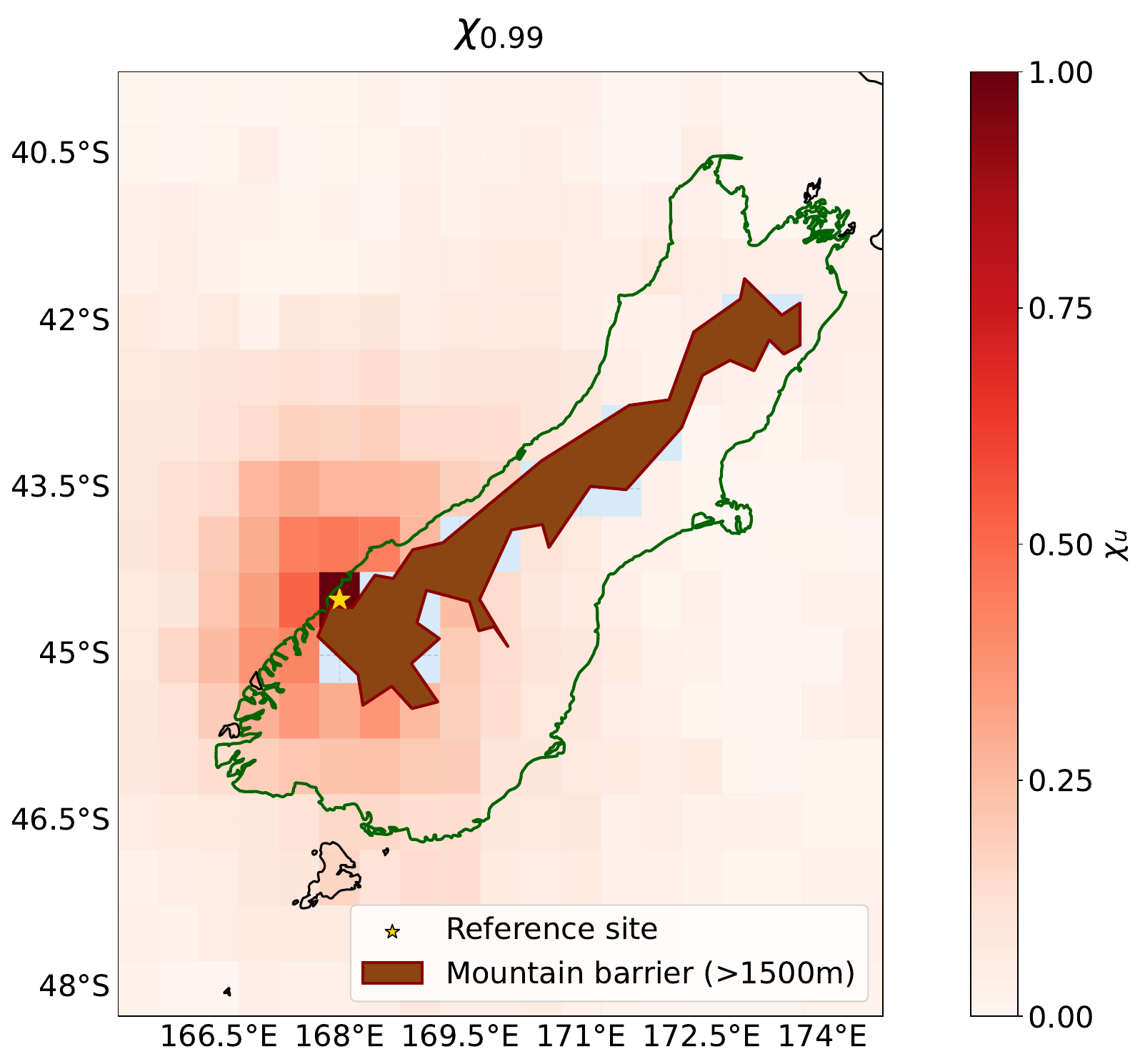}
			\\
			\includegraphics[width=0.2\textwidth, clip=true, trim=0 0 30 40]{precip_emp_chiu_inland_center_140_99.pdf} &
			\includegraphics[width=0.2\textwidth, clip=true, trim=0 0 30 40]{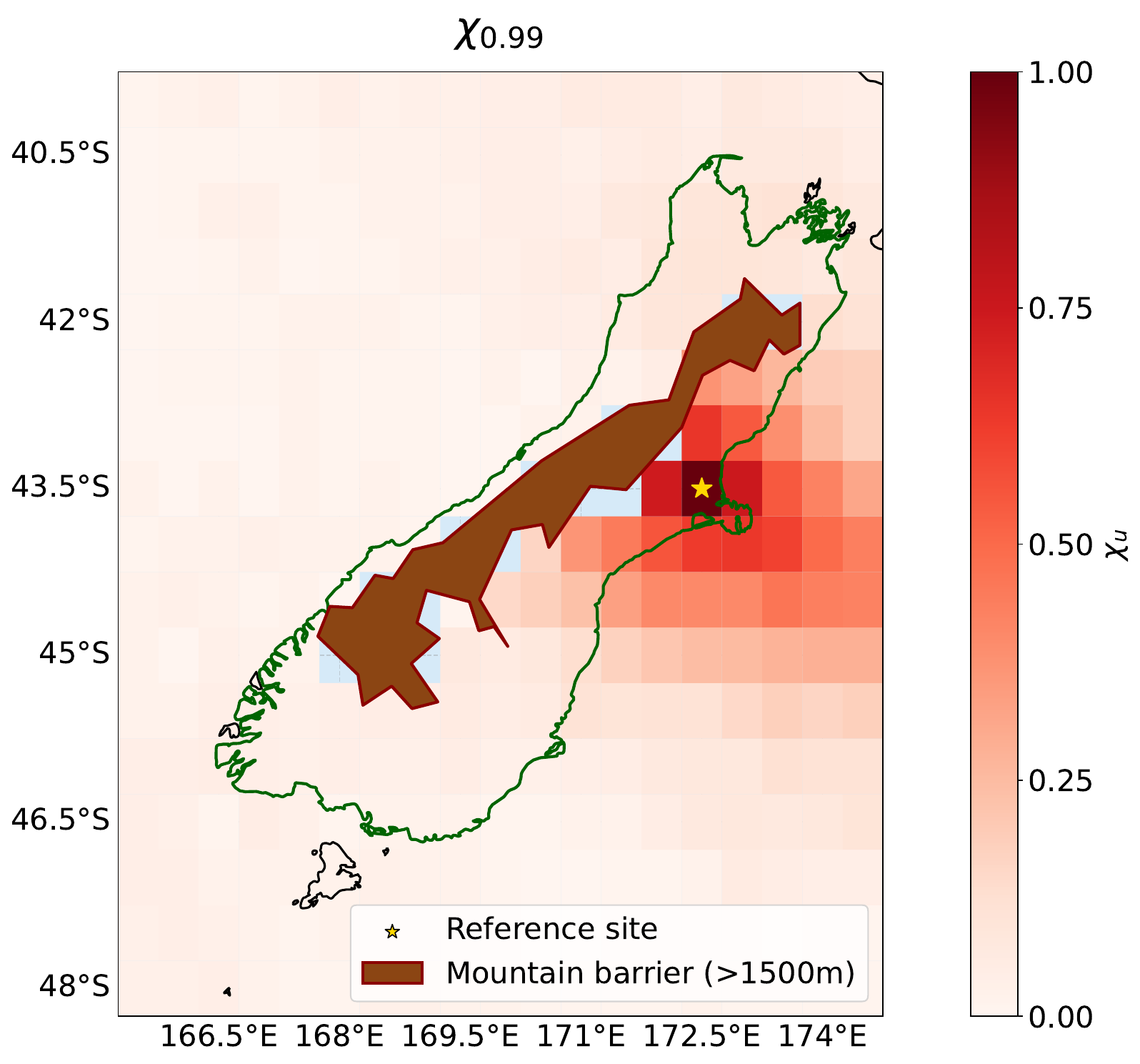} &
			\includegraphics[width=0.2\textwidth, clip=true, trim=0 0 30 40]{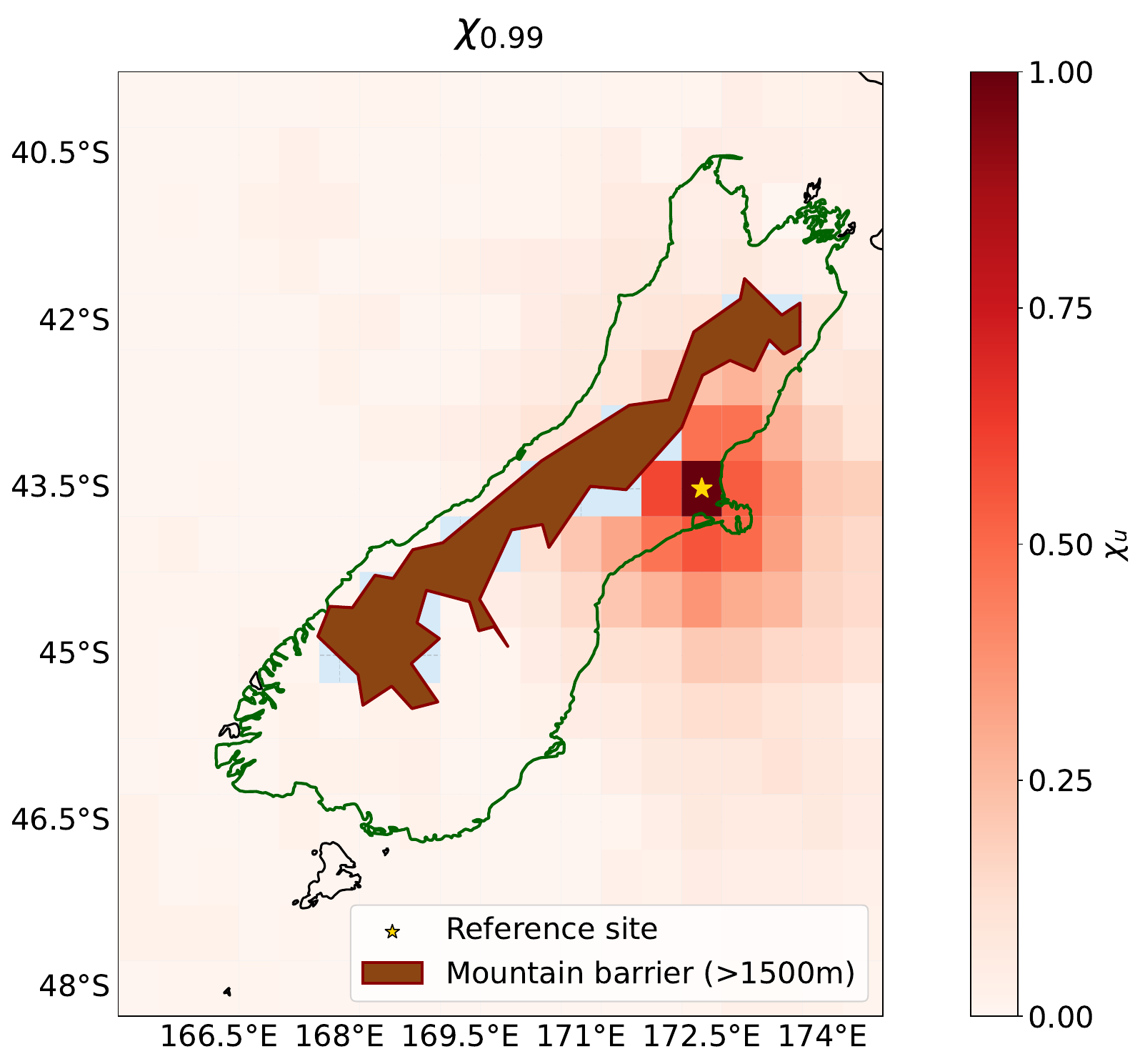} &
			\includegraphics[width=0.2\textwidth, clip=true, trim=0 0 30 40]{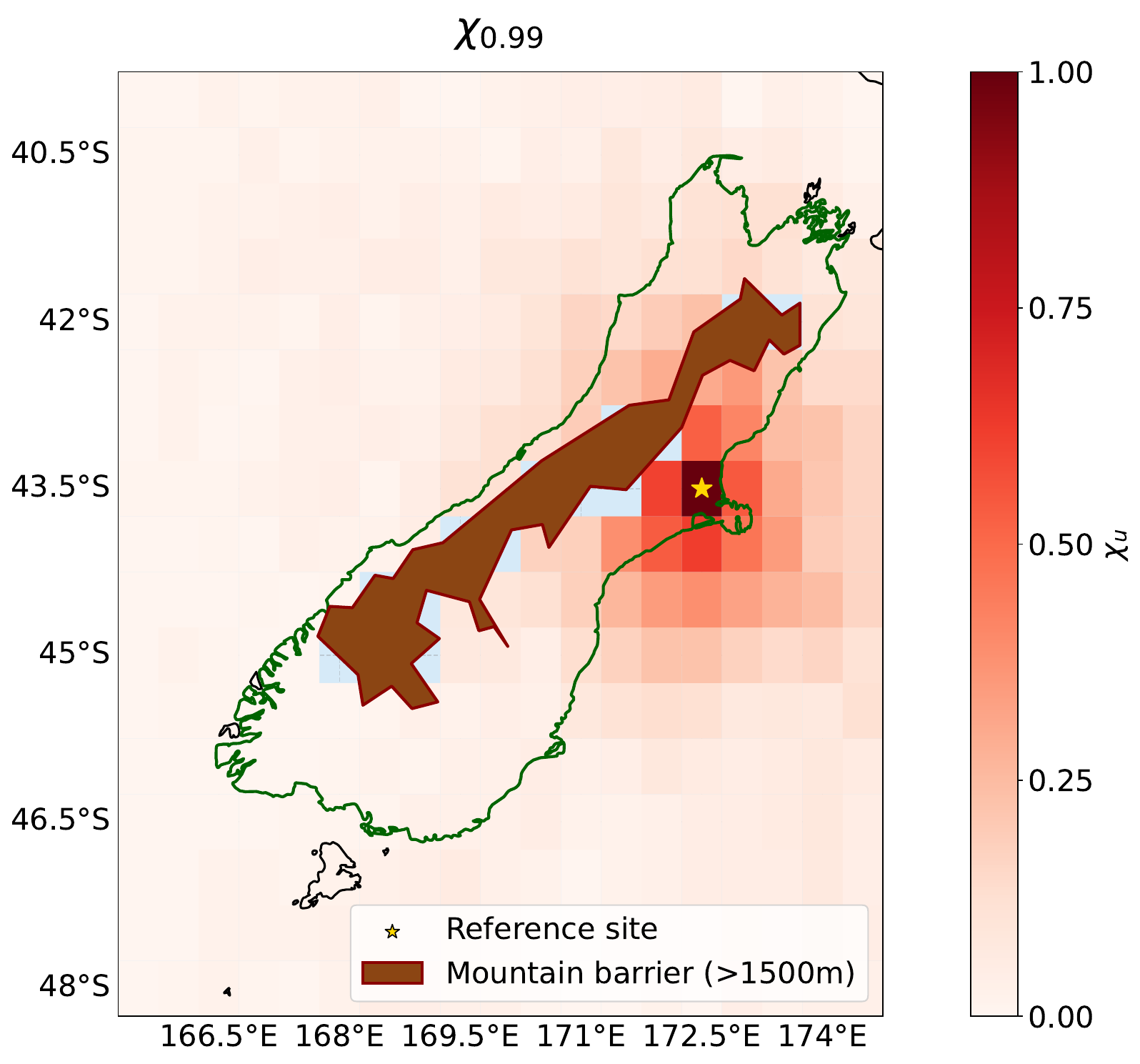}
			\\
			\includegraphics[width=0.2\textwidth, clip=true, trim=0 0 30 40]{precip_emp_chiu_inland_center_184_99.pdf} &
			\includegraphics[width=0.2\textwidth, clip=true, trim=0 0 30 40]{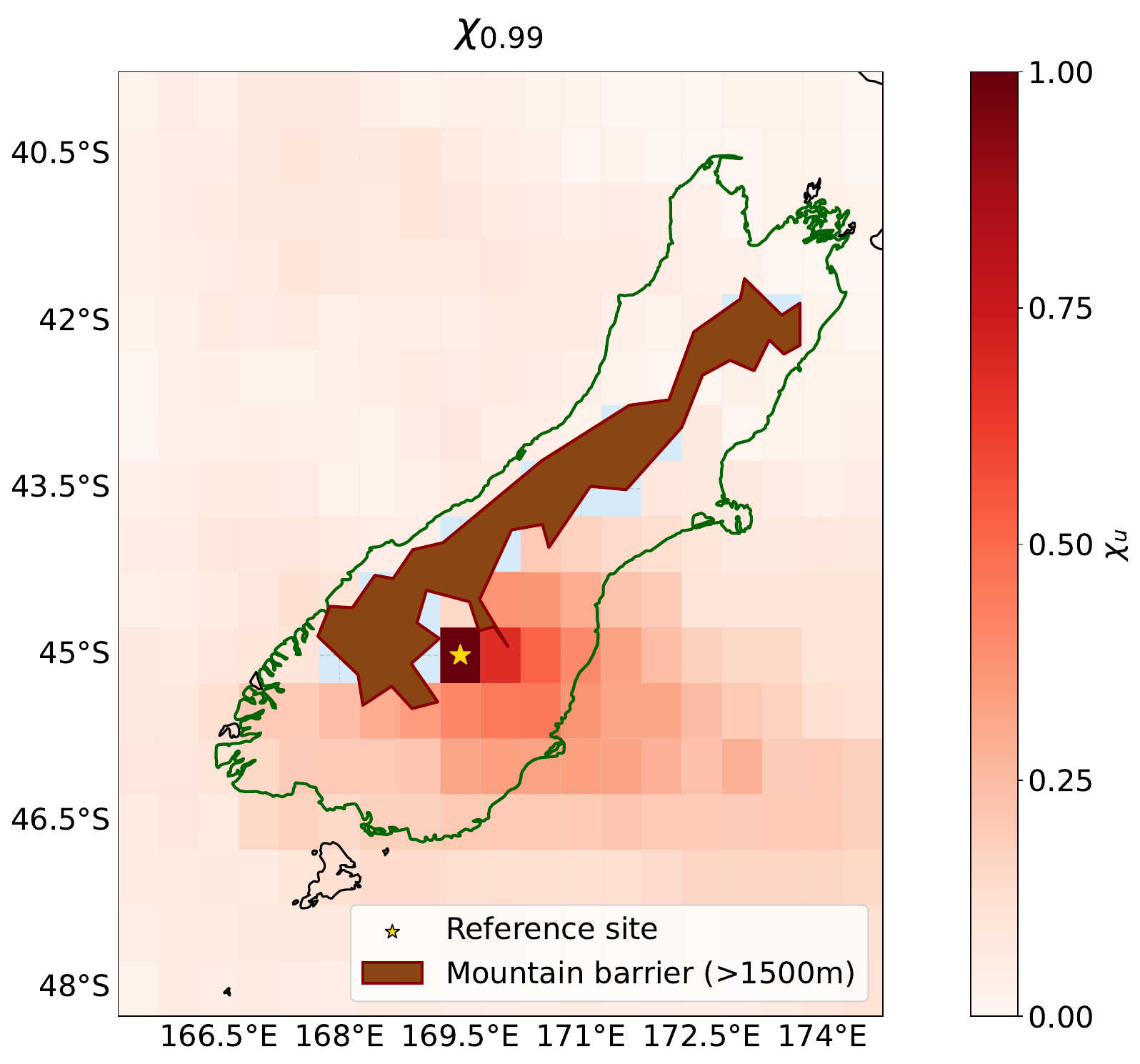} &
			\includegraphics[width=0.2\textwidth, clip=true, trim=0 0 30 40]{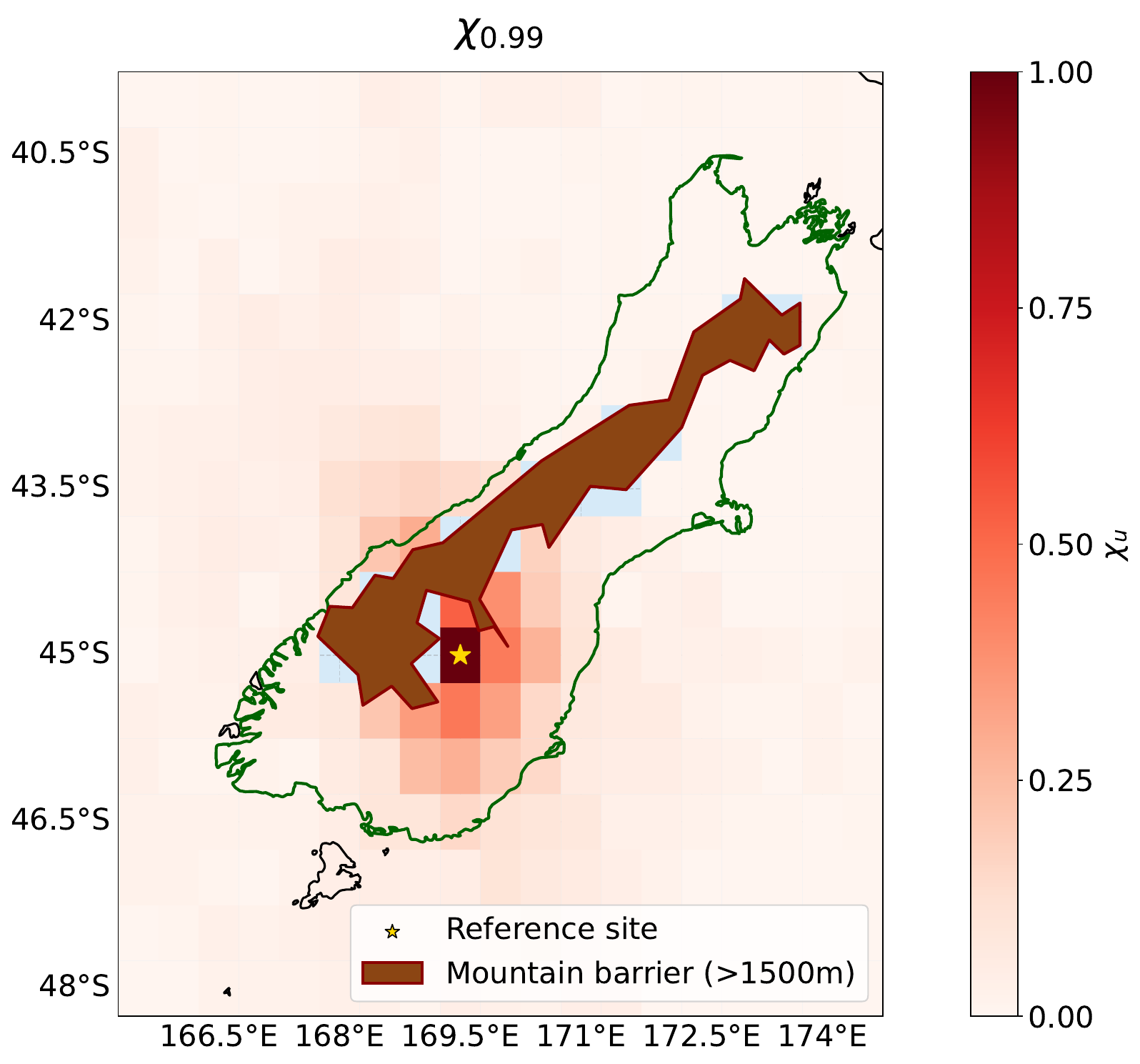} &
			\includegraphics[width=0.2\textwidth, clip=true, trim=0 0 30 40]{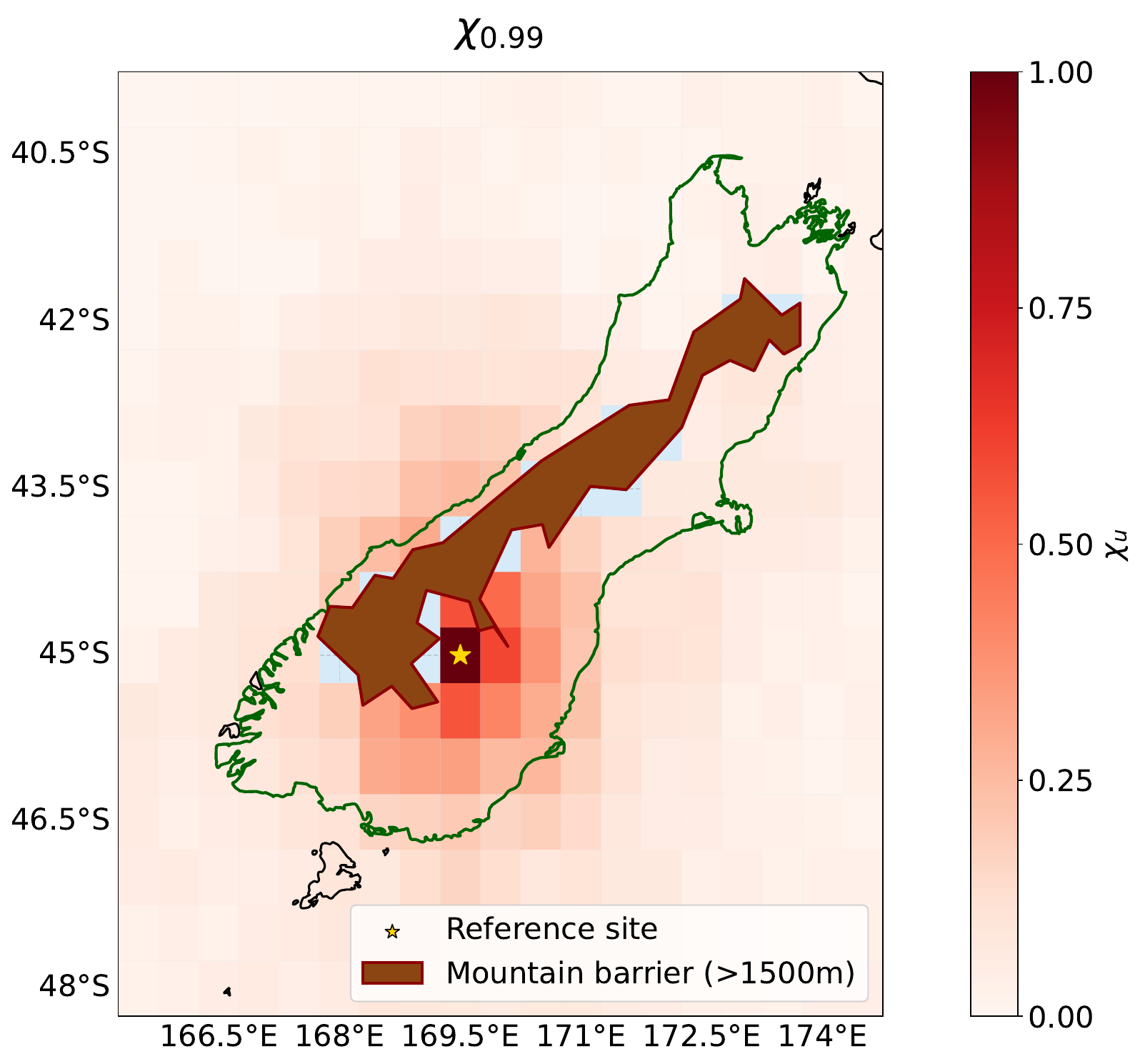}
		\end{tabular}
		\caption{Empirical and model-based $\chi_{0.99}$ surfaces around a reference site, for all three models.}
		\label{fig:precip_chiu_model_comparison}
	\end{figure}

	Figure~\ref{fig:precip_chiu_model_comparison} provides a complementary spatial view at the more extreme threshold $u = 0.99$. The empirical surface is directionally asymmetric, with attenuation in the direction across the Southern Alps. The visibility model reproduces both this attenuation and the associated anisotropic structure. In contrast, the Shi et al.\ and Huser-Wadsworth models produce substantially smoother surfaces whose dependence is primarily organized by spatial separation and therefore do not reproduce the barrier-induced directional pattern. Additional $\chi_u$ at $u = 0.90$ and $u = 0.95$ surfaces across reference sites are provided in Section~\ref{supp:precip_comparison_chiu} of the Supplementary Material. 
	
	Taken together, the comparison shows that the visibility model should not be interpreted as uniformly superior according to aggregate predictive scores. Rather, its principal advantage is structural; it is the only model among the three that explicitly represents the physical barrier and reproduces the resulting attenuation and anisotropy in extremal dependence. The holdout decomposition and the $\chi_u$ comparisons provide complementary evidence that this improvement is attributable to the dependence structure, while the weaker marginal predictive performance indicates that further refinement of the marginal GEV specification could improve the overall predictive accuracy of the visibility model. 

\section{North Atlantic Significant Wave Heights} \label{sec:wave_application}
	
	We next consider significant wave height around the Faroe Islands. We restrict the analysis to the winter storm season, defined as October through March, when severe storms are frequent in the North Atlantic and North Sea. The data are obtained from the E.U. Copernicus Marine Service Information; \url{https://doi.org/10.48670/moi-00022}. We analyze significant wave heights in the vicinity of the Faroe Islands over the spatial domain $[9^\circ \text{W}, 5^\circ \text{W}] \times [60^\circ \text{N}, 64^\circ \text{N}]$. The Global Ocean Waves Reanalysis (GOWR) dataset provides global coverage at a spatial resolution of $0.2^\circ \times 0.2^\circ$ and a temporal resolution of 3 hours. The available record spans from January 1, 1980, to present. Within the study region, the spatial grid contains $d = 432$ locations. 
	
	Rather than treating every 3-hourly observation as a separate extreme, we aggregate the observations into 5-day blocks. This blocking scheme is intended to reduce temporal dependence and to represent individual storm episodes more directly, since a single storm can generate elevated wave heights over several consecutive days. Specifically, let $H_t(\bm{s}_i, \tau)$ denote the significant wave height at spatial location $\bm{s}_i$ and 3-hourly time point $\tau$ during storm season $t$. For 5-day block $b$, we define the block maximum
	\begin{align}
		M_{t, b}(\bm{s}_i) = \max_{\tau \in \mathcal{B}_{t, b}} H_t(\bm{s}_i, \tau), 
		\label{eq:wave_block_max}
	\end{align}
	where $\mathcal{B}_{t, b}$ denotes the set of 3-hourly observations contained in block $b$. The resulting collection $\{M_{t, b}(\bm{s}_i)\}$ provides the storm-scale extreme observations used in the subsequent peaks-over-threshold analysis. 
	%
    The choice of the 5-day block length is supported by the storm-scale exploratory analysis reported in Supplementary Section~\ref{supp:wave_storm_duration}.
	
	\subsection{Marginal GPD Model} \label{subsec:gpd_margin}
	
	We use a peaks-over-threshold approach for the marginal distribution of the 5-day block maxima. At each spatial location $\bm{s}$, we first estimate the marginal $0.95$ quantile, $q_{0.95}(\bm{s})$, using a pooled spatial quantile-regression model. The model incorporates covariates describing the large-scale spatial trend, smooth spatial variation, bathymetric depth, distance to the island centroid, exposure angle, and selected interaction terms. The fitted spatial surface provides a smooth and physically interpretable estimate of the high threshold across the study region, while borrowing information across neighboring locations. The selection and construction of the spatial covariates, together with additional candidate covariates and alternative specifications considered in the analysis, are described in Section~\ref{supp:wave_gpd_covariates} of the Supplementary Material.
	
	For each 5-day block and spatial location, we define the exceedance
	\begin{align}
		Y_{t, b}(\bm{s}) = M_{t, b}(\bm{s}) - \hat{q}_{0.95}(\bm{s}), \qquad M_{t, b}(\bm{s}) > \hat{q}_{0.95}(\bm{s}).
		\label{eq:wave_exceedance}
	\end{align}
	This procedure produces 35{,}776 positive exceedances across the 432 spatial locations. Conditional on exceeding the spatially varying threshold, we model the exceedances using a generalized Pareto distribution, 
	\begin{align}
		Y_{t, b}(\bm{s}) \mid M_{t, b}(\bm{s}) > \hat{q}_{0.95}(\bm{s}) \sim \operatorname{GPD}(\sigma(\bm{s}), \xi(\bm{s})), 
		\label{eq:wave_gpd}
	\end{align}
	with scale parameter $\sigma(\bm{s}) > 0$ and shape parameter $\xi(\bm{s})$. The corresponding GPD distribution function is 
	\begin{align}
		G(y \mid \bm{s}) = 1 - \left[ 1 + \frac{\xi(\bm{s}) y}{\sigma(\bm{s})} \right]^{-1 / \xi(\bm{s})}, 
		\label{eq:wave_gpd_cdf}
	\end{align}
	for values of $y$ satisfying $1 + \xi(\bm{s}) y / \sigma(\bm{s}) > 0$, with the usual exponential limit when $\xi(\bm{s}) = 0$. 
	
	We allow both GPD parameters to vary spatially, using separate covariate specifications for the scale and shape parameters.. The scale parameter uses the same covariates as the spatial quantile-regression model for the $0.95$ threshold and is modeled through the log link $\log\sigma(\bm{s}) = \bm{x}_{\sigma}(\bm{s})^\top \bm{\beta}_{\sigma}$. In contrast, the shape parameter uses a separate set of selected spatial covariates and is modeled using the rescaled-logit transformation described in Section~\ref{subsec:gev_margin}, $g_{\xi}(\xi(\bm{s})) = \bm{x}_\xi(\bm{s})^\top \bm{\beta}_{\xi}$. Thus, the marginal model allows both the threshold and the magnitude of extreme wave events to vary systematically with the local environment and geometry of the Faroe Islands, while allowing the shape of the upper tail to have a distinct spatial structure.
	
	\subsection{Results} \label{subsec:wave_result}
	
	While we have found that joint inference of marginal and dependence parameters almost always yields better performance than plug-in procedures \citep[e.g.][]{Shi2026}, in this case we found MCMC mixing challenging for the full joint model.  Therefore, as a computational convenience, we use a two-step MCMC procedure to estimate the marginal and dependence model parameters separately. In the first step, the coefficients governing the spatially varying GPD scale and shape parameters are estimated using the No-U-Turn (NUTS), with 2{,}000 warmup iterations followed by 20{,}000 post-warmup iterations. The chain is thinned by retaining every fifth iteration, yielding 4{,}000 posterior samples. In the second step, we fix these parameters at their first-step posterior mean estimates, $\hat{\sigma}(\bm{s})$ and $\hat{\xi}(\bm{s})$, and estimate the dependence parameters, including those governing the anisotropic Wendland kernels, the spatial varying tail dependence structure, and the latent visibility Gaussian process in (\refeq{eq:nonconvex_GP}). The Wendland kernel uses 41 knots, while the spatially varying tail dependence parameter is represented using Gaussian kernel functions, with both the kernel coefficients and bandwidth estimated in the second step. For the full dataset, we run the MCMC algorithm for 100{,}000 iterations, discard the first 75{,}000 as burn-in, and thin the remaining iterations to obtain 2{,}500 posterior samples. The second step inference therefore focuses on the spatial extreme dependence structure conditional on the fitted marginal parameters.
	
\begin{figure}[ht]
		\centering
		\begin{tabular}{cc cc}
			\includegraphics[width=0.2\textwidth, clip=true, trim=0 0 30 40]
			{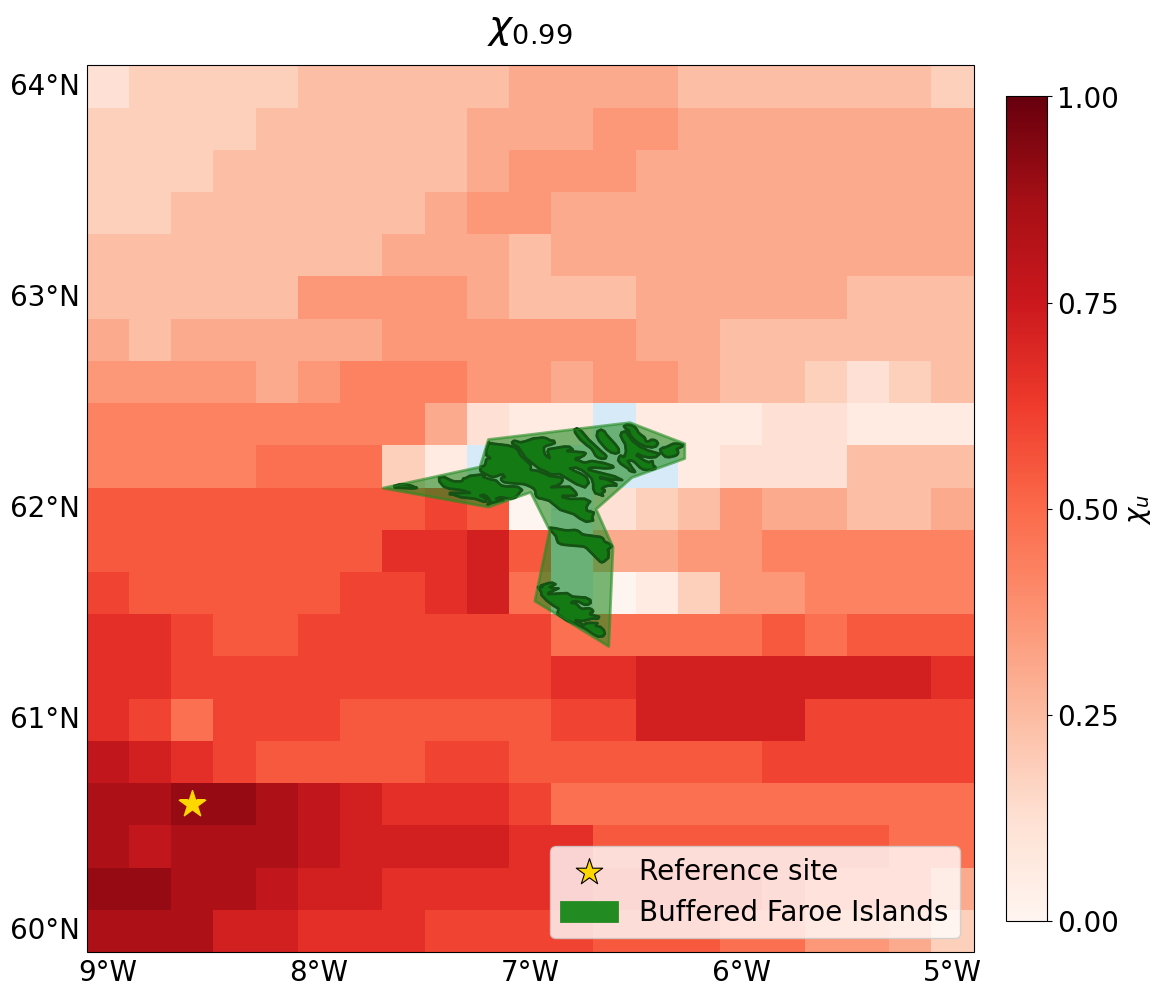} & 
			\includegraphics[width=0.2\textwidth, clip=true, trim=0 0 30 40]
			{wave_emp_chiu_center_211_99.png} &
			\includegraphics[width=0.2\textwidth, clip=true, trim=0 0 30 40]
			{wave_emp_chiu_center_222_99.png} & 
			\includegraphics[width=0.2\textwidth, clip=true, trim=0 0 30 40]
			{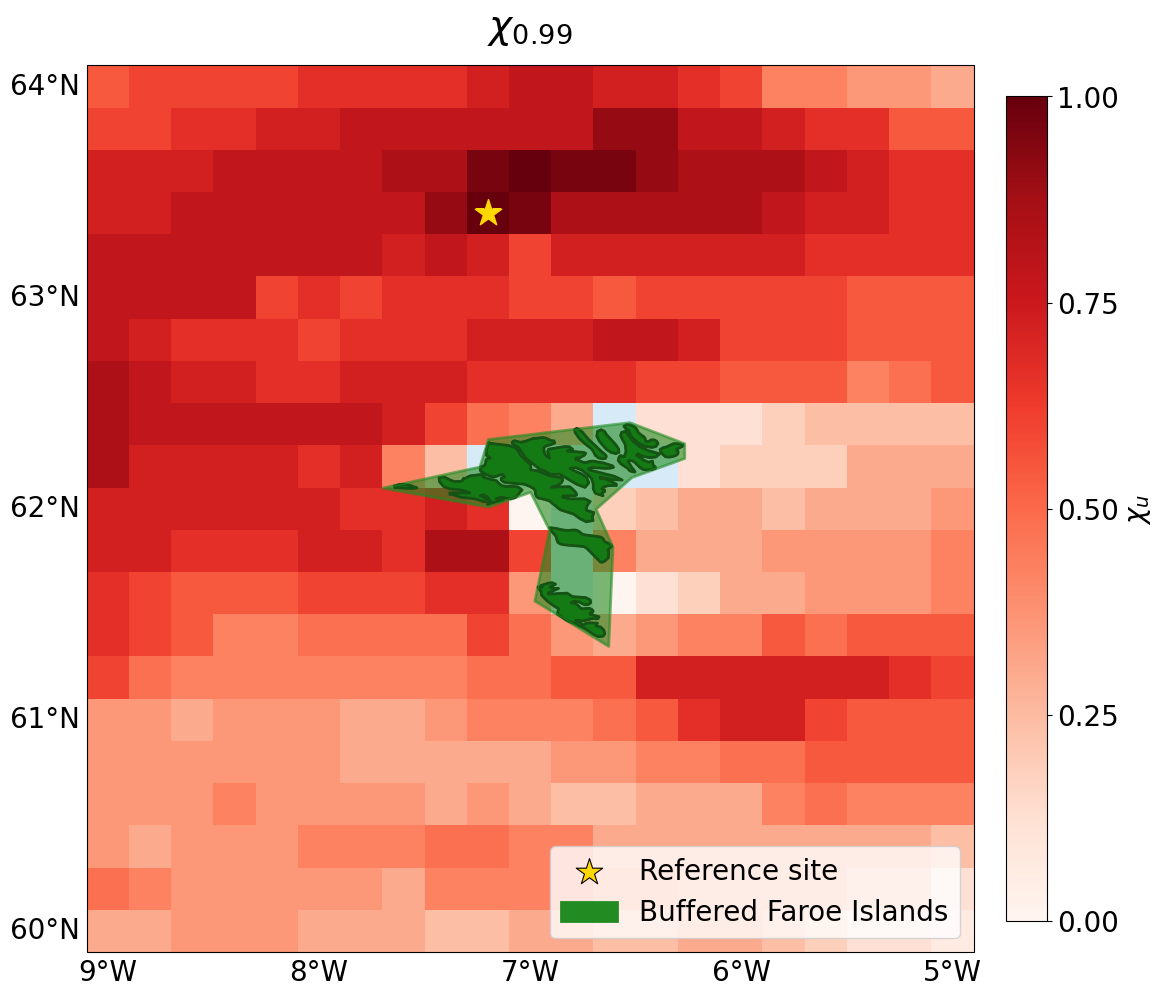} 
			\\
			\includegraphics[width=0.2\textwidth, clip=true, trim=0 0 30 40]
			{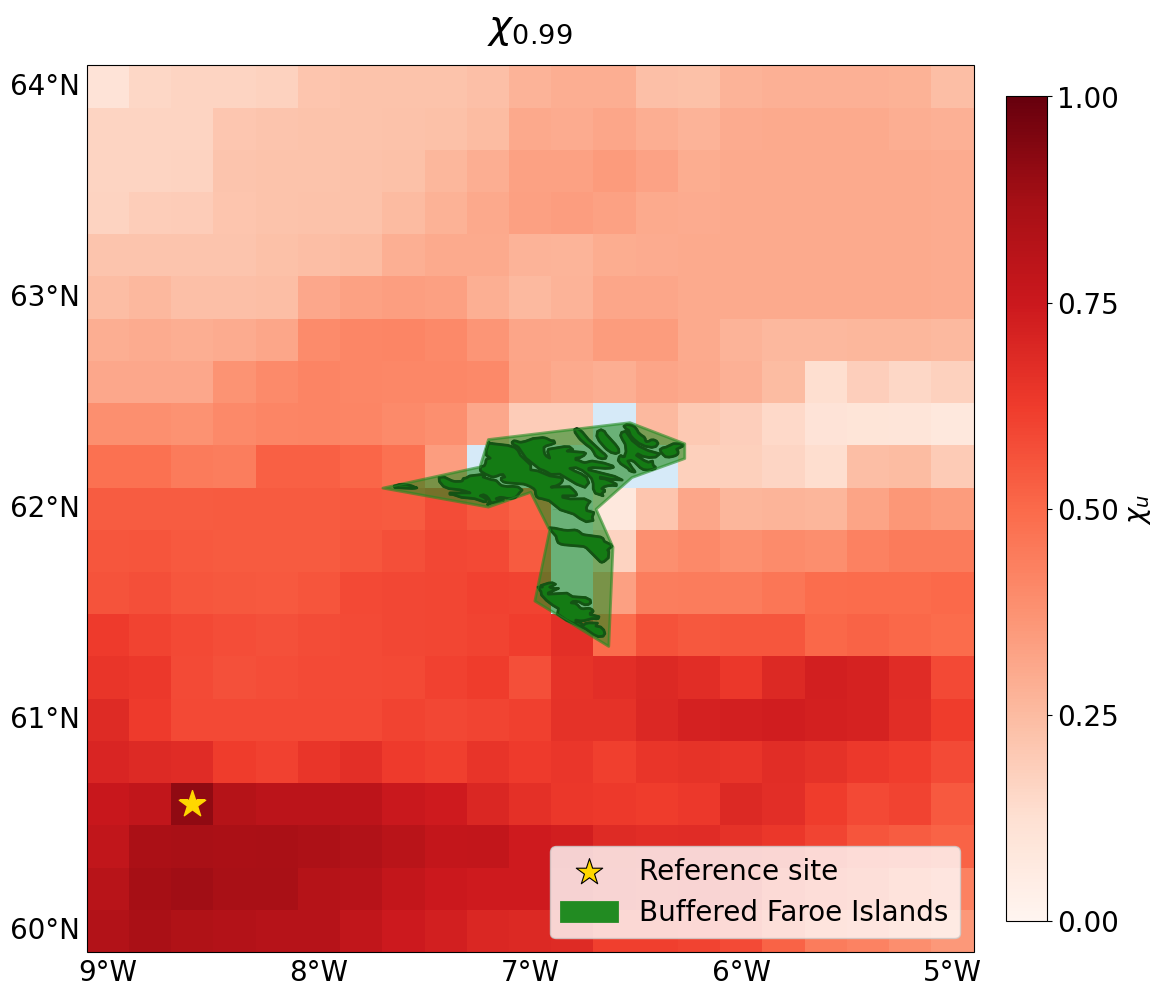} & 
			\includegraphics[width=0.2\textwidth, clip=true, trim=0 0 30 40]
			{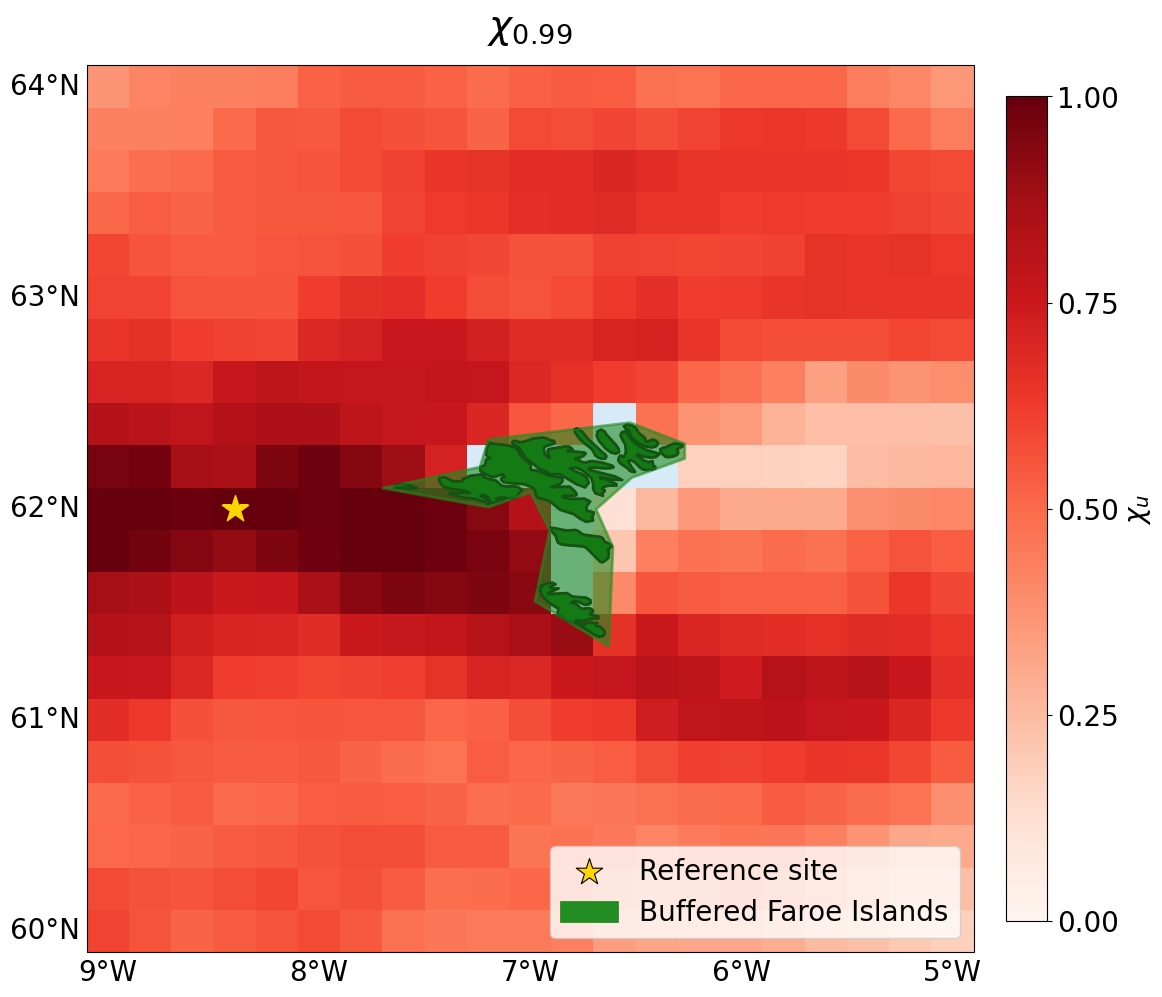} &
			\includegraphics[width=0.2\textwidth, clip=true, trim=0 0 30 40]
			{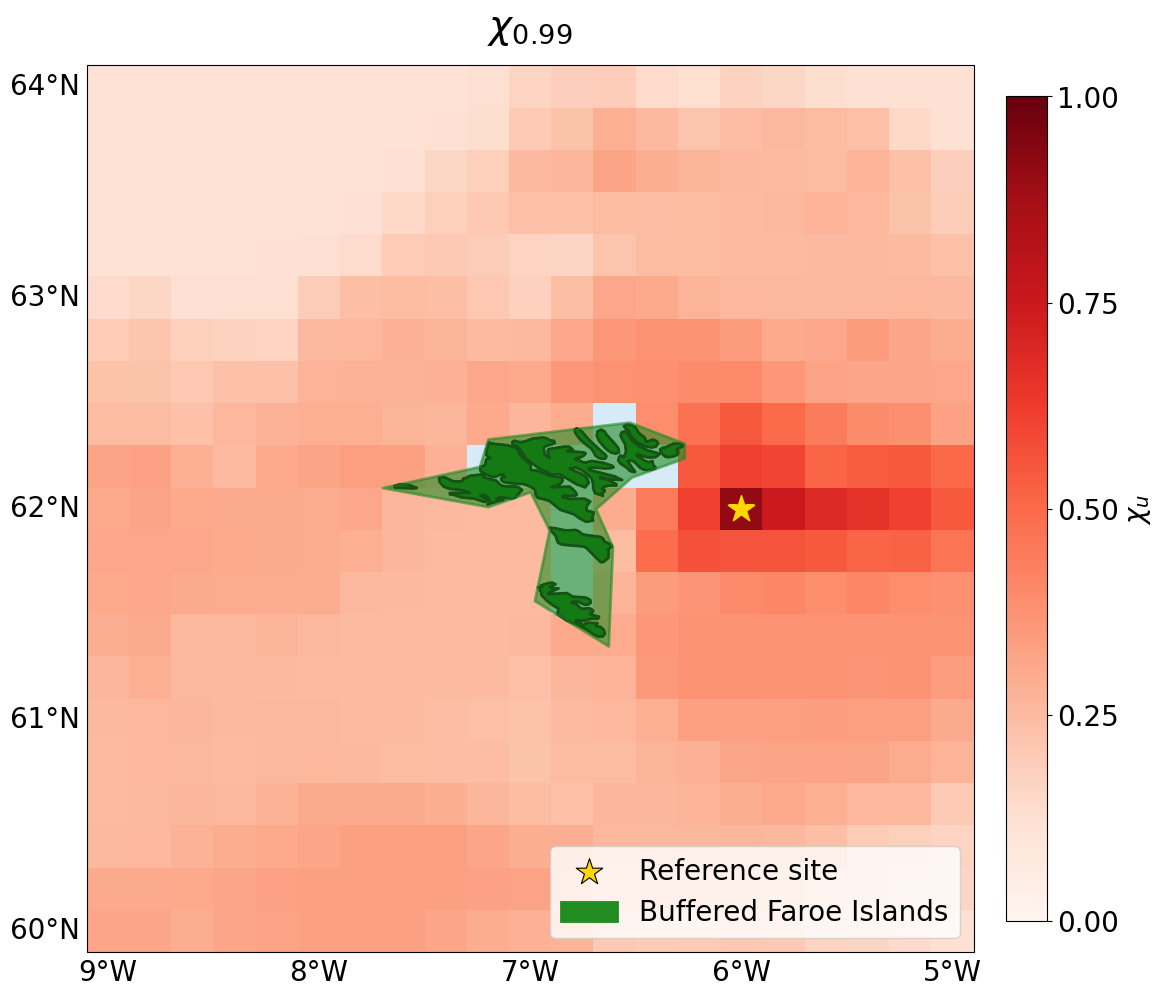} & 
			\includegraphics[width=0.2\textwidth, clip=true, trim=0 0 30 40]
			{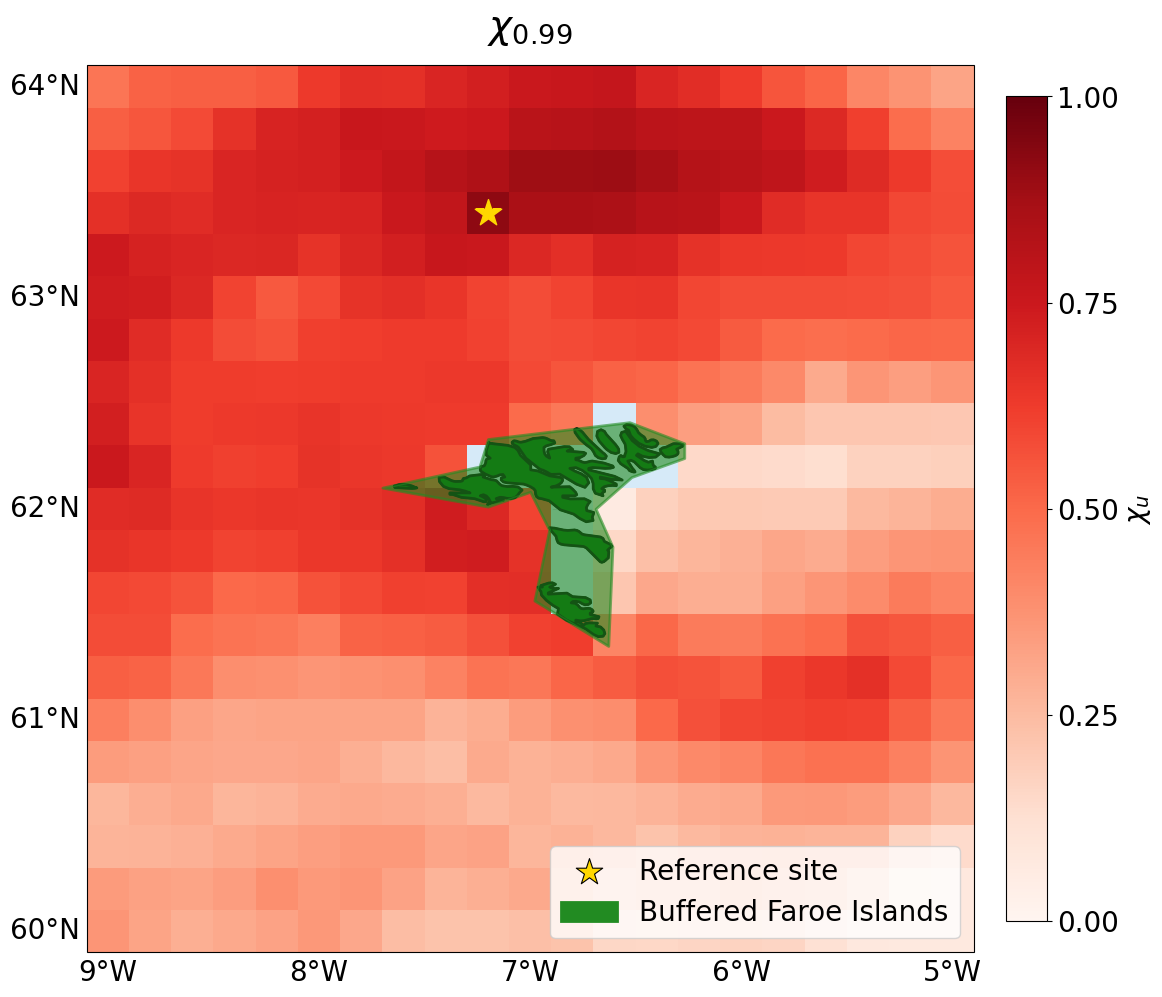} 
		\end{tabular}
		\caption{Pairwise empirical (top row) and visibility-model-based conditional (bottom row) $\chi_{0.99}$ for significant wave heights. Columns correspond to four reference sites located southwest, west, east, and north of the Faroe Islands, respectively; the reference sites are marked by gold stars.}
		\label{fig:wave_chiu_emp_est}
	\end{figure}

	We assess the ability of the visibility model to reproduce the empirical spatial structure of extremal dependence by comparing empirical and model-based estimates of the pairwise tail dependence $\chi_u$. We focus on threshold at $u = 0.99$ and four reference sites selected to represent distinct positions relative to the Faroe Islands: an open-ocean site southwest of the islands and sites located to the west, east, and north of the archipelago. The reference sites are marked by gold stars in Figure~\ref{fig:wave_chiu_emp_est}. Results for the lower thresholds at $u = 0.96$ and $u = 0.98$ are provided in Section~\ref{supp:wave_chiu} of the Supplementary Material. 
	
	Figure~\ref{fig:wave_chiu_emp_est} compares the empirical and visibility-model-based conditional $\chi_u$ surfaces. The empirical estimates exhibit substantial directional asymmetry, indicating that extremal dependence is not determined solely by Euclidean distance. For the southwest and western reference sites, dependence is generally stronger along the exposed western side of the archipelago and weaker across the islands. Conversely, for the eastern reference site, dependence is strongest toward the open ocean to the east and attenuated for sites separated from the reference location by land. The northern reference site exhibits a similar pattern, with stronger dependence along unobstructed ocean pathways than across land-obstructed directions. 
	
	These patterns indicate that the islands do more than increase the effective separation between locations. Sites at comparable Euclidean distances can exhibit different levels of extremal dependence depending on whether the direct path between them is obstructed by land. We refer to this phenomenon as barrier-induced extremal asymmetry. The contrast between unobstructed and obstructed directions is particularly informative at higher thresholds, although the empirical estimates also become more noisier as $u$ increases. 
	
	The model-based conditional $\chi_u$ surfaces reproduce the main directional features of the empirical estimates while providing smoother spatial patterns. In particular, the visibility model captures the attenuation of extremal dependence across the islands and the stronger dependence along unobstructed ocean pathways across the four reference sites. These results illustrate how the visibility graph allows the dependence structure to reflect the geometry of the domain, rather than being governed by Euclidean separation alone. 
	
	Overall, the results suggest that accounting for the geometry of the Faroe Islands is important for representing the extremal dependence of significant wave heights in the surrounding ocean. The visibility model captures the principal barrier-induced directional asymmetry evident in the empirical estimates. 
	
	Posterior means and 95\% equal-tail credible intervals for the marginal and dependence parameters are reported in the Supplementary Material (Table~\ref{tab:wave_full_posterior_mean}), together with the corresponding MCMC trace plots.
	
	\subsection{Model Comparison} \label{subsec:wave_comparison}
	
	To assess out-of-sample predictive performance, we divide the $d = 432$ spatial locations into a training set of 369 sites and a holdout set of 63 sites. Like in Section~\ref{subsec:precip_comparison}, we withhold three spatially contiguous blocks located in different parts of the study region (Figure~\ref{fig:train_test_split}(b)). 
	
	We again compare the visibility model with the spatial scale-aware model of \citet{Shi2026log} and the model of \citet{Huser2019}. All three models incorporate a log-Laplace nugget effect \citep{Shi2026log} for computational tractability and use marginal GPD parameters fixed at estimates obtained from the first-step MCMC. The visibility model follows the specification used in the full-data analysis, with an anisotropic Wendland kernel and a visibility-based Gaussian process. Its spatially varying tail dependence parameter is constructed using visibility-based Gaussian kernel functions with bandwidth fixed at $\sqrt{1/6}$. Both kernel constructions are 41 knots, including knots outside the study domain to account for edge effects.
	
	The Shi et al.\ model also uses an anisotropic Wendland kernel with 41 knots and a Gaussian kernel basis with 41 knots to model the spatially varying tail dependence parameter. To account for edge effects, knots are also placed outside the domain. For the latent Gaussian process, Shi et al.\ use a nonstationary, locally isotropic Mat{\'e}rn covariance function \citep{paciorek2006, risser2015}, constructed using 13 knots, rather than the visibility-based Gaussian process used in the visibility model. The Huser-Wadsworth model uses an isotropic stationary Mat{\'e}rn covariance structure and a spatially constant tail dependence parameter, together with the log-Laplace nugget effect. 
	
	For the visibility model and Shi et al.\ models, we run the MCMC chains for 100{,}000 iterations and discard the first 50{,}000 as burn-in. For the Huser-Wadsworth model, we run 50{,}000 iterations and discard the first 25{,}000. We thin the retained iterations by a factor of 10 for all three models, yielding 2{,}500 posterior samples per model. Additional posterior summaries and MCMC diagnostics are provided in Section~\ref{supp:wave_model_comparison} of the Supplementary Material. 
	We again evaluate predictive performance using the widely applicable information criteria (WAIC) on the training data, together with the predictive log-likelihood and the threshold-weighted continuous ranked probability score (twCRPS) on the holdout data. 
	
	Table~\ref{tab:faroe-model-comparison-diff} summarizes the resulting scores. The visibility model achieves the lowest WAIC among the three models. The difference is substantial relative to both Shi et al.\ and Huser-Wadsworth, indicating a better training-data fit after accounting for model complexity under this criterion. For the holdout predictive log-likelihood, the visibility and Shi et al.\ models achieves similar scores, whereas the Huser-Wadsworth model performs substantially worse. The Shi et al.\ model has the lowest twCRPS, although its difference from the visibility model is small; the Huser-Wadsworth model has a higher twCRPS-smooth, indicating poorer predictive performance under this score. Thus, the visibility model performs particularly well in terms of WAIC and is competitive with Shi et al.\ for holdout prediction, while the Huser-Wadsworth model performs less favorably on the two holdout scores. 
	
	\begin{table}[ht]
		\centering
		\caption{Held-out predictive performance at the 63 Faroe test sites (pooled over all three
			holdout blocks), expressed as differences from the visibility model (Diff = model $-$ Visibility). All models used marginal GPD parameters fixed at values obtained from a separate conditionally independent model fit. For WAIC and twCRPS-smooth
			(lower is better), a negative difference favors the row model over the visibility model;
			for log-score (higher is better), a positive difference favors the row model.
			\textbf{Bold} entries represent the best score and \textit{italic} entries are distinguishable from the visibility model at the $2\times$SE level.}
		\label{tab:faroe-model-comparison-diff}
			\begin{tabular}{lrrr}
				\toprule
				& WAIC &  Pred LL & twCRPS-smooth \\
				\midrule
				Visibility  &  \textbf{59{,}951} & \textbf{$-$1{,}059.2} & 0.1708 \\
				Shi et al. & 110{,}952 & $-$1{,}061.6 & \textbf{0.1704} \\
				Huser--Wadsworth & 84{,}932 & $-$3{,}286.8 & 0.1728 \\
				\midrule
				Shi $-$ Visibility  & \textit{+51,001 $\pm$ 10,627} & $-2.5 \pm 261.0$ & $-0.0004 \pm 0.0002$ \\
				HW $-$ Visibility & \textit{+24,981 $\pm$ 4,599} & \textit{-2,227.7 $\pm$ 269.8} & \textit{+0.0020 $\pm$ 0.0004} \\
				\bottomrule
			\end{tabular}%
	\end{table}


    We next examine the spatial structure of extremal dependence using the conditional $\chi_u$. The empirical estimates reveal pronounced directional asymmetry around the Faroe Islands. Extremal dependence is generally attenuated in directions obstructed by the islands, whereas stronger dependence is observed along unobstructed ocean pathways. The visibility model captures these barrier-related spatial patterns across different reference sites. In contrast, the Shi et al.\ and Huser--Wadsworth models do not reproduce the directional attenuation as clearly, reflecting their lack of an explicit representation of the visibility constraints imposed by the archipelago. These fundings illustrate the importance of accounting for physical barriers when modeling spatial extremal dependence. Comparisons of empirical and model-based conditional $\chi_u$ surfaces at $u \in \{0.96, 0.98, 0.99\}$ are provided in Section~\ref{supp:wave_comparison_chiu} of the Supplementary Material. The Supplementary Material also reports posterior means and 95\% equal-tailed credible intervals, spatial plots of the tail dependence parameter and Wendland-kernel boundary effects for the visibility and Shi et al.\ models, and QQ plots for the holdout sites.

\section{Discussion} \label{sec:discussion}
	
	We incorporate a visibility graph into a spatial random scale mixture model for environmental processes observed over domains with physical barriers, allowing extremal dependence to reflect domain geometry rather than relying solely on Euclidean separation. The simulation study and the applications to precipitation over the South Island of New Zealand and significant wave heights around the Faroe Islands illustrate the potential value of explicitly accounting for such barriers. 
	
	
	In the South Island application, the empirical extremal dependence surface exhibits differences between locations on the same side of the Southern Alps and those separated by the mountain range. The visibility model captures this directional attenuation more clearly than models primarily driven by Euclidean distance. However, improved representation of the spatial dependence structure does not necessarily yield uniformly better predictive scores, highlighting the importance of marginal model specification for overall predictive performance. In the Faroe Islands application, the visibility model reproduces the principal directional features of the empirical conditional tail dependence surfaces and achieves competitive holdout performance. Together, these applications demonstrate the value of incorporating domain geometry, while also showing that its benefits depend on the application and evaluation criterion.
	
	
	An important limitation is that the current visibility graph may attenuate dependence too strongly between sites that are not directly visible to one another. Visibility restricts contributions from knots in both the Wendland and Gaussian kernel constructions, while also influencing the latent Gaussian process. Consequently, nonvisible sites may share fewer sources of dependence, potentially amplifying attenuation across barriers. This is particularly apparent in the South Island application, where the model-based $\chi_u$ surfaces suggest weaker dependence across the Southern Alps than the empirical estimates. More generally, direct visibility does not necessarily imply conditional independence or the absence of extremal dependence, since environmental processes may remain connected through indirect pathways or atmospheric transport. 
	
	
	
	A promising direction for future work could allow the strength of the barrier effect to be estimated rather than determined entirely by the binary visibility graph. For example, visibility could be incorporated into the latent Gaussian process while the Wendland and Gaussian kernel contributions are allowed to extend across nonvisible site-knot pairs, possibly with a distance- or barrier-dependent attenuation. Alternatively, a weighted visibility graph could represent varying degrees of obstruction, allowing dependence across barriers to be reduced without being effectively excluded. Such extensions would provide greater flexibility in modeling residual dependence across barriers, although their implications for identifiability and covariance positive definiteness would require further investigation.
	
	Several additional extensions are also worth considering. First, the current applications use different marginal extreme value formulations, seasonal GEV maxima for precipitation and threshold-based GPD exceedances for wave heights. Future work could investigate alternative marginal specifications and their interaction with the dependence model, particularly because the precipitation comparison shows that differences in marginal fit can influence aggregate predictive scores. Second, more extensive predictive assessments across different barrier geometries and environmental processes would help clarify when visibility-based modeling offers the greatest benefit. Finally, the current wave height analysis uses a two-step inference strategy that fixes the marginal parameters at first-stage posterior means during dependence estimation. A Joint inference approach that propagates marginal parameter variability into the dependence component, as we took in our block-maxima model, would be preferable to the two-stage approach that took in our POT model, albeit at additional computational cost. 
	
	
	\bibliography{main}


	\clearpage
	\singlespacing
	\let\oldsection\section
	\renewcommand{\section}{\FloatBarrier\oldsection}
	\renewcommand{\topfraction}{0.9}
	\renewcommand{\bottomfraction}{0.9}
	\renewcommand{\textfraction}{0.1}
	\renewcommand{\floatpagefraction}{0.7}
	\setcounter{topnumber}{3}
	\setcounter{totalnumber}{5}
	\makeatletter
	\setlength{\@fptop}{0pt}
	\setlength{\@fpsep}{10pt plus 1fil}
	\setlength{\@fpbot}{0pt plus 1fil}
	\makeatother
	
	\begin{center}
		{\Large\bfseries Supplementary Material}
	\end{center}
	
	\setcounter{section}{0}
	\setcounter{figure}{0}\setcounter{table}{0}\setcounter{equation}{0}
	\renewcommand{\thesection}{S\arabic{section}}
	\renewcommand{\thefigure}{S\arabic{figure}}
	\renewcommand{\thetable}{S\arabic{table}}
	\renewcommand{\theequation}{S\arabic{equation}}
	\renewcommand{\theHsection}{S\arabic{section}}
	\renewcommand{\theHfigure}{S\arabic{figure}}
	\renewcommand{\theHtable}{S\arabic{table}}
	\renewcommand{\theHequation}{S\arabic{equation}}
	
	\section{Simulation for Linear Model in GEV Parameters and Spatially Varying Tail Dependence} \label{supp:sim_varying_tail_gev}
	
	We consider that spatially varying GEV parameters are modeled using thin-plate spline (TPS) basis functions to make smoothness across the domain. The location parameter is modeled on the original scale, $\mu(\bm{s}) = \bm{x}(\bm{s})^\top \bm{\beta}$, the scale parameter is modeled using on the log scale, and the shape parameter is modeled using a rescaled-logit transformation described in Section~\ref{subsec:gev_margin}. Specifically, we set the true regression coefficients to $\bm{\beta} = (\beta_0, \beta_1, \beta_2, \beta_3)^\top = (1, 1.5, 2, 2.5)^\top$ for each of the three GEV parameters. For the spatially varying tail dependence parameter $\delta(\bm{s})$, we generate the 22 latent coefficients $\{\phi_k\}$ at the same knots locations $(\bm{s}_k)$ used in the anisotropic Wendland kernel construction. These are then interpolated to all site locations using Gaussian kernel functions with bandwidth $h = 3$. We run the MCMC chain for 33,000 iterations, discard the first 3,000 iteration as burn-in, and thin the chain by retaining every third draw.
	
	Overall, the MCMC performs well on the spatial structure and parameters. While a few 95\% credible intervals fail to contain the true values, the true values lie close to the corresponding interval boundaries, suggesting only mild undercoverage. One notable exception is the Gaussian kernel bandwidth $h$, whose posterior distribution appears not to have fully stabilized, indicating that additional MCMC iterations may be required for this parameter. 
	
	\begin{figure}[htbp]
		\centering
		\begin{tabular}{ccc}
			\includegraphics[width=0.3\textwidth]{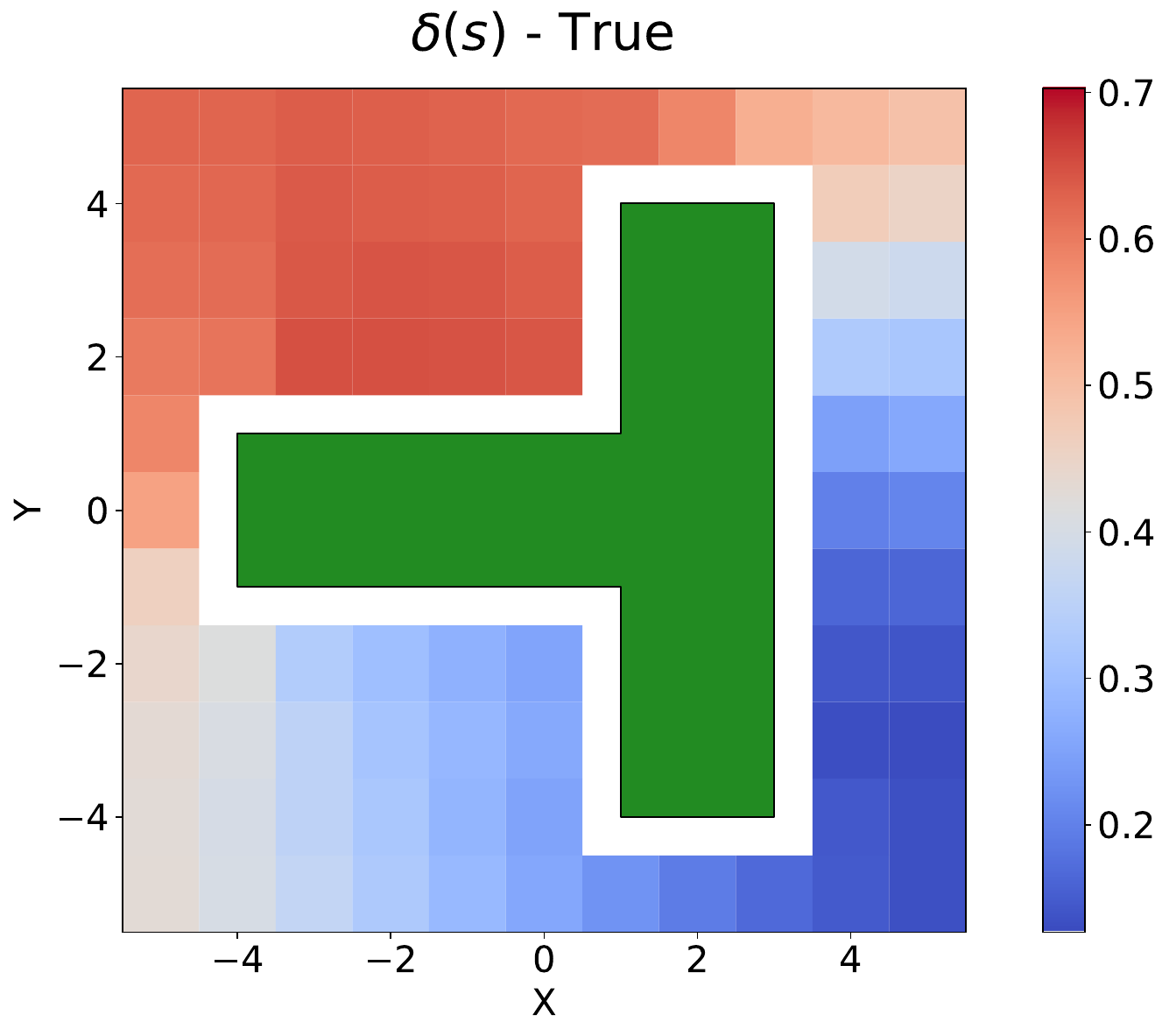} & 
			\includegraphics[width=0.3\textwidth]{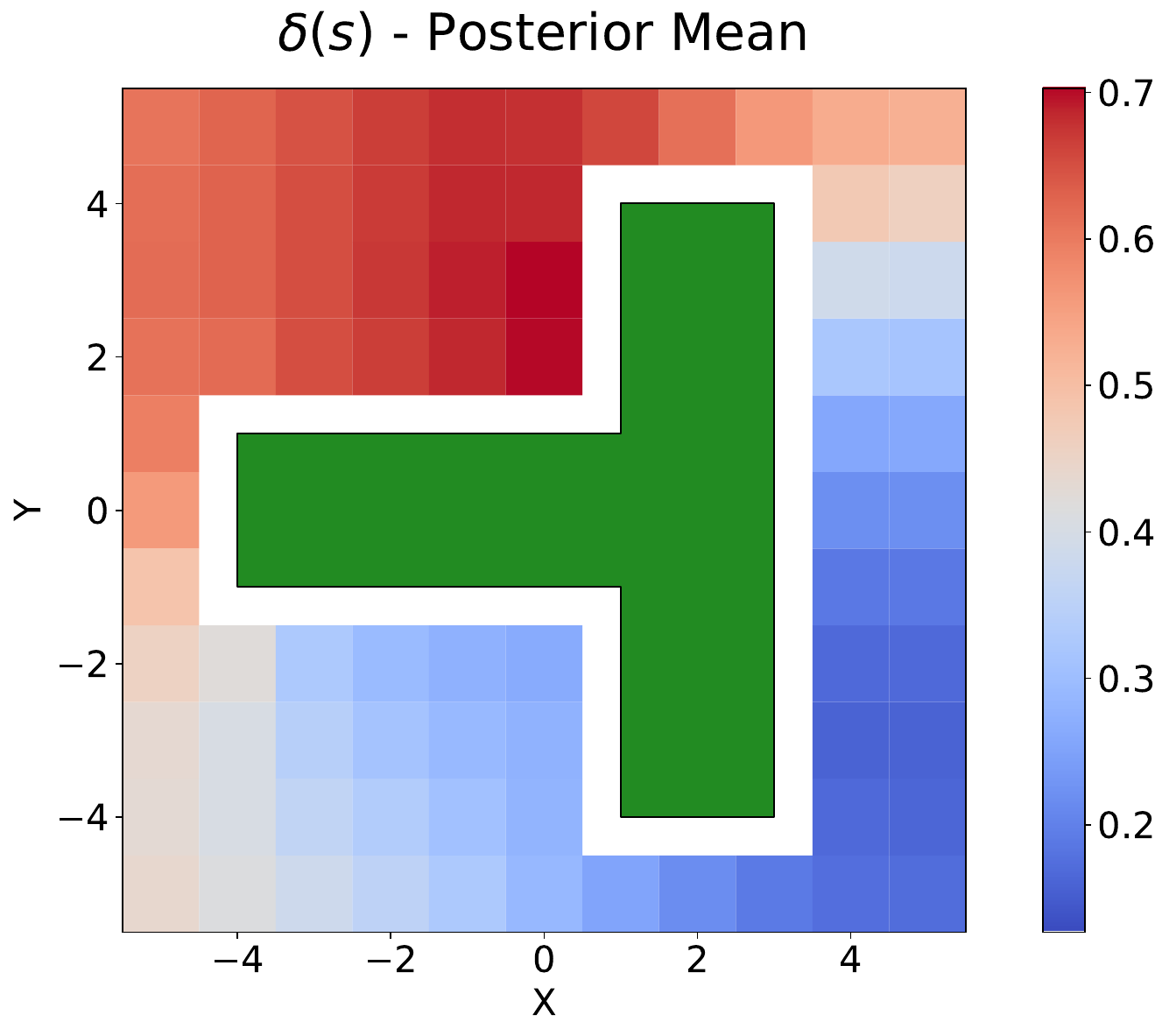}
		\end{tabular}
		\caption{True and posterior mean surfaces of the spatially varying tail dependence parameter $\delta(\bm{s})$.}
		\label{fig:sim_surface_tail}
	\end{figure}
	
	\section{Scoring Criteria for Model Comparison} \label{supp:scoring_criteria}
	
	This section describes the three criteria used to compare the visibility model with the models of \citet{Shi2026, Shi2026log} and \citet{Huser2019} in both real-data applications (Sections~\ref{sec:precip_application} and~\ref{sec:wave_application}): training the widely applicable information criterion (WAIC), the holdout predictive log-likelihood (PredLL), and the threshold-weighted CRPS (twCRPS). WAIC is evaluated using the training data, whereas PredLL and twCRPS assess predictive performance at the holdout sites. All three criteria are computed using the spatial train/test holdout design described in the main text for each application. 
	
	\paragraph{Holdout predictive log-likelihood.}
	Let $\bm{Z}$ denote the latent Gaussian field underlying the Gaussian copula representation, with $\bm{Z} \sim N(\bm{0}, \bm{\Sigma})$. At posterior draw $b = 1, \ldots, B$, the conditional distribution of the latent field at the holdout sites, given the latent field at the training sites is $\bm{Z}_{\text{test}} \mid \bm{Z}_{\text{train}} \sim N\big( \bm{\mu}_{\text{cond}}^{(b)}, \bm{\Sigma}_{\text{cond}}^{(b)} \big)$, where $\bm{\theta}^{(b)}$ denotes the corresponding model parameters. The holdout predictive density is obtained by integrating over the posterior uncertainty represented by the retained draws. Since the observations are linked to the latent Gaussian field through the model's marginal transformations, the conditional density can be evaluated using the Gaussian conditional distribution together with the relevant transformation and Jacobian terms. The pooled predictive log-likelihood is
	\begin{align*}
		\text{PredLL} = \sum_{t = 1}^{T} \log\left[ \frac{1}{B} \sum_{b=1}^{B} p\big(\bm{Y}_t^{\text{test}} \mid \bm{Z}_{\text{train}}^{(b)}, \bm{\theta}^{(b)} \big)\right], 
	\end{align*}
	where the posterior predictive density at each draw is evaluated conditional on the training data. The logarithm of the posterior-averaged predictive density is computed numerically using the log-sum-exp identity for stability. Larger PredLL values indicate better predictive performance. 
	
	\paragraph{Widely applicable information criterion.}
	WAIC \citep{watanabe2010, gelman2014} estimates predictive performance while accounting for model complexity. Let $\ell_{b, t}$ denote the log pointwise predictive density for data unit $t$ at posterior draw $b$, with the unit defined consistently with the likelihood factorization used in the analysis. The log pointwise predictive density and effective number of parameters are estimated as
	\begin{align*}
		\text{lppd} = \sum_{t=1}^{T} \log\frac{1}{B} \sum_{b=1}^{B} e^{\ell_{b, t}}, \quad 
		p_{\text{WAIC}} = \sum_{t=1}^{T} \text{Var}_b(\ell_{b, t}), \quad
		\text{WAIC} = -2 \left( \text{lppd} - p_{\text{WAIC}} \right).
	\end{align*}
	Smaller WAIC values indicate better estimated out-of-sample predictive performance. For the precipitation application, the data units correspond to years, and $\ell_{b, t}$ is the joint log-density of the spatial observations in year $t$ under posterior draw $b$. 
	
	\paragraph{Threshold-weighted continuous ranked probability score.}
	The threshold-weighted continuous ranked probability score (twCRPS) evaluates predictive distributions while emphasizing the upper tail, which is of particular interest for extreme events \citep{gneiting2011, lerch2017}. For a predictive random variable $X$, observation $y$, and nonnegative chaining function $v$ that emphasizes the upper tail, the score is defined as 
	\begin{align*}
		\text{twCRPS} = \mathbb{E}| v(X) - v(y) | - \tfrac{1}{2}\mathbb{E} | v(X) - v(X') |,
	\end{align*}
	where $X$ and $X'$ are independent draws from the predictive distribution. In practice, the score is estimated from posterior predictive samples using the sorted-order identity for the CRPS. Lower twCRPS values indicate better predictive performance, with greater emphasis placed on discrepancies in the upper tail according to the chosen function $v$. 
	
	\section{Precipitation Application} \label{supp:precip_application}
	
	\subsection{Modeling the GEV Parameters via Barrier-Informed Covariates} \label{supp:precip_gev_covariates}
	
	The GEV margins are modeled as linear (or link-transformed linear) functions of location, so that each parameter varies smoothly across the domain while remaining informed by the geometry of the barrier. Rather than using raw geographic coordinates, which have no direct physical interpretation with respect to the barrier, we constructed five covariates specifically designed to capture how the Southern Alps shape extreme precipitation:
	\begin{itemize}
		\item \textbf{Windward indicator}, $\mathds{1}_{\text{windward}}(\bm{s})$: a binary indicator of which side of the range a site falls on. We first fit a straight line running along the main axis of the Southern Alps and classify each site by the sign of the two-dimensional cross product between the line's direction vector and the vector from the line to the site; sites on the moisture-facing (northwest) side are windward, and sites on the far side are leeward. Coastal (non-inland) sites are additionally required to lie within $0.25^\circ$ of the coastline before being classified this way, so that open-ocean sites well away from the range are not spuriously assigned a side.
		
		\item \textbf{Standardized distance to the barrier centroid}, $\text{dist}(\bm{s})$: the Euclidean distance (in degree) from site $\bm{s}$ to the centroid of the mountain range polygon, standardized to zero mean and unit variance across sites. This captures the expected decay of orographic enhancement (or rain-shadow suppression) with distance from the range.
		
		\item \textbf{Bearing to the barrier centroid}, $\sin\theta(\bm{s})$ and $\cos\theta(\bm{s})$: writing $\theta(\bm{s}) = \operatorname{atan2}(\Delta\text{lon}, \Delta\text{lat})$ for the compass bearing from site $\bm{s}$ toward the range centroid, we include its sine and cosine as a smooth, periodic pair of covariates. Because the Southern Alps run diagonally from northwest to southeast rather than along a cardinal direction, a single linear covariate in latitude or longitude cannot represent the associated gradient in extreme precipitation; the $(\sin\theta, \cos\theta)$ pair lets the model place its steepest gradient along whatever compass direction the data actually support, without assuming it in advance.
		
		\item \textbf{Windward $\times$ distance interaction}, $\mathds{1}_{\text{windward}}(\bm{s}) \cdot \text{dist}(\bm{s})$: allows the rate of decay with distance from the range to differ between the wet, windward side and the dry, leeward side, which is physically expected since orographic enhancement and rain-shadow suppression need not decay at the same rate.
	\end{itemize}
	
	\begin{figure}[htbp]
		\centering
		\includegraphics[width=0.8\textwidth]{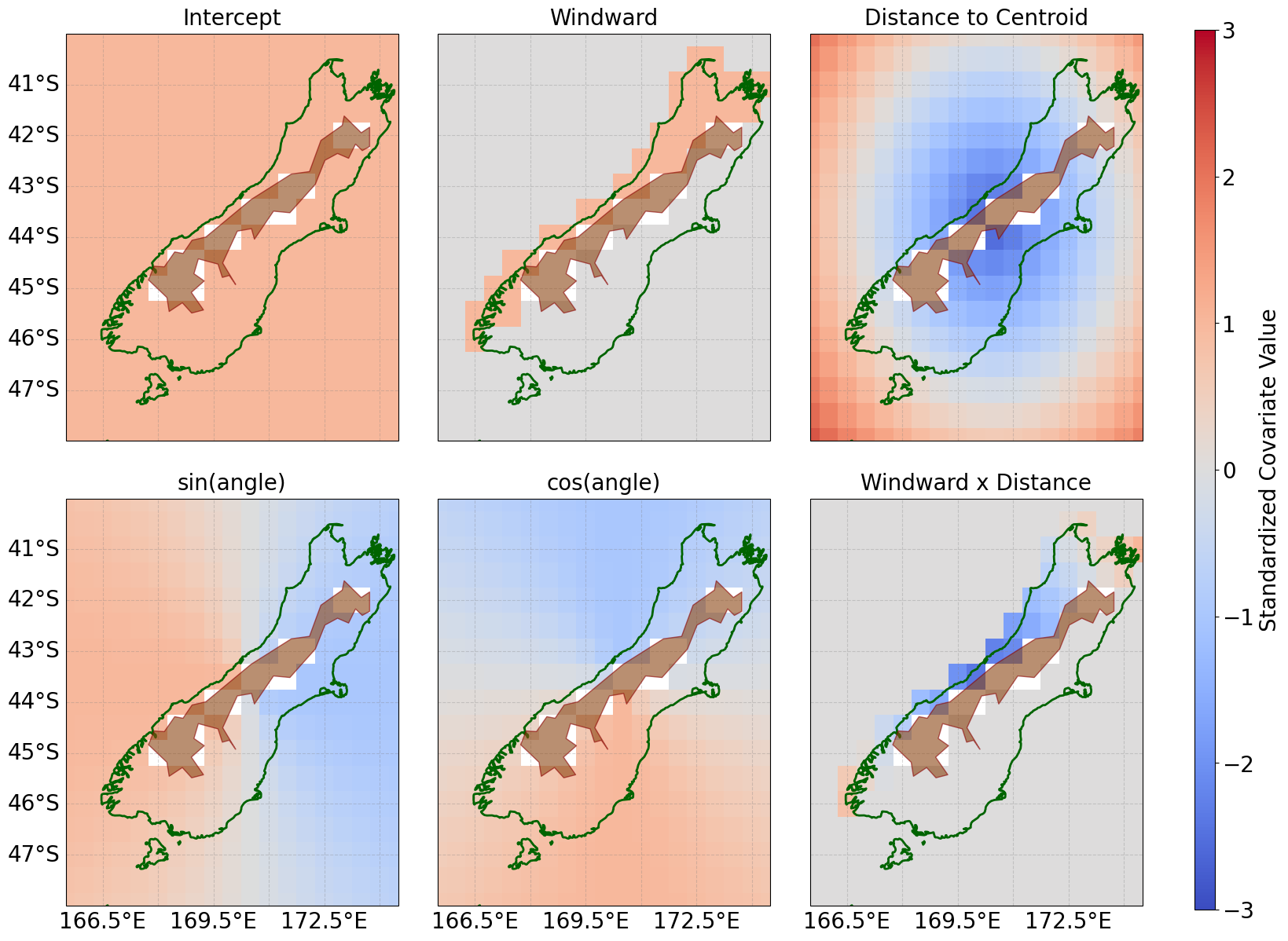}
		\caption{Spatial surfaces of the barrier-informed covariates used to model the GEV parameters for seasonal maximum precipitation over the South Island of New Zealand. The mountain barrier is defined by elevations exceeding 1500m.}
		\label{fig:precip_covariates_surface}
	\end{figure}
	
	These five covariates (plus an intercept) were not chosen a priori. We first assembled a larger candidate pool, the four covariates above, together with standardized elevation and all pairwise interactions among elevation, the windward indicator, distance, and bearing, and performed a best subset search, fitting every combination by ordinary least squares against the site-wise GEV maximum likelihood estimates (MLE; obtained independently at each of the $d = 309$ sites) and ranking candidate models of a given size by out-of-sample RMSE. This search was carried out separately for $\mu$, $\log\sigma$, and the rescaled-logit transform of $\xi$. Models that additionally included elevation or its interactions achieved only marginally lower RMSE for some parameters, while the five covariates above achieved competitive RMSE for all three GEV parameters simultaneously. See Figure~\ref{fig:precip_covariates_surface} for the resulting covariate surfaces. We therefore adopted this common five-covariate specification across $\mu(\bm{s})$, $\sigma(\bm{s})$, and $\xi(\bm{s})$, favoring parsimony and a single physically interpretable design matrix over marginal gains in in-sample fit for any one parameter. Writing $\bm{x}(\bm{s})^\top = (1, \mathds{1}_{\text{windward}}(\bm{s}), \text{dist}(\bm{s}), \sin\theta(\bm{s}), \cos\theta(\bm{s}), \mathds{1}_{\text{windward}}(\bm{s})\text{dist}(\bm{s}))$ for the common design vector, the three GEV parameters are modeled as 
	\begin{align*}
		\mu(\bm{s}) =&\; \bm{x}(\bm{s})^\top \bm{\beta}_\mu, 
		\\
		\log\sigma(\bm{s}) =&\; \bm{x}(\bm{s})^\top \bm{\beta}_\sigma, 
		\\
		\xi(\bm{s}) =&\; \tfrac{3}{2} \text{expit} \left( \bm{x}(\bm{s})^\top \bm{\beta}_\xi \right) - \tfrac{1}{2}, 
	\end{align*}
	where the log link on $\sigma$ enforces positively and the rescaled-logit link on $\xi$ constrains the shape parameter to the regularity range $\xi(\bm{s}) \in (-0.5, 1)$. As a preliminary check prior to full Bayesian estimation, each $\bm{\beta}$ was fit by the ordinary least squares against the corresponding (possibly transformed) site-wise MLEs. Top row of Figure~\ref{fig:precip_gev_posterior_mean} shows the resulting fitted surfaces alongside the site-wise MLEs (see also Section~\ref{supp:precip_gev_surfaces} below for the corresponding comparison using the full Bayesian posterior mean).
	
	\FloatBarrier
	\subsection{Model Fit Assessment} \label{supp:precip_fit}
	
	\subsubsection{MCMC Diagnostics} \label{supp:precip_mcmc}
	
	Table~\ref{tab:precip_full_posterior_mean} reports the posterior means and 95\% equal-tailed credible intervals for all margin and dependence parameters of the visibility model. The MCMC chain was run for 60,000 iterations, with the first 30,000 discarded as burn-in and the remainder thinned by a factor of 3, yielding $B = 10{,}000$ posterior samples.
	
	\begin{table}[htbp]
		\centering
		\caption{Posterior mean and 95\% equal-tail credible interval, visibility model.}
		\label{tab:precip_full_posterior_mean}
		\resizebox{\textwidth}{!}{
			\begin{tabular}{c ccc ccc c}
				\toprule
				\multicolumn{8}{c}{Margin Model} \\
				\hline
				Parameter & $\beta_0$ & $\beta_1$ & $\beta_2$ & $\beta_3$ & $\beta_4$ & $\beta_5$ & \\
				\hline
				$\mu$ & $\underset{(15.935,\, 19.337)}{17.646}$ & $\underset{(-0.934,\, 0.857)}{-0.021}$ & $\underset{(0.111,\, 1.082)}{0.615}$ & $\underset{(-1.924,\, -0.379)}{-1.144}$ & $\underset{(-1.920,\, 0.154)}{-0.886}$ & $\underset{(-0.929,\, 1.178)}{0.116}$ &  \\
				\hline
				$\sigma$ & $\underset{(2.756,\, 2.879)}{2.817}$ & $\underset{(0.109,\, 0.223)}{0.166}$ & $\underset{(-0.129,\, -0.066)}{-0.098}$ & $\underset{(-0.176,\, -0.090)}{-0.134}$ & $\underset{(-0.203,\, -0.087)}{-0.144}$ & $\underset{(-0.164,\, -0.070)}{-0.118}$ &  \\
				\hline
				$\xi$ & $\underset{(0.130,\, 0.377)}{0.246}$ & $\underset{(0.144,\, 0.352)}{0.247}$ & $\underset{(-0.063,\, 0.017)}{-0.023}$ & $\underset{(-0.354,\, -0.211)}{-0.283}$ & $\underset{(0.034,\, 0.226)}{0.130}$ & $\underset{(0.145,\, 0.328)}{0.239}$ &  \\
				\midrule
				\multicolumn{8}{c}{Dependence Model} \\
				\hline
				\multirow{8}{*}{Tail} & $\phi_1$ & $\phi_2$ & $\phi_3$ & $\phi_4$ & $\phi_5$ & $\phi_6$ & $\phi_7$ \\
				\cline{2-8}
				& $\underset{(0.006,\, 0.363)}{0.129}$ & $\underset{(0.756,\, 1.208)}{0.978}$ & $\underset{(1.865, \, 2.938)}{2.452}$ & $\underset{(0.220,\, 0.641)}{0.429}$ & $\underset{(1.132,\, 1.573)}{1.345}$ & $\underset{(0.496,\, 1.203)}{0.857}$ & $\underset{(1.249,\, 2.016)}{1.623}$ \\
				\cline{2-8}
				& $\phi_8$ & $\phi_9$ & $\phi_{10}$ & $\phi_{11}$ & $\phi_{12}$ & $\phi_{13}$ & $\phi_{14}$ \\
				\cline{2-8}
				& $\underset{(0.573,\, 0.935)}{0.751}$ & $\underset{(0.398,\, 0.597)}{0.496}$ & $\underset{(0.565,\, 0.916)}{0.741}$ & $\underset{(0.002,\, 0.238)}{0.078}$ & $\underset{(0.191,\, 0.432)}{0.312}$ & $\underset{(0.042,\, 0.677)}{0.342}$ & $\underset{(0.613,\, 1.334)}{0.965}$ \\
				\cline{2-8}
				& $\phi_{15}$ & $\phi_{16}$ & $\phi_{17}$ & $\phi_{18}$ & $\phi_{19}$ & $\phi_{20}$ & $\phi_{21}$ \\
				\cline{2-8}
				& $\underset{(0.741,\, 1.695)}{1.199}$ & $\underset{(0.008,\, 0.642)}{0.226}$ & $\underset{(0.001,\, 0.106)}{0.029}$ & $\underset{(0.543,\, 1.555)}{1.032}$ & $\underset{(0.223,\, 0.992)}{0.587}$ & $\underset{(0.000,\, 0.053)}{0.015}$ & $\underset{(0.001,\, 0.147)}{0.041}$ \\
				\cline{2-5}
				& $\phi_{22}$ & $\phi_{23}$ & $\phi_{24}$ & $h$ & & & \\
				\cline{2-5}
				& $\underset{(0.058,\, 0.789)}{0.388}$ & $\underset{(0.002,\, 0.224)}{0.068}$ & $\underset{(0.003,\, 0.248)}{0.081}$ & $\underset{(1.555,\, 1.750)}{1.648}$ & & & \\
				\hline
				\multirow{2}{*}{Wendland} & $r$ & $\lambda$ & $\alpha$ & & & & \\
				\cline{2-4}
				& $\underset{(5.403,\, 6.355)}{5.867}$ & $\underset{(0.551,\, 0.703)}{0.625}$ & $\underset{(2.826,\, 3.052)}{2.920}$ & & & & \\
				\hline
				\multirow{2}{*}{Mat{\'e}rn} & $\rho$ & $\nu$ & & & & & \\
				\cline{2-3}
				& $\underset{(8.348,\, 10.454)}{9.354}$ & $\underset{(0.720,\, 0.759)}{0.739}$ & & & & & \\
				\bottomrule
			\end{tabular}
		}
	\end{table}
	
	\begin{figure}[htbp]
		\centering
		\includegraphics[width=0.8\textwidth]{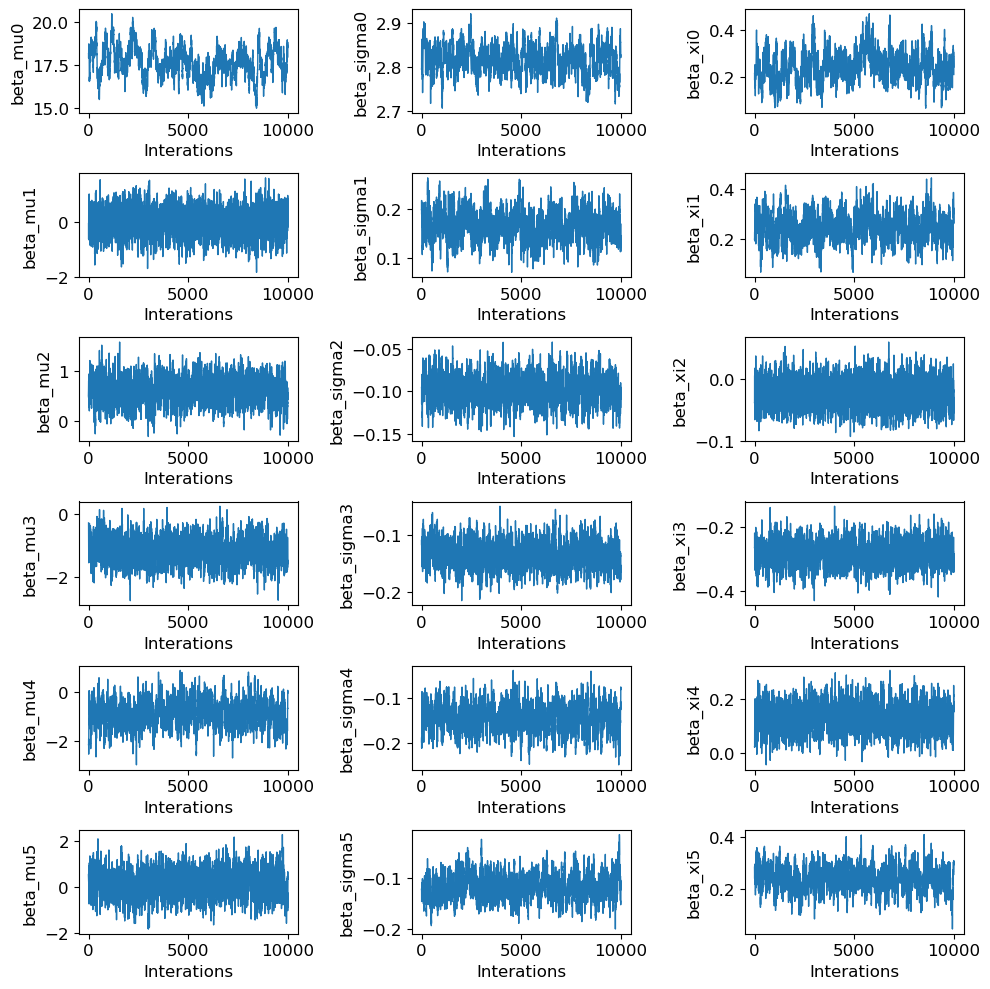}
		\caption{Trace plots for the margin regression coefficients $\bm{\beta}_\mu$ (left), $\bm{\beta}_\sigma$ (middle), and $\bm{\beta}_\xi$ (right), $B = 10{,}000$ retained posterior draws.}
		\label{fig:precip_full_trace_beta}
	\end{figure}
	
	\begin{figure}[htbp]
		\centering
		\includegraphics[width=0.8\textwidth]{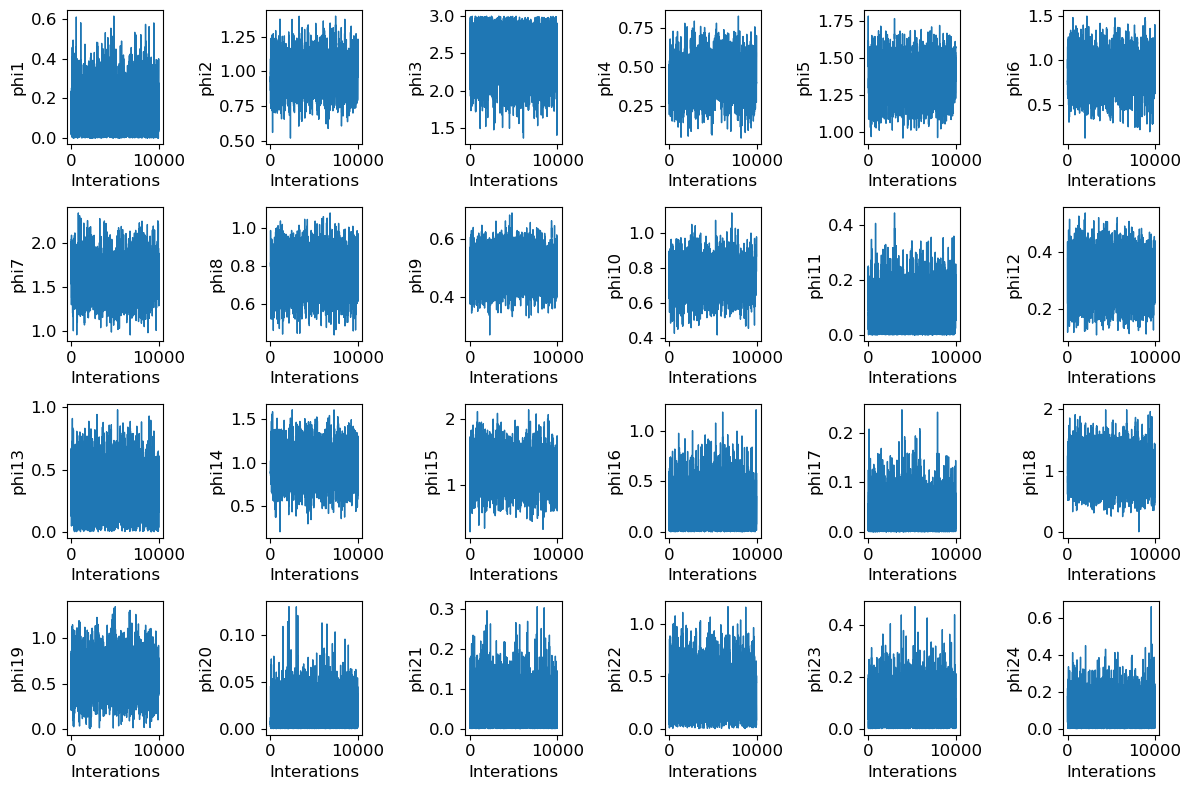}\\
		(a) \\[2mm]
		\includegraphics[width=0.8\textwidth]{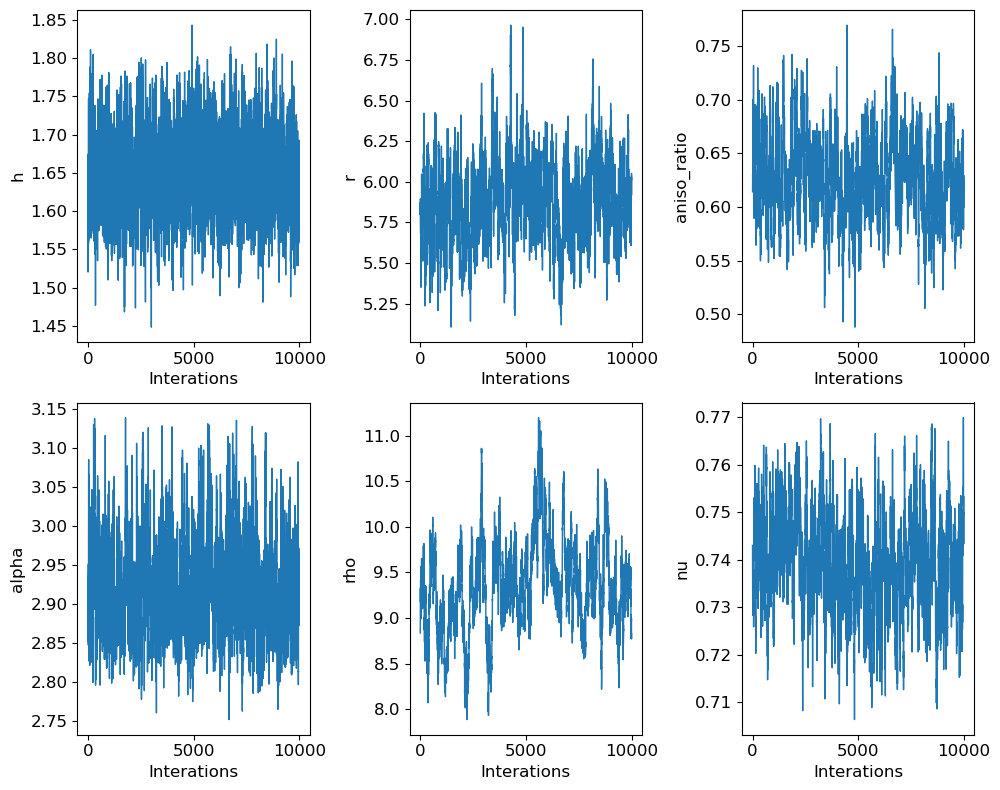}\\
		(b)
		\caption{Trace plots for (a) the tail dependence kernel coefficients $\phi_1, \ldots, \phi_{24}$ and (b) bandwidth $h$, the remaining dependence parameters $r$, $\lambda$, $\alpha$ (Wendland) and $\rho$, $\nu$ (Mat\'ern).}
		\label{fig:precip_full_trace_other}
	\end{figure}
	
	Figure~\ref{fig:precip_full_trace_beta} shows trace plots for the margin regression coefficients $\bm{\beta}_\mu$, $\bm{\beta}_\sigma$, and $\bm{\beta}_\xi$; Figure~\ref{fig:precip_full_trace_other} shows the tail-dependence Gaussian kernel coefficients $\phi_1, \ldots, \phi_{24}$ and the remaining dependence parameters ($h$, $r$, $\lambda'$, $\alpha$, $\rho$, $\nu$). All chains shown were run for $60{,}000$ iterations. After discarding the first $30{,}000$ as burn-in and thinning by a factor of $3$, the remaining $10{,}000$ draws are summarized in Table~\ref{tab:precip_full_posterior_mean}; the margin coefficients in Figure~\ref{fig:precip_full_trace_beta} oscillate stably around a constant level with no visible trend across the run, consistent with a well-mixed chain.
	
	\subsubsection{GEV Parameter Surfaces} \label{supp:precip_gev_surfaces}
	
\begin{figure}[htbp]
		\centering
		\includegraphics[width=0.7\textwidth]{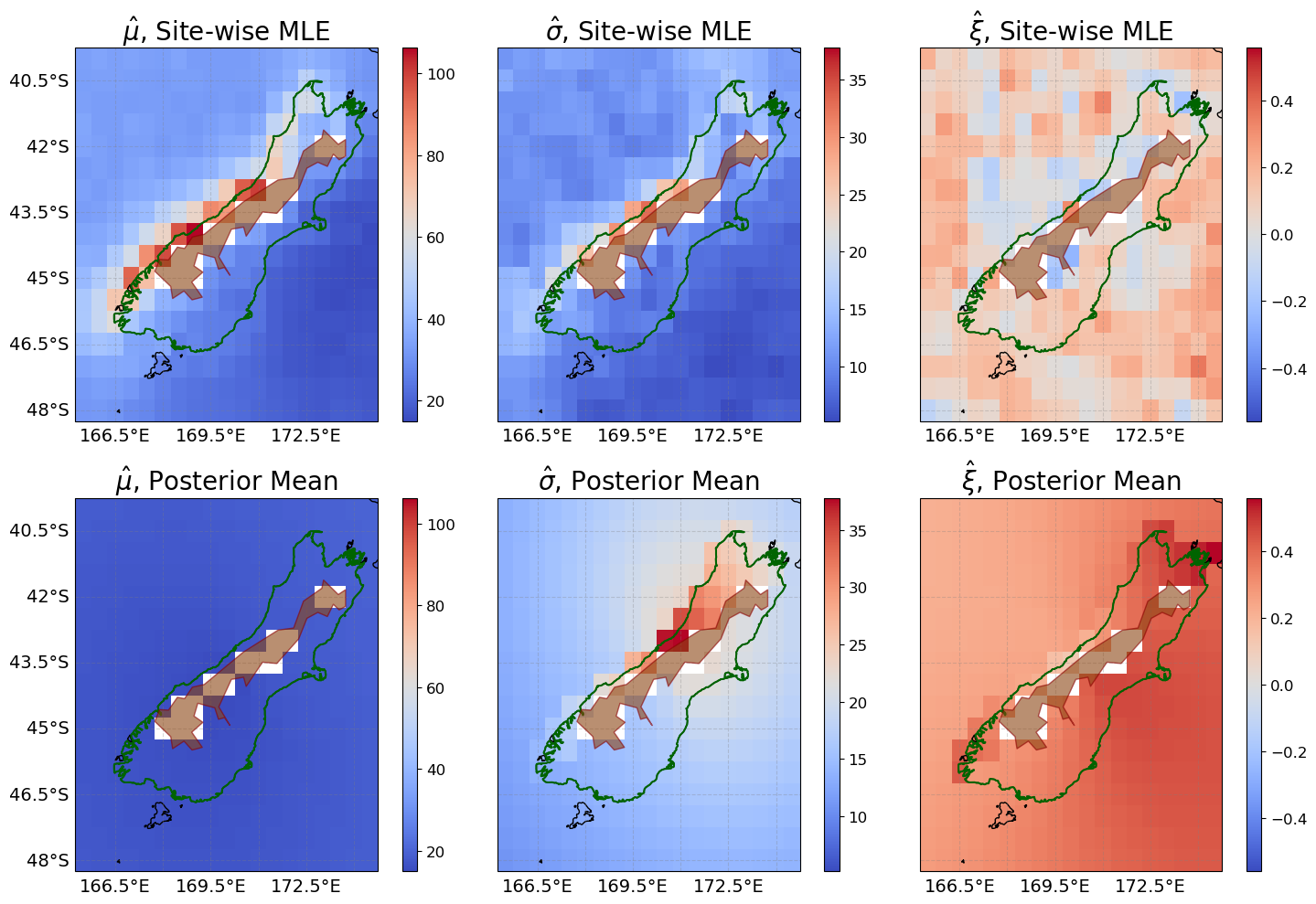}
		\caption{Site-wise GEV maximum likelihood estimates (top row) and posterior mean surfaces under the fitted visibility model (bottom row), for the location $\mu(\bm{s})$ (left), scale $\sigma(\bm{s})$ (middle), and shape $\xi(\bm{s})$ (right) parameters.}
		\label{fig:precip_gev_posterior_mean}
	\end{figure}

	Figure~\ref{fig:precip_gev_posterior_mean} shows the posterior mean surfaces of $\mu(\bm{s})$, $\sigma(\bm{s})$, and $\xi(\bm{s})$ under the fitted hierarchical model, to be compared against the site-wise MLE surfaces shown in the top row. 
	
	The scale parameter $\sigma(\bm{s})$ shows reasonable qualitative agreement between the MLE and the posterior mean surface show elevated values concentrated in a band running along the barrier, consistent with orographic enhancement of precipitation variability near the range. The location parameter $\mu(\bm{s})$ and the shape parameter $\xi(\bm{s})$ diverge much more sharply from their MLE counterparts, for two potential reasons.
	
	For $\mu(\bm{s})$, the site-wise MLE shows a narrow ridge of high values hugging the barrier itself, decaying quickly away from it on both sides. Our covariate specification (Section~\ref{supp:precip_gev_covariates}) can represent broad windward/leeward and distance-to-centroid gradients, but it is a smooth function of position relative to the barrier's centroid, not of position relative to the barrier's actual, irregular outline, it therefore cannot reproduce a narrow feature that tightly follows the range's specific shape the way the raw MLE does. The posterior mean instead shows a single smooth gradient across the whole domain, which is the best this covariate set can do; it captures the broad windward-to-leeward contrast but not the localized ridge.
	
	For $\xi(\bm{s})$, the divergence is more fundamental and, we think, expected rather than a shortcoming. GEV shape parameter estimates are notoriously imprecise at a single site: with only $T = 77$ annual maxima, the site-wise MLE for $\xi$ has substantial sampling variance, and the MLE panel in Figure~\ref{fig:precip_gev_posterior_mean} shows essentially no coherent spatial structure -- it looks close to spatial noise. The hierarchical model, by construction, pools information across all $d = 309$ sites through the shared covariate structure, trading this noisy, unusable site-by-site signal for a stable, low-variance estimate of how $\xi$ varies smoothly with position. That the posterior mean surface looks nothing like the MLE surface here is precisely the intended effect of this regularization, not evidence that the margin model has failed. An individually noisy, spatially incoherent MLE surface for $\xi$ is not a meaningful target to try to reproduce in the first place.
	
	\subsubsection{Wendland Kernel} \label{supp:precip_wendland}
	
	\begin{figure}[htbp]
		\centering
		\includegraphics[width=0.7\textwidth]{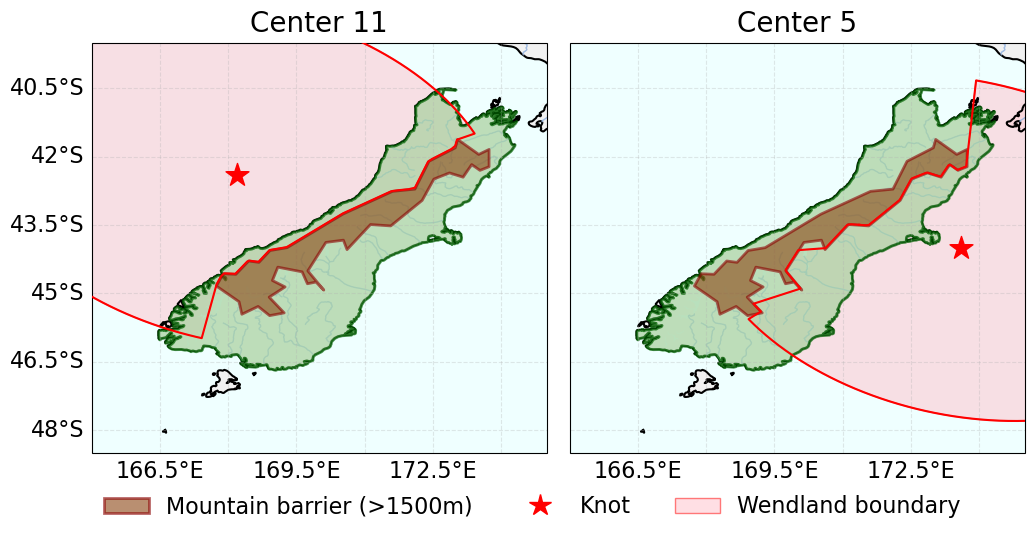}
		\caption{Visibility-restricted Wendland kernel support for two representative knots: Center 11 (left), an inland knot near the range whose support reaches northwest along the storm corridor; Center 5 (right), a leeward knot whose support reaches southeast, away from the range, truncated sharply at the barrier in both cases.}
		\label{fig:precip_wendland_examples}
	\end{figure}
	
	Figure~\ref{fig:precip_wendland_examples} shows the visibility-restricted Wendland kernel support for two representative knots. Center 11 sits inland, close to the range; its support extends broadly to the northwest, the direction from which Alps-blocking atmospheric rivers deliver moisture, while being sharply truncated along the barrier to its southeast. Center 5 sits on the leeward (eastern) side of the island; its support extends to the southeast, away from the range, and is cut off entirely along the mountain barrier rather than smoothly decaying across it. In both cases, the kernel's reach is not an isotropic disk but is elongated along, and truncated by, the same northwest-southeast corridor that the AR season's storm systems travel. The visibility restriction, not an assumed distance decay, is what shapes the support.
	
	\subsubsection{Tail Dependence Surface} \label{supp:precip_delta}
	
\begin{figure}[htbp]
		\centering
		\includegraphics[width=0.5\textwidth]{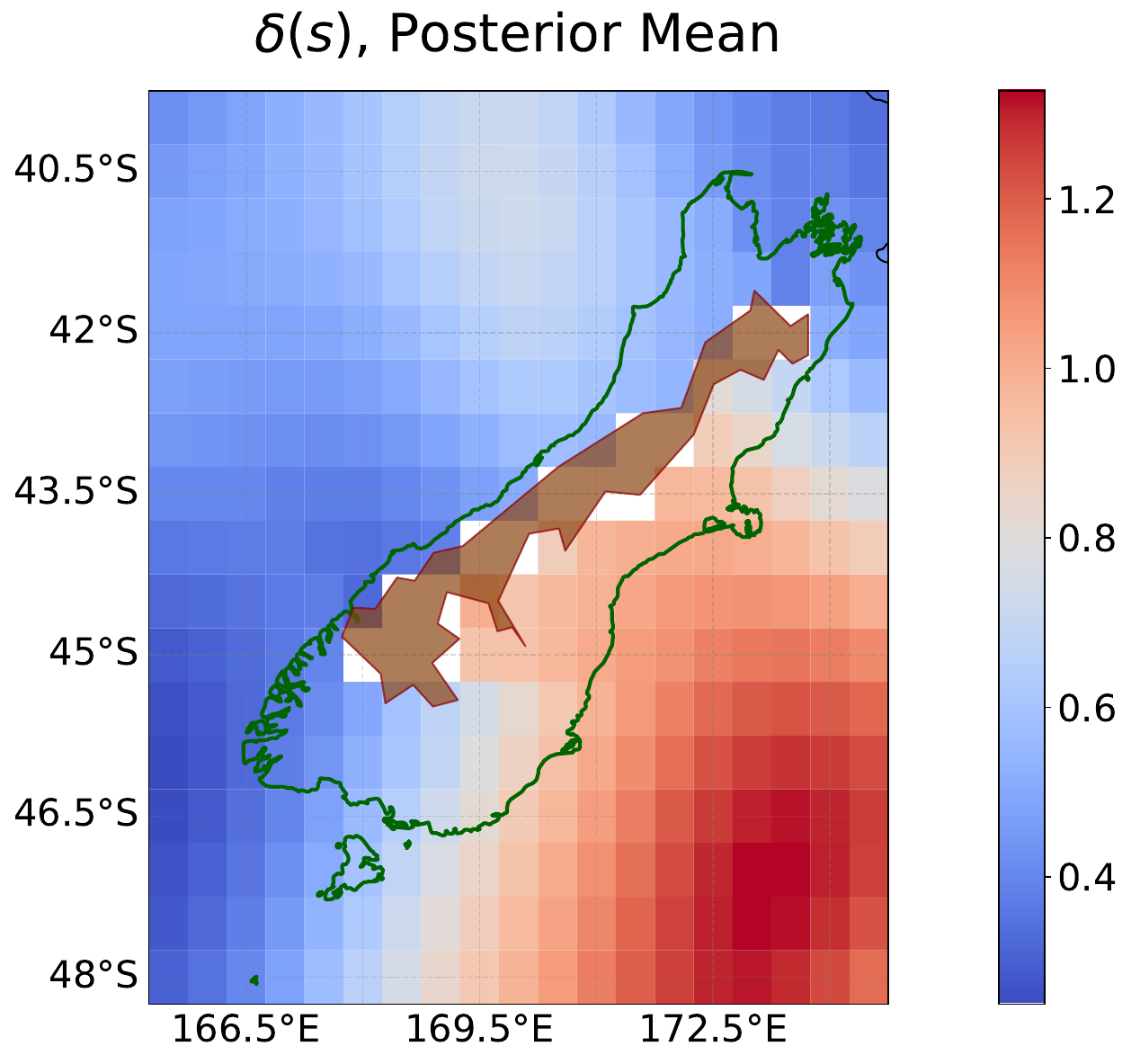}
		\caption{Posterior mean of the site-level tail-dependence parameter $\delta(\bm{s})$ under the fitted visibility model.}
		\label{fig:precip_full_delta_posterior_mean}
	\end{figure}

	Figure~\ref{fig:precip_full_delta_posterior_mean} shows the posterior mean of the site-level tail-dependence parameter $\delta(\bm{s})$. The surface separates cleanly along the barrier: values are low ($\approx 0.3$--$0.5$) on the windward (northwest) side and substantially higher ($\approx 1.0$--$1.3$) on the leeward (southeast) side, with the transition tracking the range itself rather than running along some other axis of the domain. This is notable because, unlike the GEV margin parameters (Section~\ref{supp:precip_gev_covariates}), $\delta(\bm{s})$ has no explicit windward/leeward covariate in its construction -- it is a purely spatial Gaussian-kernel smooth, $\delta(\bm{s}) = \bm{G}(\bm{s})^\top\bm{\phi}$, restricted by the visibility graph in the same way as the Wendland kernel and copula precision matrix (Section~\ref{supp:precip_wendland}). The windward--leeward contrast is therefore inferred from the data through the likelihood, not assumed by the model's structure.
	
	A higher $\delta(\bm{s})$ corresponds to stronger asymptotic dependence -- extremes at nearby sites being more likely to co-occur -- while a lower $\delta(\bm{s})$ moves the process toward asymptotic independence. That the leeward side shows the higher value is not obviously the naive guess (the windward side sees the more intense precipitation on average), but it is physically plausible: on the windward slopes, the very largest precipitation totals are shaped substantially by fine-scale, local orographic forcing -- exact terrain, elevation, and exposure -- so which specific sites experience the most extreme totals during a given atmospheric river event can vary considerably even over short distances, pushing the windward side toward more site-specific, less mutually dependent extremes. On the leeward side, by contrast, extreme totals are comparatively rare and are more likely to occur only during unusually strong or moist events capable of producing widespread rainfall well beyond the range -- a broader, synoptic-scale mechanism that would tend to produce more spatially coherent, simultaneous extremes across leeward sites. We offer this as a plausible physical interpretation of the pattern rather than a mechanism we have separately verified.
	
	\subsubsection{Extremal Dependence $\chi_u$ at Lower Thresholds} \label{supp:precip_chiu}
	
\begin{figure}[htbp]
		\centering
		\begin{tabular}{cc cc}
			\multicolumn{4}{c}{\large$\chi_{0.90}$} \\
			\includegraphics[width=0.2\textwidth, clip=true, trim=0 0 30 40]{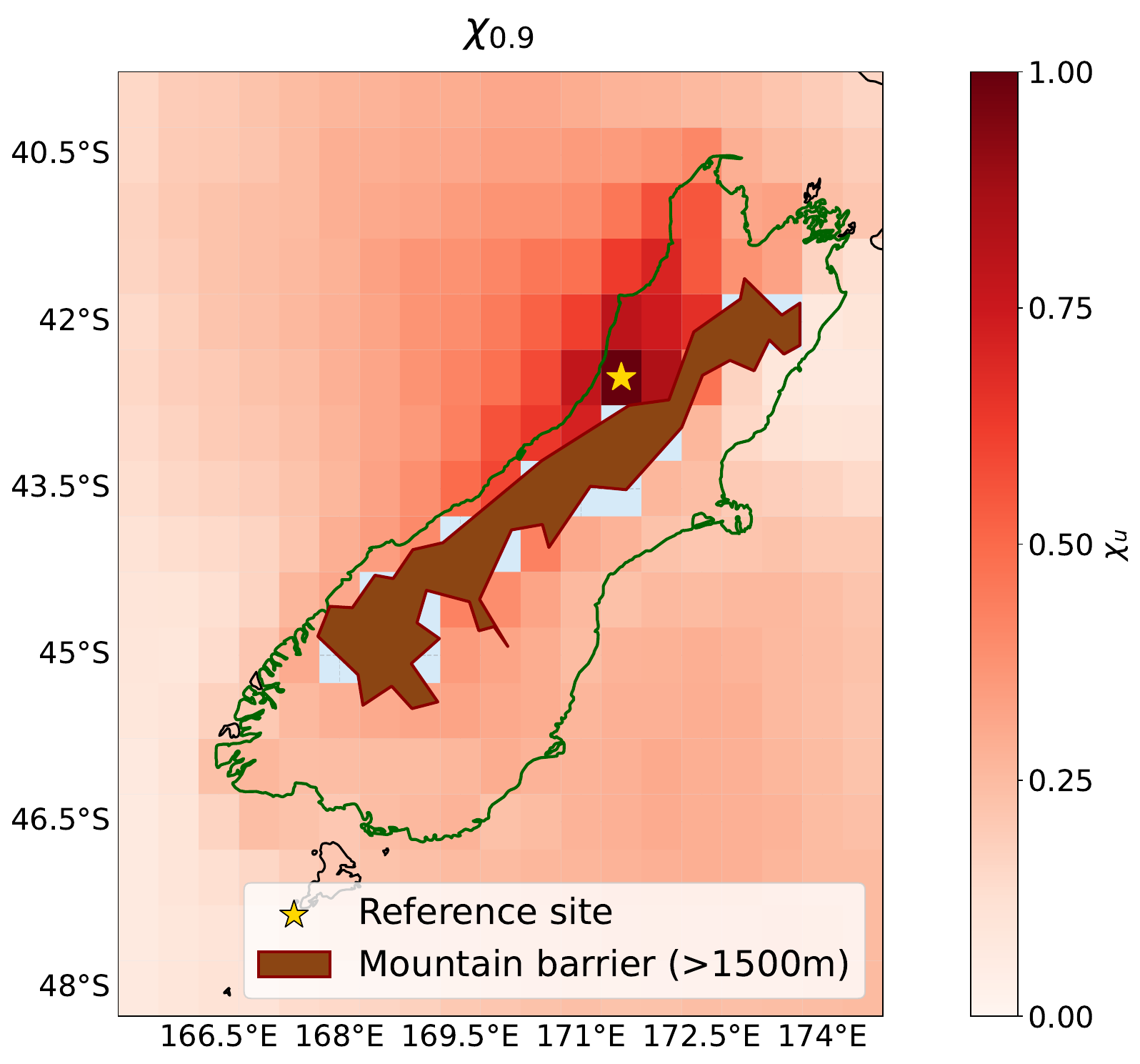} & 
			\includegraphics[width=0.2\textwidth, clip=true, trim=0 0 30 40]{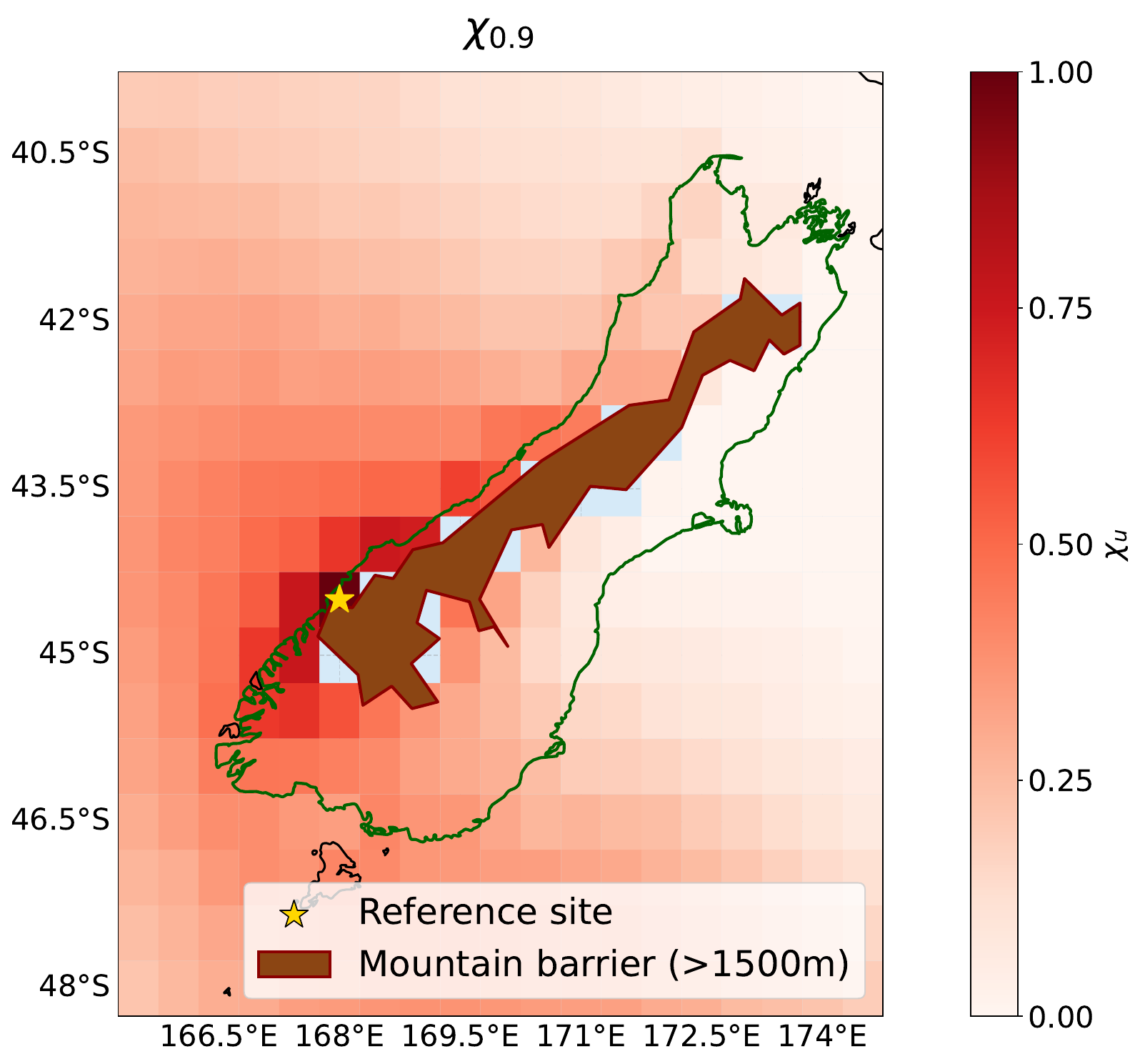} & 
			\includegraphics[width=0.2\textwidth, clip=true, trim=0 0 30 40]{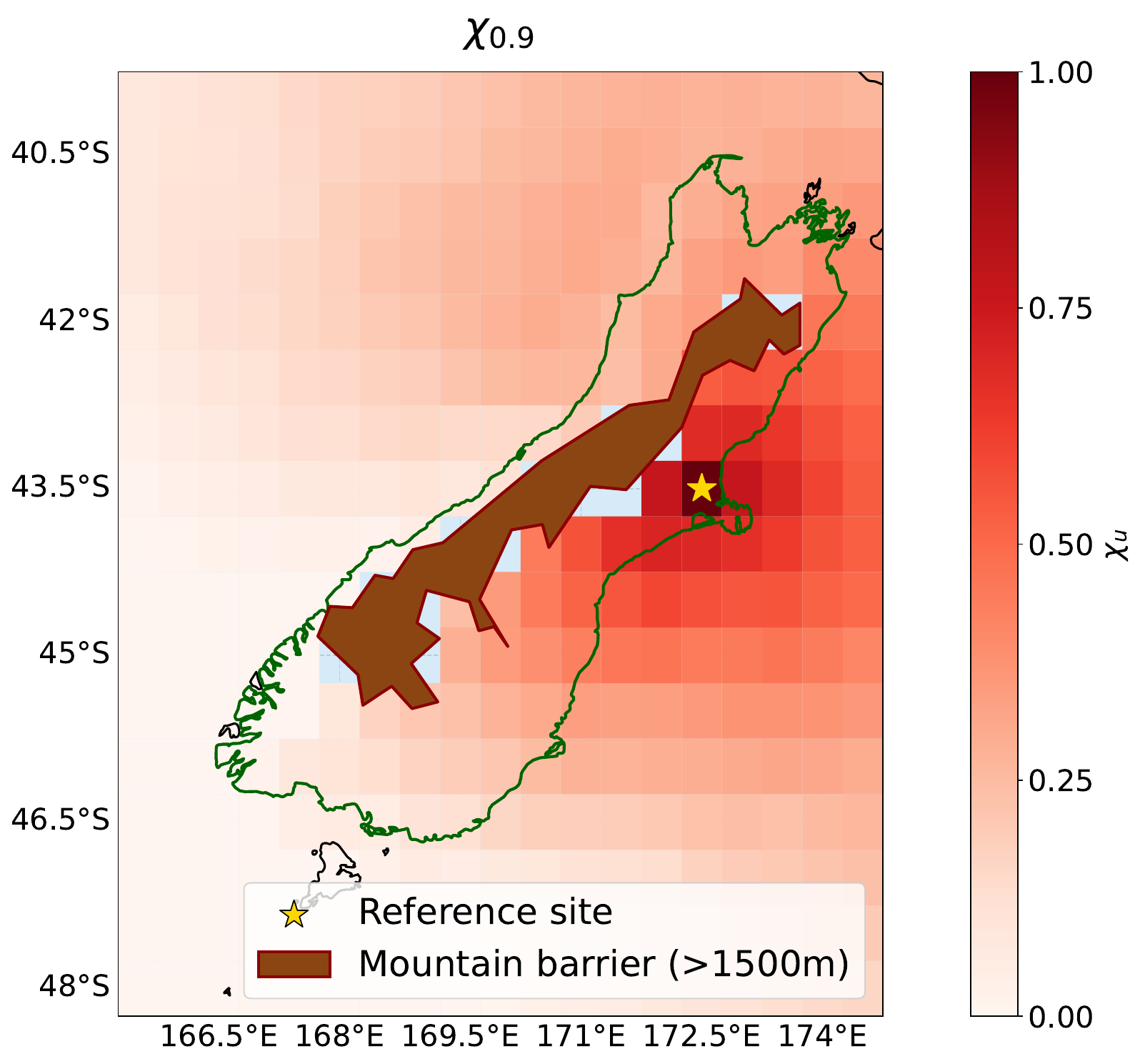} & 
			\includegraphics[width=0.2\textwidth, clip=true, trim=0 0 30 40]{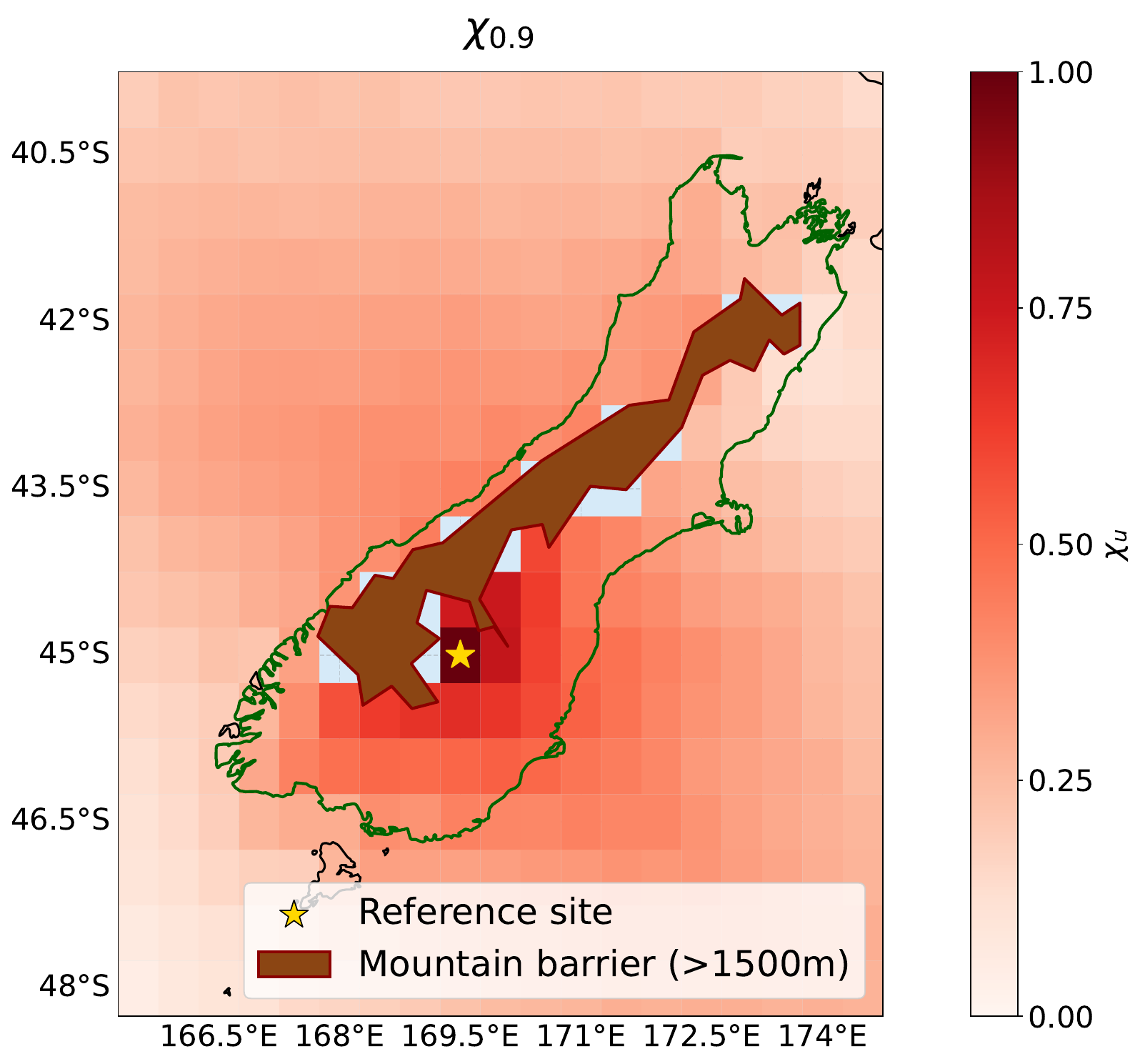}
			\\
			\includegraphics[width=0.2\textwidth, clip=true, trim=0 0 30 40]{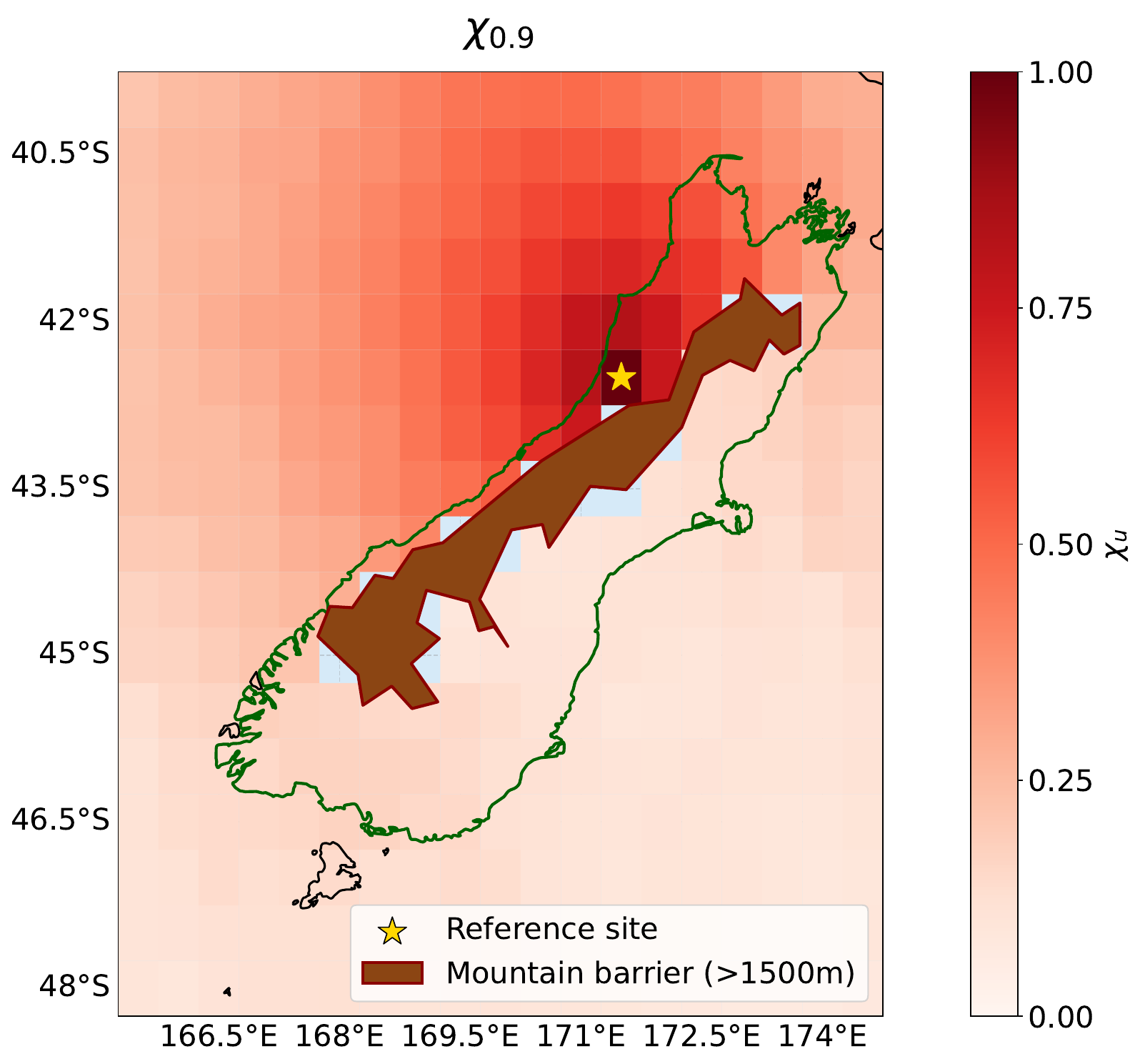} & 
			\includegraphics[width=0.2\textwidth, clip=true, trim=0 0 30 40]{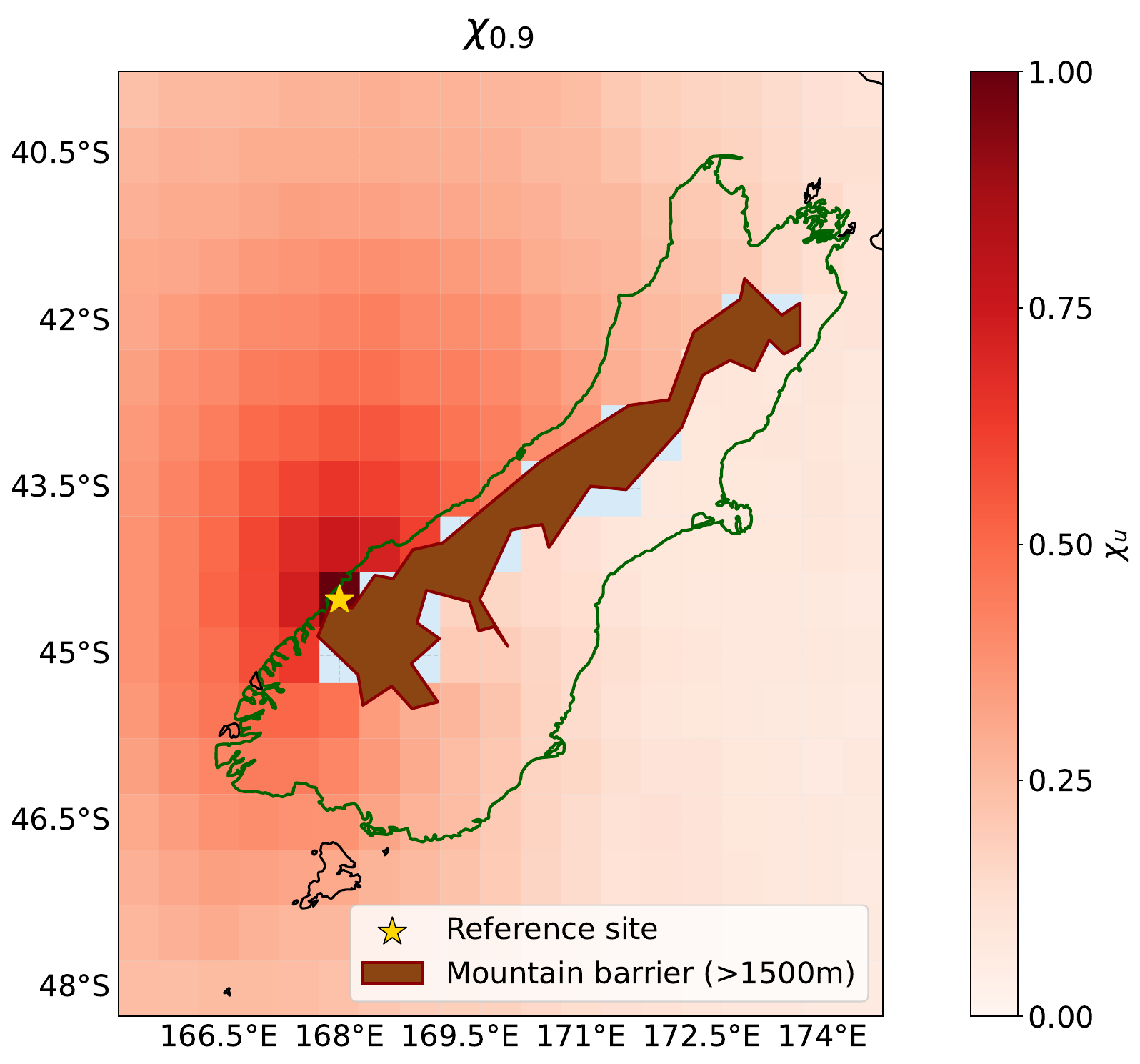} & 
			\includegraphics[width=0.2\textwidth, clip=true, trim=0 0 30 40]{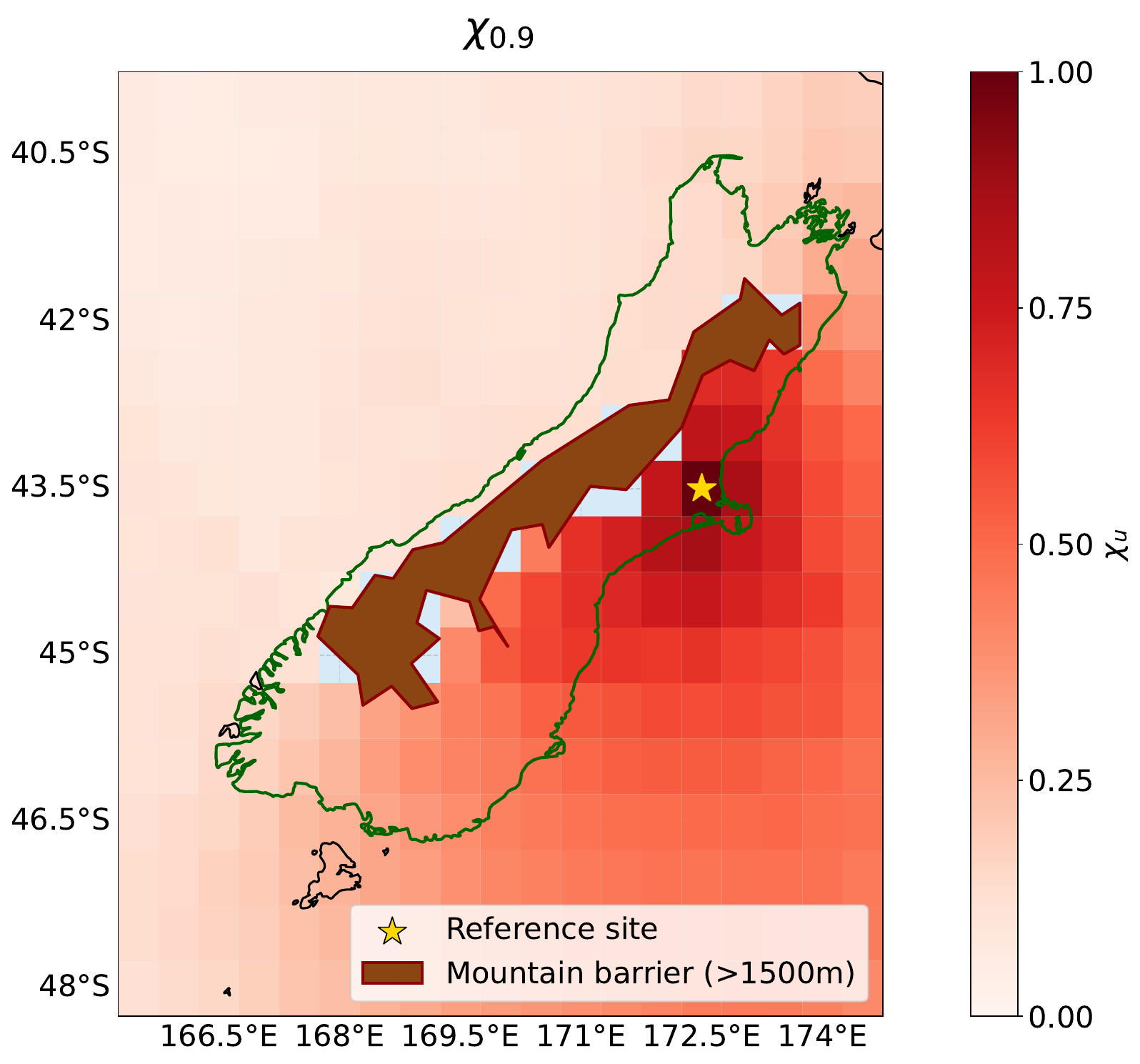} & 
			\includegraphics[width=0.2\textwidth, clip=true, trim=0 0 30 40]{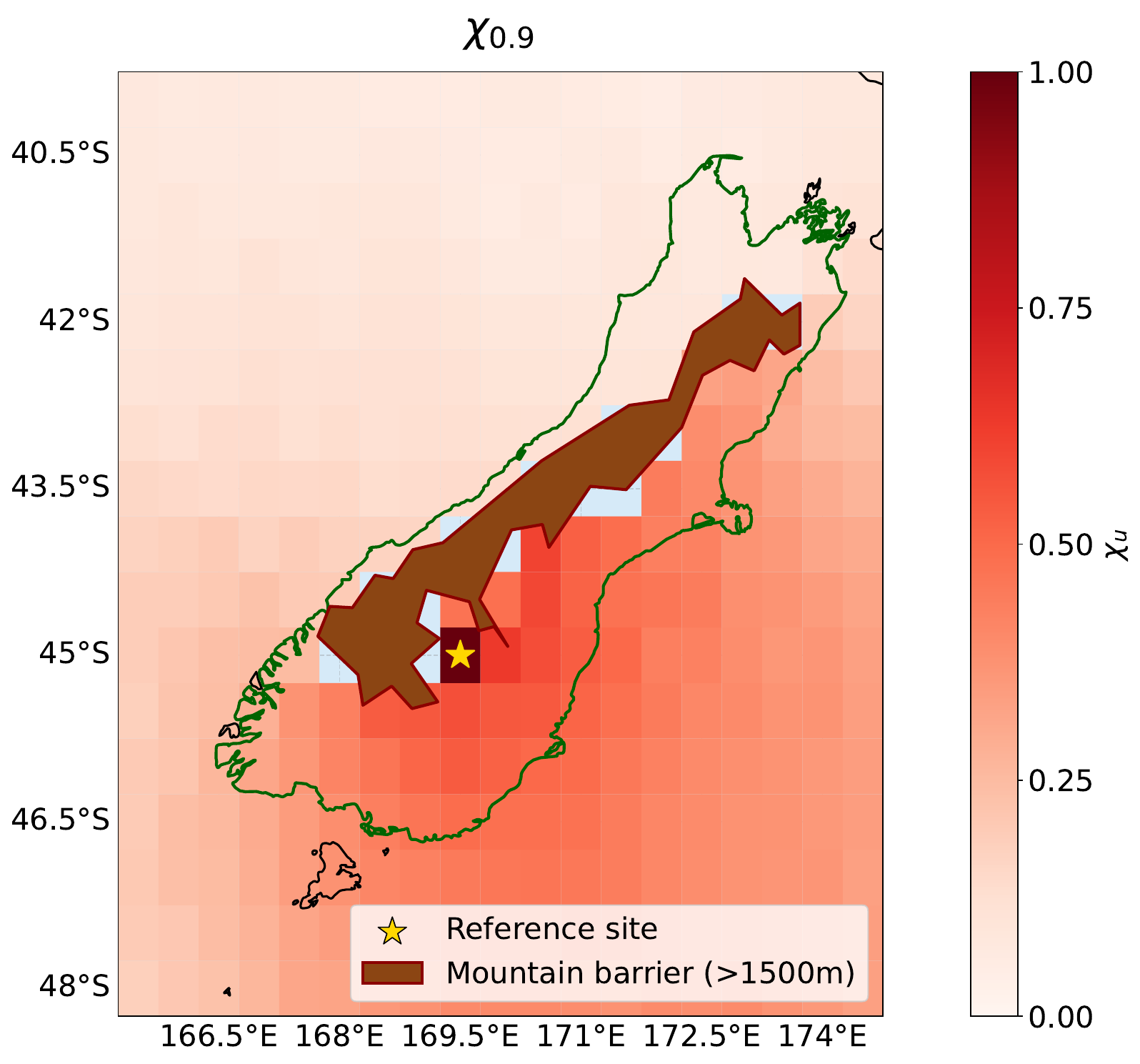}
			\\
			\multicolumn{4}{c}{\large$\chi_{0.95}$} \\
			\includegraphics[width=0.2\textwidth, clip=true, trim=0 0 30 40]{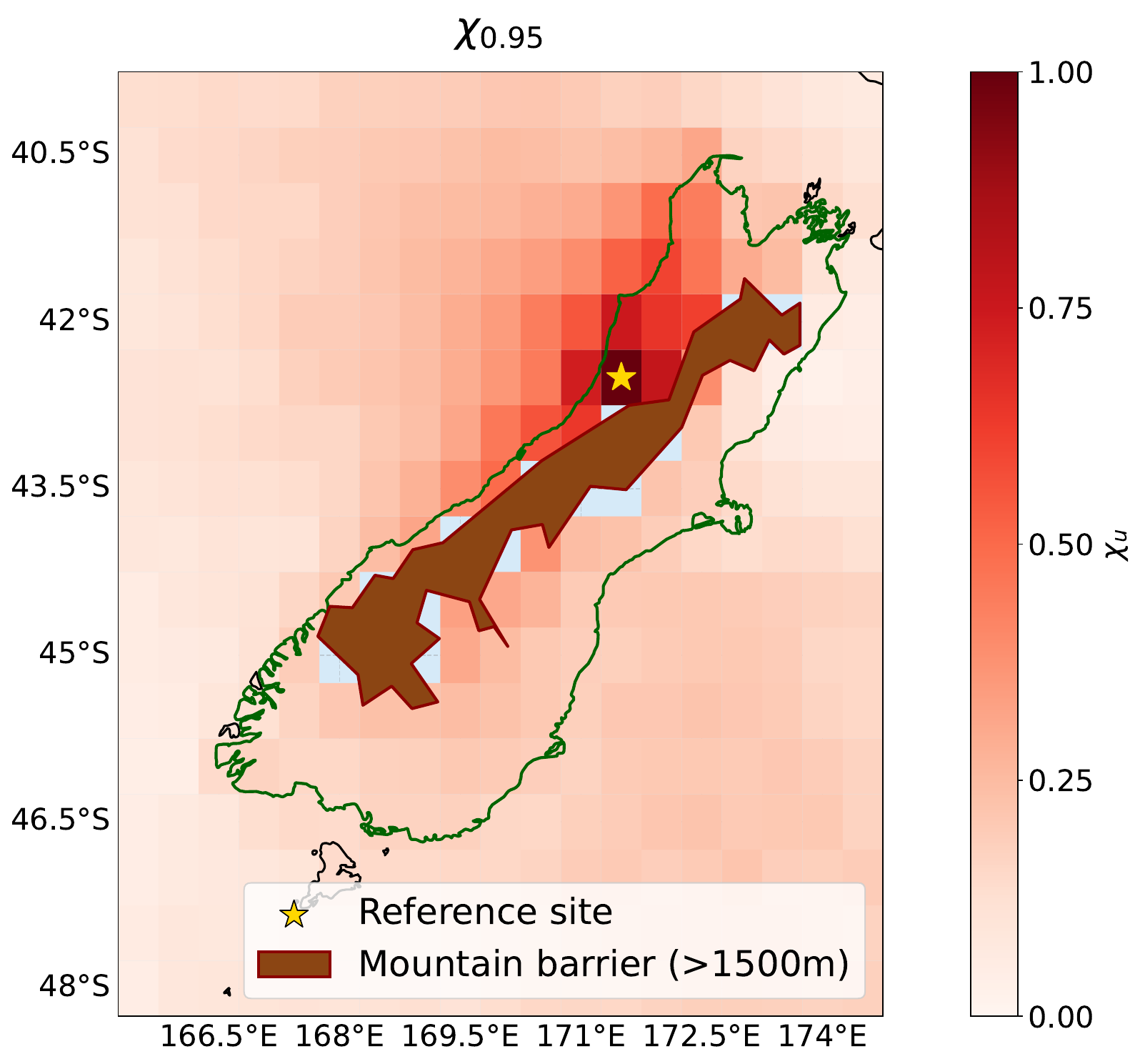} & 
			\includegraphics[width=0.2\textwidth, clip=true, trim=0 0 30 40]{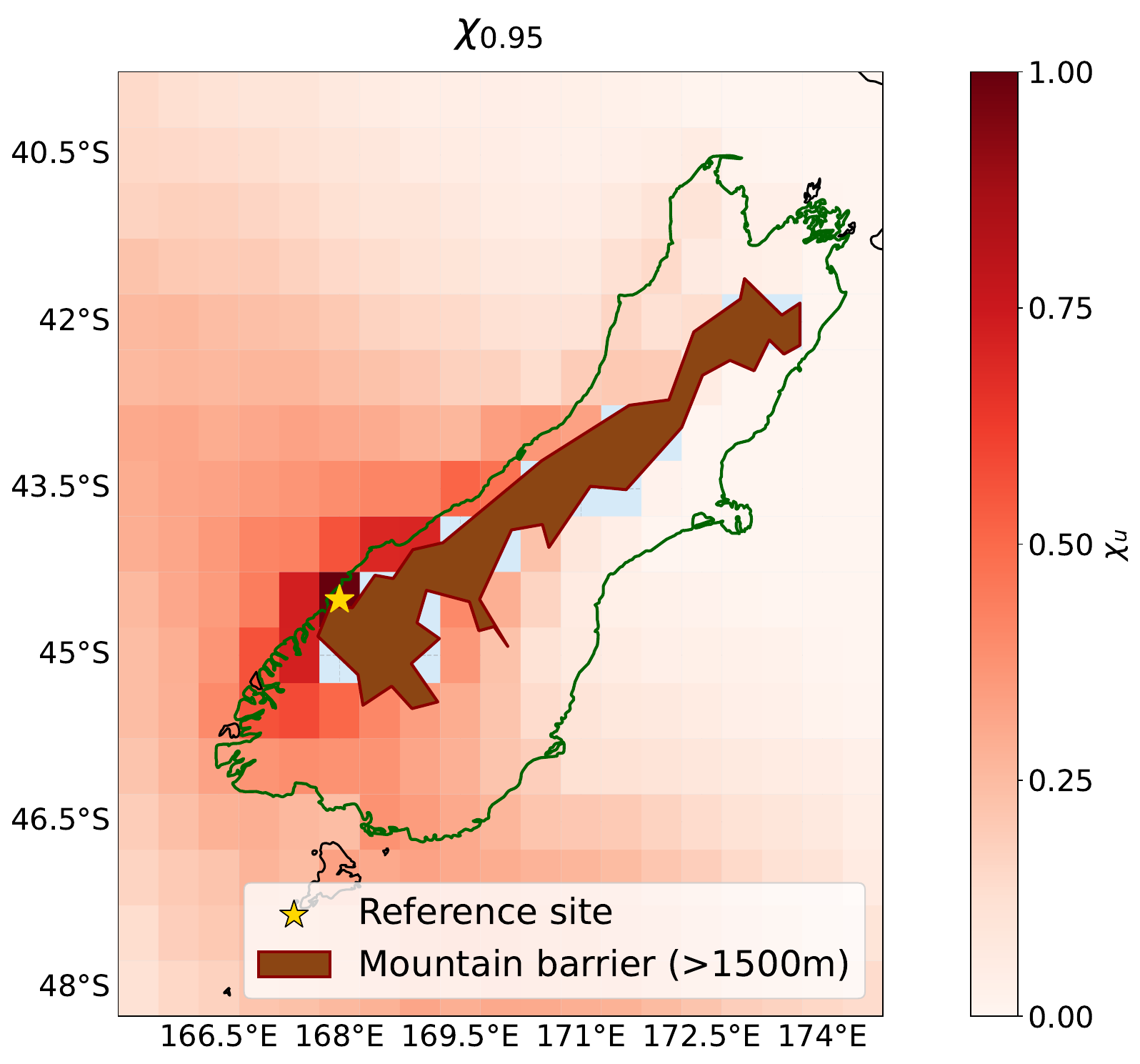} & 
			\includegraphics[width=0.2\textwidth, clip=true, trim=0 0 30 40]{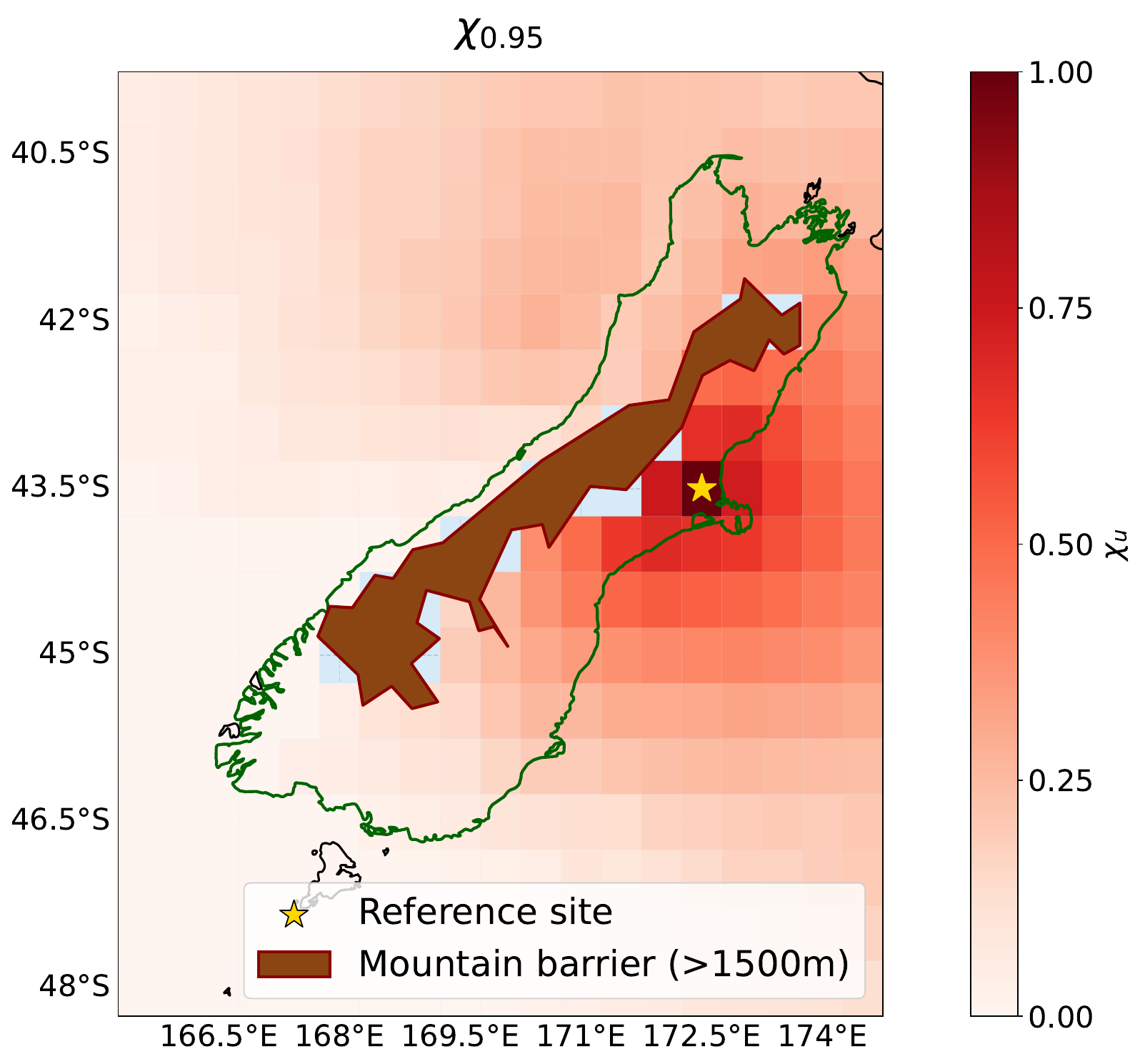} & 
			\includegraphics[width=0.2\textwidth, clip=true, trim=0 0 30 40]{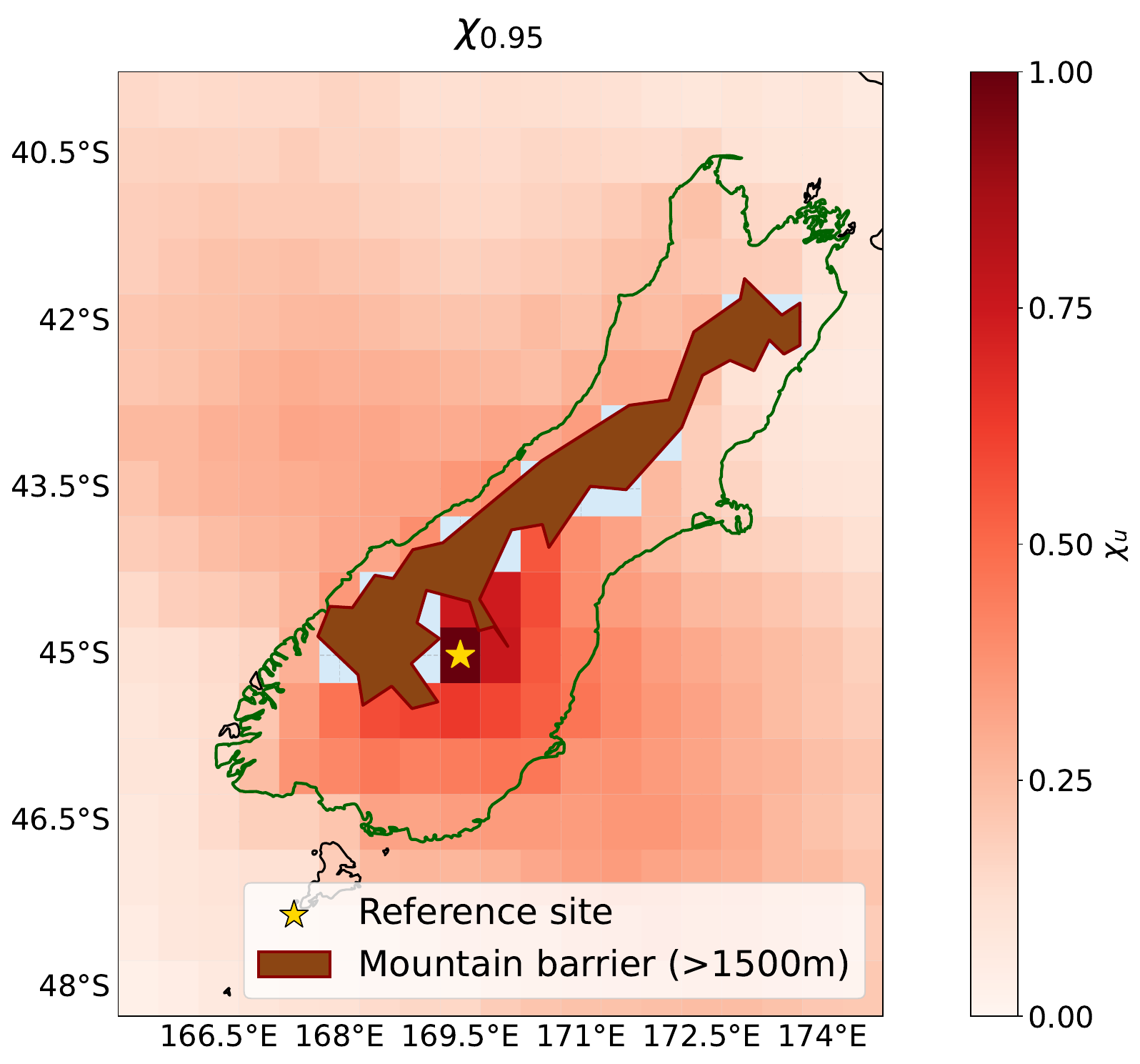}
			\\
			\includegraphics[width=0.2\textwidth, clip=true, trim=0 0 30 40]{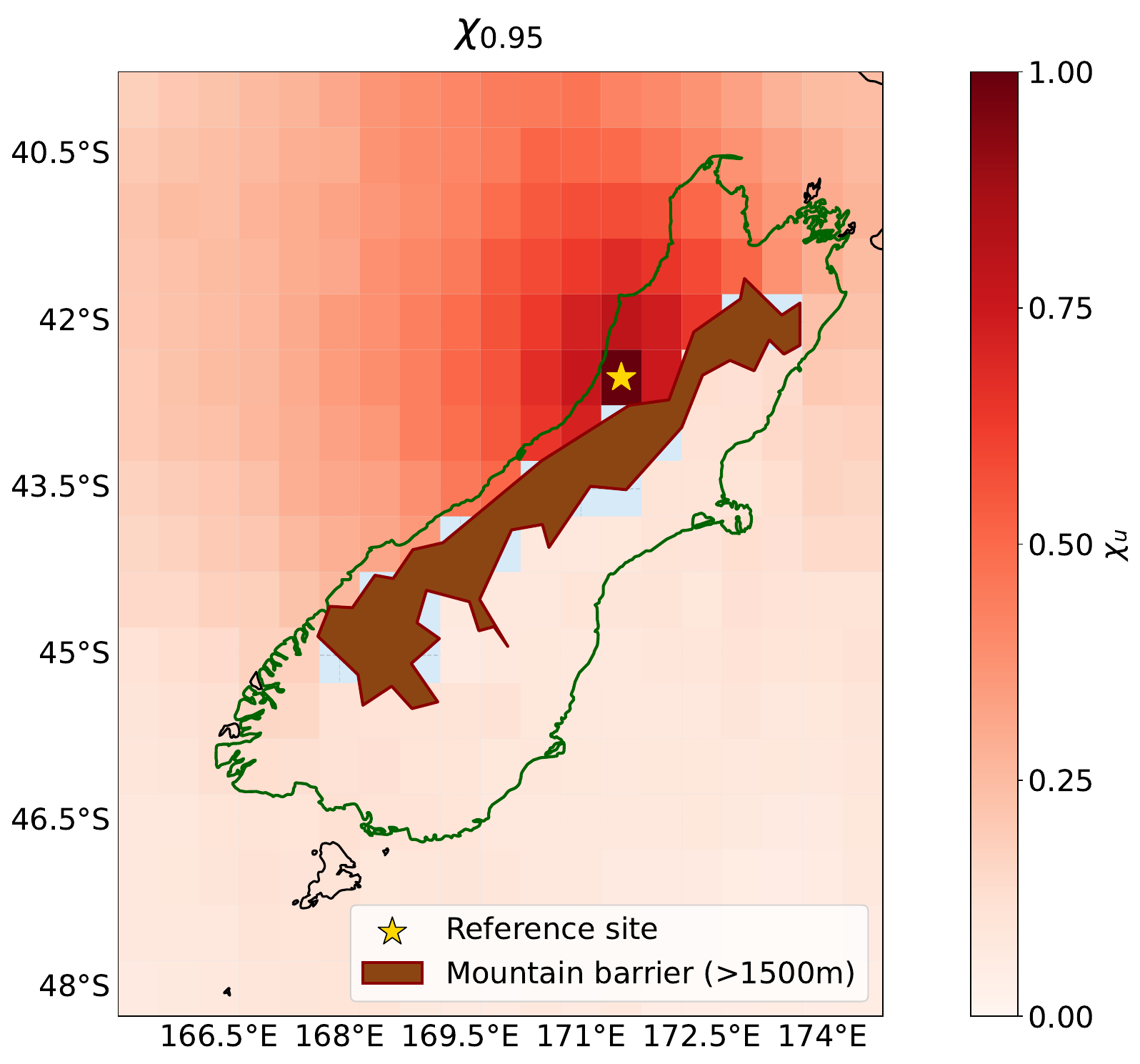} & 
			\includegraphics[width=0.2\textwidth, clip=true, trim=0 0 30 40]{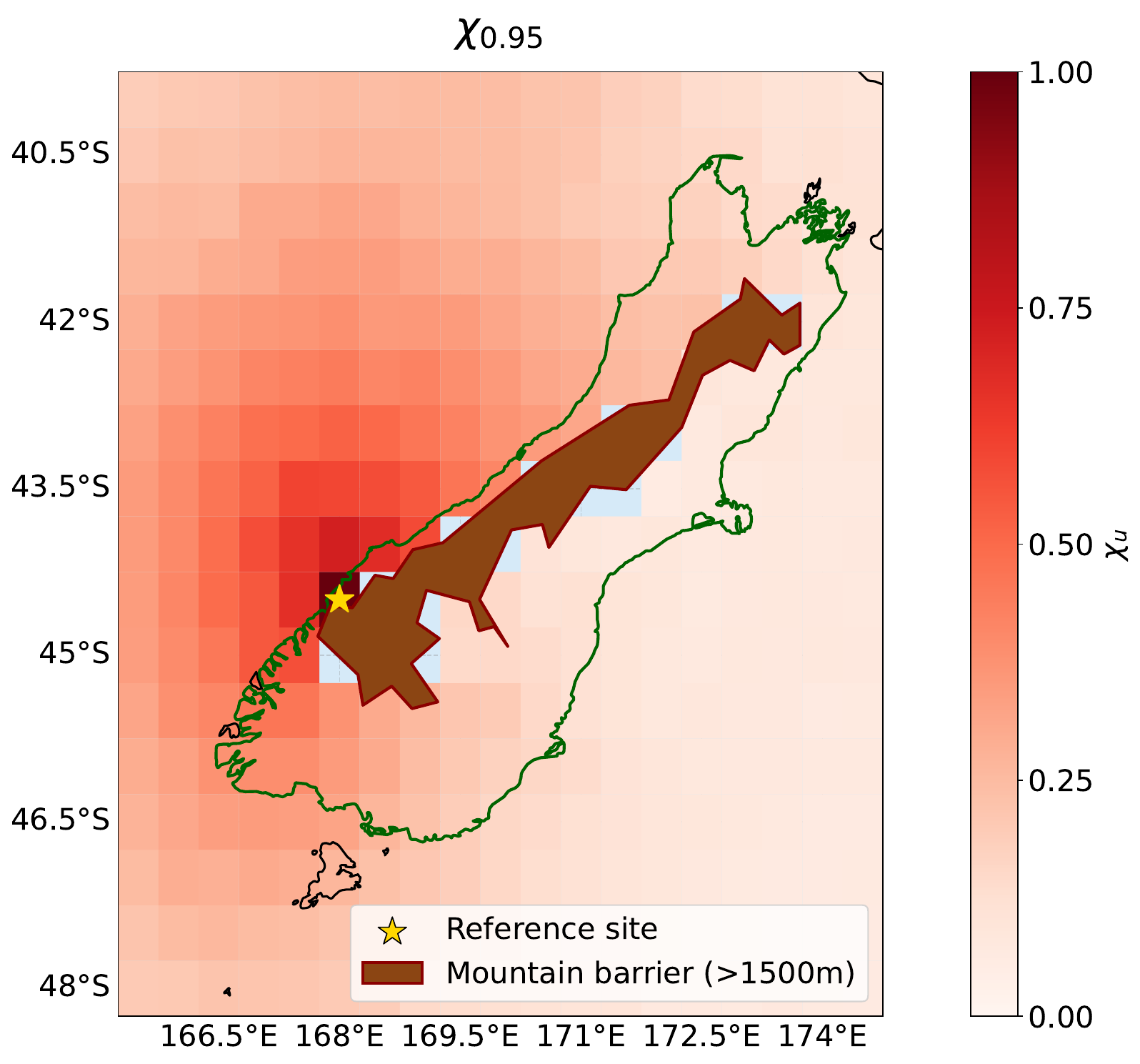} & 
			\includegraphics[width=0.2\textwidth, clip=true, trim=0 0 30 40]{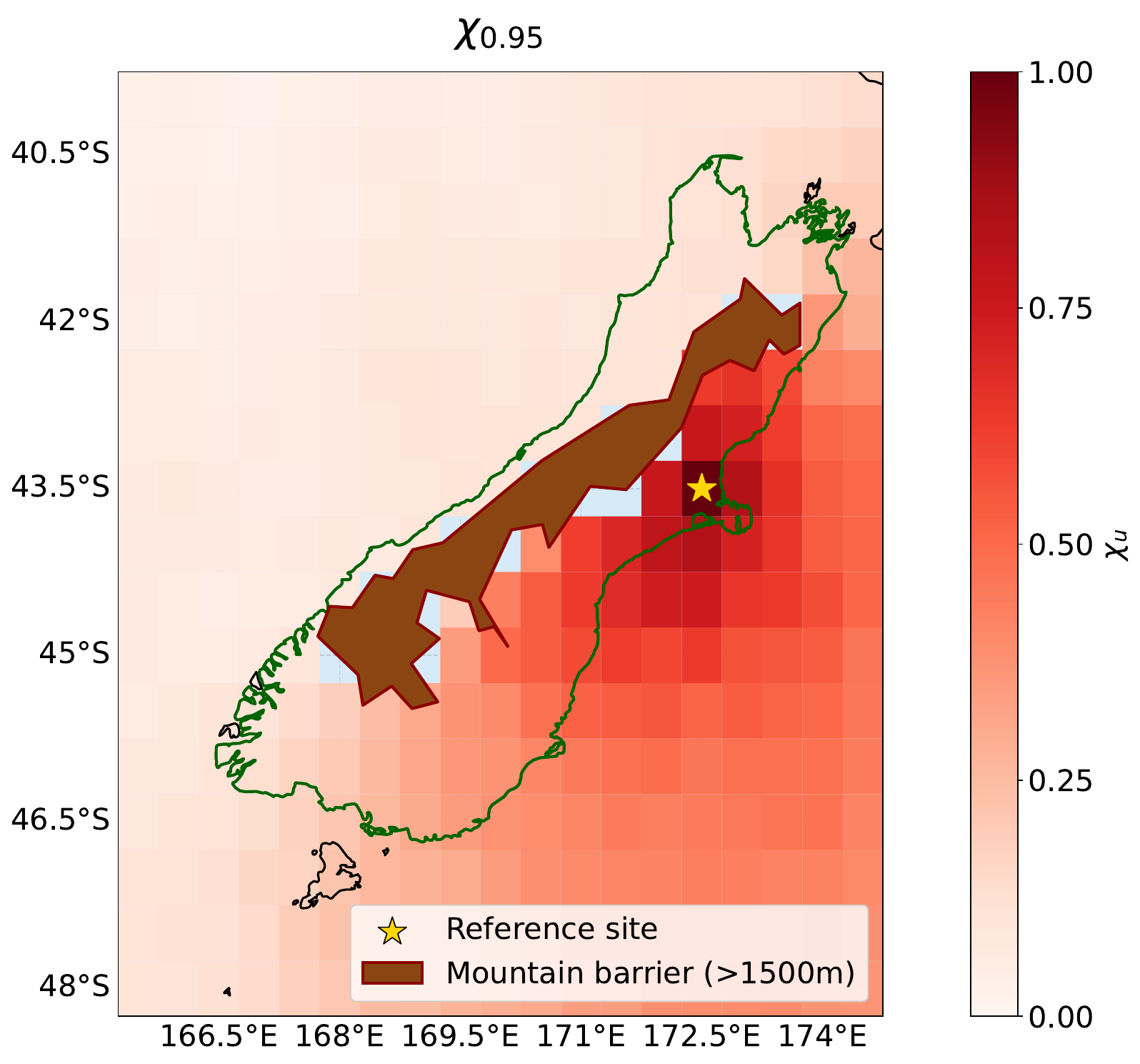} & 
			\includegraphics[width=0.2\textwidth, clip=true, trim=0 0 30 40]{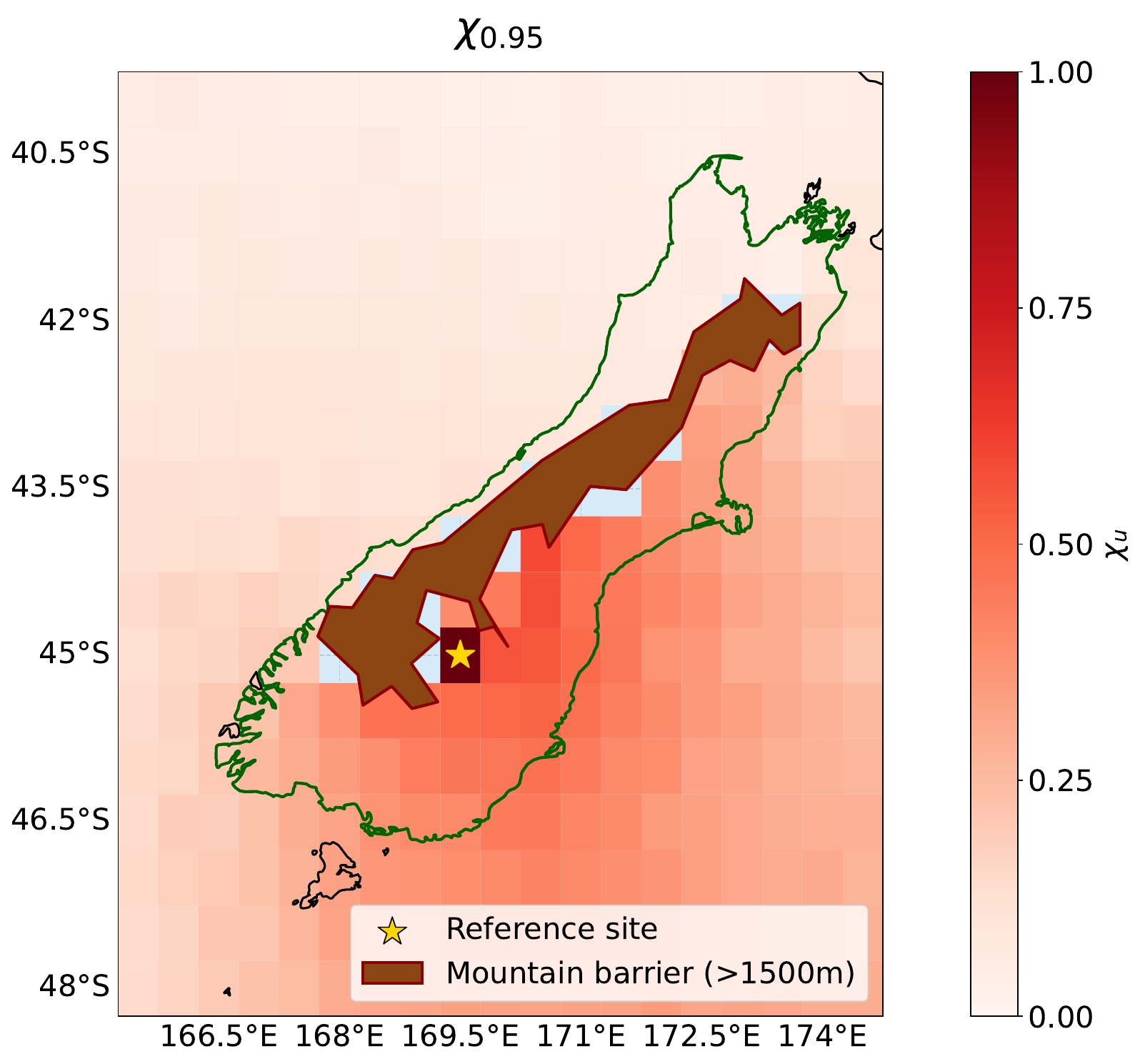}
		\end{tabular}
		\caption{Empirical (top row within each threshold) and visibility model-based (bottom row within each threshold) $\chi_{0.90}$ and $\chi_{0.95}$ surfaces for seasonal maximum precipitation over the South Island of New Zealand. Columns from left to right correspond to reference sites near Hokitika, Milford Sound, Christchurch, and Queenstown.}
		\label{fig:precip_chiu_emp_est_more}
	\end{figure}

	Figure~\ref{fig:precip_chiu_emp_est_more} extends the comparison of empirical and model-based $\chi_u$ surfaces in the Figure~\ref{fig:precip_chiu_emp_est} by considering the lower thresholds $u = 0.90$ and $u = 0.95$. For each threshold, the top row shows the empirical estimates and the bottom row shows the corresponding visibility model estimates for four reference sites near Hokitika, Milford Sound, Christchurch, and Queenstown, ordered from left to right. These additional results allow us to assess whether the spatial patterns observed at $u = 0.99$ persist at less extreme thresholds. 
	
	The empirical surfaces exhibit directional variation in extremal dependence, with attenuation across the Southern Alps and stronger dependence along unobstructed regions. The visibility model reproduces these broad spatial features across the four reference sites and both thresholds, although the local patterns and magnitudes are not identical to the empirical estimates. The comparison suggests that incorporating the barrier geometry helps the model represent spatial variation in extremal dependence that would be difficult to capture using distance alone. Together with the results at $u = 0.99$ in the main text, these maps illustrate the model's ability to represent barrier-related dependence patterns across a range of thresholds. 
	
	\FloatBarrier
	\subsection{Comparison with Alternative Models} \label{supp:precip_model_comparison}
	
	\subsubsection{Posterior Summaries} \label{supp:precip_comparison_posterior}
	
	Table~\ref{tab:precip_margin_three_model} reports the posterior means and 95\% equal-tailed credible intervals for the shared margin regression coefficients across all three models (all three use the same five-covariate specification from Section~\ref{supp:precip_gev_covariates}).
	
	\begin{table}[htbp]
		\centering
		\caption{Posterior mean and 95\% equal-tail credible interval, margin regression coefficients, all three models.}
		\label{tab:precip_margin_three_model}
		\resizebox{\textwidth}{!}{
			\begin{tabular}{c c cccccc}
				\toprule
				Parameter & Model & $\beta_0$ & $\beta_1$ & $\beta_2$ & $\beta_3$ & $\beta_4$ & $\beta_5$ \\
				\midrule
				\multirow{3}{*}{$\mu$} & Visibility & $\underset{(17.011,\, 21.025)}{18.964}$ & $\underset{(-1.178,\, 0.786)}{-0.164}$ & $\underset{(-0.058,\, 0.971)}{0.438}$ & $\underset{(-1.719,\, 0.038)}{-0.856}$ & $\underset{(-0.345,\, 1.716)}{0.686}$ & $\underset{(-0.750,\, 1.428)}{0.322}$ \\
				& Shi et al. & $\underset{(24.821,\, 28.314)}{26.589}$ & $\underset{(0.452,\, 2.256)}{1.388}$ & $\underset{(0.032,\, 1.315)}{0.680}$ & $\underset{(0.258,\, 2.701)}{1.459}$ & $\underset{(-4.777,\, -2.987)}{-3.885}$ & $\underset{(-2.170,\, -0.150)}{-1.176}$ \\
				& Huser--Wadsworth & $\underset{(16.529,\, 20.352)}{18.340}$ & $\underset{(-1.917,\, -0.076)}{-0.983}$ & $\underset{(1.063,\, 1.955)}{1.514}$ & $\underset{(-1.215,\, 0.957)}{-0.166}$ & $\underset{(-4.163,\, -2.589)}{-3.364}$ & $\underset{(-1.226,\, 0.929)}{-0.152}$ \\
				\midrule
				\multirow{3}{*}{$\sigma$} & Visibility & $\underset{(2.743,\, 2.853)}{2.799}$ & $\underset{(0.129,\, 0.251)}{0.193}$ & $\underset{(-0.120,\, -0.058)}{-0.089}$ & $\underset{(-0.133,\, -0.041)}{-0.088}$ & $\underset{(-0.121,\, -0.012)}{-0.065}$ & $\underset{(-0.176,\, -0.085)}{-0.131}$ \\
				& Shi et al. & $\underset{(2.557,\, 2.637)}{2.597}$ & $\underset{(0.218,\, 0.308)}{0.264}$ & $\underset{(-0.092,\, -0.035)}{-0.064}$ & $\underset{(-0.068,\, 0.032)}{-0.020}$ & $\underset{(-0.220,\, -0.118)}{-0.169}$ & $\underset{(-0.171,\, -0.086)}{-0.129}$ \\
				& Huser--Wadsworth & $\underset{(2.330,\, 2.424)}{2.378}$ & $\underset{(0.166,\, 0.275)}{0.221}$ & $\underset{(-0.061,\, -0.007)}{-0.034}$ & $\underset{(-0.094,\, 0.032)}{-0.032}$ & $\underset{(-0.349,\, -0.261)}{-0.306}$ & $\underset{(-0.155,\, -0.059)}{-0.107}$ \\
				\midrule
				\multirow{3}{*}{$\xi$} & Visibility & $\underset{(0.003,\, 0.223)}{0.112}$ & $\underset{(0.122,\, 0.341)}{0.227}$ & $\underset{(-0.041,\, 0.027)}{-0.007}$ & $\underset{(-0.300,\, -0.178)}{-0.239}$ & $\underset{(0.150,\, 0.293)}{0.221}$ & $\underset{(0.175,\, 0.339)}{0.259}$ \\
				& Shi et al. & $\underset{(-0.459,\, -0.347)}{-0.404}$ & $\underset{(0.070,\, 0.210)}{0.142}$ & $\underset{(-0.037,\, 0.022)}{-0.007}$ & $\underset{(-0.098,\, 0.001)}{-0.049}$ & $\underset{(-0.089,\, 0.036)}{-0.026}$ & $\underset{(0.050,\, 0.186)}{0.117}$ \\
				& Huser--Wadsworth & $\underset{(-0.288,\, -0.117)}{-0.206}$ & $\underset{(0.093,\, 0.218)}{0.155}$ & $\underset{(-0.036,\, 0.014)}{-0.011}$ & $\underset{(-0.039,\, 0.035)}{-0.002}$ & $\underset{(-0.096,\, 0.001)}{-0.047}$ & $\underset{(-0.001,\, 0.118)}{0.058}$ \\
				\bottomrule
			\end{tabular}
		}
	\end{table}
	
	\begin{table}[htbp]
		\centering
		\caption{Posterior mean and 95\% equal-tail credible interval, dependence hyperparameters, all three models. Visibility and Shi et al.\ both model tail dependence as spatially varying, $\delta(\bm{s}) = \bm{G}(\bm{s})^\top \bm{\phi}$, a 24-knot Gaussian-kernel expansion with shared bandwidth $h$ (individual knot coefficients $\phi_1,\ldots,\phi_{24}$ shown in Figure~\ref{fig:precip_comparison_delta_posterior_mean} rather than tabulated); Huser--Wadsworth instead uses a single constant $\delta$, so its tail dependence row is not comparable cell-by-cell with the other two.}
		\label{tab:precip_dependence_three_model_other}
		\begin{tabular}{ll ccc}
			\toprule
			\multicolumn{2}{c}{Parameter} & Visibility & Shi et al. & Huser--Wadsworth \\
			\midrule
			\multicolumn{2}{l}{Tail dependence, bandwidth $h$} & $\underset{(1.450,\, 1.676)}{1.555}$ & $\underset{(1.090,\, 1.235)}{1.162}$ & -- \\
			\multicolumn{2}{l}{Tail dependence, constant $\delta$} & -- & -- & $\underset{(0.000,\, 0.017)}{0.005}$ \\
			\midrule
			\multirow{3}{*}{Wendland} & $r$ & $\underset{(2.379,\, 2.471)}{2.418}$ & $\underset{(3.459,\, 3.921)}{3.672}$ & -- \\
			& $\lambda$ & $\underset{(4.371,\, 6.088)}{5.132}$ & -- & -- \\
			& $\alpha$ & $\underset{(1.213,\, 1.255)}{1.234}$ & -- & -- \\
			\midrule
			\multirow{2}{*}{Mat\'ern} & $\rho$ & $\underset{(7.970,\, 10.422)}{9.106}$ & $\underset{(3.288,\, 4.215)}{3.732}$ & $\underset{(4.535,\, 6.116)}{5.322}$ \\
			& $\nu$ & $\underset{(0.697,\, 0.743)}{0.719}$ & $\underset{(0.820,\, 0.899)}{0.857}$ & $\underset{(0.787,\, 0.848)}{0.816}$ \\
			\bottomrule
		\end{tabular}
	\end{table}
	
	\begin{table}[htbp]
		\centering
		\caption{Posterior mean and 95\% credible interval of the tail-dependence knot coefficients $\phi_1,\ldots,\phi_{24}$, Visibility and Shi et al.\ models. Huser--Wadsworth is omitted here since it has no spatially varying tail dependence coefficients, using a single constant $\delta$ instead (Table~\ref{tab:precip_dependence_three_model_other}).}
		\label{tab:precip_dependence_two_model_phi}
			\begin{tabular}{c cc @{\hskip 8mm} ccc}
				\toprule
				Parameter & Visibility & Shi et al. & Parameter & Visibility & Shi et al. \\
				\midrule
				$\phi_{1}$ & $\underset{(0.002,\, 0.219)}{0.067}$ & $\underset{(0.001,\, 0.096)}{0.027}$ & $\phi_{13}$ & $\underset{(0.072,\, 0.695)}{0.370}$ & $\underset{(0.006,\, 0.405)}{0.155}$ \\
				$\phi_{2}$ & $\underset{(0.518,\, 0.940)}{0.730}$ & $\underset{(0.352,\, 0.862)}{0.605}$ & $\phi_{14}$ & $\underset{(1.108,\, 1.809)}{1.444}$ & $\underset{(0.282,\, 0.766)}{0.511}$ \\
				$\phi_{3}$ & $\underset{(1.701,\, 2.873)}{2.285}$ & $\underset{(1.079,\, 1.854)}{1.449}$ & $\phi_{15}$ & $\underset{(0.725,\, 1.604)}{1.155}$ & $\underset{(0.923,\, 1.677)}{1.293}$ \\
				$\phi_{4}$ & $\underset{(0.058,\, 0.458)}{0.254}$ & $\underset{(0.001,\, 0.076)}{0.021}$ & $\phi_{16}$ & $\underset{(0.006,\, 0.512)}{0.172}$ & $\underset{(0.324,\, 1.223)}{0.761}$ \\
				$\phi_{5}$ & $\underset{(1.181,\, 1.656)}{1.412}$ & $\underset{(0.870,\, 1.493)}{1.170}$ & $\phi_{17}$ & $\underset{(0.001,\, 0.081)}{0.022}$ & $\underset{(0.000,\, 0.051)}{0.014}$ \\
				$\phi_{6}$ & $\underset{(0.025,\, 0.575)}{0.273}$ & $\underset{(0.026,\, 0.460)}{0.220}$ & $\phi_{18}$ & $\underset{(0.216,\, 1.100)}{0.643}$ & $\underset{(0.621,\, 1.343)}{0.978}$ \\
				$\phi_{7}$ & $\underset{(0.329,\, 1.079)}{0.741}$ & $\underset{(0.970,\, 1.550)}{1.251}$ & $\phi_{19}$ & $\underset{(0.191,\, 0.809)}{0.488}$ & $\underset{(0.002,\, 0.209)}{0.068}$ \\
				$\phi_{8}$ & $\underset{(0.399,\, 0.739)}{0.569}$ & $\underset{(0.001,\, 0.128)}{0.039}$ & $\phi_{20}$ & $\underset{(0.000,\, 0.028)}{0.007}$ & $\underset{(0.011,\, 0.369)}{0.162}$ \\
				$\phi_{9}$ & $\underset{(0.386\, 0.584)}{0.487}$ & $\underset{(0.002,\, 0.191)}{0.056}$ & $\phi_{21}$ & $\underset{(0.001,\, 0.121)}{0.035}$ & $\underset{(0.003,\, 0.267)}{0.089}$ \\
				$\phi_{10}$ & $\underset{(0.429,\, 0.779)}{0.609}$ & $\underset{(0.866,\, 1.305)}{1.080}$ & $\phi_{22}$ & $\underset{(0.006,\, 0.336)}{0.123}$ & $\underset{(0.044,\, 0.445)}{0.238}$ \\
				$\phi_{11}$ & $\underset{(0.006,\, 0.315)}{0.131}$ & $\underset{(0.037,\, 0.377)}{0.206}$ & $\phi_{23}$ & $\underset{(0.001,\, 0.173)}{0.052}$ & $\underset{(0.004,\, 0.262)}{0.100}$ \\
				$\phi_{12}$ & $\underset{(0.143,\, 0.359)}{0.251}$ & $\underset{(0.000,\, 0.039)}{0.010}$ & $\phi_{24}$ & $\underset{(0.001,\, 0.144)}{0.043}$ & $\underset{(0.003,\, 0.253)}{0.083}$ \\
				\bottomrule
			\end{tabular}
	\end{table}
	
	The three model's dependence structures are not directly comparable term-by-term. The visibility model and \citet{Shi2026} both use a 24-knot Gaussian kernel tail dependence surface and Wendland kernel (isotropic for \citet{Shi2026}, visibility-restricted and anisotropic for the visibility model), while \citet{Huser2019} uses a single global scalar tail dependence parameter $\delta$. Table~\ref{tab:precip_dependence_three_model_other} and \ref{tab:precip_dependence_two_model_phi} reports each model's own dependence parameters separately.
	
	\subsubsection{GEV Parameter Surfaces: All Three Models} \label{supp:precip_comparison_gev}
	
\begin{figure}[htbp]
		\centering
		\includegraphics[width=0.7\textwidth]{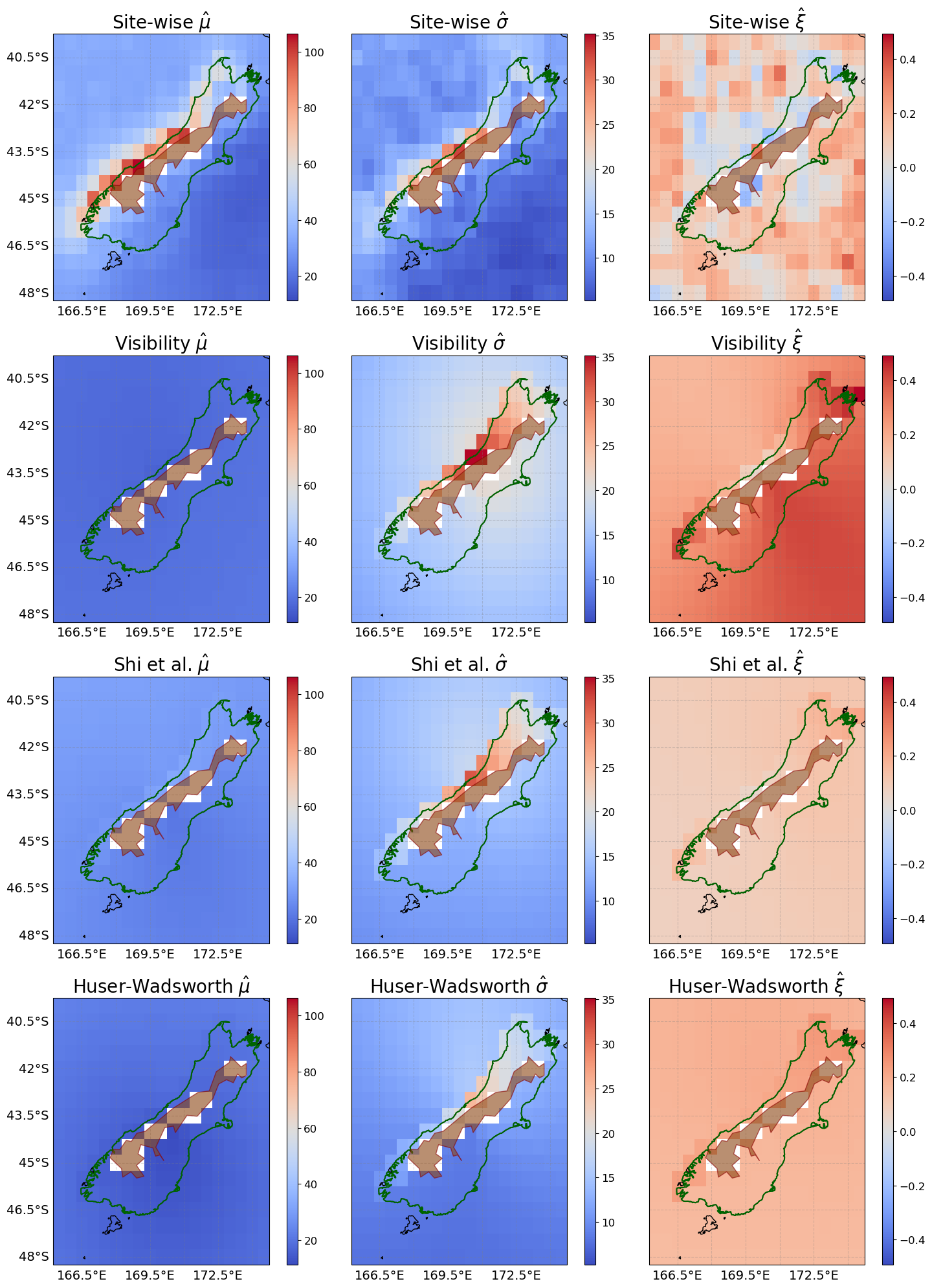}
		\caption{Posterior mean GEV location (left), scale (middle), and shape (right) surfaces: visibility model (top), Shi et al.\ (middle), Huser--Wadsworth (bottom).}
		\label{fig:precip_gev_three_model}
	\end{figure}

	Figure~\ref{fig:precip_gev_three_model} compares the fitted posterior mean GEV surfaces across the three models. Unlike Section~\ref{supp:precip_gev_surfaces}. The three models share the same five-covariate margin specification (Section~\ref{supp:precip_gev_covariates}). 
	
	The scale surfaces $\hat\sigma$ agree closely across all three models. Each shows the same qualitative pattern, with values elevated in a band along the barrier and low elsewhere. The location and shape surfaces, $\hat\mu$ and $\hat\xi$, agree far less. For $\hat\mu$ in particular, the visibility model's surface is lowest in the west and highest in the east, while \citet{Shi2026} and \citet{Huser2019} surface shows nearly the reverse gradient, highest in the northwest and lowest in the southeast. Since all three models use an identical covariate specification for the margins, this disagreement cannot come from a difference in what the margin model is allowed to represent; it must instead reflect the fact that the margin and dependence parameters are fit jointly, not separately, and are therefore not fully separable in the likelihood. A margin parameter that is only weakly identified by the marginal data alone (as $\xi$ generally is, and as $\mu$ can be here given the modest signal in a five-covariate linear predictor) is correspondingly more free to be pulled toward different values depending on which dependence structure the rest of the model assumes, whereas $\sigma$ appears to be identified strongly enough by the data itself to be comparatively insensitive to that choice.
	
	\subsubsection{QQ Plot Diagnostics at Holdout Sites} \label{supp:precip_comparison_qq}
	
	\begin{figure}[htbp]
		\centering
		{\bf Visibility} \hfill {\bf Shi et al.} \hfill {\bf Huser--Wadsworth}\\[1mm]
		\includegraphics[width=0.3\textwidth]{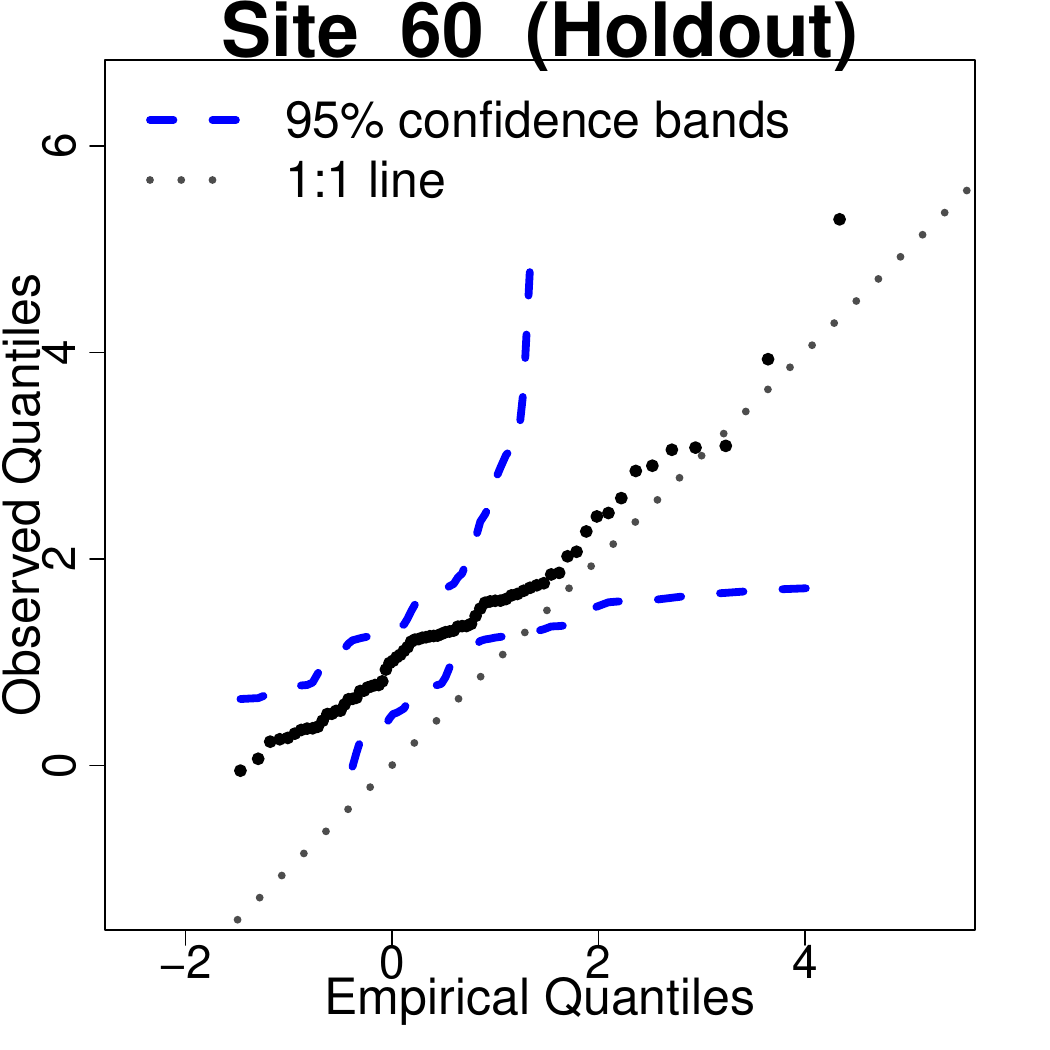}\hfill
		\includegraphics[width=0.3\textwidth]{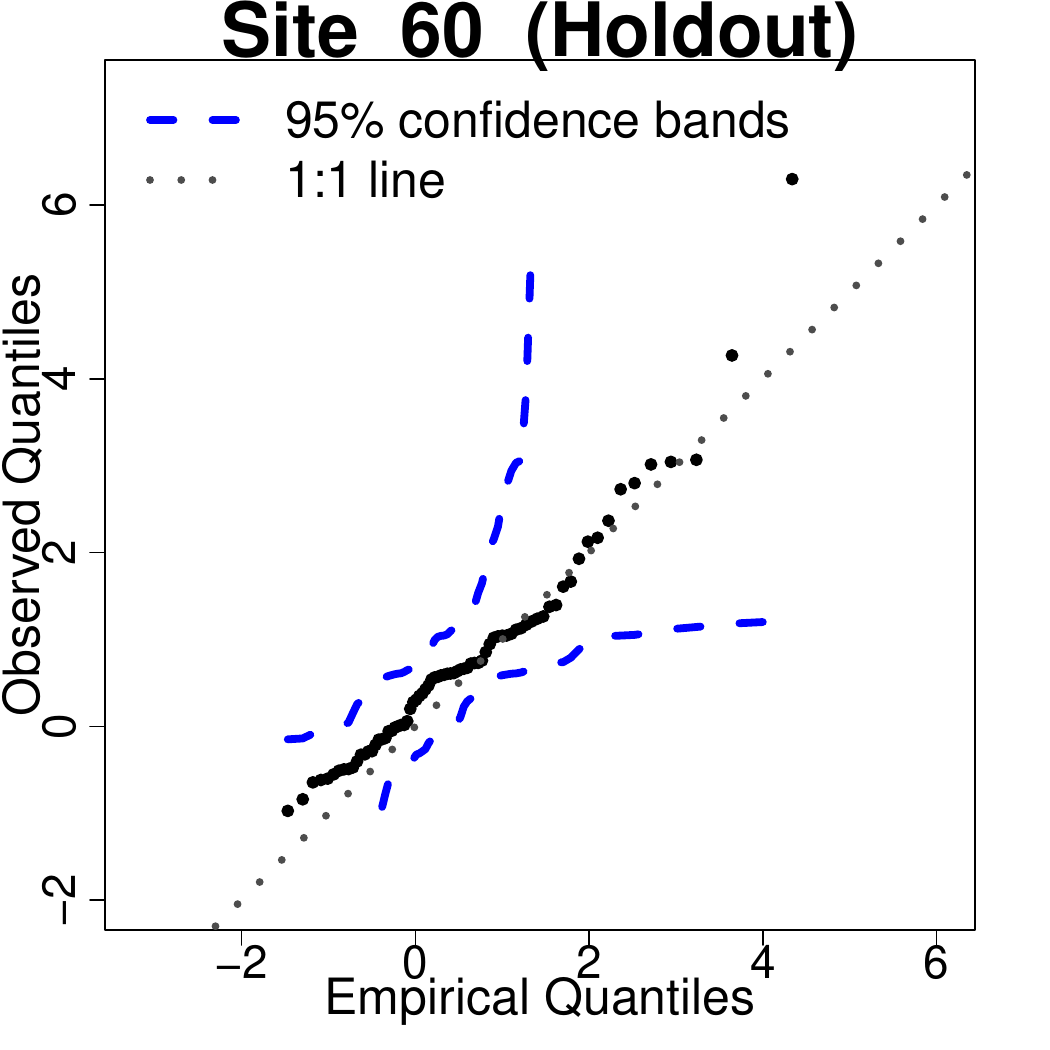}\hfill
		\includegraphics[width=0.3\textwidth]{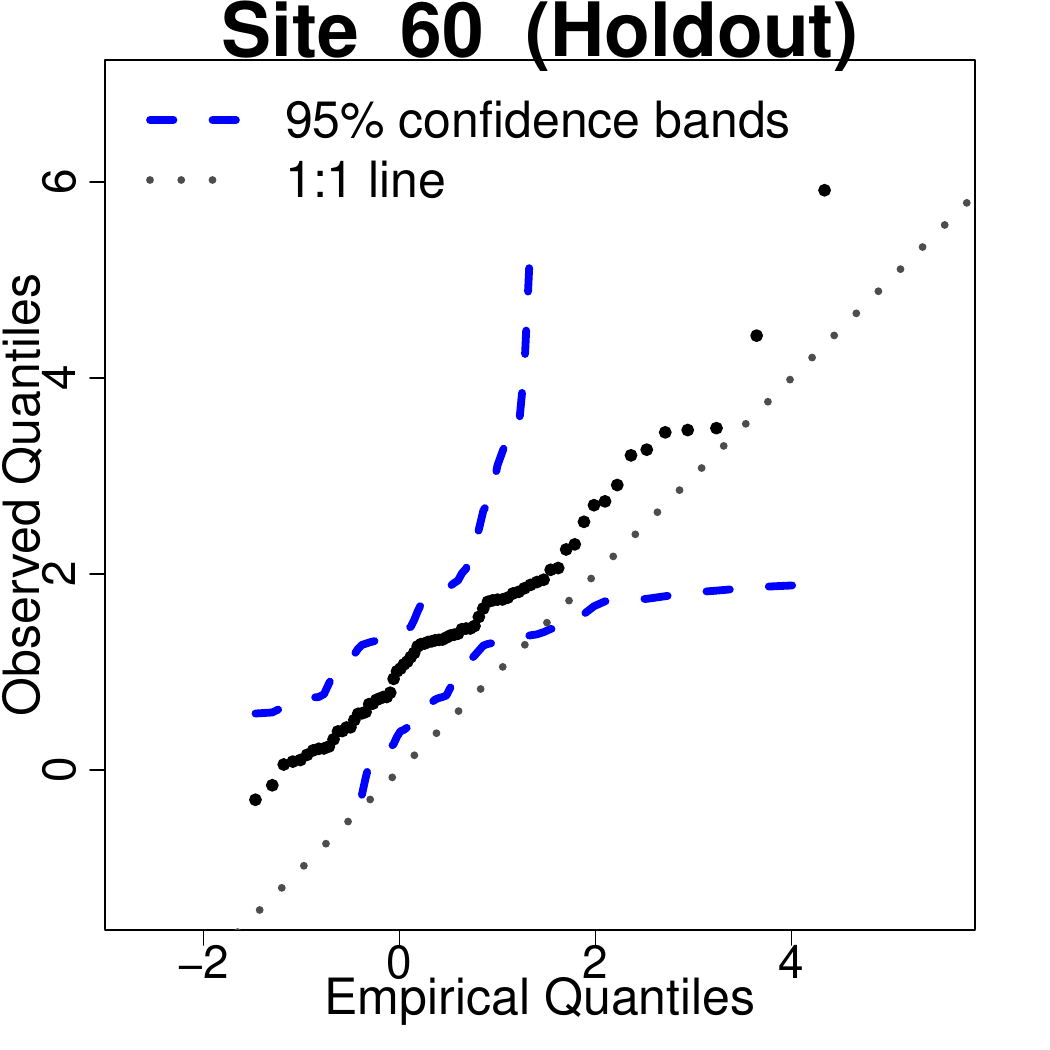}\\[2mm]
		\includegraphics[width=0.3\textwidth]{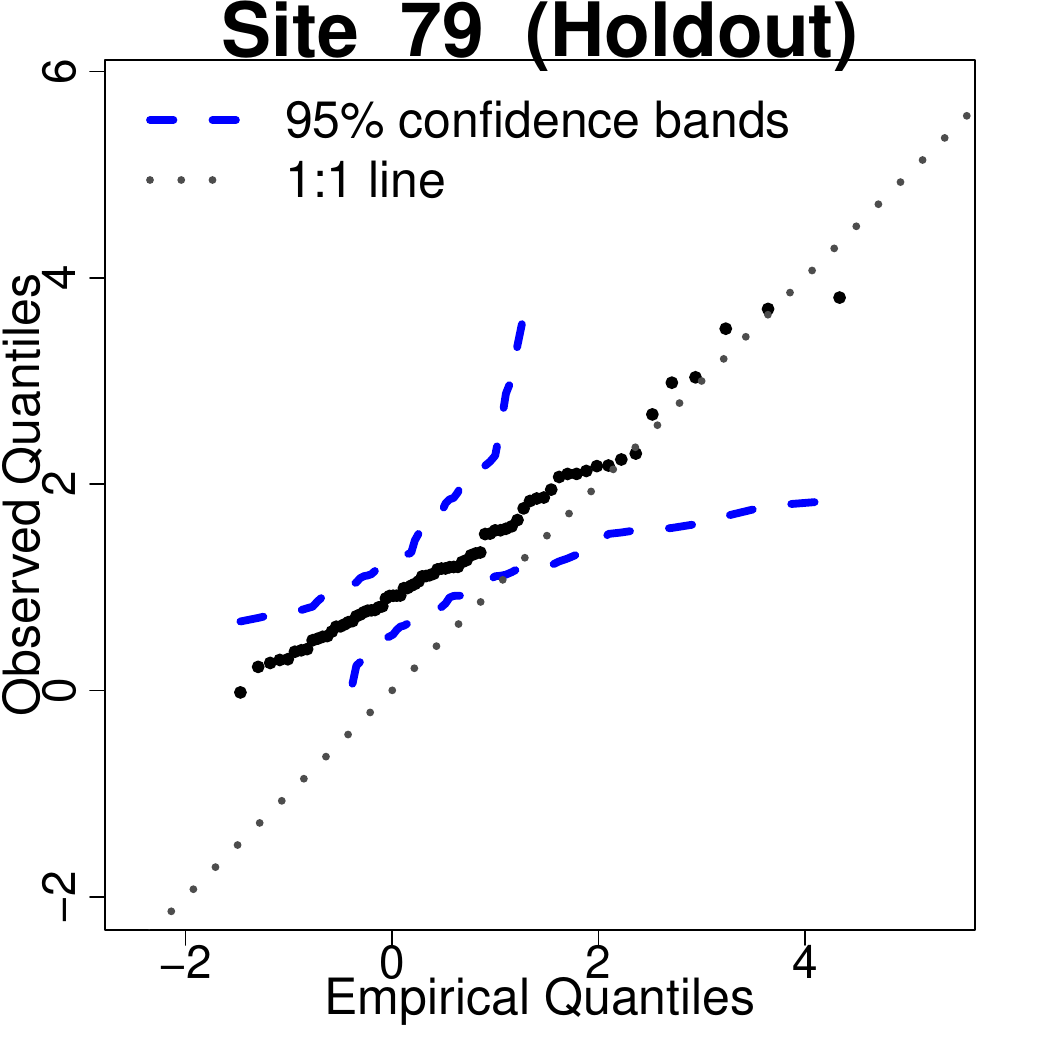}\hfill
		\includegraphics[width=0.3\textwidth]{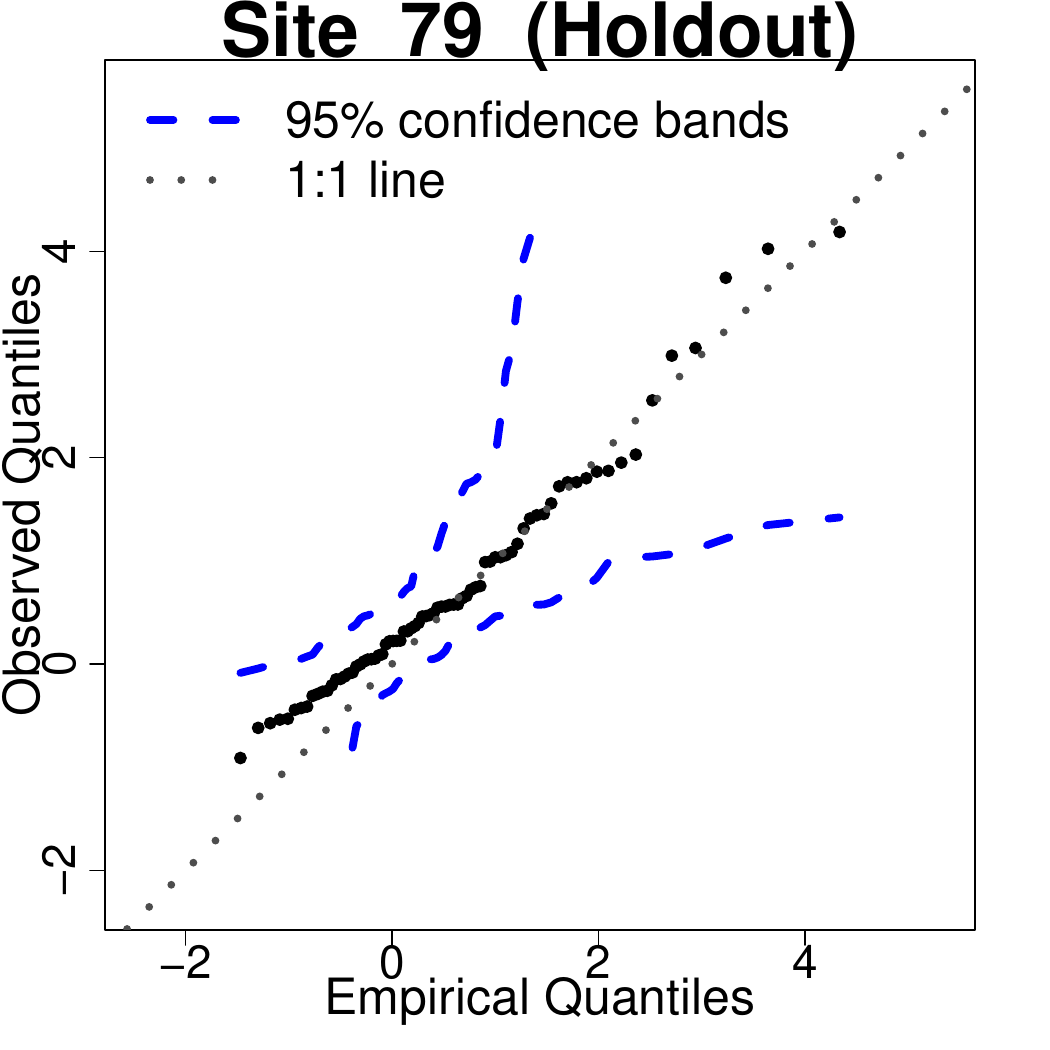}\hfill
		\includegraphics[width=0.3\textwidth]{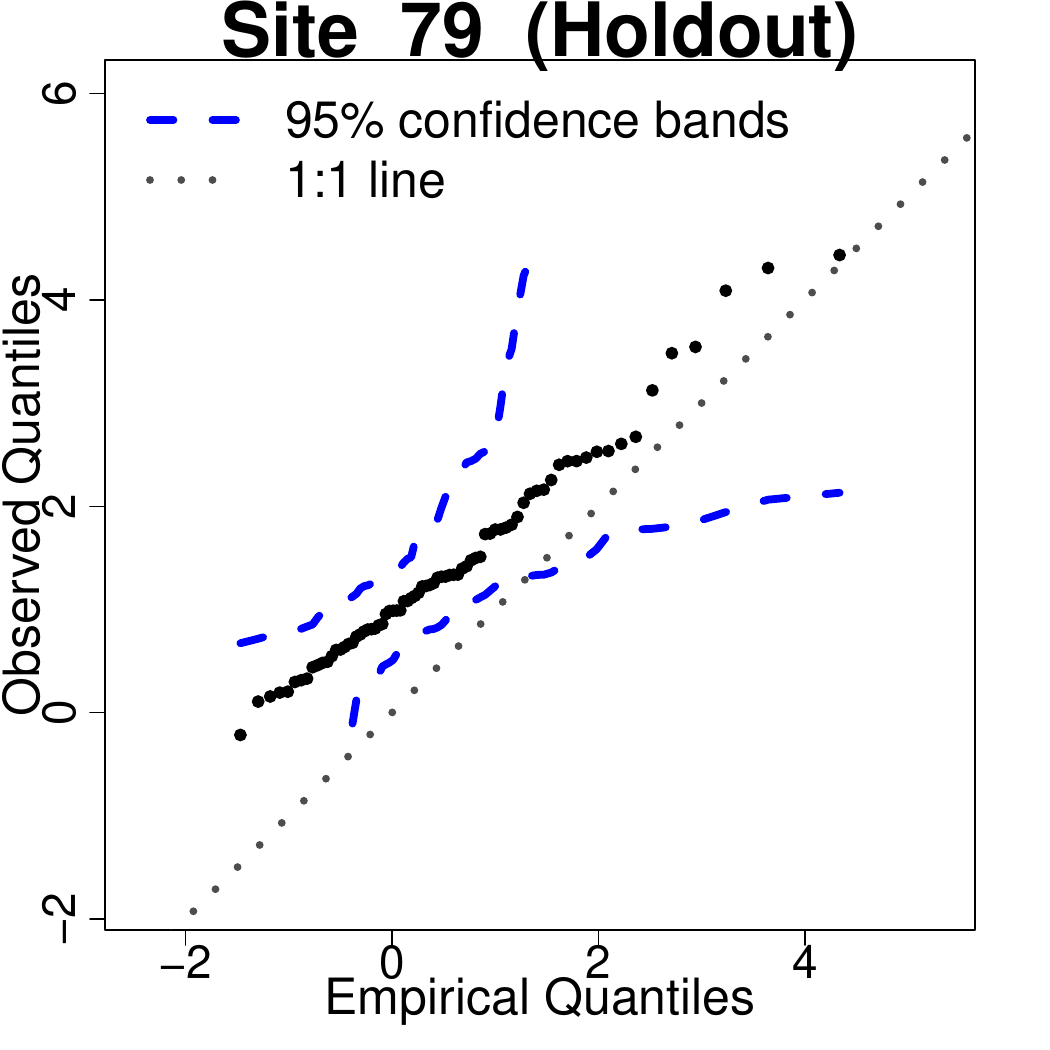}\\[2mm]
		\includegraphics[width=0.3\textwidth]{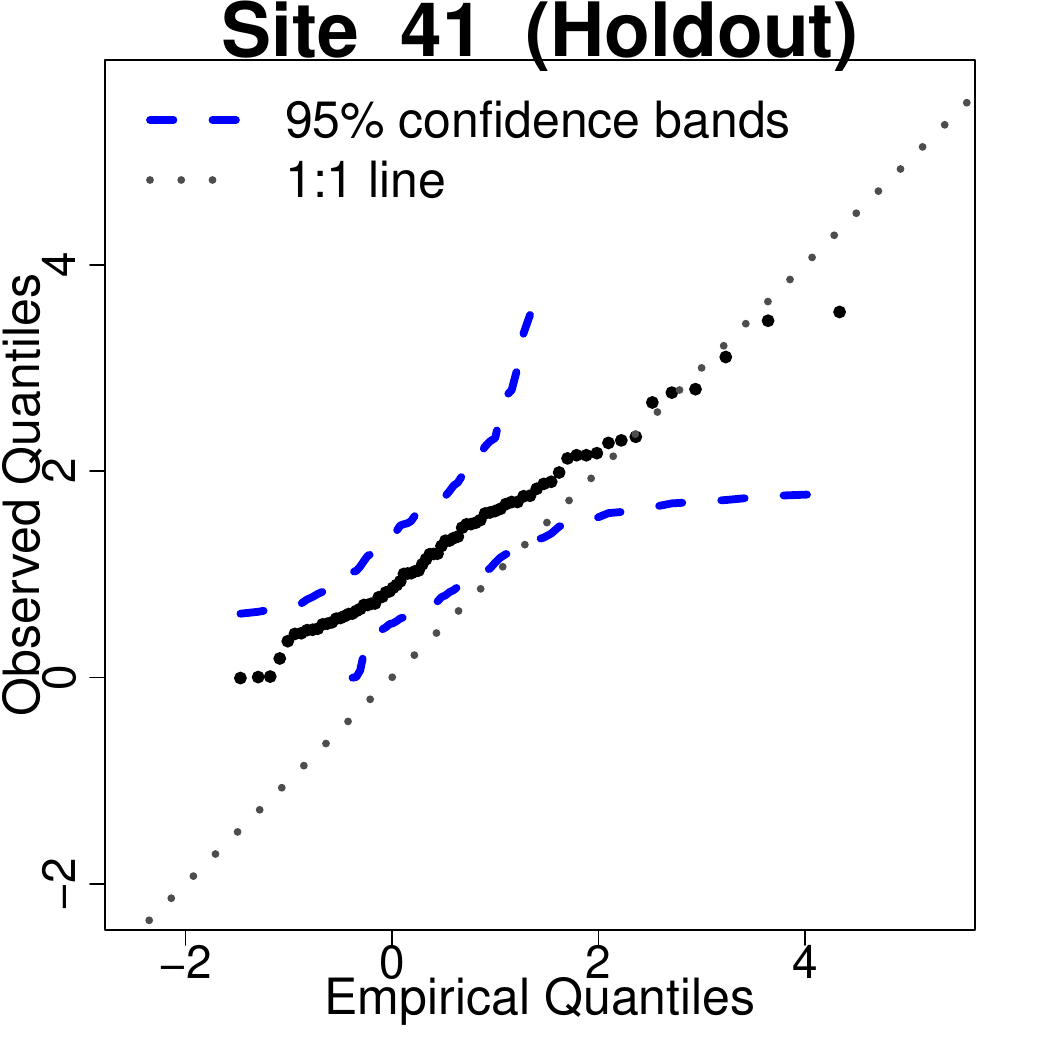}\hfill
		\includegraphics[width=0.3\textwidth]{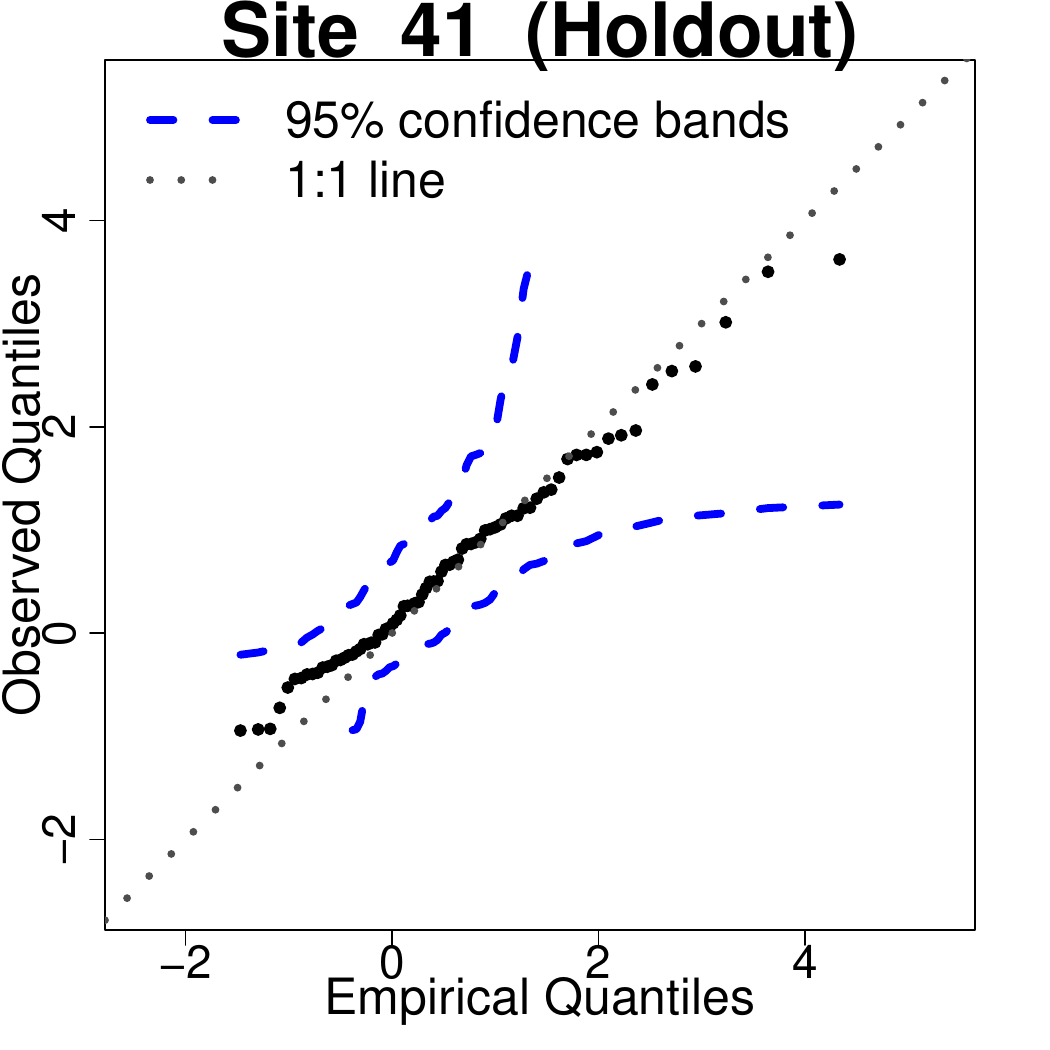}\hfill
		\includegraphics[width=0.3\textwidth]{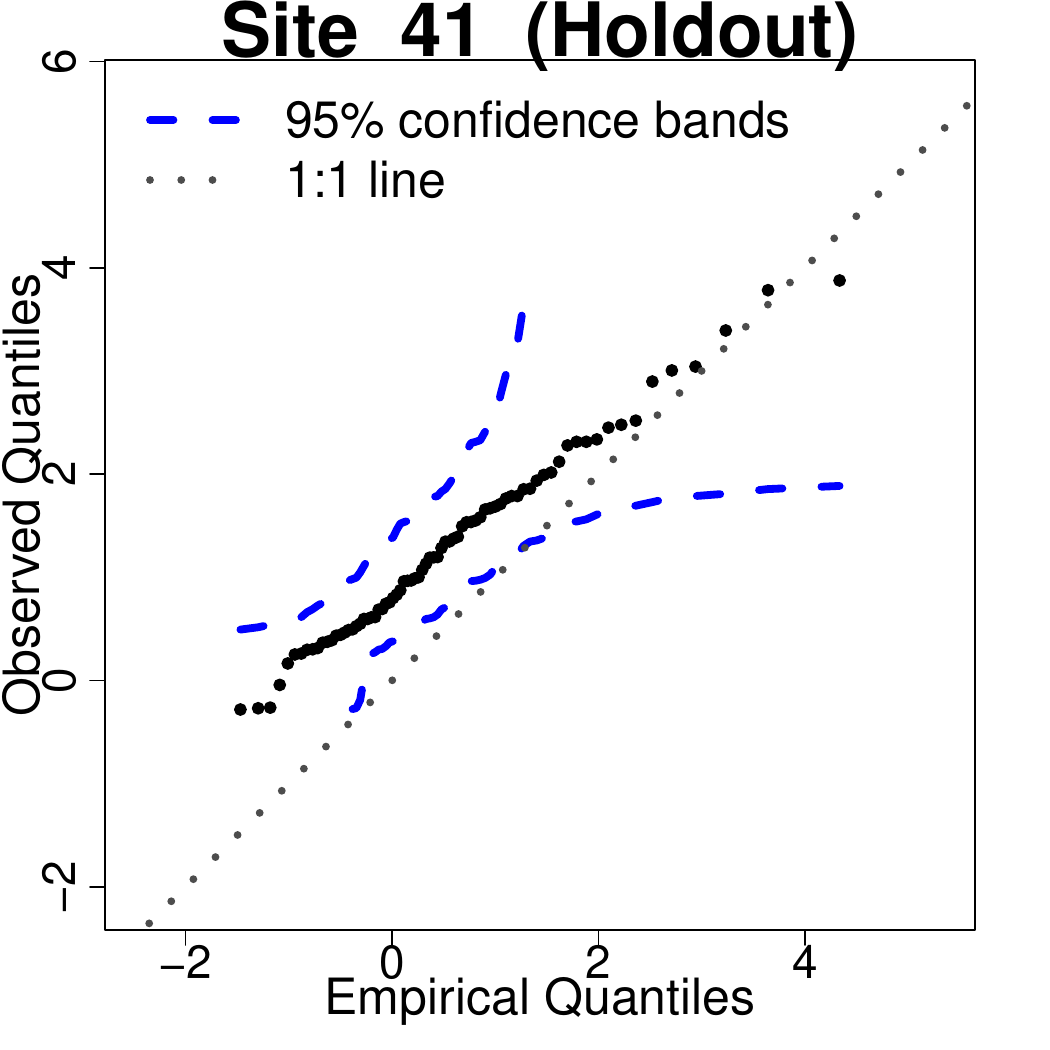}\\[2mm]
		\includegraphics[width=0.3\textwidth]{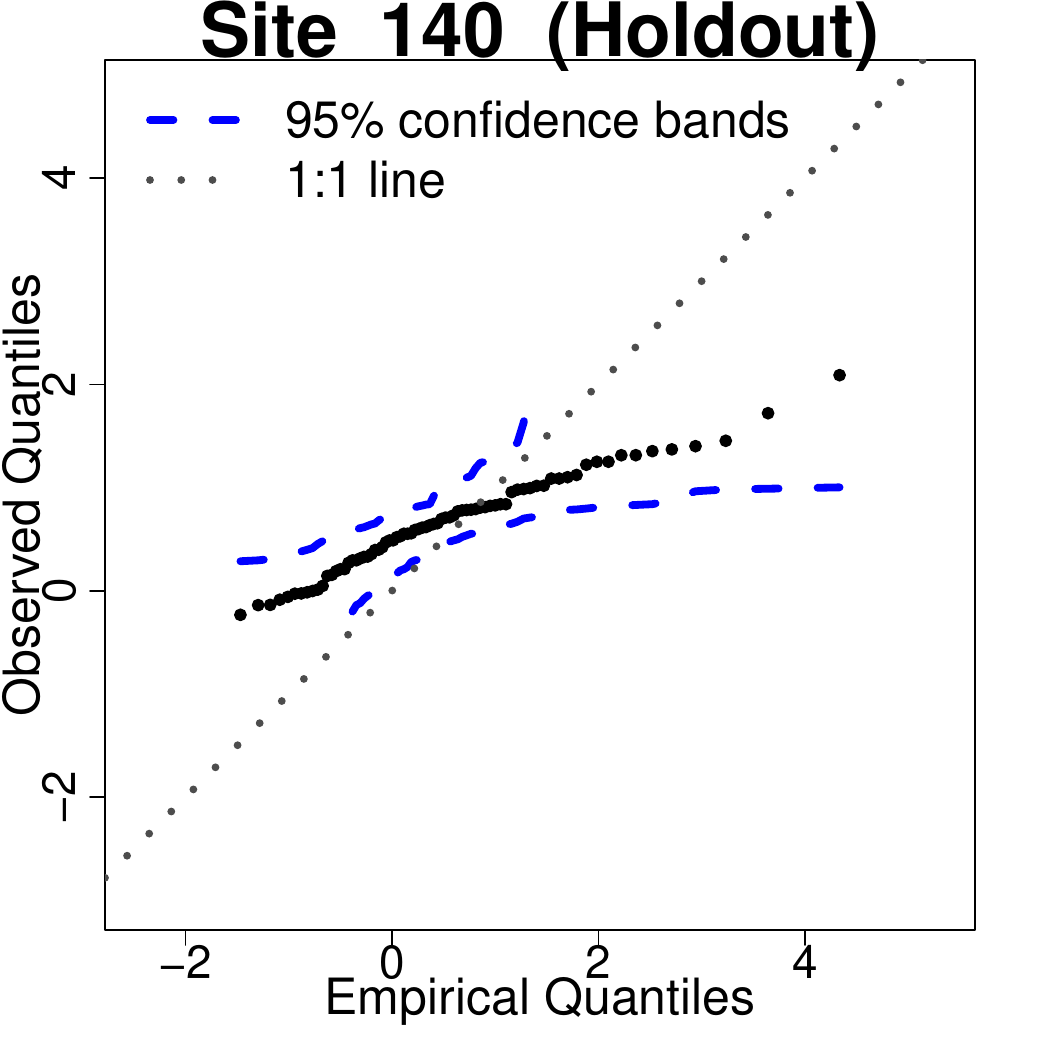}\hfill
		\includegraphics[width=0.3\textwidth]{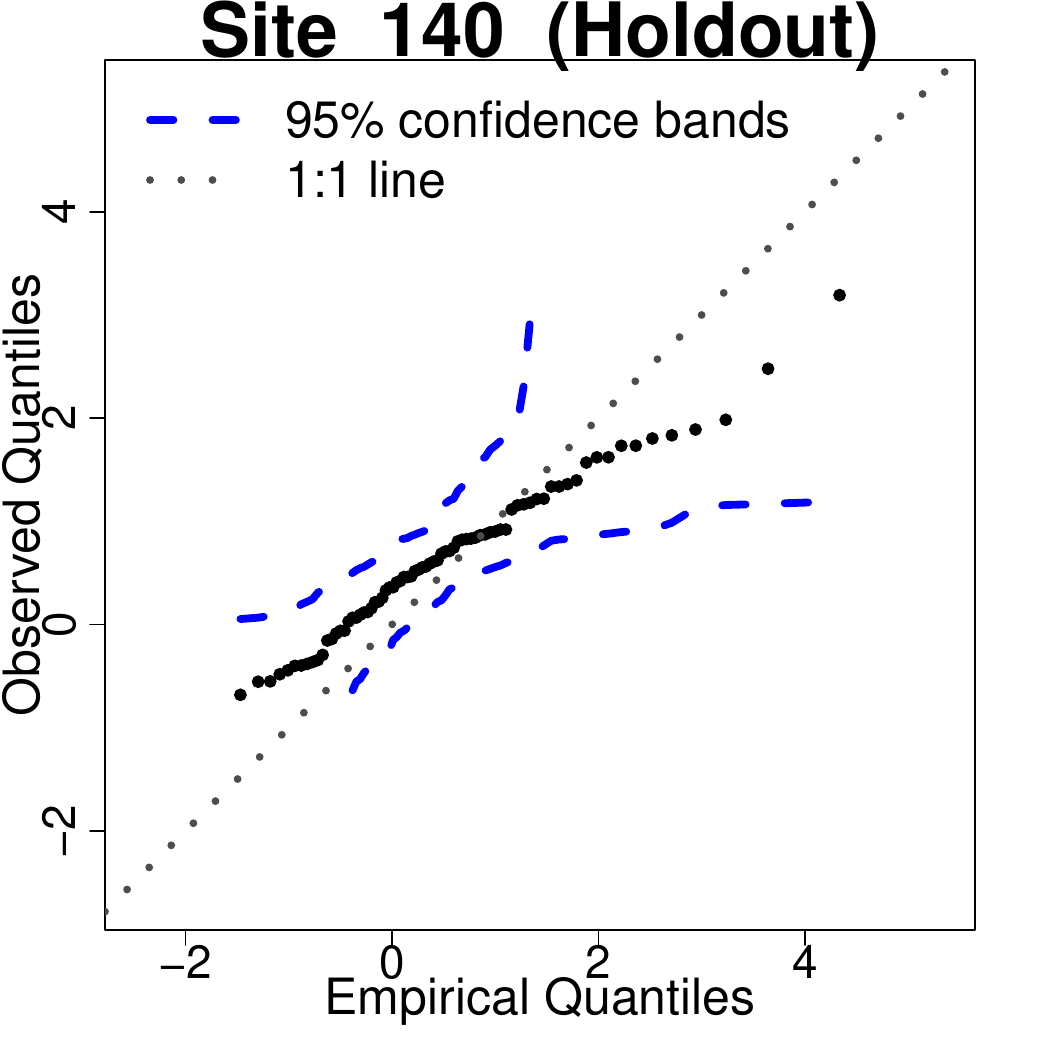}\hfill
		\includegraphics[width=0.3\textwidth]{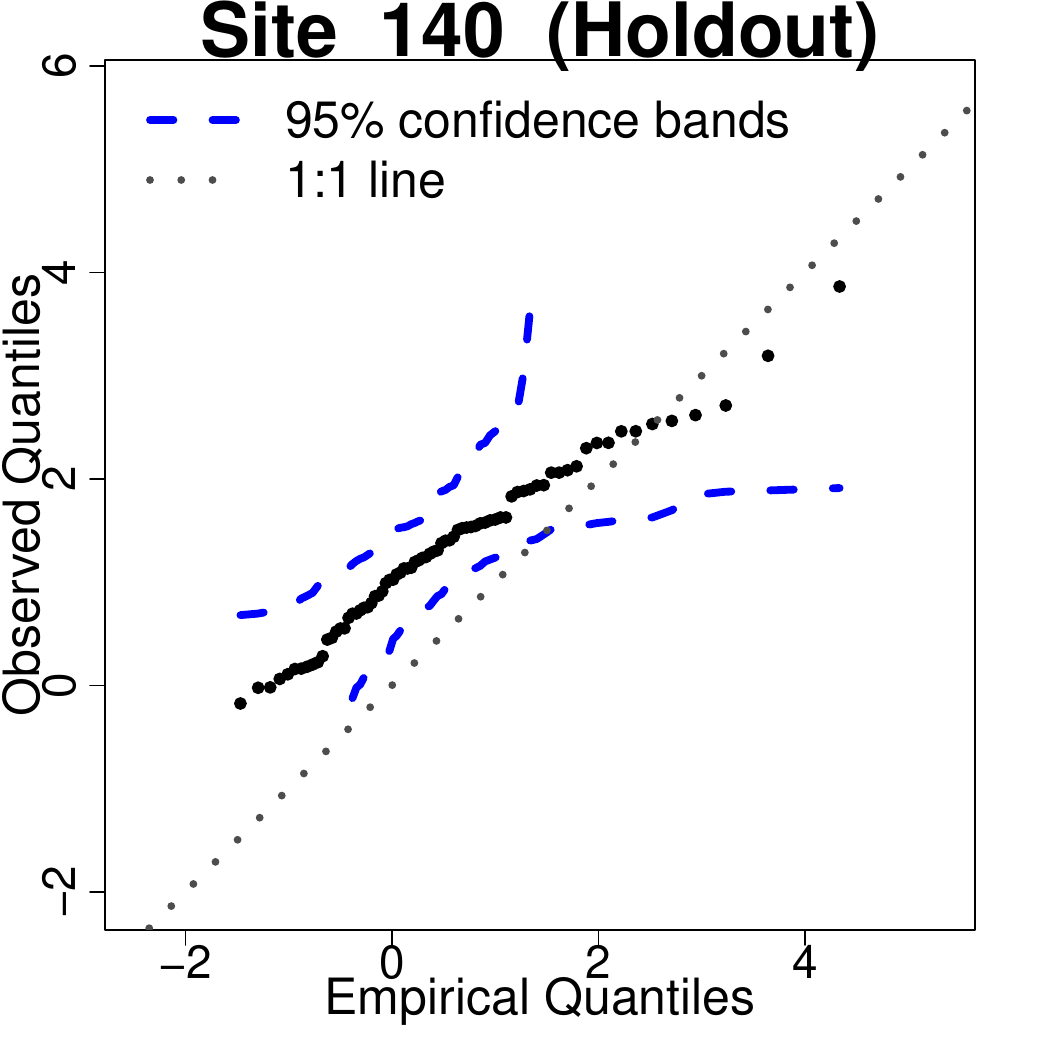}
		\caption{GEV QQ plots by site (rows: 60, 79 -- well-calibrated; 41, 140 -- poorly-calibrated) and model (columns), holdout sites.}
		\label{fig:precip_qq_holdout}
	\end{figure}
	
	Figure~\ref{fig:precip_qq_holdout} shows GEV QQ plots (posterior-averaged CDF against the observed holdout annual maxima, with 95\% confidence bands) at four holdout sites, deliberately chosen to mix two well-calibrated sites (60, 79) with two poorly-calibrated sites (41, 140), for all three models. The posterior mean GEV surfaces are not close to the site-wise MLE surfaces at every site (Section~\ref{supp:precip_comparison_gev}), yet the QQ plots diagnostics show that this does not translate into a uniformly poor fit, sites 60 and 79 track the 1:1 line closely under three models, while sites 41 and 140 show visible curvature. We read this as the signature of a spatially smoothed hierarchical margin, accurate at some sites, less so at others, rather than a wholesale failure of the GEV specification.
	
	\subsubsection{Wendland Kernel: Visibility and Shi et al. Model} \label{supp:precip_comparison_wendland}
	
	\begin{figure}[htbp]
		\centering
		\includegraphics[width=0.7\textwidth]{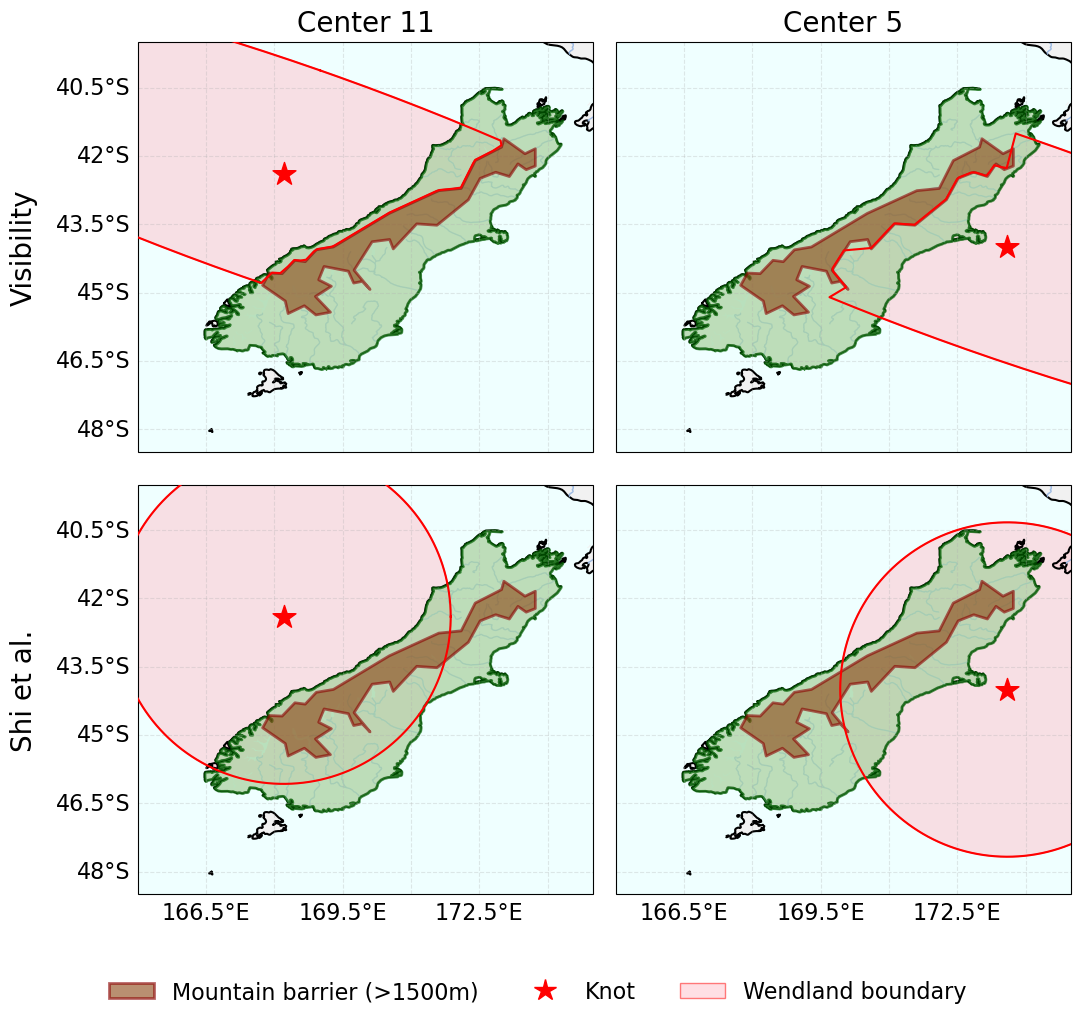}
		\caption{Wendland kernel support for two representative knots (Center 11, left; Center 5, right) under the visibility model (top row) and the Shi et al.\ model (bottom row), both fit using only the $276$ training sites.}
		\label{fig:precip_compairson_wendland_examples}
	\end{figure}
	
	Figure~\ref{fig:precip_compairson_wendland_examples} compares the Wendland kernel support at the same two knots (Center 11, Center 5) under the visibility model and the Shi et al.\ model, both fit using only the $276$ training sites from the spatial holdout described in Section~\ref{subsec:precip_comparison} of the main text (the same split used throughout the model comparison). The contrast is exactly what the two models' constructions predict, the visibility model's support is truncated at the barrier in both cases. For Center 11, it is bounded on multiple sides, reflecting that its line of sight is restricted both by the range and by the extent of the domain itself, producing a narrow wedge rather than a simple circle segment. Shi et al.'s kernel, by contrast, is isotropic and blind to the barrier by construction, so it produces a full, untruncated circle at both knots regardless of whether the barrier lies within that radius. This is a direct visual confirmation of the structural difference, rather than a fitted result. The shape of each panel follows immediately from which model generated it, independent of which sites were used to fit it.
	
	\subsubsection{Tail Dependence Surface: Visibility and Shi et al. Model} \label{supp:precip_comparison_tail_surface}
	
	\begin{figure}[htbp]
		\centering
		\includegraphics[width=0.7\textwidth]{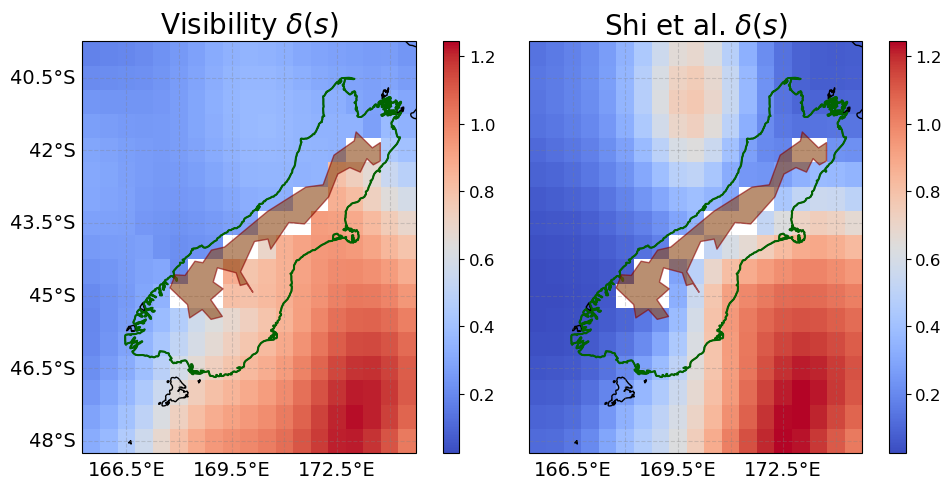}
		\caption{Posterior mean of the site-level tail-dependence parameter $\delta(\bm{s})$ under the visibility model (left) and the Shi et al.\ model (right), both fit using only the $276$ training sites.}
		\label{fig:precip_comparison_delta_posterior_mean}
	\end{figure}
	
	Figure~\ref{fig:precip_comparison_delta_posterior_mean} compares the posterior mean of $\delta(\bm{s})$ under the visibility model and the Shi et al.\ model, again both fit to the $276$ training sites only. The two surfaces are qualitatively similar, both show the same low-windward/high-leeward gradient described in Section~\ref{supp:precip_delta} for the full-data fit. This is notable because, unlike the visibility model, Shi et al.'s Gaussian kernel has no visibility restriction built into it at all, its $\delta(\bm{s})$ is a purely barrier-blind spatial smooth. That a similar windward-leeward asymmetry emerges under both models, and under a reduced (training-only) set of sites, is evidence that this pattern is being driven by the data through the likelihood rather than being an artifact of the visibility restriction specifically built into the visibility model's kernel, or of which sites happen to be included in the fit.
	
	\subsubsection{Extremal Dependence: All Three Models} \label{supp:precip_comparison_chiu}
	
	\begin{figure}[htbp]
		\centering
		\begin{tabular}{cccc}
			Empirical & Visibility & Shi et al. & Huser--Wadsworth \\
			\includegraphics[width=0.2\textwidth, clip=true, trim=0 0 30 40]{precip_emp_chiu_inland_center_105_90.pdf} &
			\includegraphics[width=0.2\textwidth, clip=true, trim=0 0 30 40]{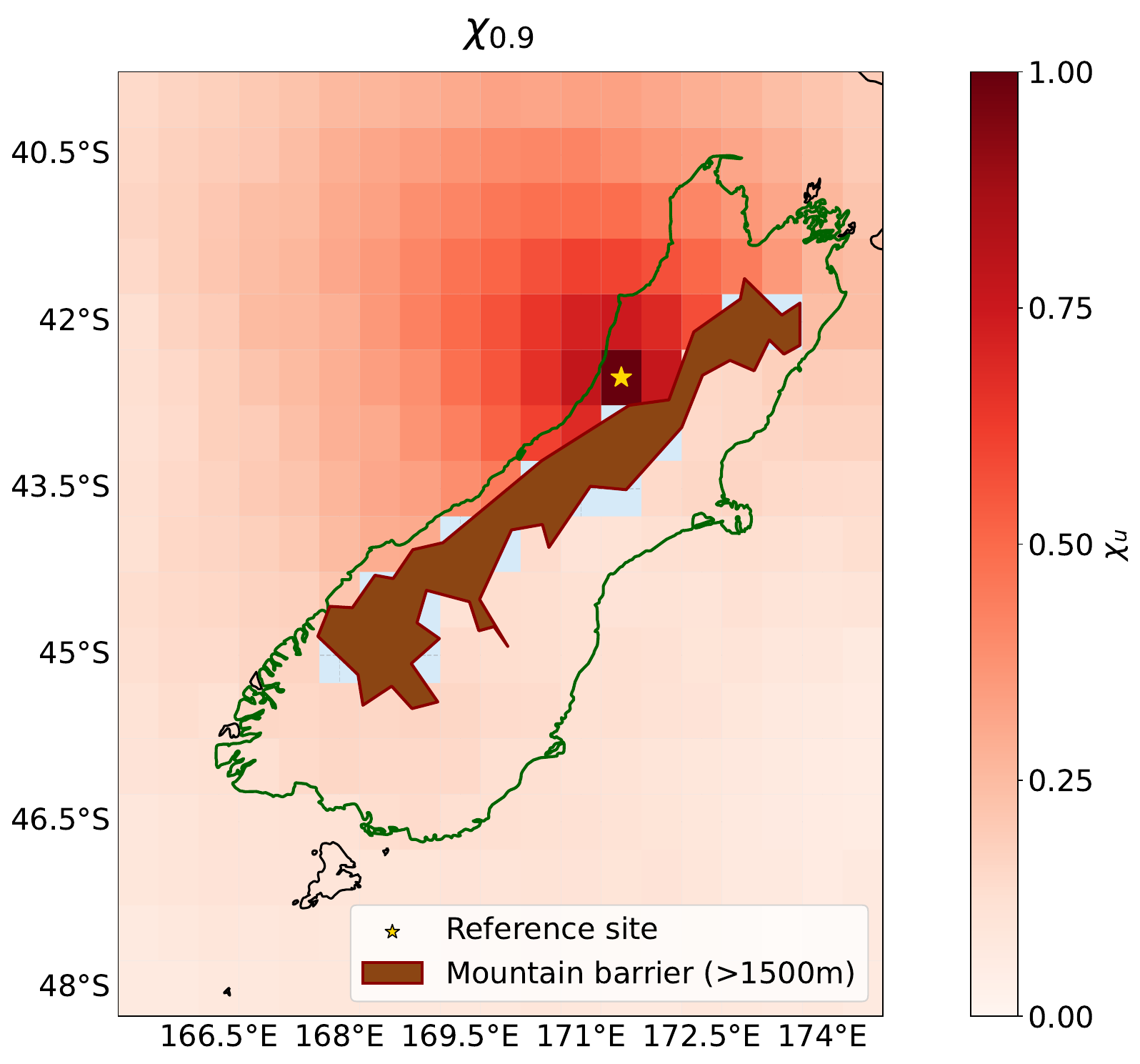} &
			\includegraphics[width=0.2\textwidth, clip=true, trim=0 0 30 40]{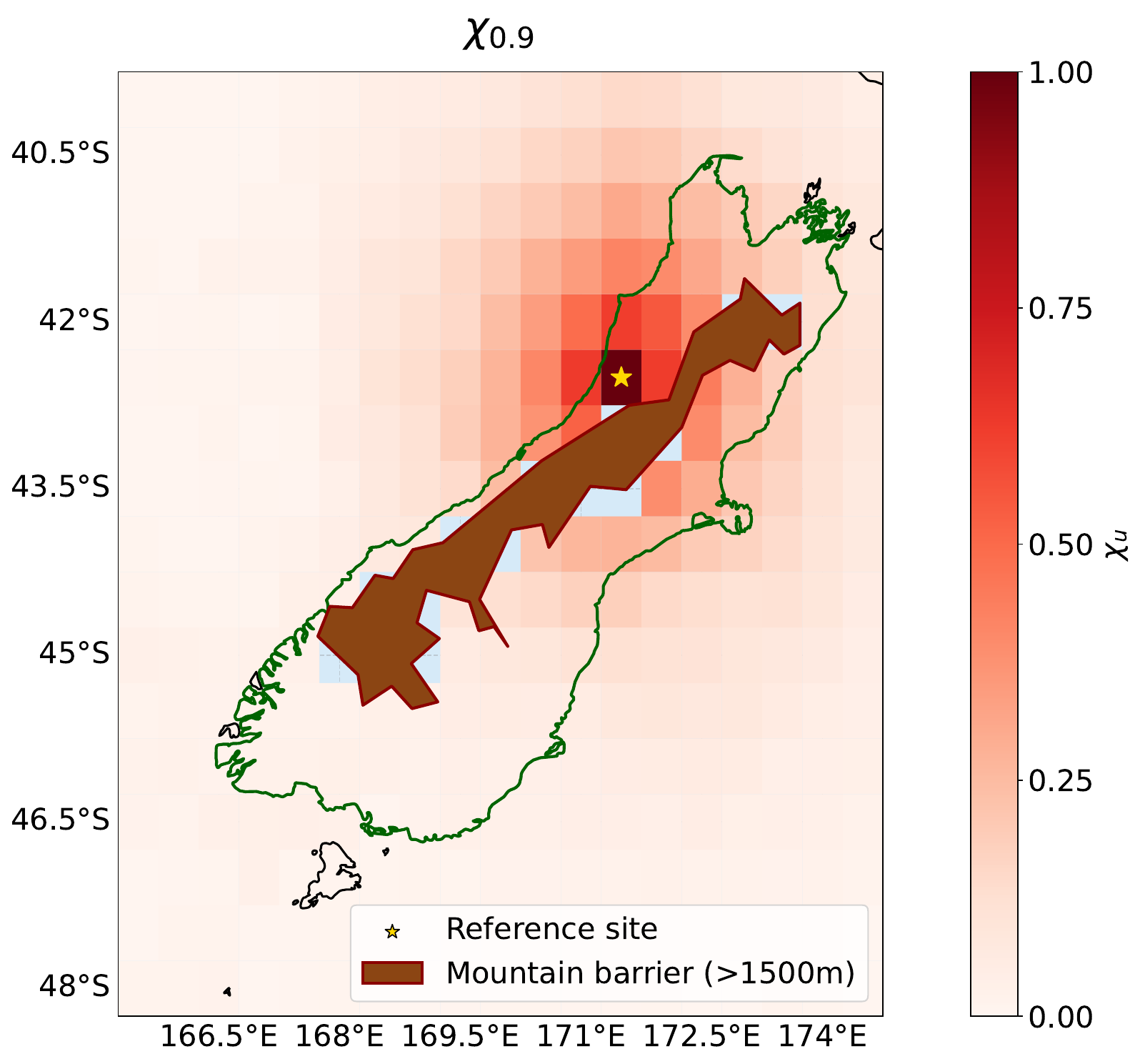} &
			\includegraphics[width=0.2\textwidth, clip=true, trim=0 0 30 40]{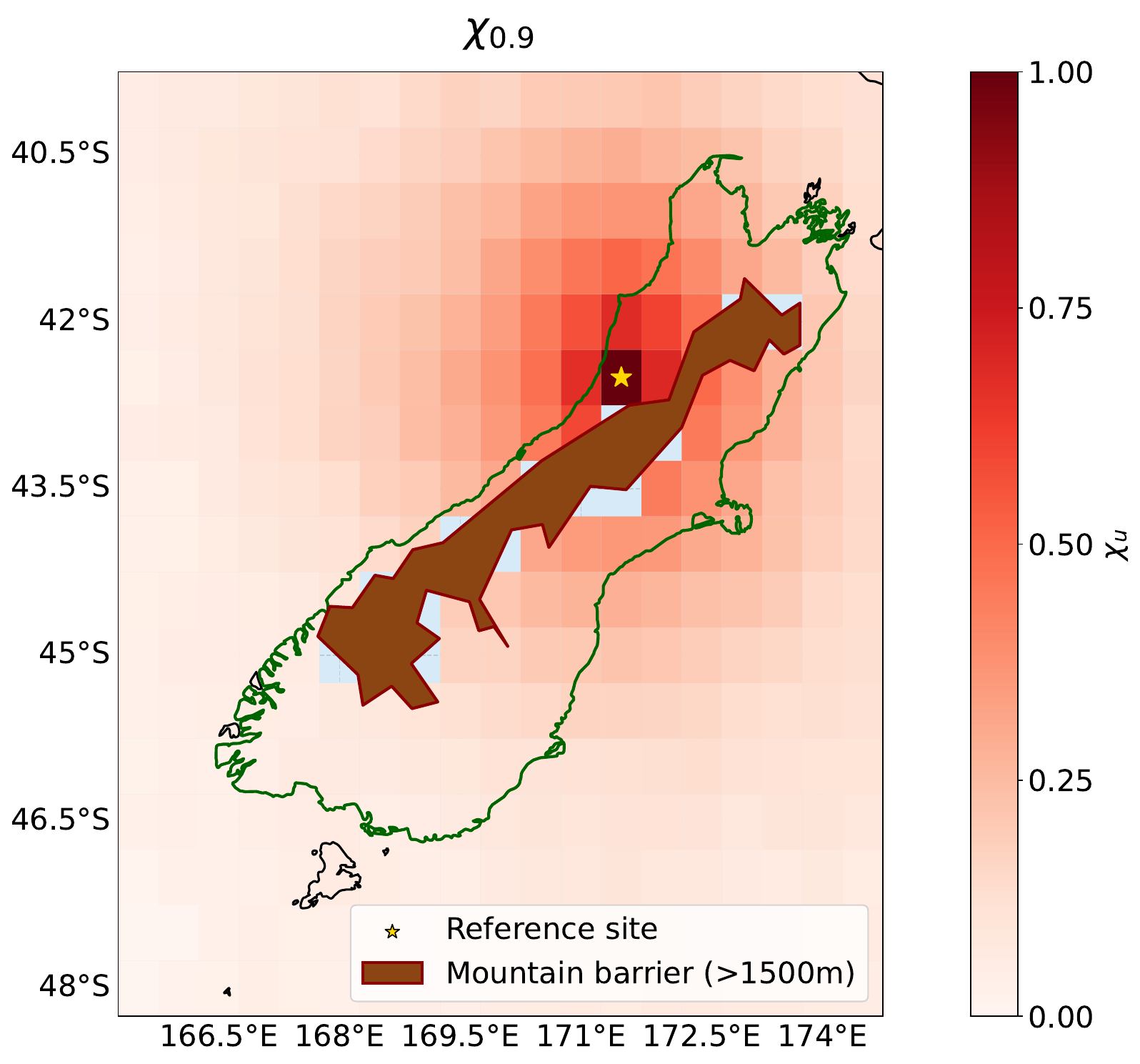}
			\\
			\includegraphics[width=0.2\textwidth, clip=true, trim=0 0 30 40]{precip_emp_chiu_inland_center_167_90.pdf} &
			\includegraphics[width=0.2\textwidth, clip=true, trim=0 0 30 40]{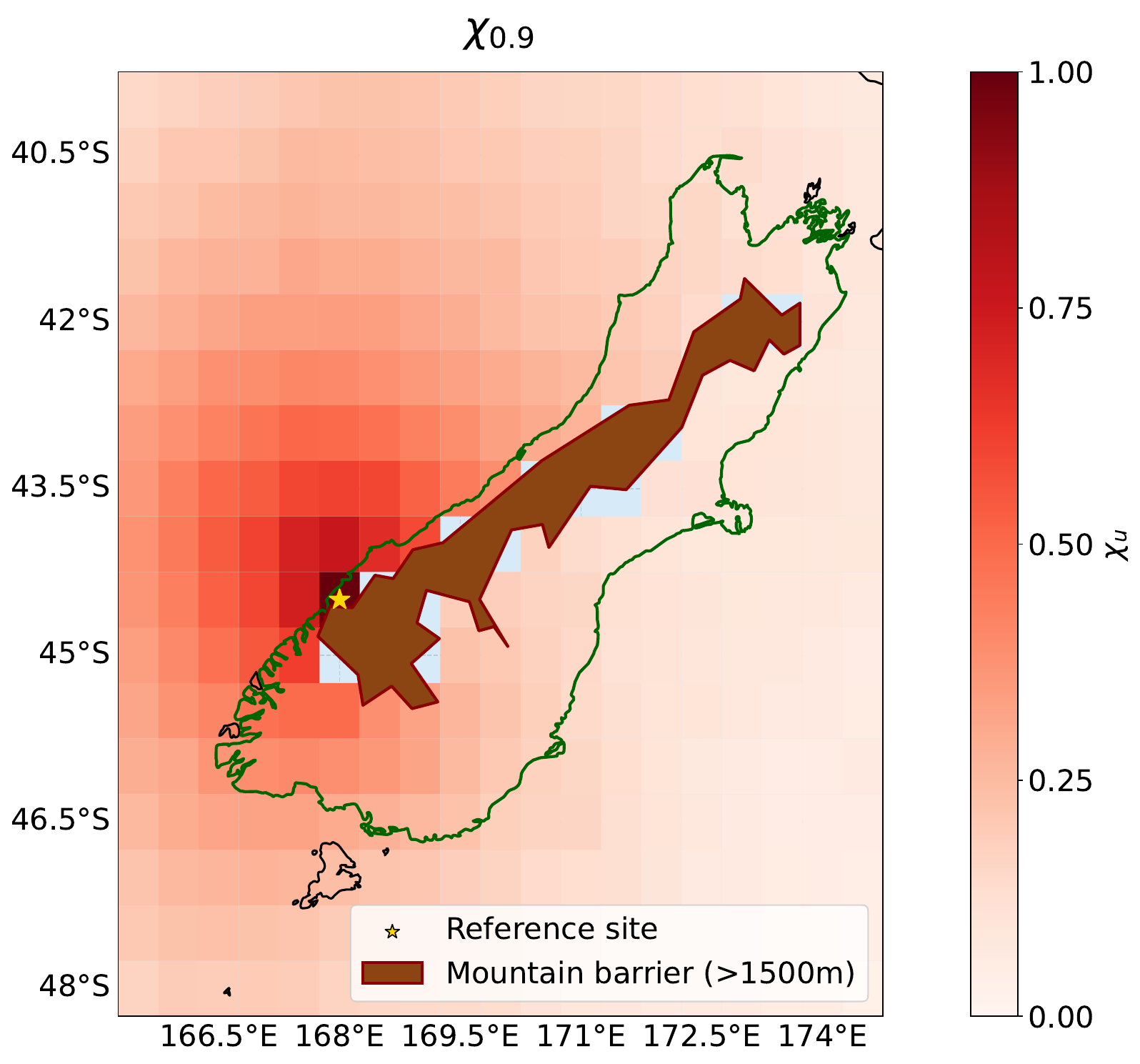} &
			\includegraphics[width=0.2\textwidth, clip=true, trim=0 0 30 40]{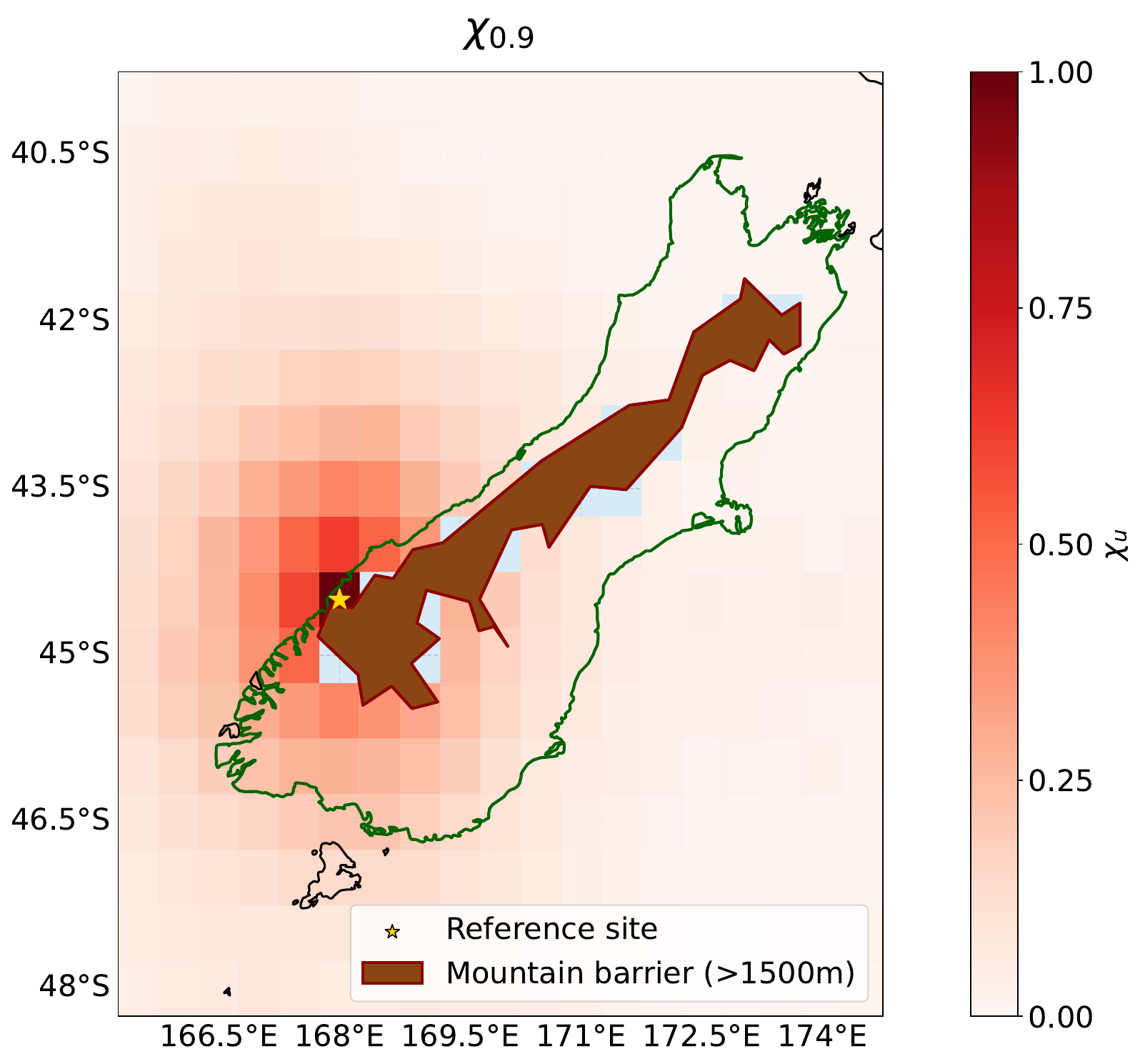} &
			\includegraphics[width=0.2\textwidth, clip=true, trim=0 0 30 40]{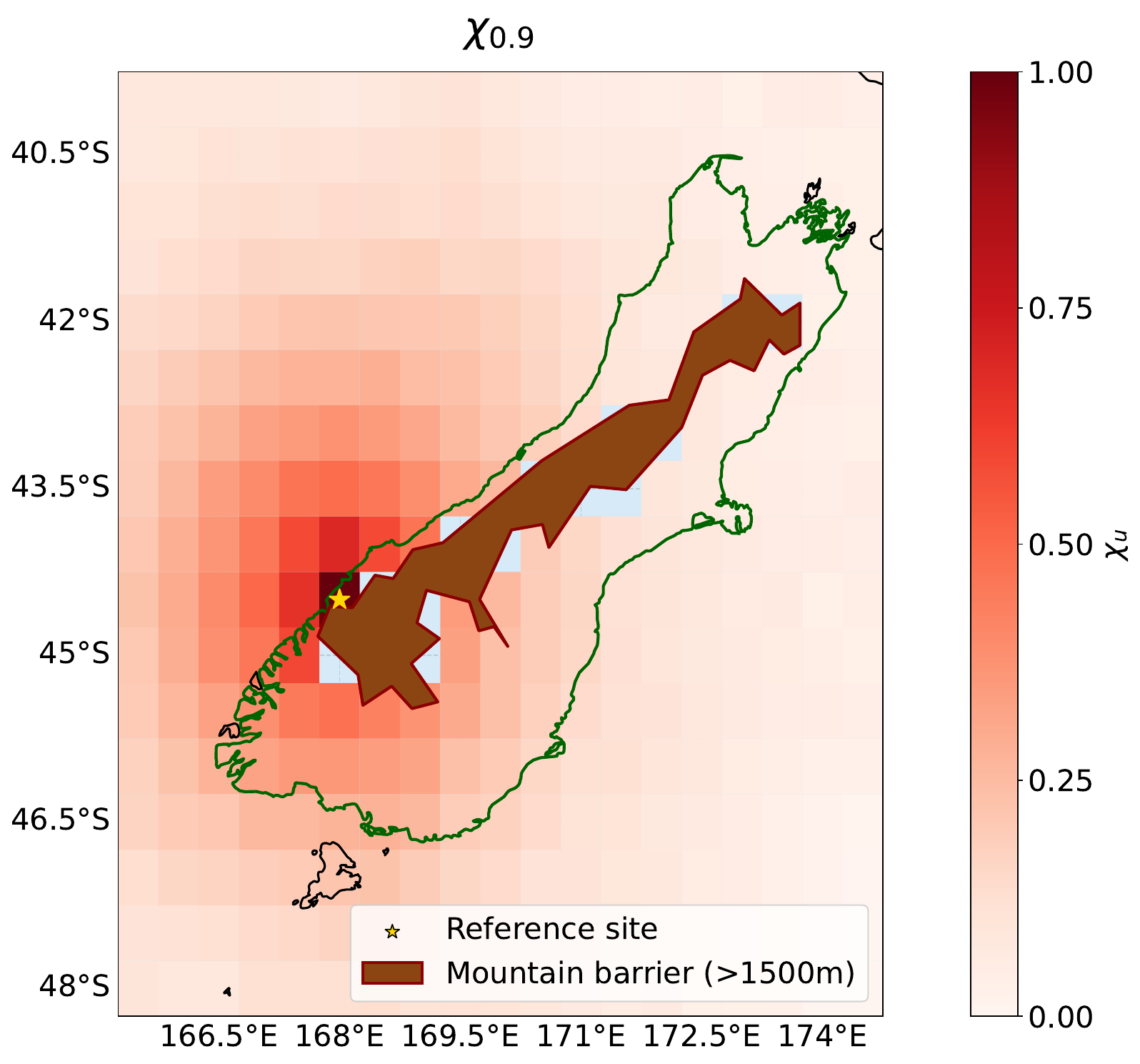}
			\\
			\includegraphics[width=0.2\textwidth, clip=true, trim=0 0 30 40]{precip_emp_chiu_inland_center_140_90.pdf} &
			\includegraphics[width=0.2\textwidth, clip=true, trim=0 0 30 40]{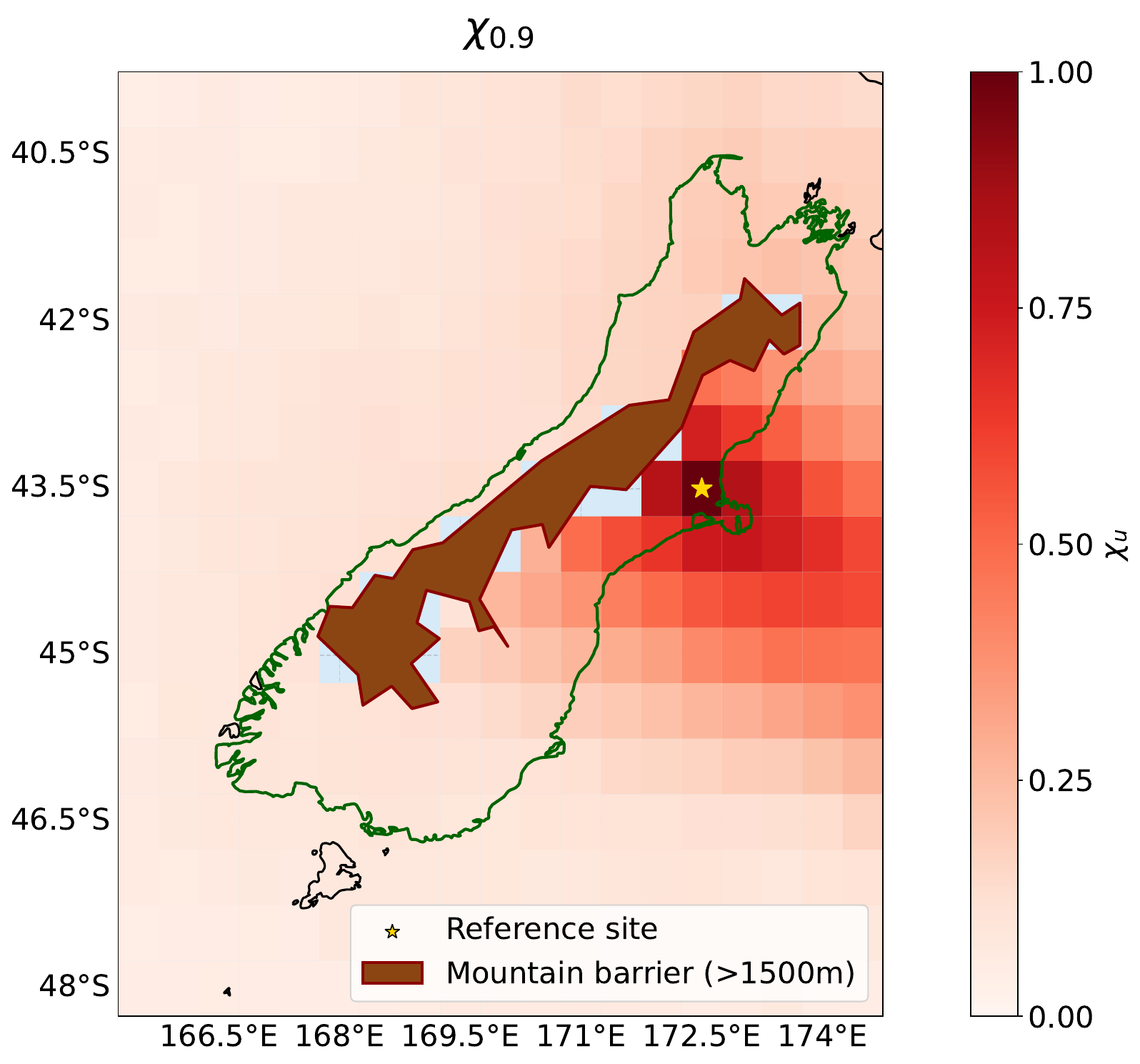} &
			\includegraphics[width=0.2\textwidth, clip=true, trim=0 0 30 40]{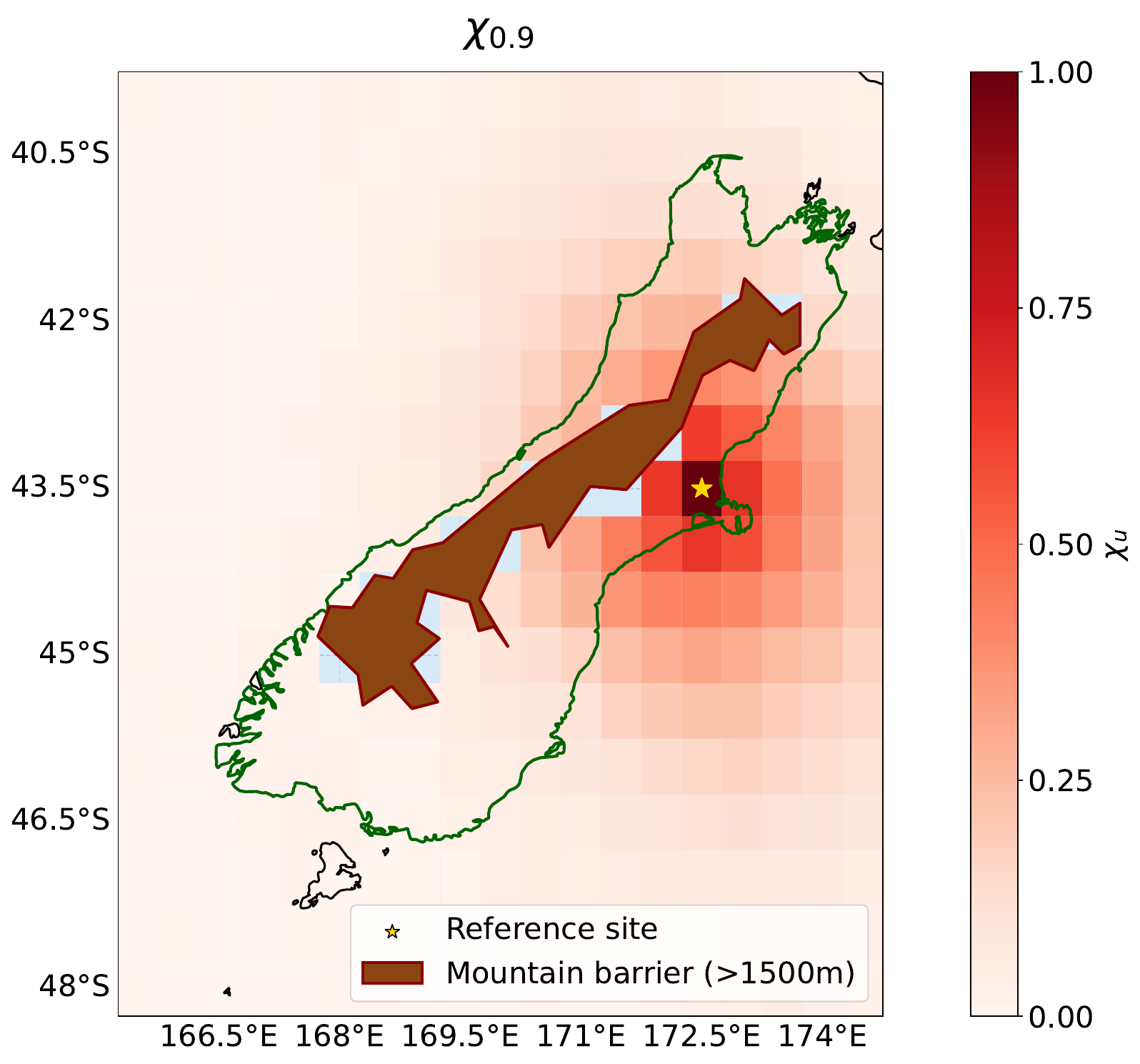} &
			\includegraphics[width=0.2\textwidth, clip=true, trim=0 0 30 40]{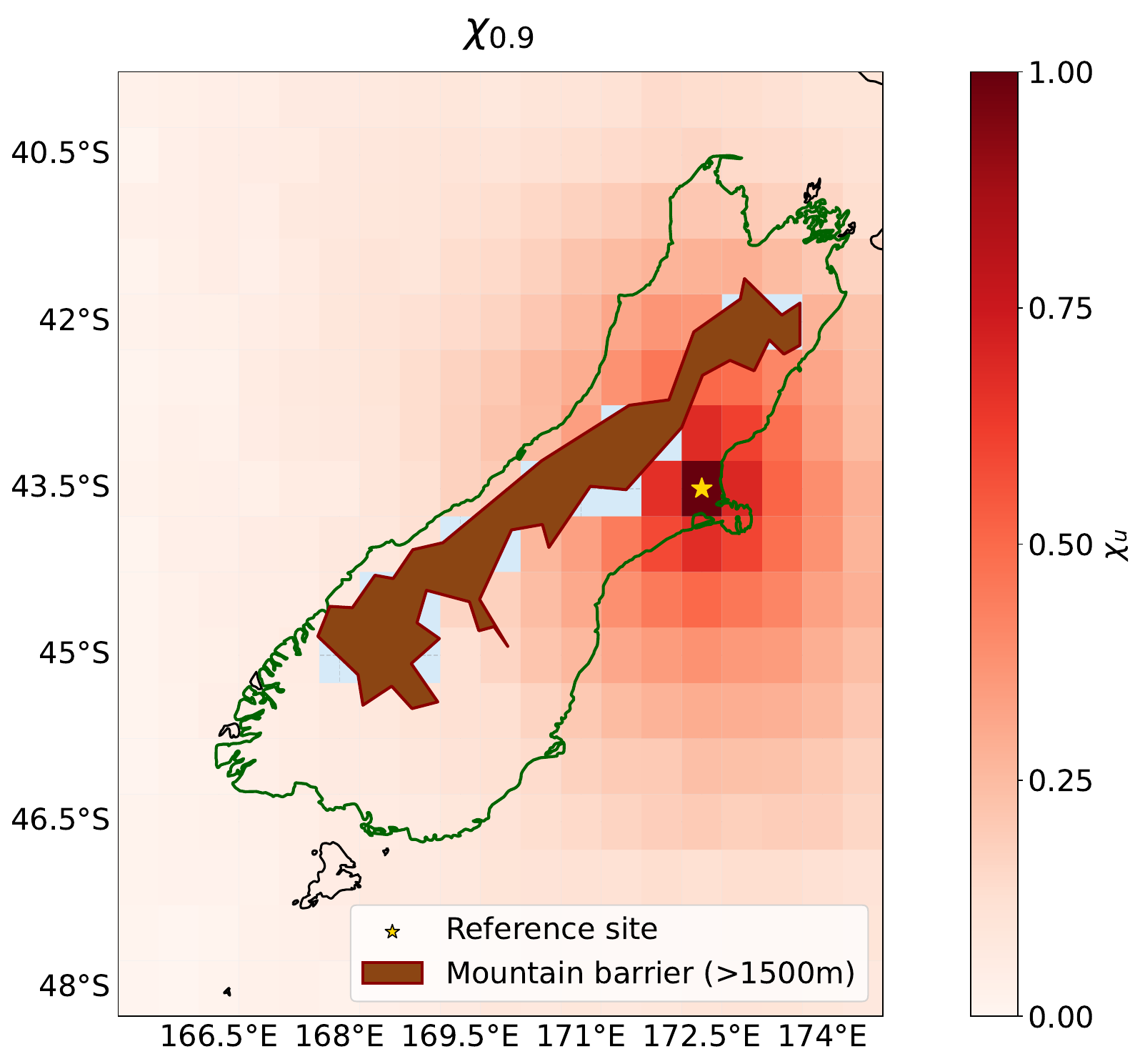}
			\\
			\includegraphics[width=0.2\textwidth, clip=true, trim=0 0 30 40]{precip_emp_chiu_inland_center_184_90.pdf} &
			\includegraphics[width=0.2\textwidth, clip=true, trim=0 0 30 40]{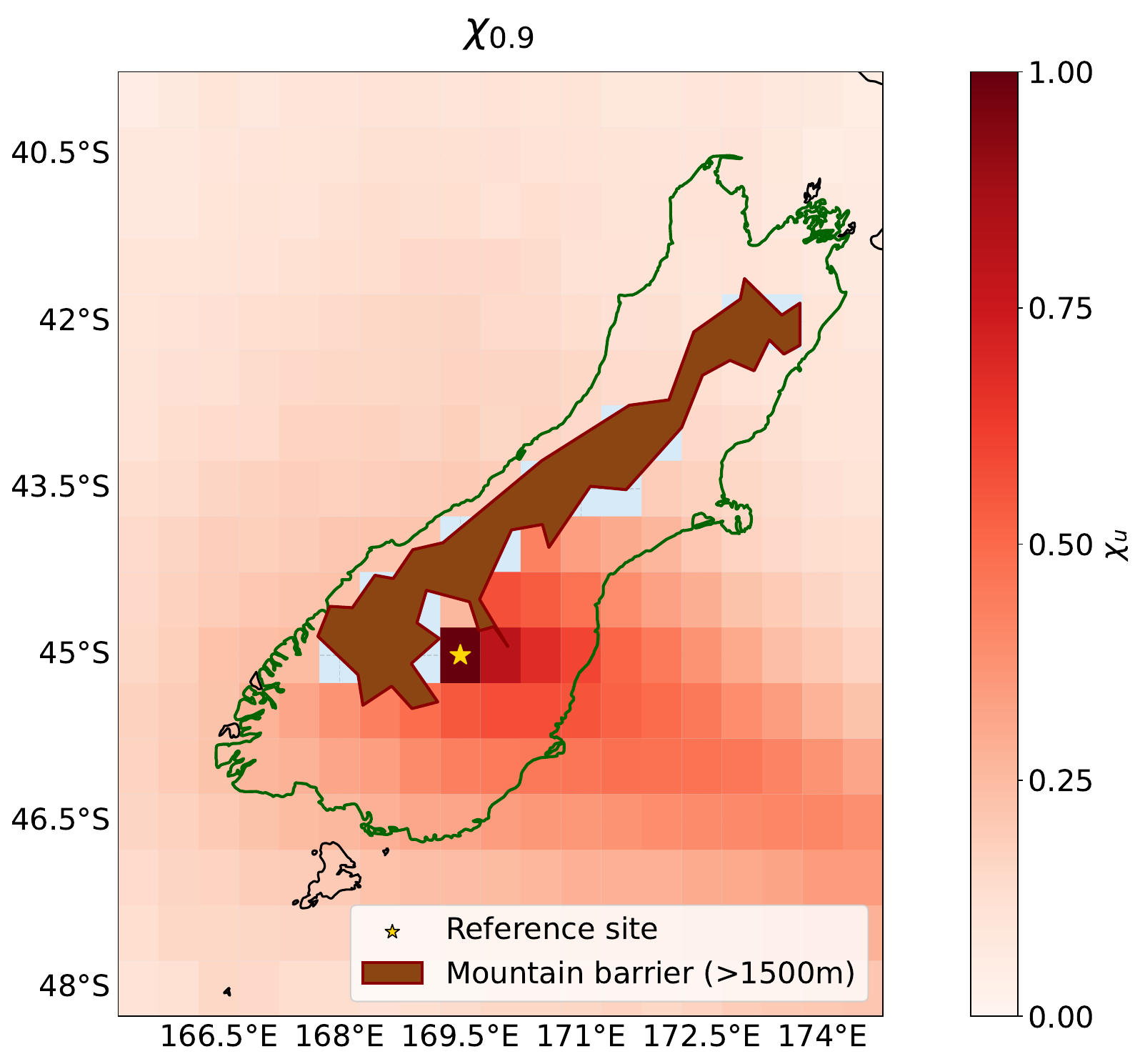} &
			\includegraphics[width=0.2\textwidth, clip=true, trim=0 0 30 40]{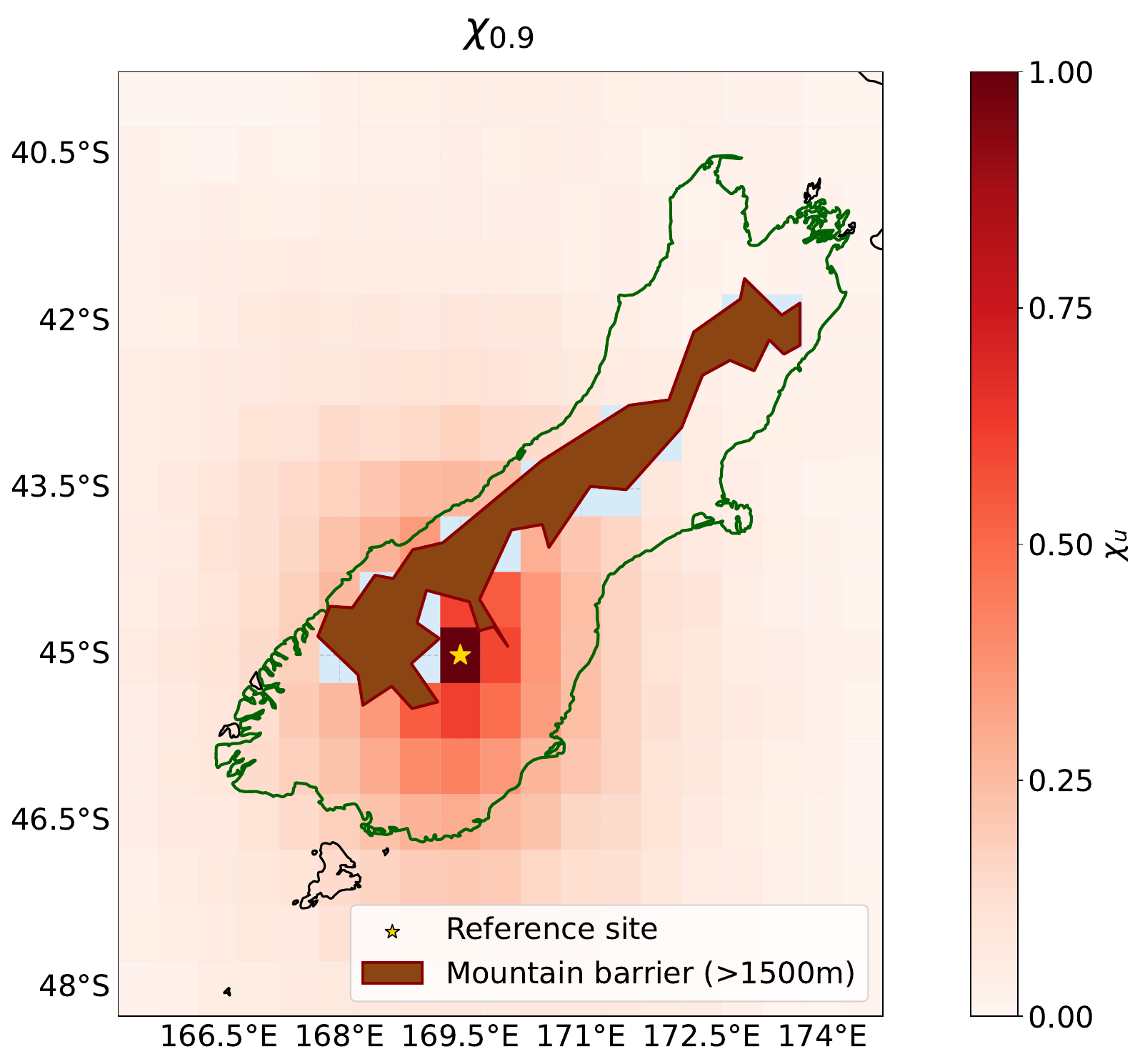} &
			\includegraphics[width=0.2\textwidth, clip=true, trim=0 0 30 40]{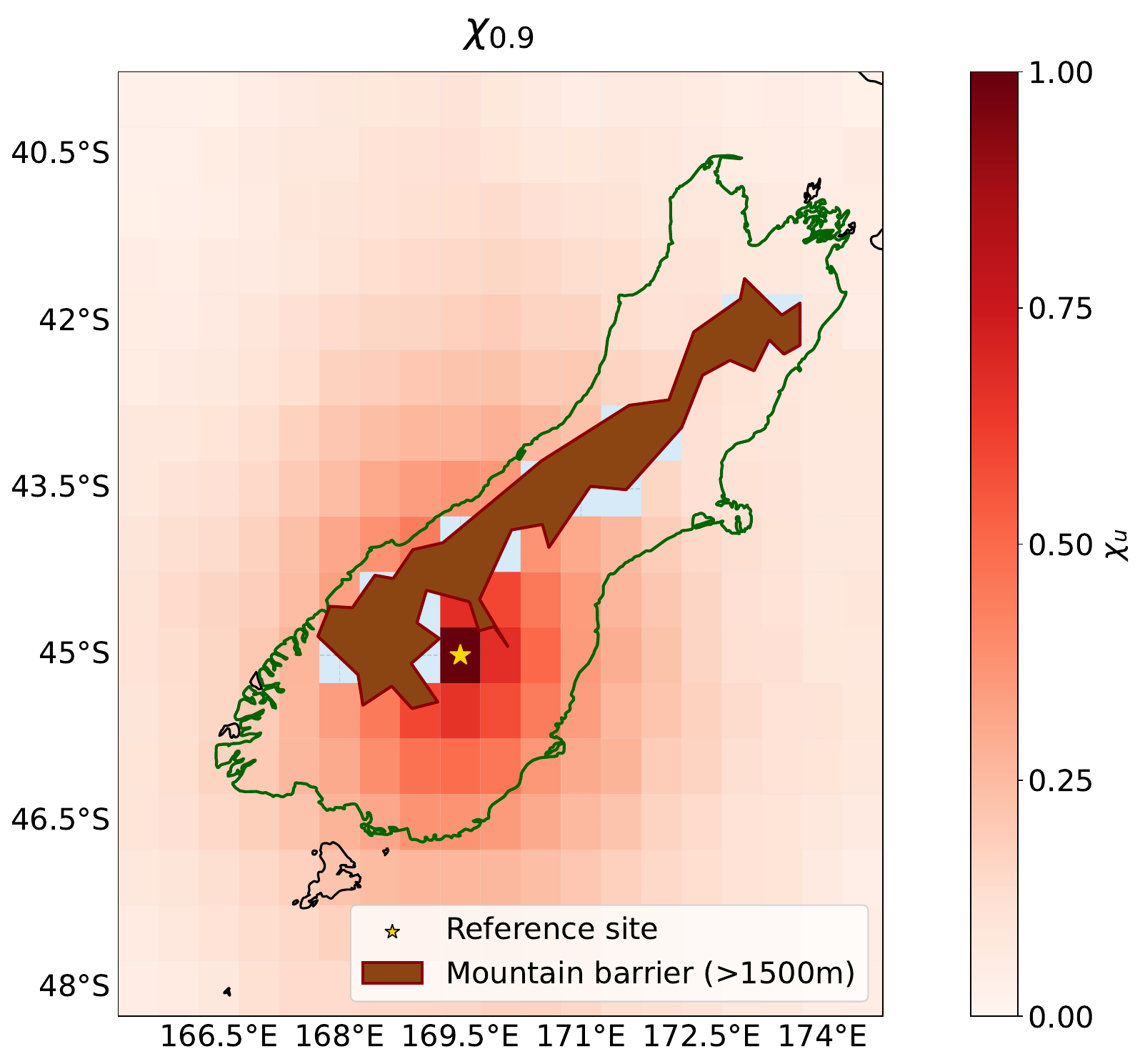}
		\end{tabular}
		\caption{Empirical and model-implied $\chi_{0.90}$ surfaces around a reference site, for all three models.}
		\label{fig:precip_chiu_model_comparison_90}
	\end{figure}
	
	\begin{figure}[htbp]
		\centering
		\begin{tabular}{cccc}
			Empirical & Visibility & Shi et al. & Huser--Wadsworth \\
			\includegraphics[width=0.2\textwidth, clip=true, trim=0 0 30 40]{precip_emp_chiu_inland_center_105_95.pdf} &
			\includegraphics[width=0.2\textwidth, clip=true, trim=0 0 30 40]{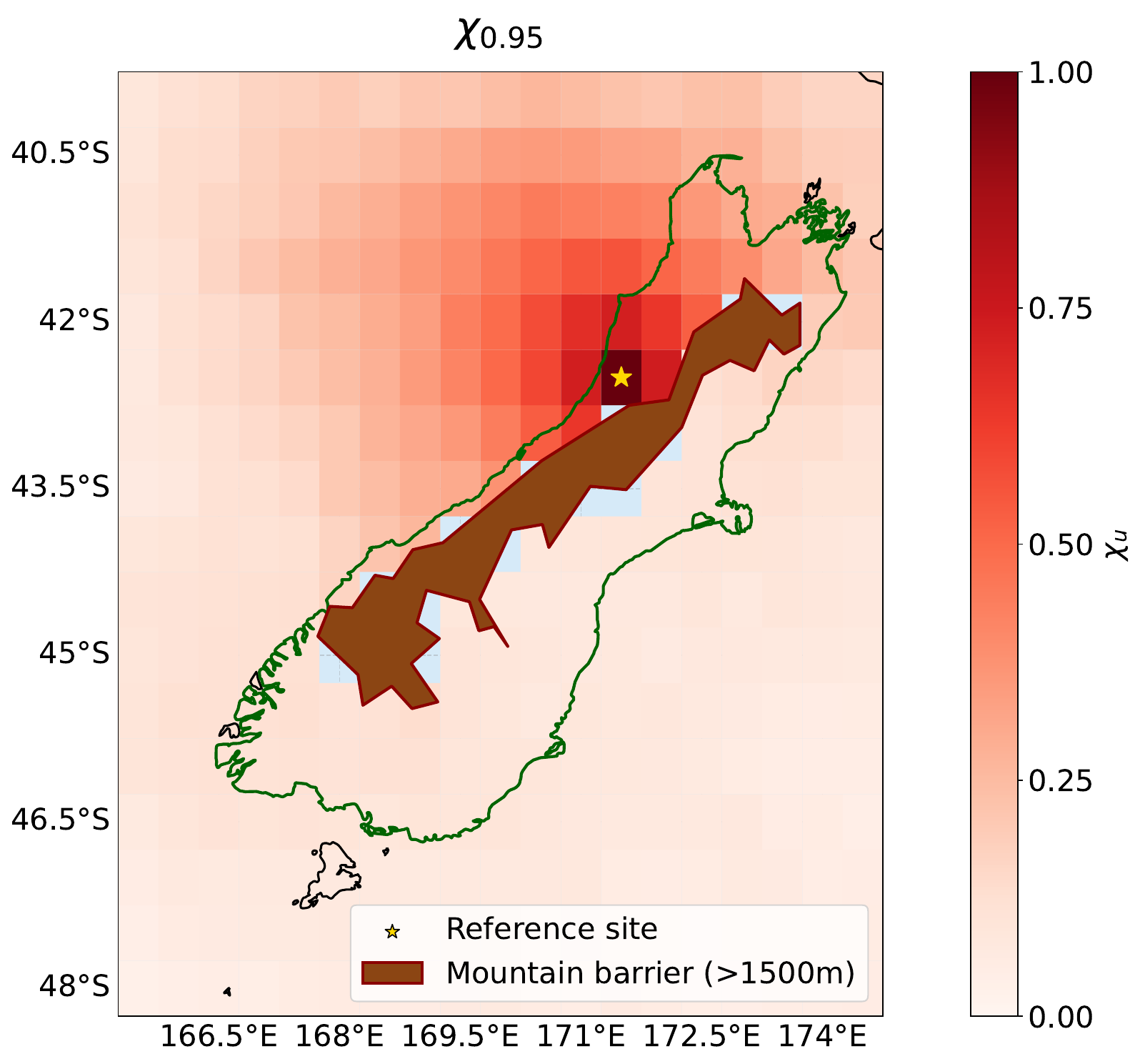} &
			\includegraphics[width=0.2\textwidth, clip=true, trim=0 0 30 40]{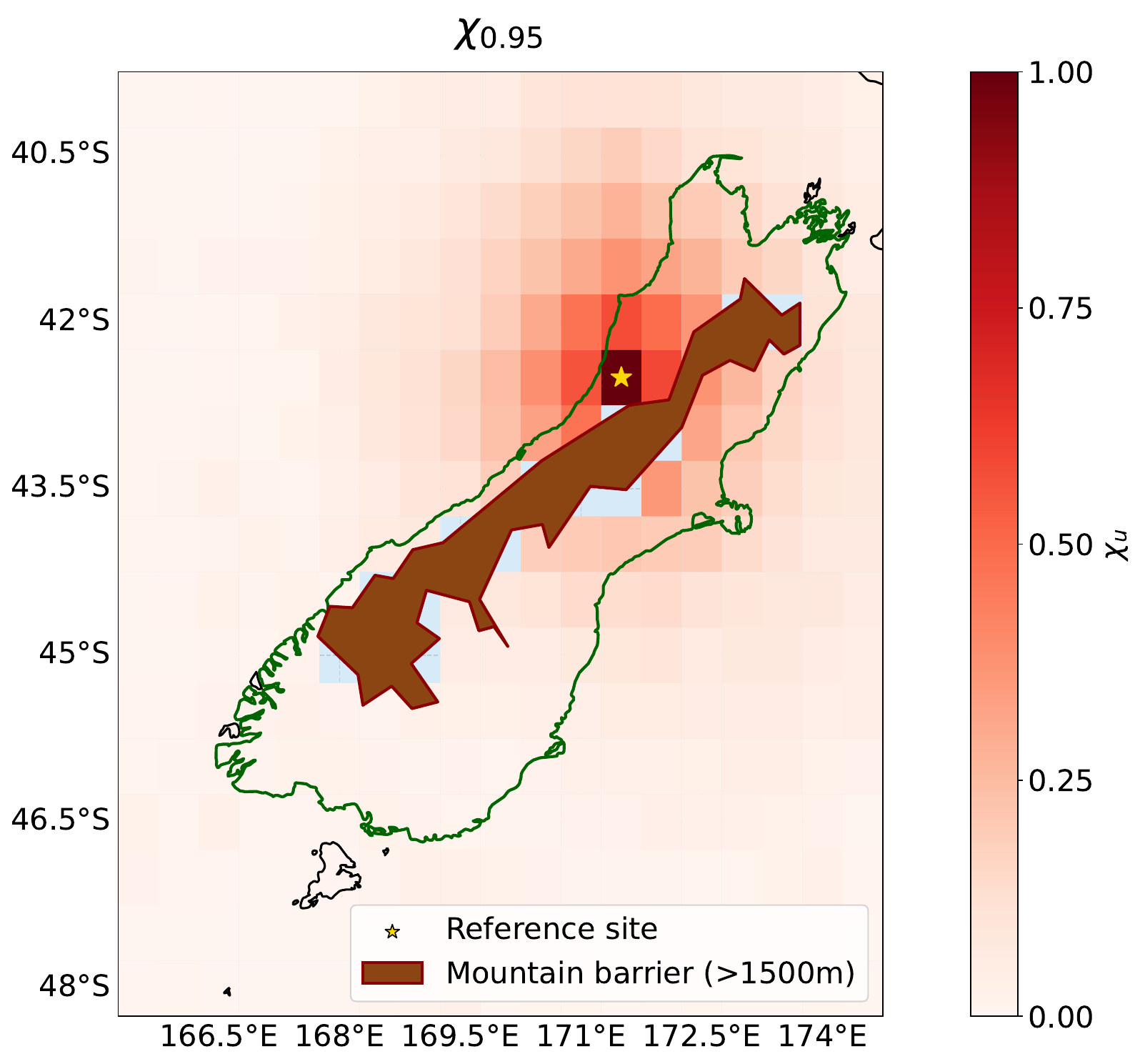} &
			\includegraphics[width=0.2\textwidth, clip=true, trim=0 0 30 40]{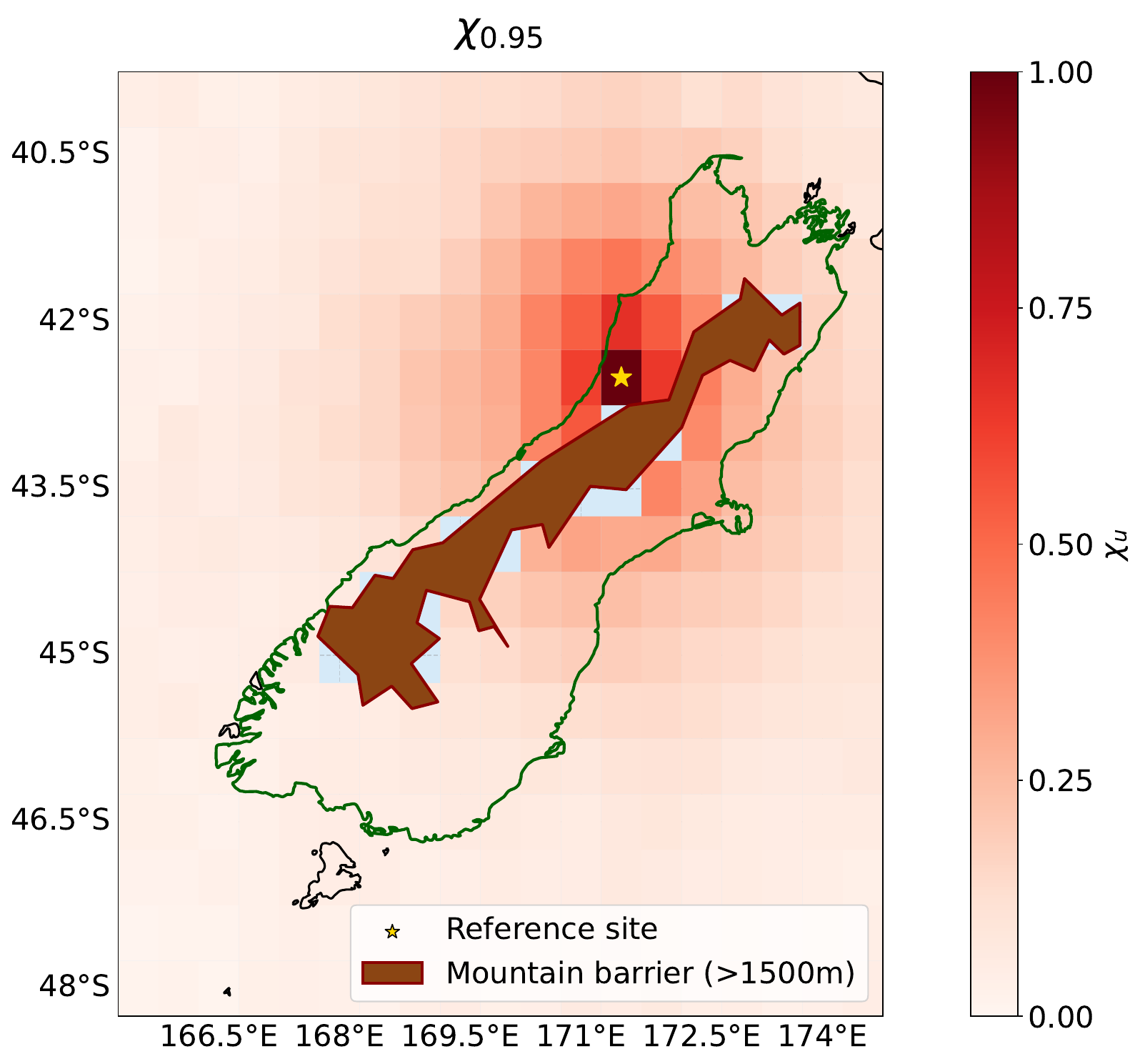}
			\\
			\includegraphics[width=0.2\textwidth, clip=true, trim=0 0 30 40]{precip_emp_chiu_inland_center_167_95.pdf} &
			\includegraphics[width=0.2\textwidth, clip=true, trim=0 0 30 40]{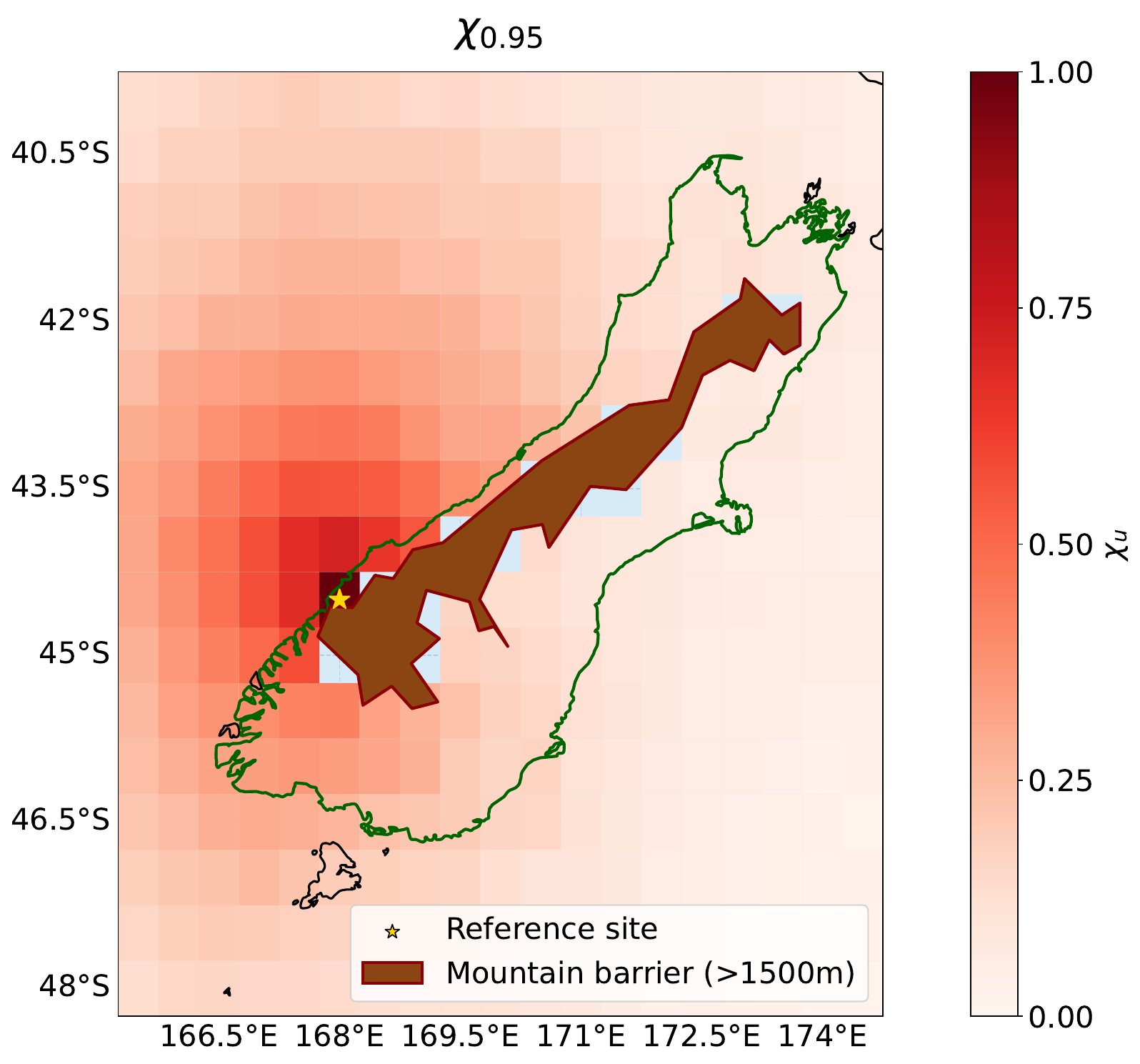} &
			\includegraphics[width=0.2\textwidth, clip=true, trim=0 0 30 40]{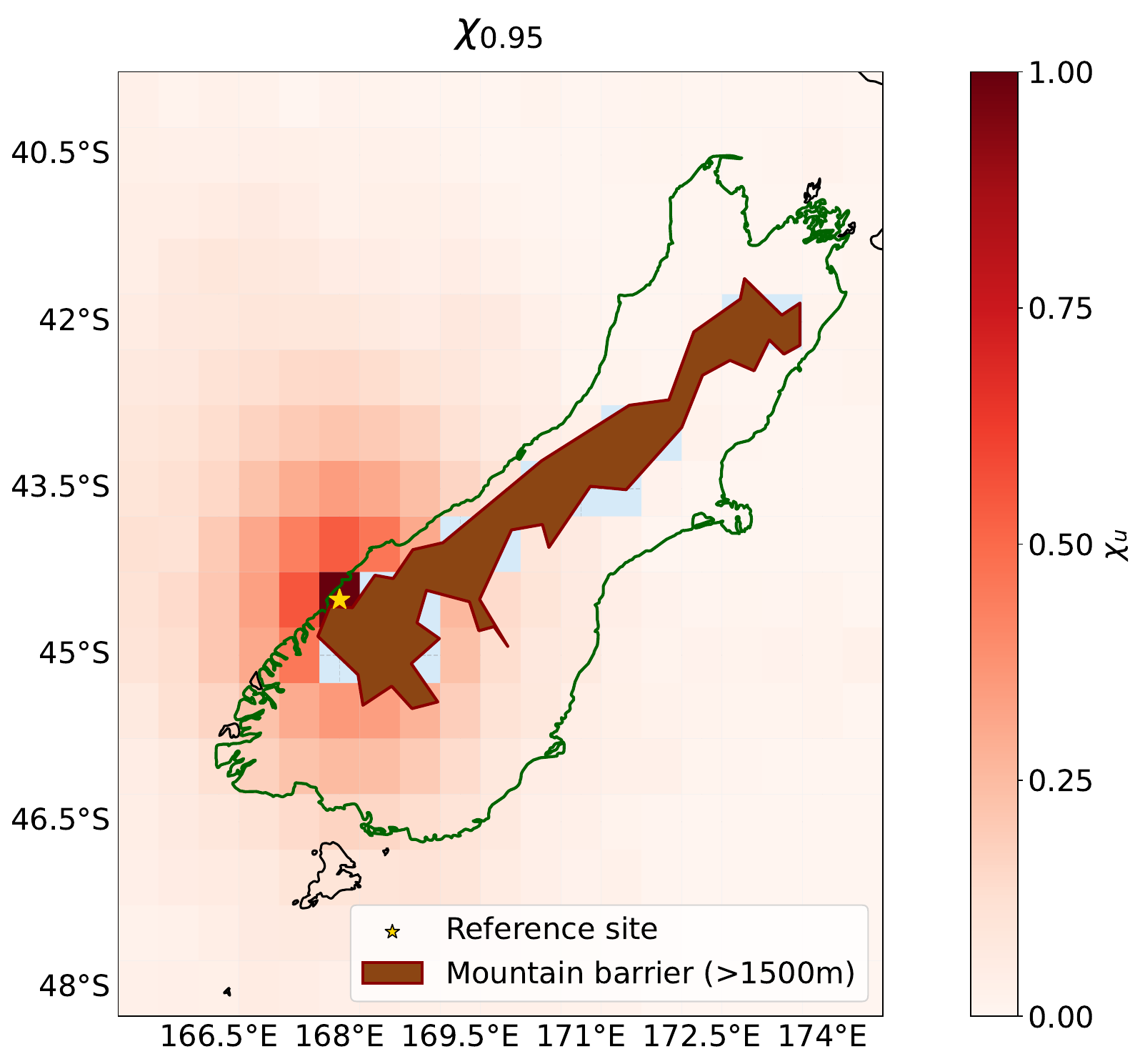} &
			\includegraphics[width=0.2\textwidth, clip=true, trim=0 0 30 40]{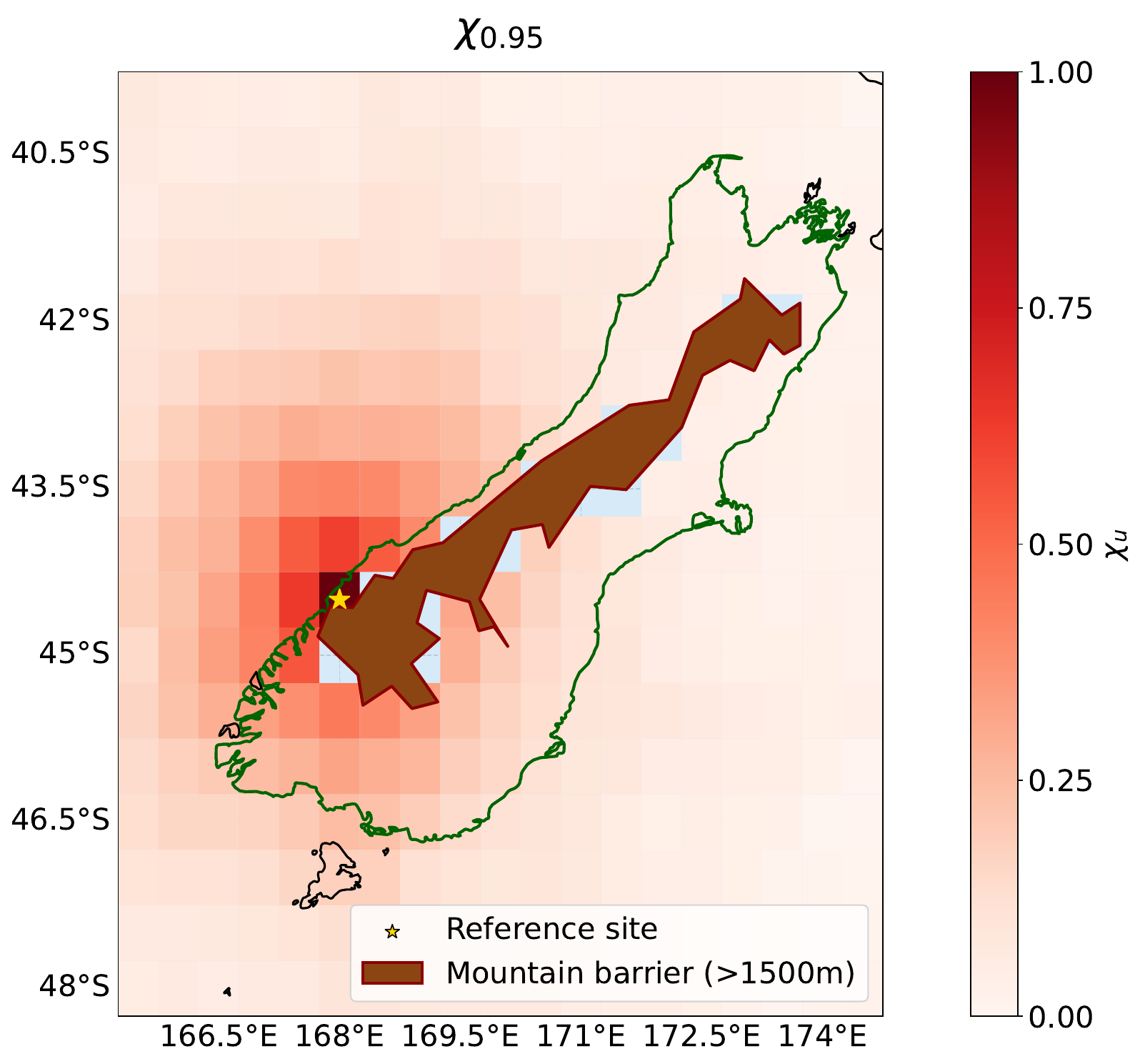}
			\\
			\includegraphics[width=0.2\textwidth, clip=true, trim=0 0 30 40]{precip_emp_chiu_inland_center_140_95.pdf} &
			\includegraphics[width=0.2\textwidth, clip=true, trim=0 0 30 40]{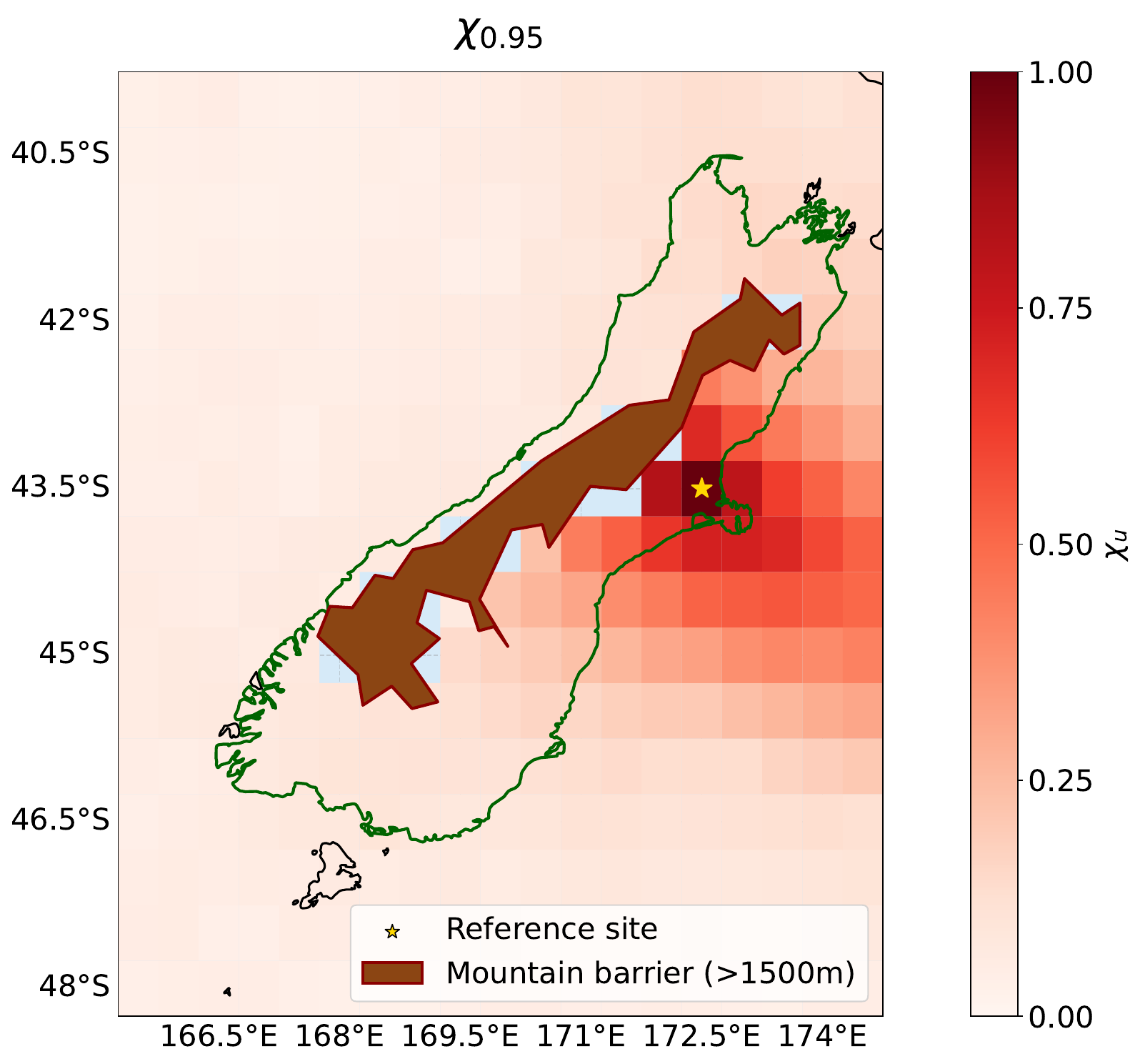} &
			\includegraphics[width=0.2\textwidth, clip=true, trim=0 0 30 40]{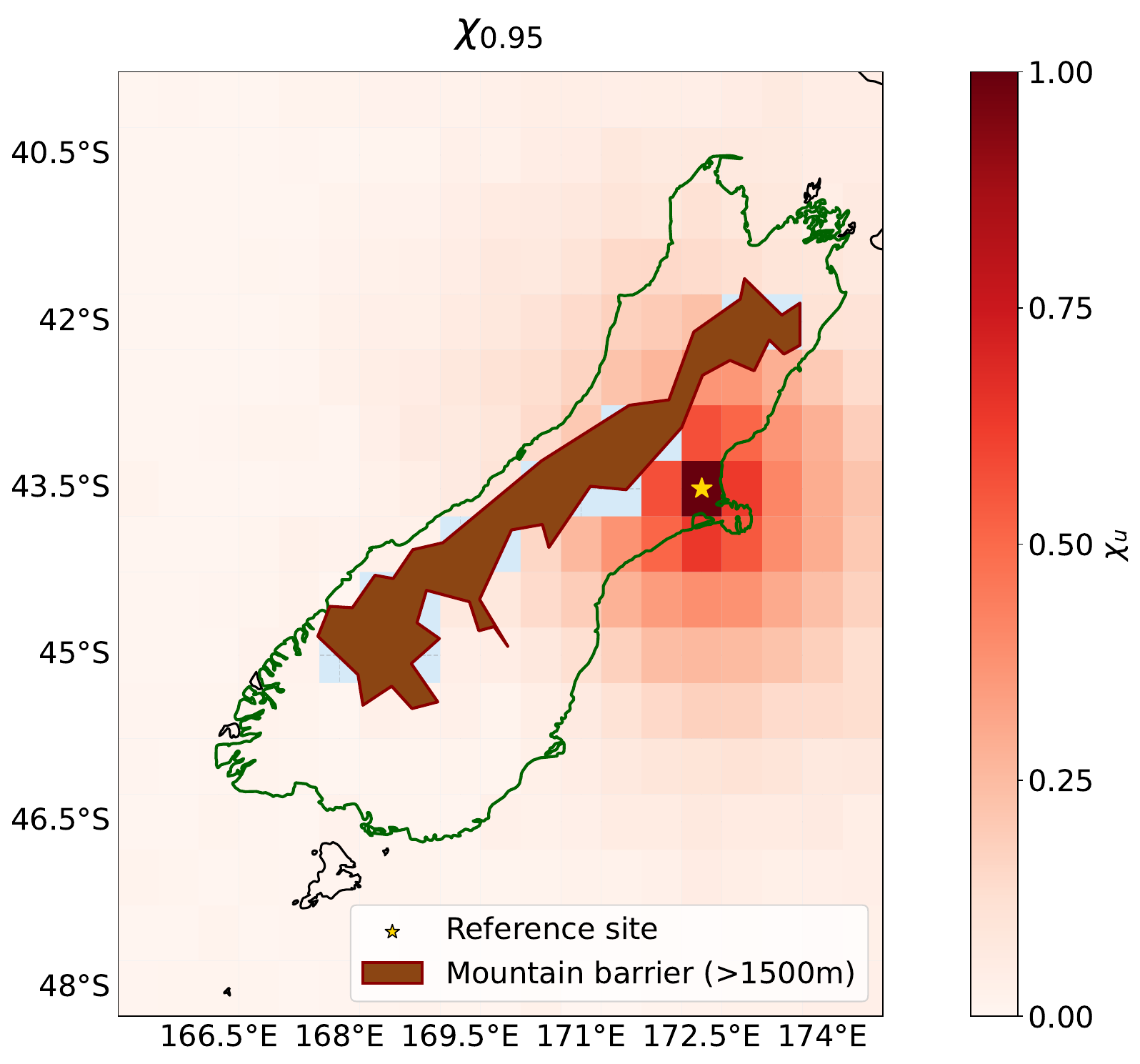} &
			\includegraphics[width=0.2\textwidth, clip=true, trim=0 0 30 40]{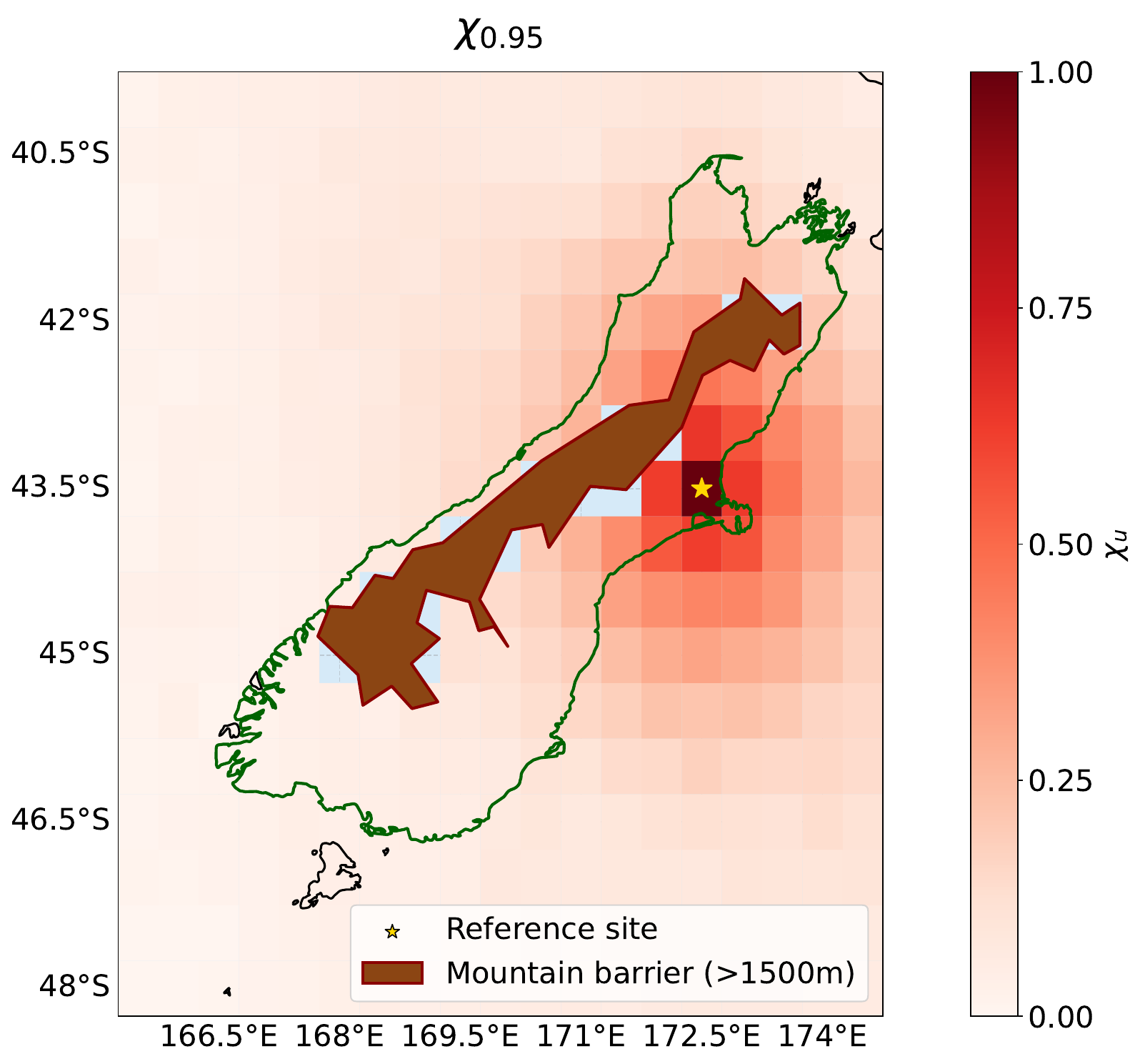}
			\\
			\includegraphics[width=0.2\textwidth, clip=true, trim=0 0 30 40]{precip_emp_chiu_inland_center_184_95.pdf} &
			\includegraphics[width=0.2\textwidth, clip=true, trim=0 0 30 40]{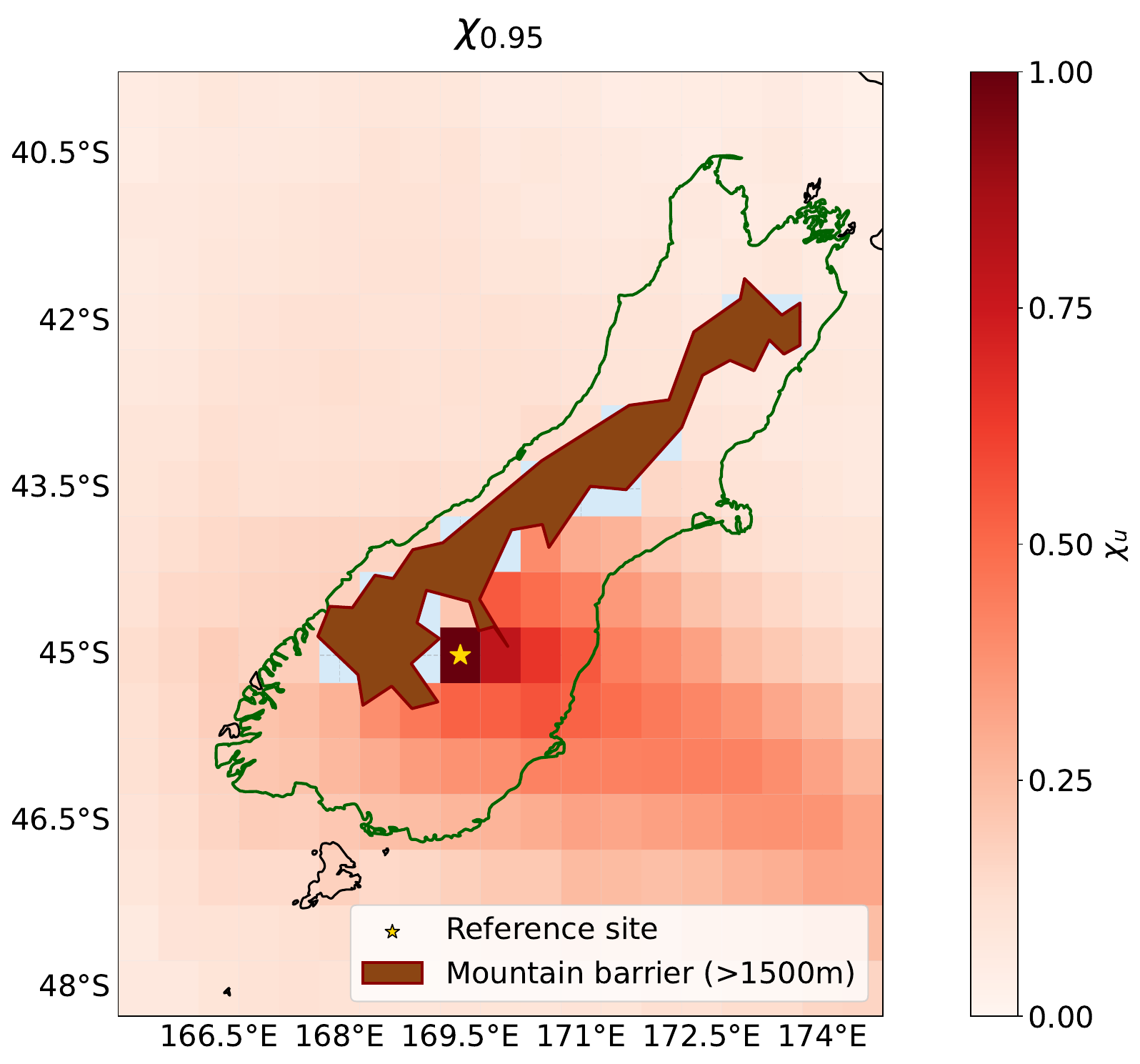} &
			\includegraphics[width=0.2\textwidth, clip=true, trim=0 0 30 40]{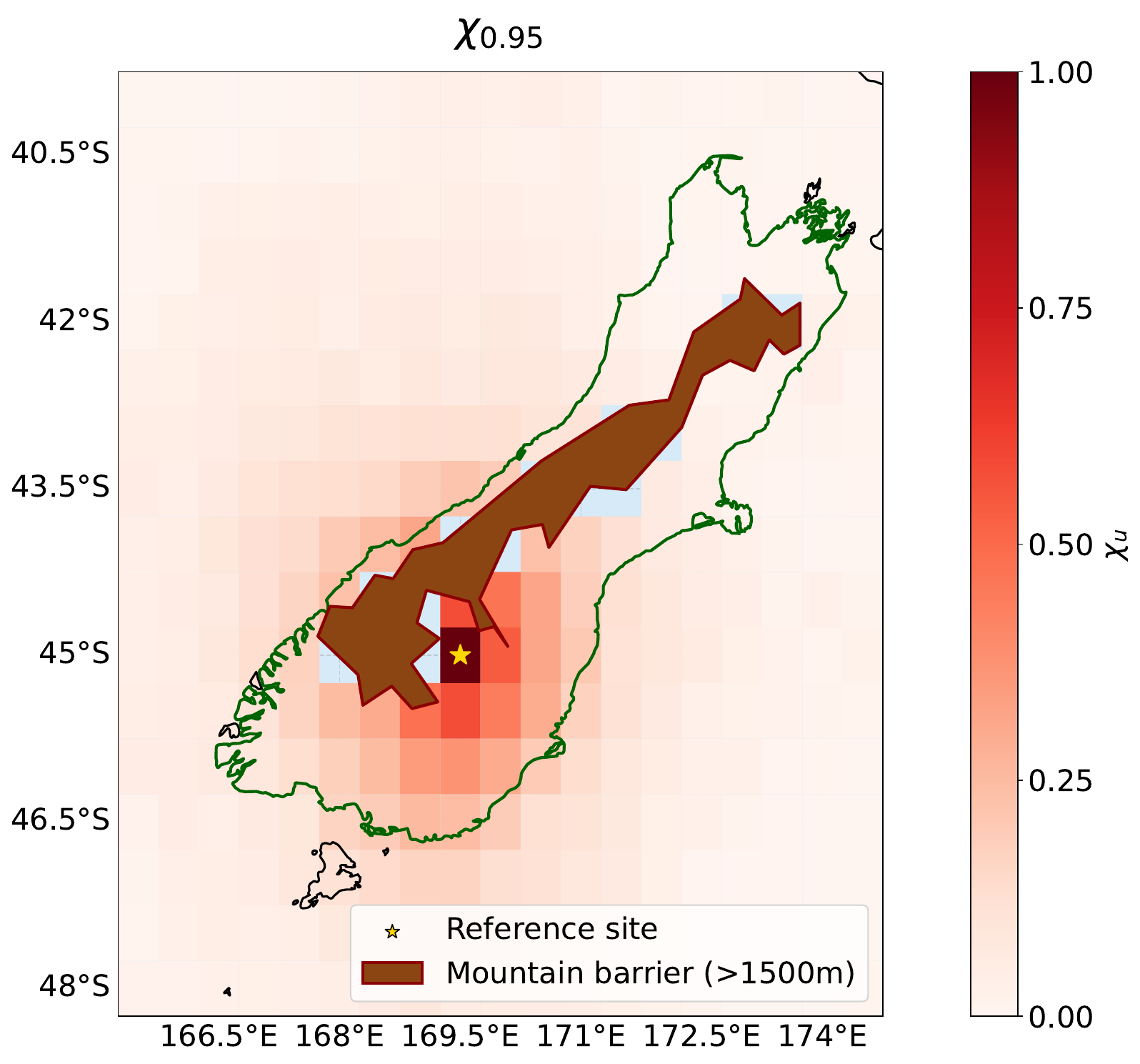} &
			\includegraphics[width=0.2\textwidth, clip=true, trim=0 0 30 40]{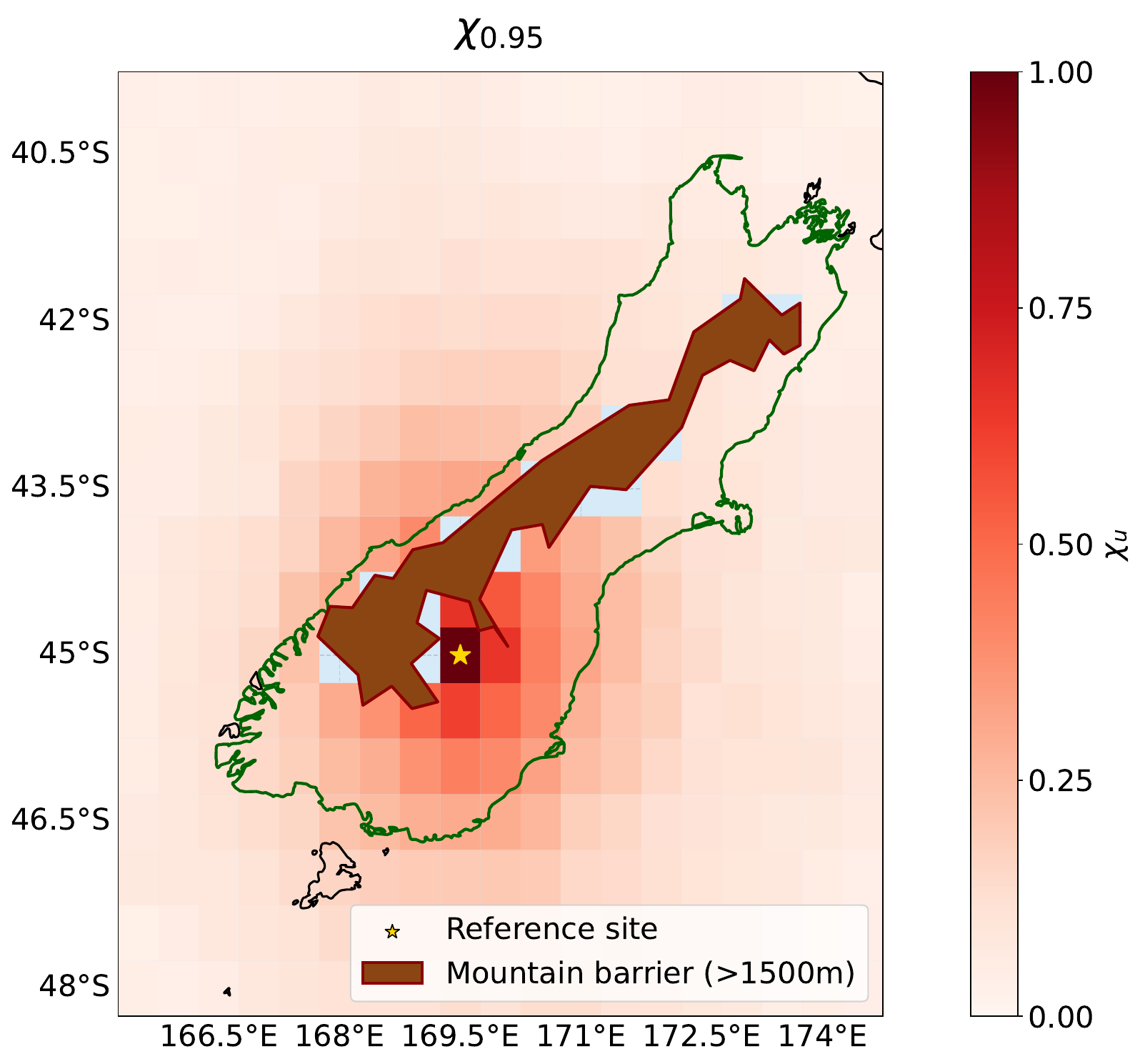}
		\end{tabular}
		\caption{Empirical and model-implied $\chi_{0.95}$ surfaces around a reference site, for all three models.}
		\label{fig:precip_chiu_model_comparison_95}
	\end{figure}
	
	Figure~\ref{fig:precip_chiu_model_comparison_90} and \ref{fig:precip_chiu_model_comparison_95} compare the empirical $\chi_{0.90}$ and $\chi_{0.95}$ surfaces at four reference sites against all three models' implied surfaces. The visibility model reproduces the empirical barrier attenuation and directional anisotropy most closely; \citet{Shi2026} built on an isotropic Wendland kernel with no barrier information, produces a smooth, radially symmetric surface that misses both features; \citet{Huser2019} single global tail dependence parameter likewise cannot represent the directional asymmetry, through its isotropic Mat{\'e}rn does capture some of the overall distance decay.
	
	\section{Significant Wave Height Application} \label{supp:wave_application}
	
	\subsection{Exploratory Analysis of Storm Duration} \label{supp:wave_storm_duration}
	
	The temporal aggregation of extreme wave heights requires distinguishing individual extreme observations from storm-scale events. A single storm can produce elevated wave heights over multiple consecutive daily or 3-hourly observations. Treating these observations as separate extremes may overrepresent individual storms and introduce temporal dependence that is not explicitly accounted for by the subsequent marginal GPD analysis. We therefore conducted an exploratory analysis of storm durations to guide the temporal aggregation used in the Faroe Islands application. 
	
\begin{figure}[htbp]
		\centering
		\includegraphics[width=0.8\textwidth]{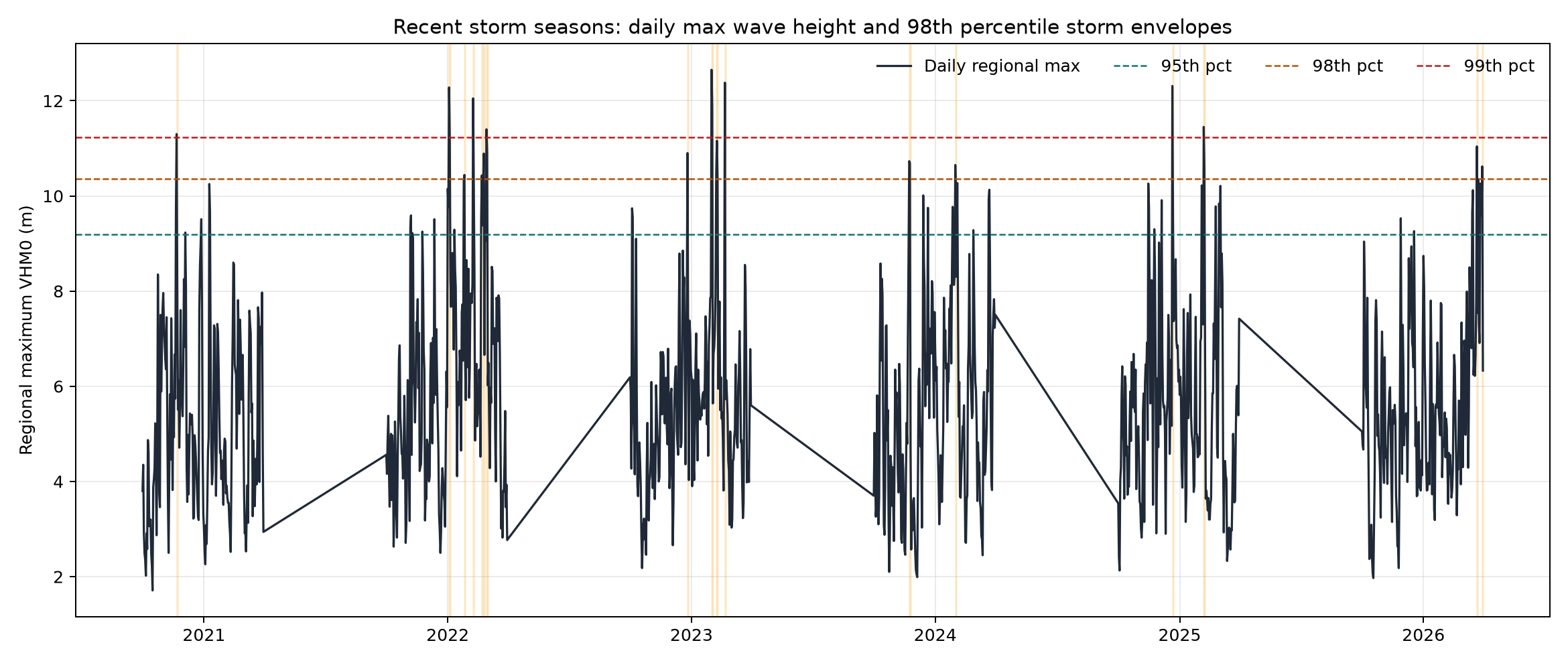}
		\caption{Regional daily maximum significant wave heights during recent storm seasons. Horizontal dashed lines indicate the 95th, 98th, and 99th percentiles used to examine high-wave episodes.}
		\label{fig:wave_storm_ts}
	\end{figure}

	We first examined the temporal evolution of regional maxima significant wave heights over recent storm seasons, together with the duration of threshold exceedance events at daily and 3-hourly resolutions. Figure \ref{fig:wave_storm_ts} illustrates the regional daily maximum wave heights during recent years, with horizontal lines indicating the 95\%, 98\%, and 99\% percentiles. The series shows that high-wave episodes can contain multiple elevated observations over a short period, supporting the use of temporal aggregation to represent storm-scale extremes rather than treating every exceedances as a separate event. 
	
\begin{figure}[htbp]
		\centering
		\includegraphics[width=0.8\textwidth]{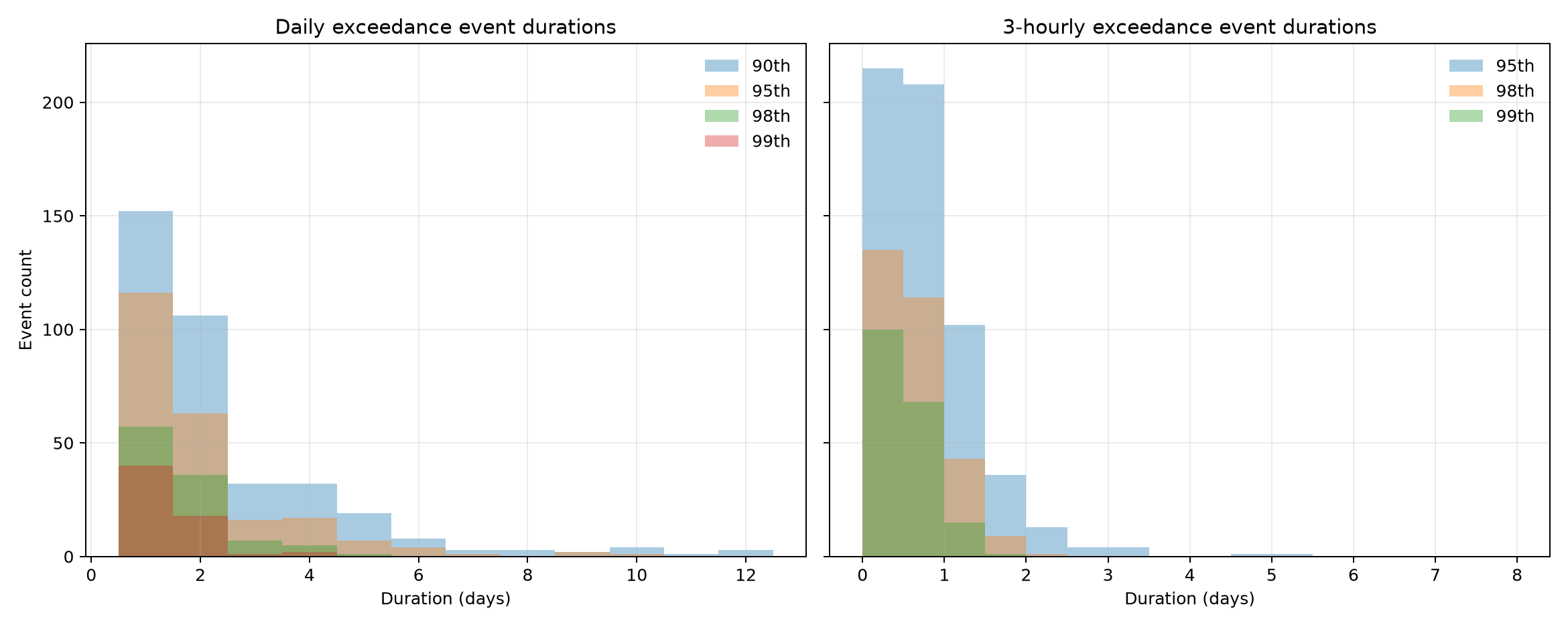}
		\caption{Distributions of threshold-exceedance event durations at daily (left) and 3-hourly (right) temporal resolutions. The histograms illustrate the generally shorter durations of events defined at higher thresholds.}
		\label{fig:wave_duration_histograms}
	\end{figure}

	We next summarized exceedance-event durations at several thresholds. An event was defined as a sequence of threshold exceedances, and its duration was measured in days. Figure~\ref{fig:wave_duration_histograms} shows the distributions of event durations at daily and 3-hourly temporal resolutions. At the daily resolution, event durations are concentrated at one to two days, with progressively shorter upper tail as the threshold increases. At the 90th percentile, some events persist for considerably longer, whereas events defined by the 98th and 99th percentiles are generally shorter. The 3-hourly results show an even stronger concentration of short-duration events, with most exceedance episodes lasting approximately one day or less. These patterns suggest that extreme wave conditions tend to occur in relatively short episodes, although some events particularly at lower thresholds, can persist over several days. 
	
	Table~\ref{tab:wave_storm_duration} provides numerical summaries of these distribution. At the daily resolution, the median, 90th percentile, and maximum event durations are 1, 4, and 10 days at the 95th percentile threshold, respectively, compared with 1, 2, and 4 days at the 99th-percentile threshold. At the 3-hourly resolution, events above the 95th percentile have a median duration of 0.625 days and a 90th percentile duration of approximately 1.46 days. Together, the histograms and summary statistics indicate that the most extreme events are typically shorter than events identified using lower thresholds, while also showing that a small number of episodes can extend over several days. 
	
	\begin{table}[htbp]
		\centering
		\caption{Observed exceedance-event durations at daily and 3-hourly temporal resolutions for the significant wave height data. Durations are reported in days.}
		\label{tab:wave_storm_duration}
		\begin{tabular}{llrrrr}
			\toprule
			Resolution & Threshold & Number of events
			& Median & 90th percentile & Maximum \\
			\midrule
			Daily & 90th percentile & 367 & 2.0 & 5.0 & 15.0 \\
			Daily & 95th percentile & 227 & 1.0 & 4.0 & 10.0 \\
			Daily & 98th percentile & 106 & 1.0 & 3.0 & 5.0 \\
			Daily & 99th percentile & 61  & 1.0 & 2.0 & 4.0 \\
			3-hourly & 95th percentile & 584 & 0.625 & 1.463 & 5.0 \\
			\bottomrule
		\end{tabular}
	\end{table}
	
\begin{figure}[htbp]
		\centering
		\includegraphics[width=0.7\textwidth]{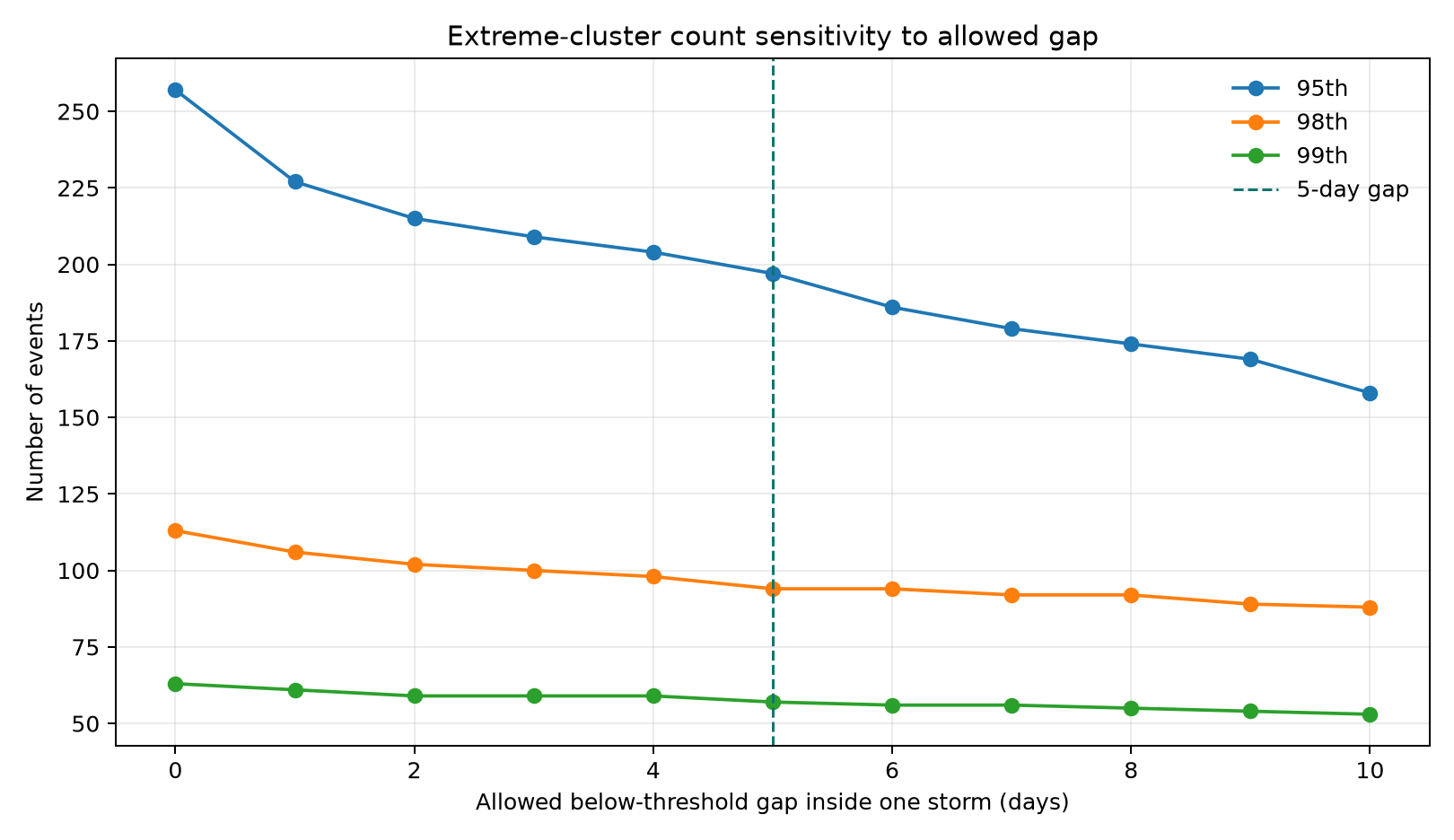}
		\caption{Sensitivity of the number of identified extreme events to the allowed below-threshold gap within a storm, for the 95th, 98th, and 99th percentile thresholds. The vertical dashed line indicates the five-day gap used as a reference in selecting the temporal aggregation scale.}
		\label{fig:wave_gap_sensitivity}
	\end{figure}

	To further assess the temporal separation required to distinguish storm events, we examined the sensitivity of extreme-event counts to the allowed below-threshold gap within a storm. Under this declustering rule, exceedance episodes separated by a sufficiently short below-threshold interval are grouped into same event. Figure~\ref{fig:wave_gap_sensitivity} shows that the number of identified events decreases as the allowed gap increases, with the largest reductions occurring at shorter gaps. At a five-day gap, the event counts have declined substantially relative to the zero-gap definition, while increasing the gap further produces more modest reductions, particularly at the higher thresholds. This pattern suggests that a five-day interval captures much of the clustering in the observed extreme-wave episodes without requiring substantially longer temporal separation. 
	
	Based on these exploratory results, we use nonoverlapping 5-day block maxima in the subsequent spatial extremes analysis. This choice provides a practical compromise between aggregation multiple observations associated with the same storm and retaining a sufficient number of temporal replicates for spatial tail modeling. The block length should not be interpreted as guaranteeing independence between successive blocks; rather, it is a pragmatic storm-scale reduction supported by the observed event durations and declustering sensitivity analysis. 
	
	\subsection{Modeling the GPD Parameters via Barrier-Informed Covariates and the High Threshold} \label{supp:wave_gpd_covariates}
	
	We use spatial regression models to describe the high-threshold surface and the spatial variation in the generalized Pareto distribution (GPD) parameters for significant wave heights around the Faroe Islands. The covariates are designed to represent both broad spatial variation and local environmental and geometric features that may influence extreme wave conditions. In particular, we consider large-scale spatial trends, smooth spatial variation, bathymetric depth, distance to the island centroid, exposure angle, and selected interaction terms. These covariates allow the fitted surfaces to reflect differences between exposed and sheltered locations, as well as spatial variation associated with the islands and surrounding seabed. 
	
\begin{figure}[htbp]
		\centering
		\includegraphics[width=0.8\textwidth]{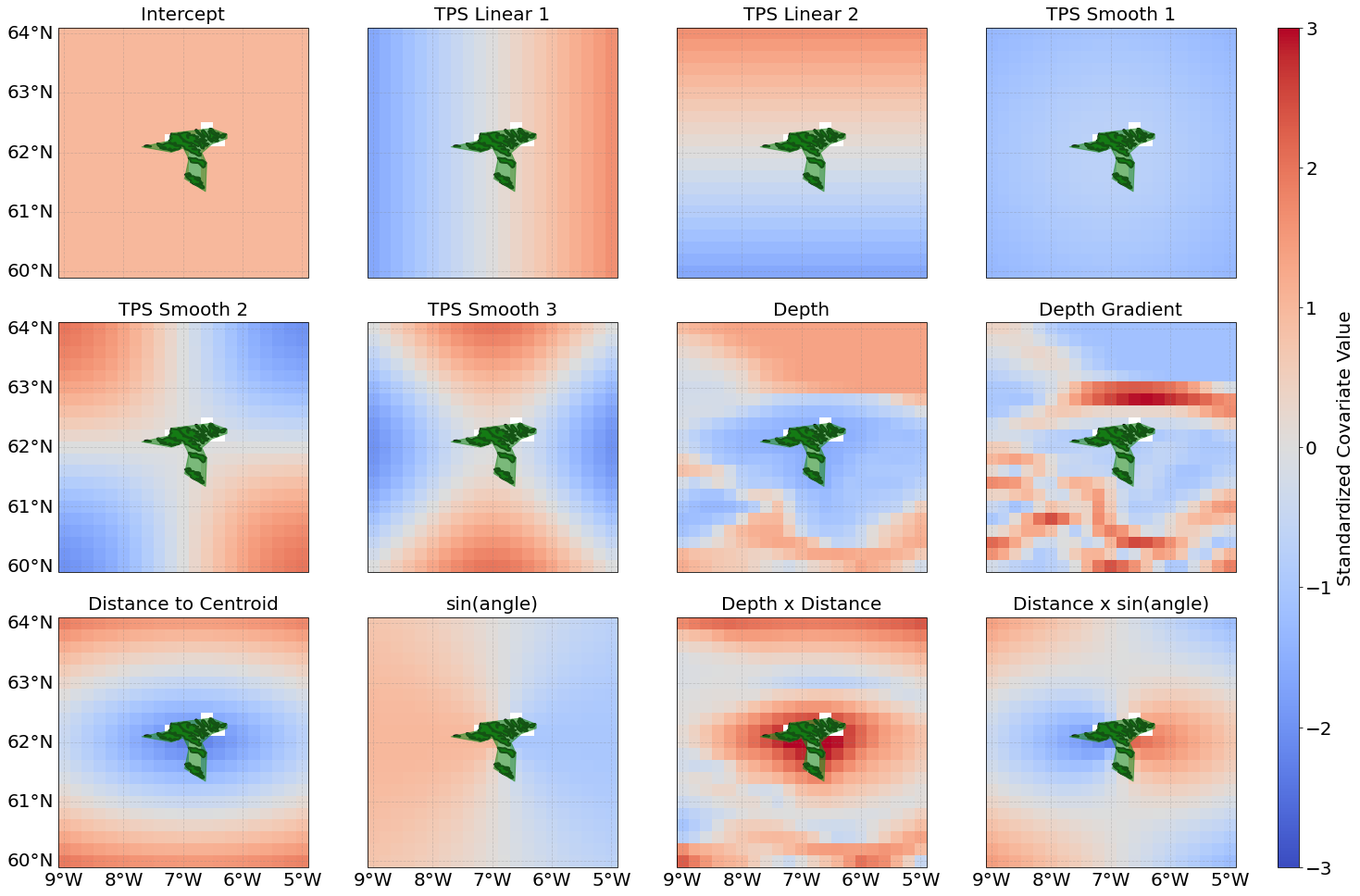}
		\caption{Spatial covariates used to model the spatially varying 0.95 quantile of significant wave height and the GPD scale and shape parameters around the Faroe Islands.}
		\label{fig:wave_covariates_surface}
	\end{figure}

	\paragraph{High-threshold surface.}
	At each location $\bm{s}$, we estimate the marginal 0.95 quantile $q_{0.95}(\bm{s})$ using a pooled spatial quantile-regression model. Rather than estimating a separate threshold independently at each location, the regression model borrows information across sites to produce a spatially coherent threshold surface. Its covariates describe the large-scale spatial trend, smooth spatial variation, bathymetric depth, distance to the island centroid, exposure angle, and selected interactions. The resulting surface accommodates spatial variation in the level above which wave-height observations are treated as extreme. Figure~\ref{fig:wave_covariates_surface} displays the covariate surfaces used in this regression.
	
	\paragraph{GPD scale parameter.}
	The GPD scale parameter $\sigma(\bm{s})$ is modeled using the same covariates as the high-threshold quantile-regression model. Specifically, letting $\bm{x}_{\sigma}(\bm{s})$ denote the corresponding design vector, we use$\log\sigma(\bm{s}) = \bm{x}_\sigma(\bm{s})^\top \bm{\beta}_\sigma$. Using a common covariate specification for the threshold and scale models provides a consistent representation of the spatial structure of high wave heights and the magnitudes of exceedances above the estimated threshold. The log link ensures that the GPD scale parameter is positive. 
	
	\paragraph{GPD shape parameters.}
	The GPD shape parameter $\xi(\bm{s})$ is modeled using a separate covariate specification. Let $\bm{x}_\xi(\bm{s})$ denote its design vector. We use the rescaled-logit link $\xi(\bm{s}) = \tfrac{3}{2} \operatorname{expit}\left( \bm{x}_\xi(\bm{s})^\top \bm{\beta}_\xi \right) - \tfrac{1}{2}$, which constrains the shape parameter to $\xi(\bm{s}) \in (-0.5, 1)$. Unlike the threshold and scale models, the shape parameter model does not use the same covariate set; its specification allows the spatial variation in the tail shape to be represented separately from the spatial variation in the threshold and exceedance scale. The covariates contain 4 thin-plate spline (TPS) functions, bathymetric depth, bathymetric depth gradient, distance to the island centroid, and the sine of exposure angle (Figure~\ref{fig:wave_covariates_surface}).
	
	\FloatBarrier
	\subsection{Model Fit Assessment} \label{supp:wave_fit}
	
	\subsubsection{MCMC Diagnostics} \label{supp:wave_mcmc}
	
	We report posterior summaries and MCMC diagnostics for the visibility mode fitted to the full significant wave height dataset, comprising $d = 432$ spatial locations. The model is estimated using the two-step MCMC procedure described in Section~\ref{subsec:wave_result}. In the first-stage, we estimate the spatially varying GPD scale and shape parameters using the threshold exceedances. The posterior mean estimates of these marginal parameters are then held fixed in the second step, which estimates the spatial extremal dependence structure, including the parameters governing the anisotropic Wendland kernel, the spatially varying tail dependence parameter, and the latent visibility-based Gaussian process. The dependence model also incorporates the log-Laplace nugget effect \citet{Shi2026log}.
	
	For the full-data analysis, the second step MCMC chain was run for 50{,}000 iterations. We discard the first 25{,}000 iterations as burn-in and thinned the remaining draws by a factor of 10, yielding $B = 2{,}500$ retained posterior samples. The anisotropic Wendland kernel uses 41 knots, and the spatially varying tail dependence parameter is constructed using Gaussian kernel functions with 41 knots and fixed bandwidth $h = \sqrt{1/6}$. 
	
	Table~\ref{tab:wave_full_posterior_mean} reports the posterior means and 95\% equal-tailed credible intervals for the marginal and dependence parameters. The marginal summaries correspond to the first stage estimates, whereas the dependence summaries are obtained from the second-step MCMC. Because the marginal posterior means are fixed during the second step, the dependence inference does not propagate the remaining uncertainty in the marginal parameters. 
	
	\begin{table}[htbp]
		\centering
		\caption{Posterior means and 95\% equal-tailed credible intervals for the marginal GPD and dependence parameters of the visibility model fitted to the full Faroe Islands significant wave height dataset. The marginal parameter estimates are obtained from the first-stage MCMC, while the dependence parameter estimates are obtained from the second-stage MCMC with the marginal parameters fixed at their first-stage posterior means.}
		\label{tab:wave_full_posterior_mean}
		\resizebox{\textwidth}{!}{
			\begin{tabular}{c ccc cccc}
				\toprule
				\multicolumn{8}{c}{Margin Model} \\
				\midrule
				\multirow{4}{*}{$\sigma$} & $\beta_0$ & $\beta_1$ & $\beta_2$ & $\beta_3$ & $\beta_4$ & $\beta_5$ & $\beta_6$ \\ \cline{2-8}
				& $\underset{(0.187,\, 0.227)}{0.207}$ & $\underset{(0.027,\, 0.161)}{0.094}$ & $\underset{(-0.069,\, -0.035)}{-0.052}$ & $\underset{(0.052,\, 0.292)}{0.173}$ & $\underset{(0.012,\, 0.030)}{0.021}$ & $\underset{(-0.178,\, -0.030)}{-0.104}$ & $\underset{(-0.057,\, -0.004)}{-0.031}$ \\
				\cline{2-8}
				& $\beta_7$ & $\beta_8$ & $\beta_9$ & $\beta_{10}$ & & & \\ \cline{2-5}
				& $\underset{(0.098,\, 0.385)}{0.242}$ & $\underset{(0.083,\, 0.276)}{0.180}$ & $\underset{(-0.030,\, 0.003)}{-0.014}$ & $\underset{(-0.042,\, 0.014)}{-0.014}$ & & & \\ 
				\midrule
				\multirow{4}{*}{$\xi$} & $\beta_0$ & $\beta_1$ & $\beta_2$ & $\beta_3$ & $\beta_4$ & $\beta_5$ & $\beta_6$ \\ \cline{2-8}
				& $\underset{(-1.245,\, -1.174)}{-1.210}$ & $\underset{(-0.333,\, -0.130)}{-0.232}$ & $\underset{(-0.118,\, -0.035)}{-0.077}$ & $\underset{(-0.067,\, 0.075)}{0.003}$ & $\underset{(0.018,\, 0.147)}{0.083}$ & $\underset{(-0.050,\, -0.002)}{-0.026}$ & $\underset{(-0.136,\, 0.047)}{-0.044}$ \\ 
				\cline{2-8}
				& $\beta_7$ & & & & & & \\ \cline{2-2}
				& $\underset{(-0.472,\, -0.198)}{-0.336}$ & & & & & & \\
				\midrule
				\multicolumn{8}{c}{Dependence Model} \\
				\midrule
				\multirow{12}{*}{Tail} & $\phi_1$ & $\phi_2$ & $\phi_3$ & $\phi_4$ & $\phi_5$ & $\phi_6$ & $\phi_7$ \\ \cline{2-8}
				& $\underset{(1.440,\, 1.506)}{1.475}$ & $\underset{(1.280,\, 1.325)}{1.302}$ & $\underset{(1.128,\, 1.173)}{1.150}$ & $\underset{(1.216,\, 1.283)}{1.249}$ & $\underset{(1.743,\, 1.833)}{1.782}$ & $\underset{(1.064,\, 1.100)}{1.082}$ & $\underset{(0.453,\, 0.502)}{0.478}$ \\ 
				\cline{2-8}
				& $\phi_8$ & $\phi_9$ & $\phi_{10}$ & $\phi_{11}$ & $\phi_{12}$ & $\phi_{13}$ & $\phi_{14}$ \\ \cline{2-8}
				& $\underset{(0.413,\, 0.466)}{0.440}$ & $\underset{(1.695,\, 1.775)}{1.737}$ & $\underset{(0.897,\, 0.937)}{0.917}$ & $\underset{(0.323,\, 0.361)}{0.342}$ & $\underset{(0.803,\, 0.856)}{0.830}$ & $\underset{(1.548,\, 1.602)}{1.574}$ & $\underset{(1.105,\, 1.148)}{1.127}$ \\ 
				\cline{2-8}
				& $\phi_{15}$ & $\phi_{16}$ & $\phi_{17}$ & $\phi_{18}$ & $\phi_{19}$ & $\phi_{20}$ & $\phi_{21}$ \\ \cline{2-8}
				& $\underset{(1.006,\, 1.046)}{1.026}$ & $\underset{(0.800,\, 0.861)}{0.831}$ & $\underset{(1.255,\, 1.325)}{1.283}$ & $\underset{(1.195,\, 1.248)}{1.219}$ & $\underset{(0.761,\, 0.811)}{0.788}$ & $\underset{(0.703,\, 0.748)}{0.726}$ & $\underset{(0.245,\, 0.305)}{0.275}$ \\ 
				\cline{2-8}
				& $\phi_{22}$ & $\phi_{23}$ & $\phi_{24}$ & $\phi_{25}$ & $\phi_{26}$ & $\phi_{27}$ & $\phi_{28}$ \\ \cline{2-8}
				& $\underset{(0.454,\, 0.509)}{0.481}$ & $\underset{(1.424,\, 1.482)}{1.450}$ & $\underset{(1.309,\, 1.347)}{1.327}$ & $\underset{(1.232,\, 1.275)}{1.253}$ & $\underset{(0.004,\, 0.137)}{0.071}$ & $\underset{(1.041,\, 1.104)}{1.073}$ & $\underset{(1.409,\, 1.480)}{1.445}$ \\ 
				\cline{2-8}
				& $\phi_{29}$ & $\phi_{30}$ & $\phi_{31}$ & $\phi_{32}$ & $\phi_{33}$ & $\phi_{34}$ & $\phi_{35}$ \\ \cline{2-8}
				& $\underset{(0.905,\, 0.986)}{0.948}$ & $\underset{(0.843,\, 1.095)}{0.958}$ & $\underset{(0.913,\, 0.998)}{0.954}$ & $\underset{(0.131,\, 0.209)}{0.171}$ & $\underset{(0.305,\, 0.368)}{0.338}$ & $\underset{(-0.091,\, -0.009)}{-0.050}$ & $\underset{(0.871,\, 0.929)}{0.900}$ \\ 
				\cline{2-8}
				& $\phi_{36}$ & $\phi_{37}$ & $\phi_{38}$ & $\phi_{39}$ & $\phi_{40}$ & $\phi_{41}$ &  \\ \cline{2-7}
				& $\underset{(0.641,\, 0.717)}{0.679}$ & $\underset{(1.271,\, 1.416)}{1.343}$ & $\underset{(1.024,\, 1.084)}{1.054}$ & $\underset{(1.397,\, 1.461)}{1.428}$ & $\underset{(1.476,\, 1.549)}{1.513}$ & $\underset{(1.682,\, 1.918)}{1.793}$ &  \\ 
				\midrule
				\multirow{2}{*}{Wendland} & $r$ & $\lambda$ & $\alpha$ & & & & \\ \cline{2-4}
				& $\underset{(1.687,\, 1.703)}{1.695}$ & $\underset{(0.633,\, 0.640)}{0.637}$ & $\underset{(3.033,\, 3.049)}{3.041}$ & & & & \\
				\midrule
				\multirow{2}{*}{Mat{\'e}rn} & $\rho$ & $\nu$ & & & & & \\ \cline{2-3}
				& $\underset{(3.203,\, 3.305)}{3.250}$ & $\underset{(0.962,\, 0.980)}{0.971}$ & & & & & \\
				\midrule
				\multirow{2}{*}{Nugget} & $\alpha_0$ & & & & & & \\ \cline{2-2}
				& $\underset{(2.001,\, 2.002)}{2.001}$ & & & & & & \\
				\bottomrule
			\end{tabular}
		}
	\end{table}
	
	\begin{figure}[htbp]
		\centering
		\includegraphics[width=0.75\textwidth]
		{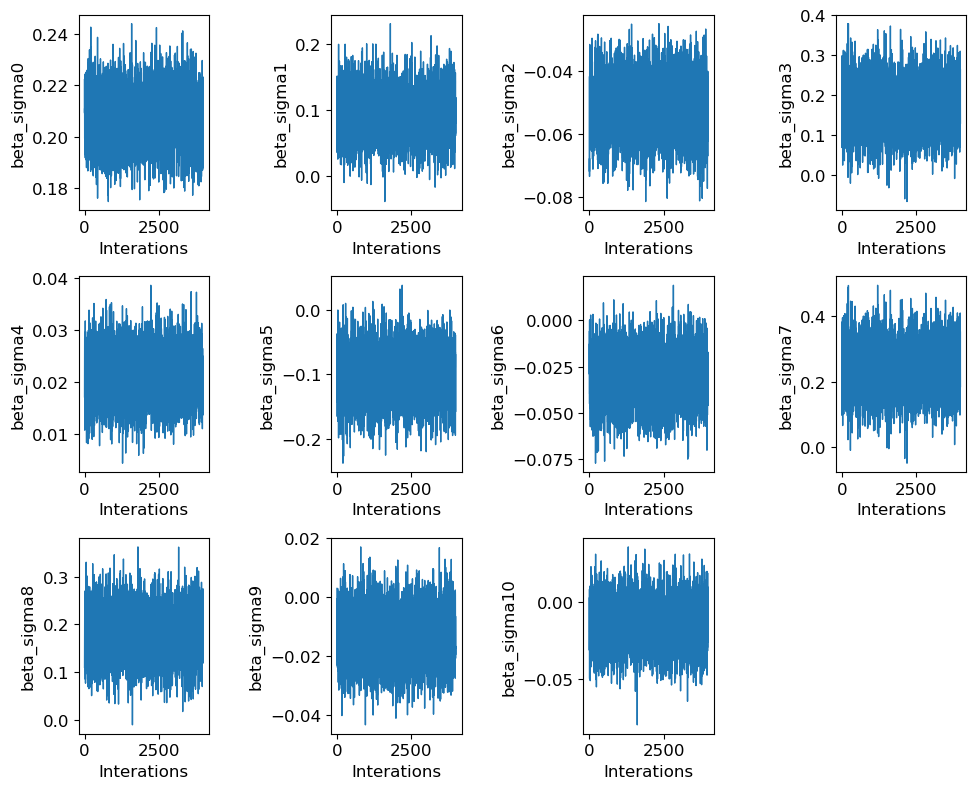}\\
		(a) \\[2mm]
		\includegraphics[width=0.75\textwidth]
		{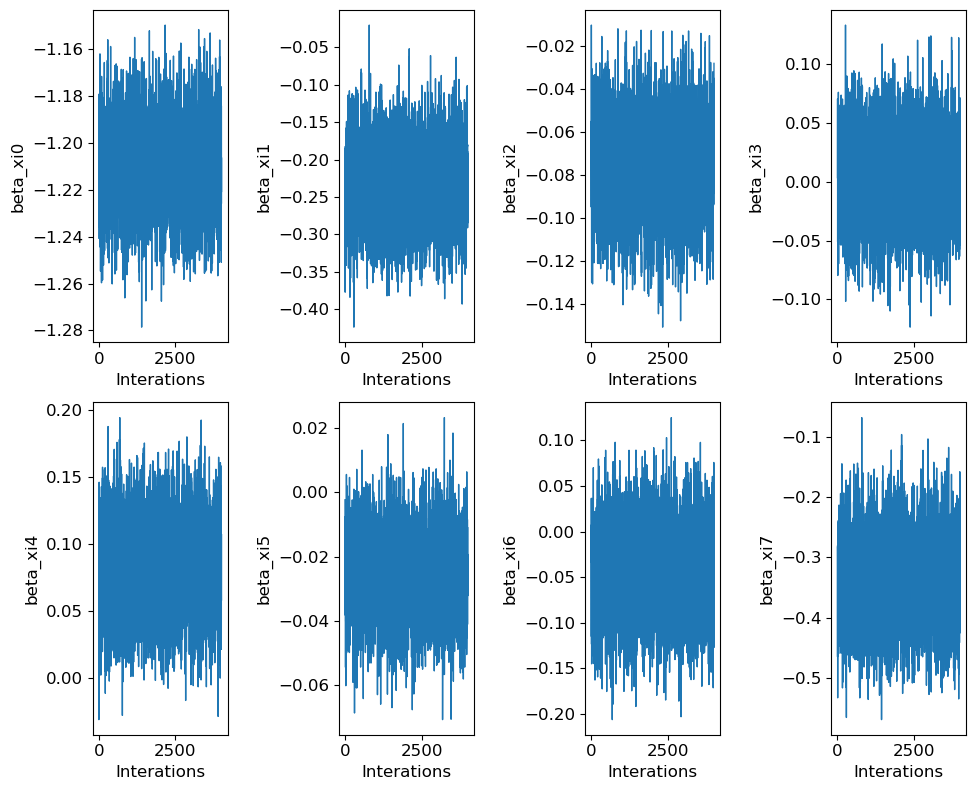}\\
		(b)
		\caption{Trace plots for the margin regression coefficients (a) $\bm{\beta}_\sigma$ and (b) $\bm{\beta}_\xi$.}
		\label{fig:wave_full_trace_gpd}
	\end{figure}
	
	\begin{figure}[htbp]
		\centering
		\includegraphics[width=0.75\textwidth]
		{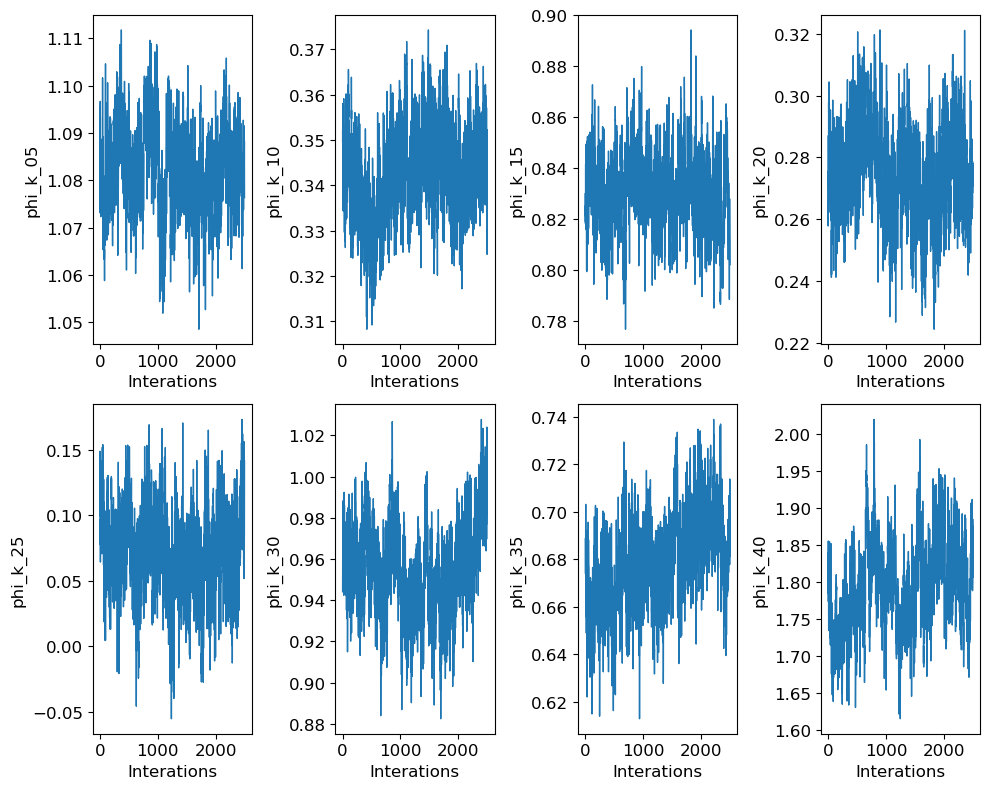}\\
		(a) \\[2mm]
		\includegraphics[width=0.75\textwidth]
		{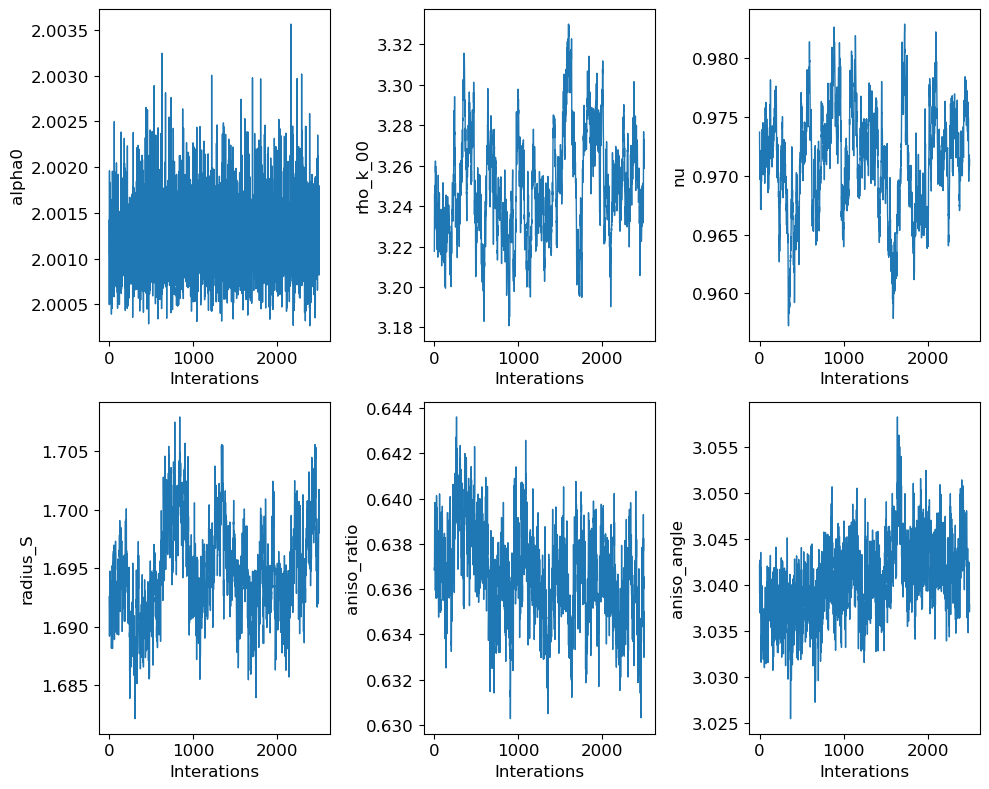}\\
		(b)
		\caption{Trace plots for the dependence parameters of the visibility model fitted to the full wave height dataset. The panels show (a) the selected Gaussian kernel coefficients for the spatially varying tail dependence parameter and (b) the remaining dependence parameters, including the log-Laplace nugget effect, Mat{\'e}rn parameter, and the Wendland parameters.}
		\label{fig:wave_full_trace_dependence}
	\end{figure}
	
	Figure~\ref{fig:wave_full_trace_gpd} and \ref{fig:wave_full_trace_dependence} show trace plots for the marginal regression coefficients and the dependence parameters, respectively. These plots are used to assess chain stability and mixing. The trace plots should be interpreted alongside the posterior summaries when evaluating the reliability of the MCMC estimates. 
	
	\subsubsection{GPD Parameters and High Threshold Surfaces} \label{supp:wave_gpd_surfaces}
	
\begin{figure}[htbp]
		\centering
		\includegraphics[width=0.7\textwidth]{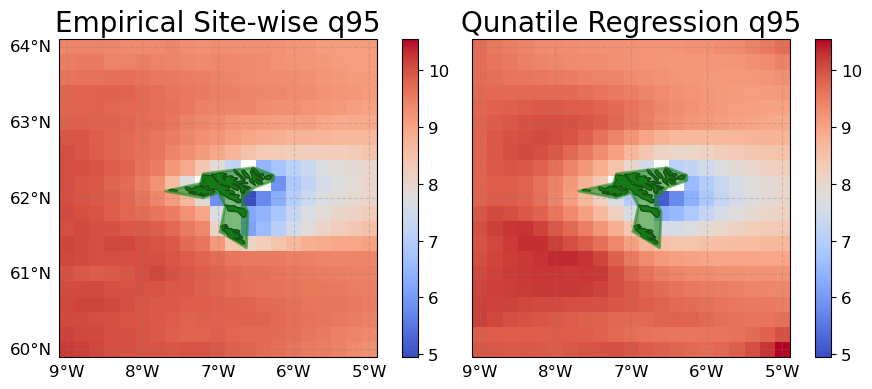}
		\caption{Empirical site-wise 0.95 quantiles of significant wave height and the fitted spatial threshold surface obtained from pooled quantile regression.}
		\label{fig:wave_q95_emp_fit}
	\end{figure}

\begin{figure}[htbp]
		\centering
		\includegraphics[width=0.7\textwidth]{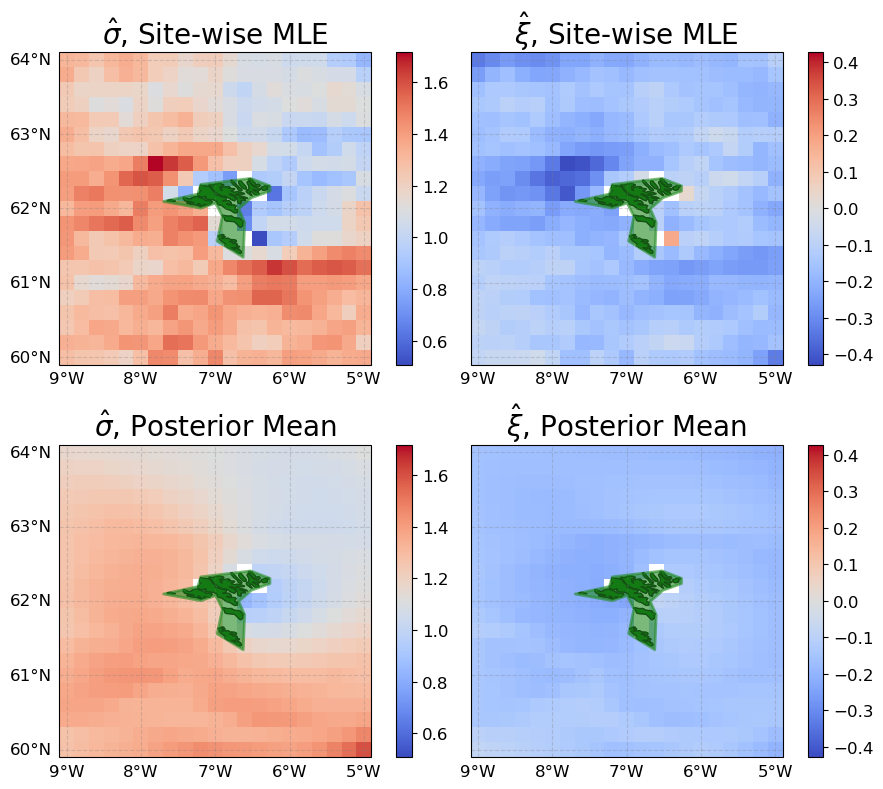}
		\caption{Site-wise maximum likelihood estimates (left) and fitted spatial surfaces (right) for the GPD scale $\sigma(\bm{s})$ and shape $\xi(\bm{s})$ parameters in the full-data Faroe Islands wave height analysis.}
		\label{fig:wave_gpd_parameter_maps}
	\end{figure}

	Figure~\ref{fig:wave_q95_emp_fit} and \ref{fig:wave_gpd_parameter_maps} compare the fitted marginal surfaces with site-wise estimates for the full-data wave height analysis. Figure~\ref{fig:wave_q95_emp_fit} compares the empirical site-wise 0.95 quantiles of significant wave height with the fitted threshold surface obtained from the pooled spatial quantile-regression model. The fitted surface captures the main spatial variation in the empirical quantiles while providing a smoother representation across the domain. This agreement supports the use of the fitted spatially varying threshold in defining exceedances for the subsequent GPD analysis. 
	
	Figure~\ref{fig:wave_gpd_parameter_maps} shows the site-wise maximum likelihood estimates (MLEs) and the corresponding fitted surfaces for the GPD scale $\sigma(\bm{s})$ and shape $\xi(\bm{s})$ parameters. Overall, the fitted surfaces agree well with the broad spatial patterns in the site-wise estimates while exhibiting less local variation. This smoothing is particularly useful for the shape parameter, whose site-wise estimates can be variable because they are obtained independently at each location. By modeling the GPD parameters through spatial covariates, the fitted model borrows information across locations and produces spatially coherent estimates of the marginal tail behavior. 
	
	\subsubsection{Wendland Kernel} \label{supp:wave_wendland}
	
\begin{figure}[htbp]
		\centering
		\includegraphics[width=0.7\textwidth]{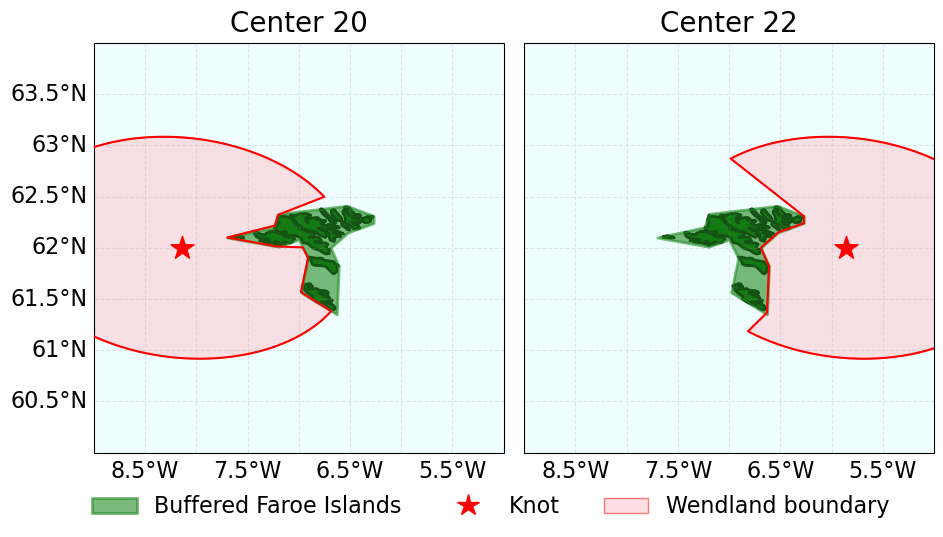}
		\caption{Visibility-restricted anisotropic Wendland kernel supports for two representative knots around the Faroe Islands.}
		\label{fig:wave_wendland_examples}
	\end{figure}

	Figure~\ref{fig:wave_wendland_examples} illustrates the visibility-restricted anisotropic Wendland kernel supports for two representative knots in the Faroe Islands domain. The posterior mean estimates of radius, anisotropy ratio, and orientation angle are $1.672$, $0.642$, and $3.026$, respectively, yielding an anisotropic, elliptical kernel support. For the knot west of the islands (left panel), the support extends into the open ocean to the west but is sharply truncated where the islands obstruct visibility. Thus, locations may lie within the nominal elliptical boundary yet receive zero kernel weight when the line segment connecting them to the knot intersects land. 
	
	The second knot, located east of the islands (right panel), provides a complementary illustration of how the visibility restriction shapes the support on the opposite side of the island chain. Together, the two examples demonstrate that the kernel support is determined not only by the anisotropic distance from each knot but also by the geometry of the ocean domain. This restriction allows the model to represent spatial extremal dependence along unobstructed ocean pathways while excluding direct connections blocked by the islands. 
	
	\subsubsection{Tail Dependence Surface} \label{supp:wave_delta}
	
\begin{figure}[htbp]
		\centering
		\includegraphics[width=0.4\textwidth]{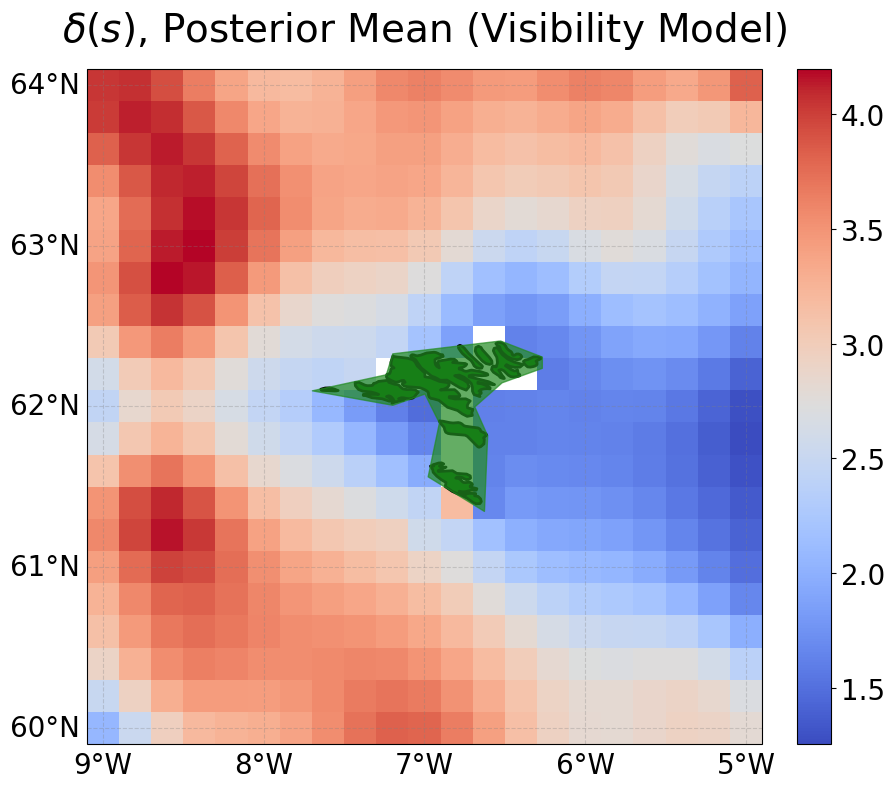}
		\caption{Posterior mean of the site-level tail dependence parameter $\delta(\bm{s})$ under the fitted visibility model for significant wave height around the Faroe Islands.}
		\label{fig:wave_full_delta_posterior_mean}
	\end{figure}

	Figure~\ref{fig:wave_full_delta_posterior_mean} shows the posterior mean of the site-level tail dependence parameter $\delta(\bm{s})$ for the full-data wave height analysis. The estimated surface exhibits substantial spatial variation across the Faroe Islands domain. The highest values occur primarily over the open ocean west and northwest of the islands, reaching approximately $3.5$-$4$ in some locations. In contrast, values are generally lower east of and around the islands, where much of the surface lie between approximately $1.5$ and $2.5$. Elevated values also occur in parts of the ocean south and southwest of the islands, indicating that the spatial pattern is more complex than a simple division between the western and eastern sides. 
	
	The spatial variation in $\delta(\bm{s})$ is represented through visibility-restricted Gaussian-kernel basis functions, rather than imposed through an explicit east-west regional effect. The contrast in the estimated surface is therefore inferred from the data through the likelihood and the dependence model. In conjunction with the visibility-restricted Wendland kernel and the latent visibility-restricted Gaussian process structure, this specification allows the model to represent spatially varying extremal dependence across the domain. 
	
	Under the model's parameterization, higher values of $\delta(\bm{s})$ correspond to stronger asymptotic dependence, whereas lower values indicate a weaker tendency toward joint extremes. The elevated values over the western open ocean are consistent with stronger extremal dependence in this region, while the lower values around and east of the islands suggest weaker dependence there. This pattern is compatible with the possibility that unobstructed ocean regions support more spatially coherent extreme wave conditions, whereas the islands contribute to spatial heterogeneity in dependence. However, the estimated surface does not follow the island boundary exactly, and the spatial contrast should not be interpreted as evidence that the islands alone cause the observed differences. 
	
	\subsubsection{Extremal Dependence $\chi_u$ at Lower Thresholds} \label{supp:wave_chiu}
	
\begin{figure}[htbp]
		\centering
		\begin{tabular}{cc cc}
			\multicolumn{4}{c}{$\chi_{0.96}$} \\
			\includegraphics[width=0.2\textwidth, clip=true, trim=0 0 30 40]
			{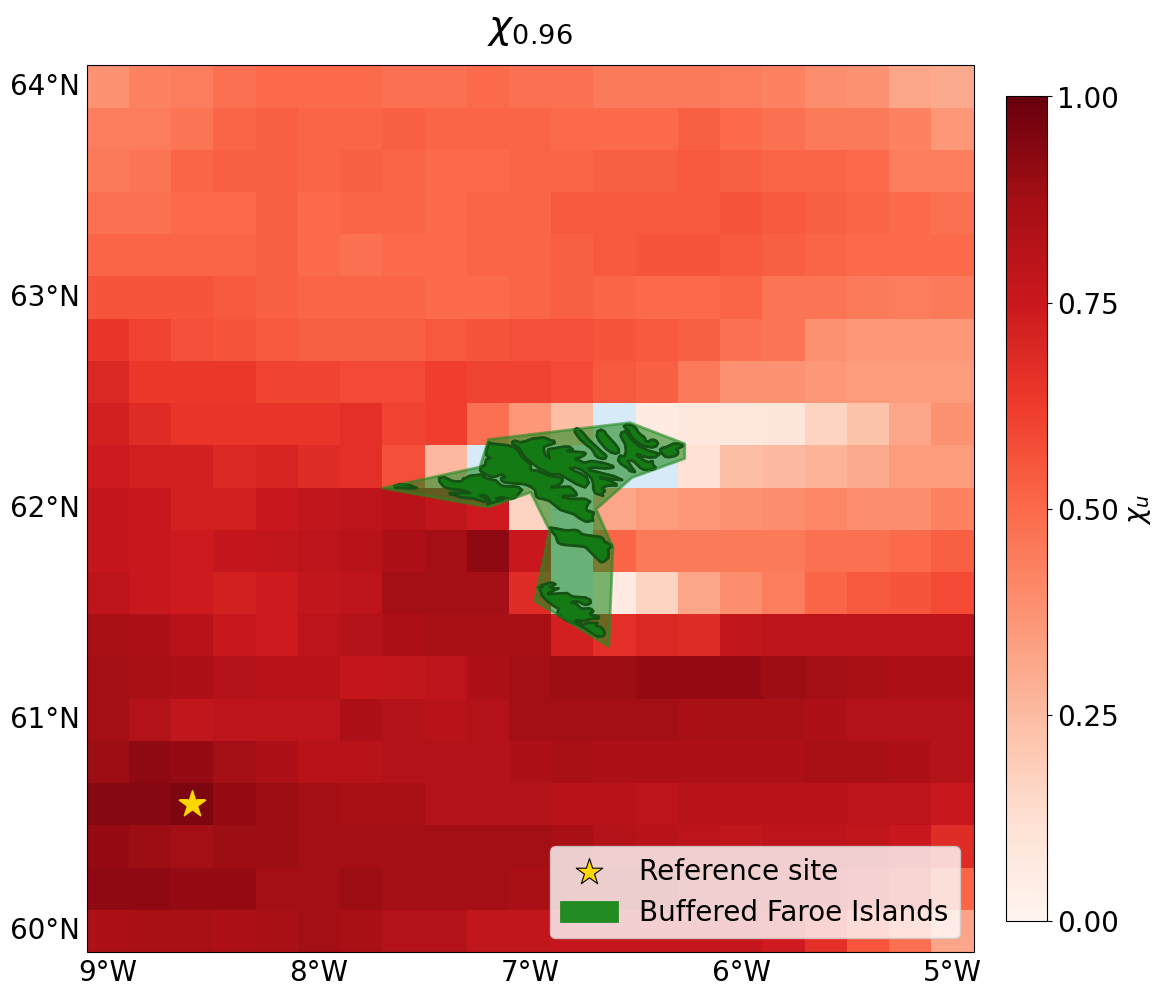} & 
			\includegraphics[width=0.2\textwidth, clip=true, trim=0 0 30 40]
			{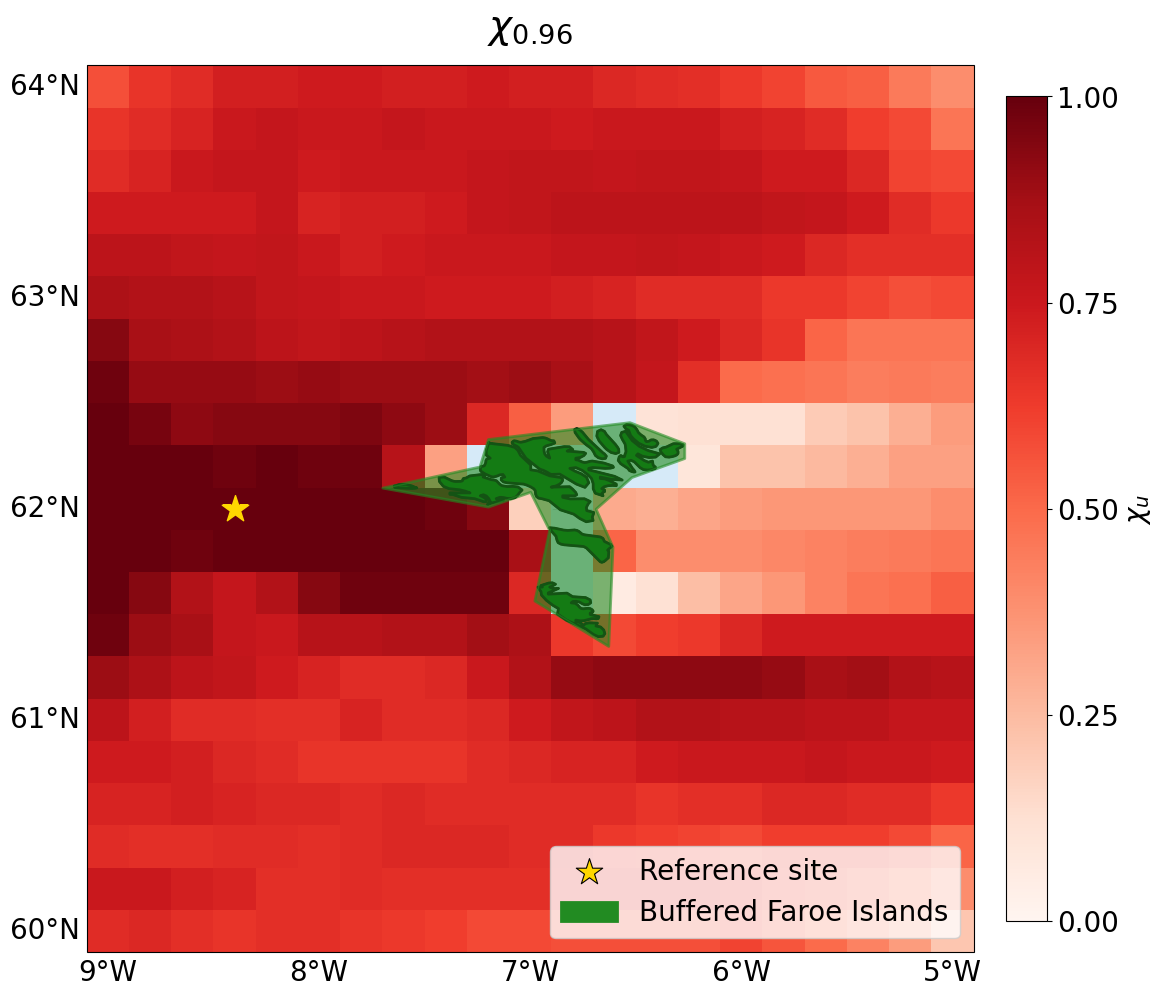} &
			\includegraphics[width=0.2\textwidth, clip=true, trim=0 0 30 40]
			{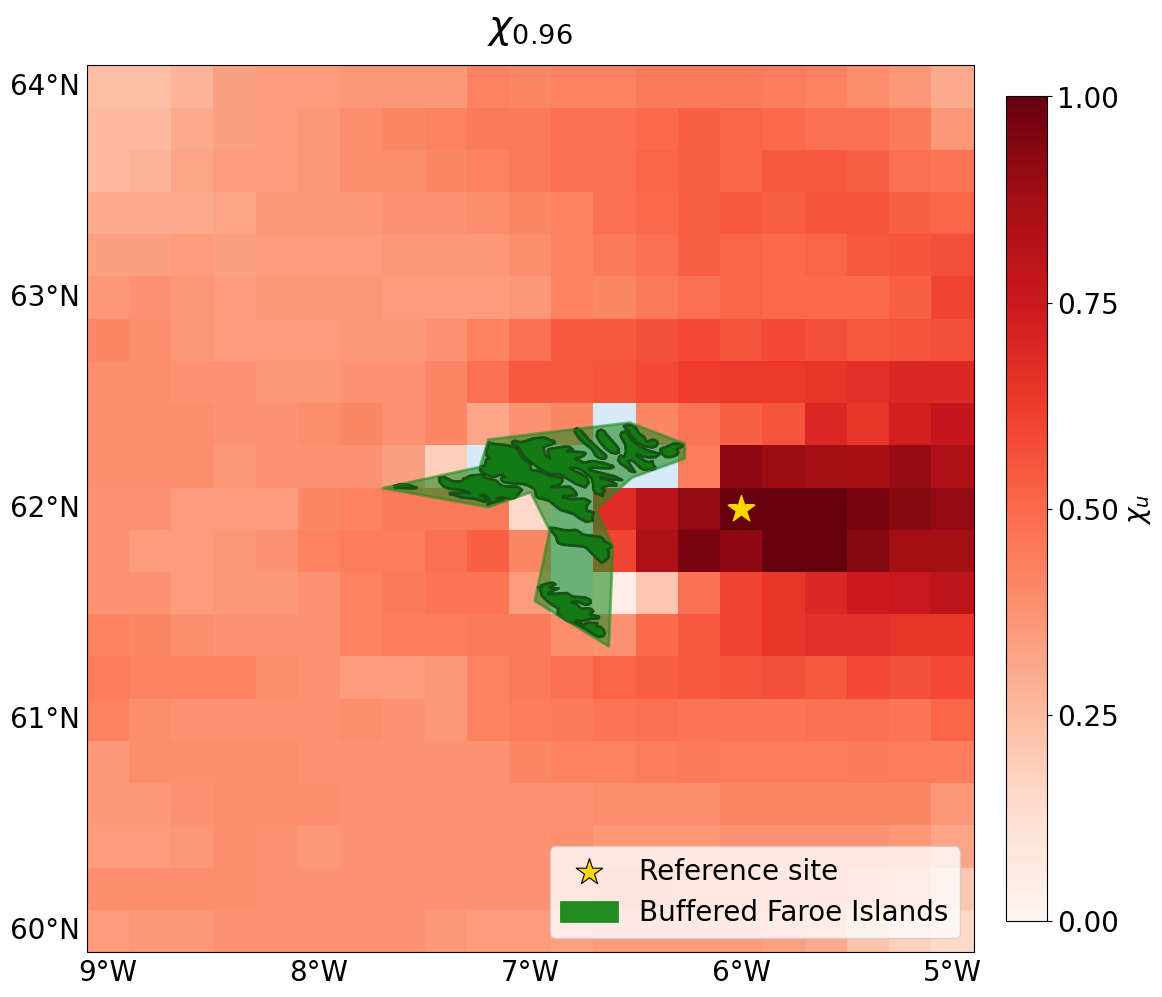} & 
			\includegraphics[width=0.2\textwidth, clip=true, trim=0 0 30 40]
			{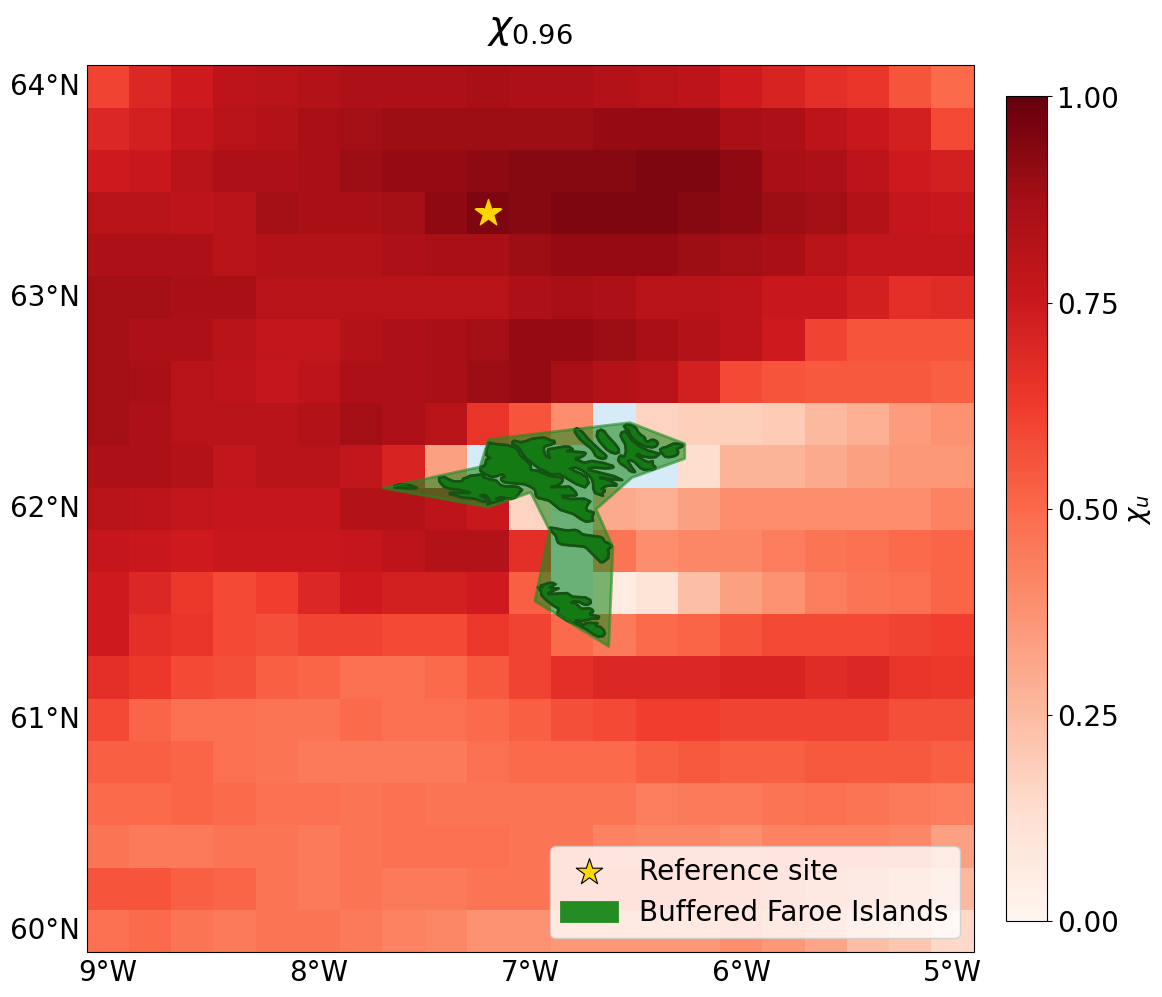} 
			\\
			\includegraphics[width=0.2\textwidth, clip=true, trim=0 0 30 40]
			{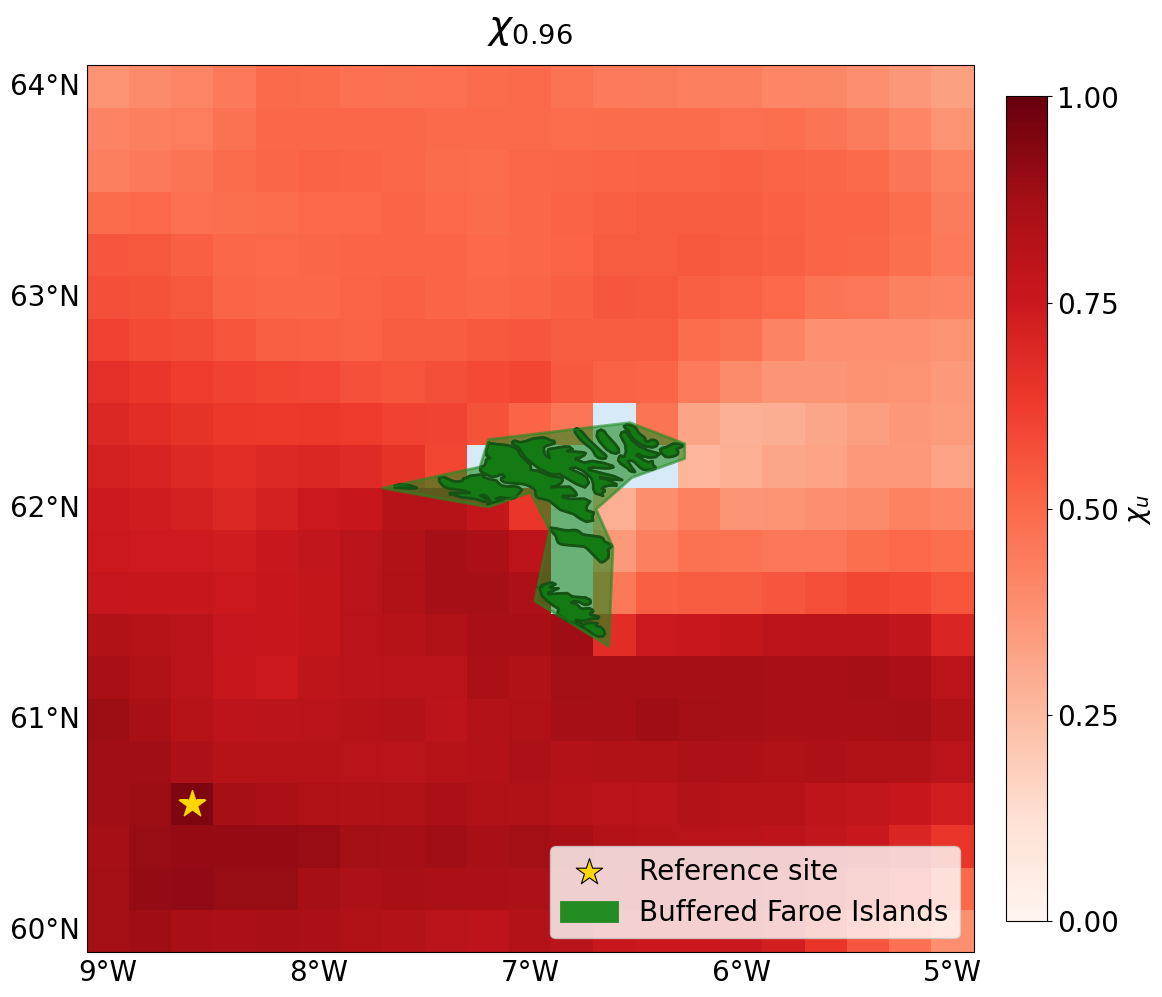} & 
			\includegraphics[width=0.2\textwidth, clip=true, trim=0 0 30 40]
			{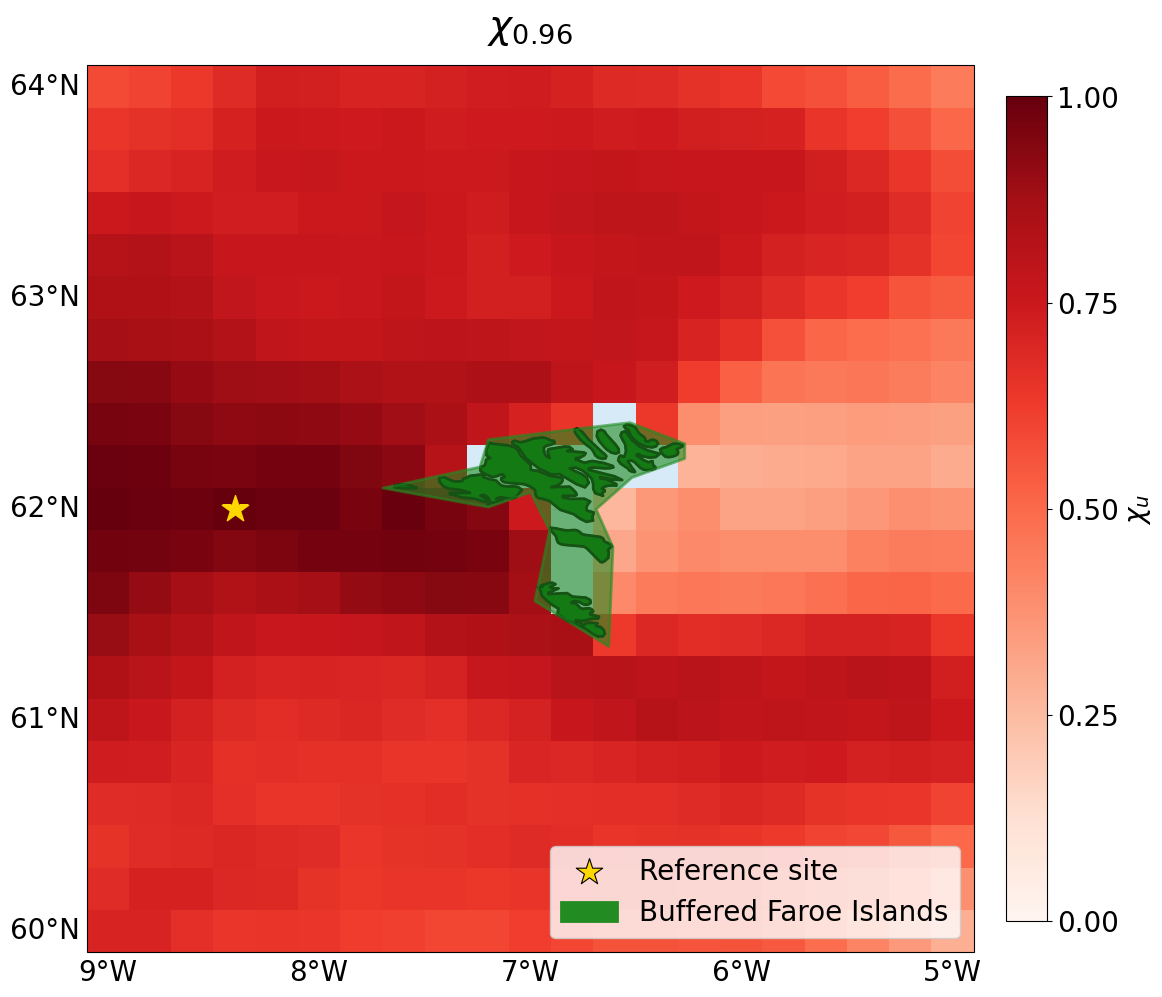} &
			\includegraphics[width=0.2\textwidth, clip=true, trim=0 0 30 40]
			{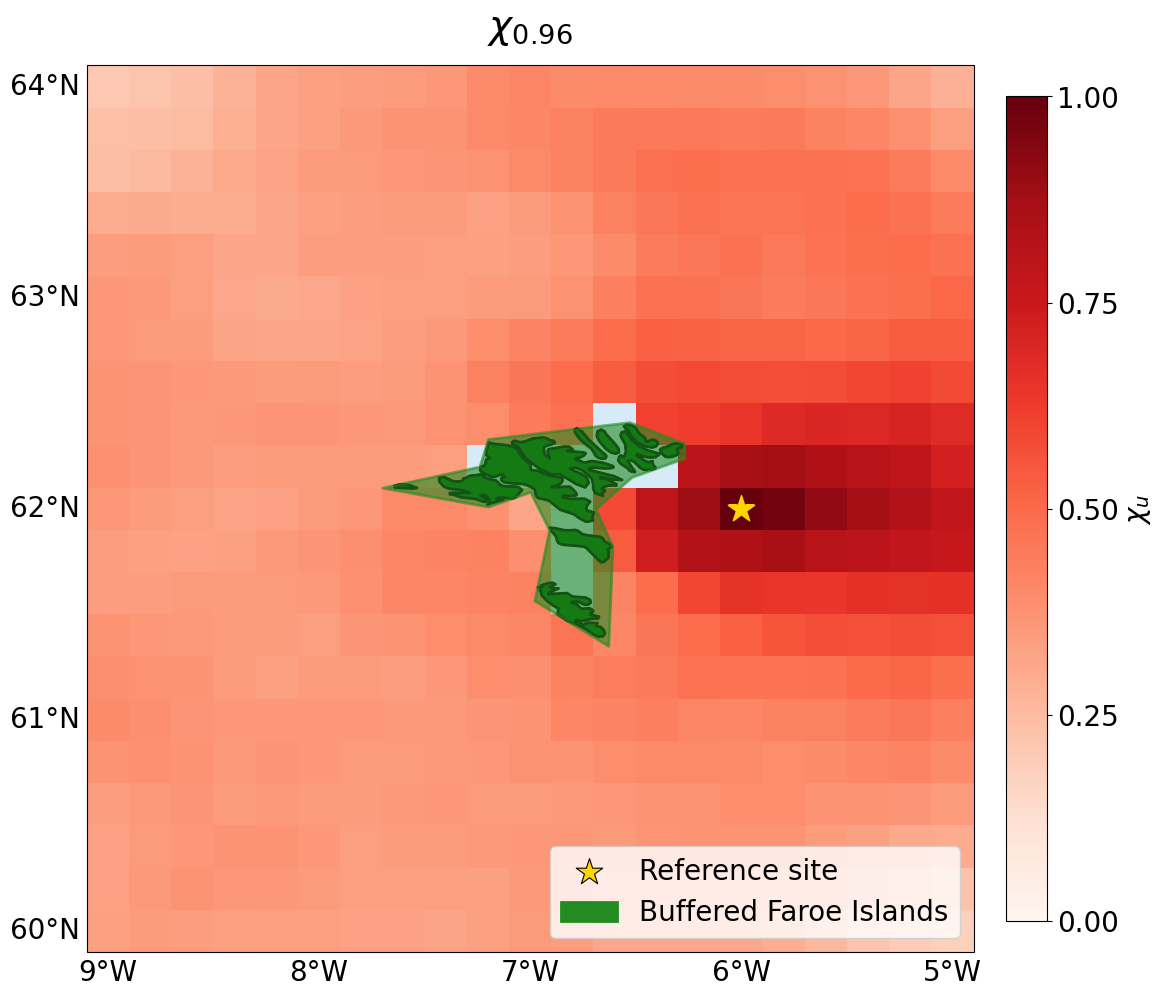} & 
			\includegraphics[width=0.2\textwidth, clip=true, trim=0 0 30 40]
			{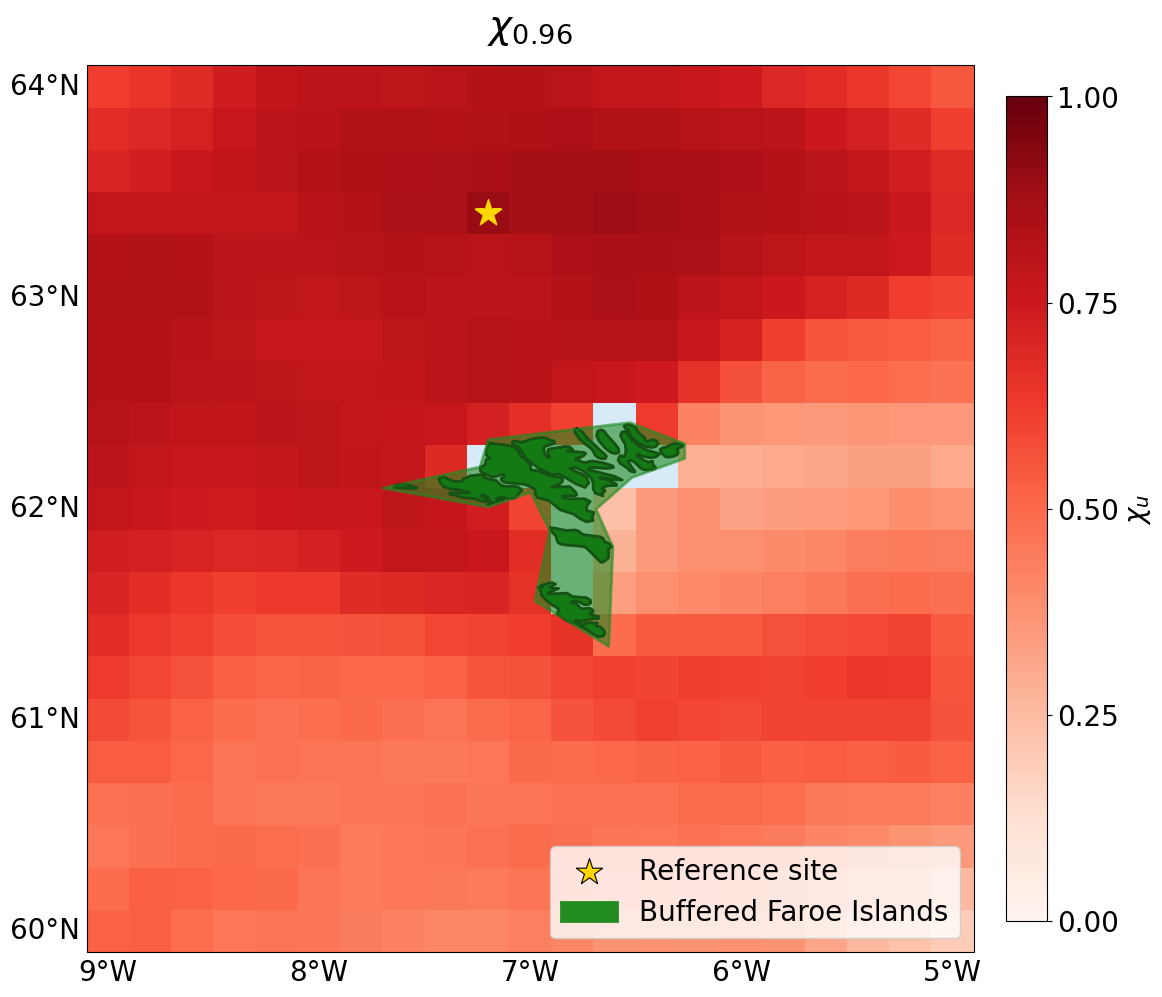} 
			\\
			\multicolumn{4}{c}{$\chi_{0.98}$} \\
			\includegraphics[width=0.2\textwidth, clip=true, trim=0 0 30 40]
			{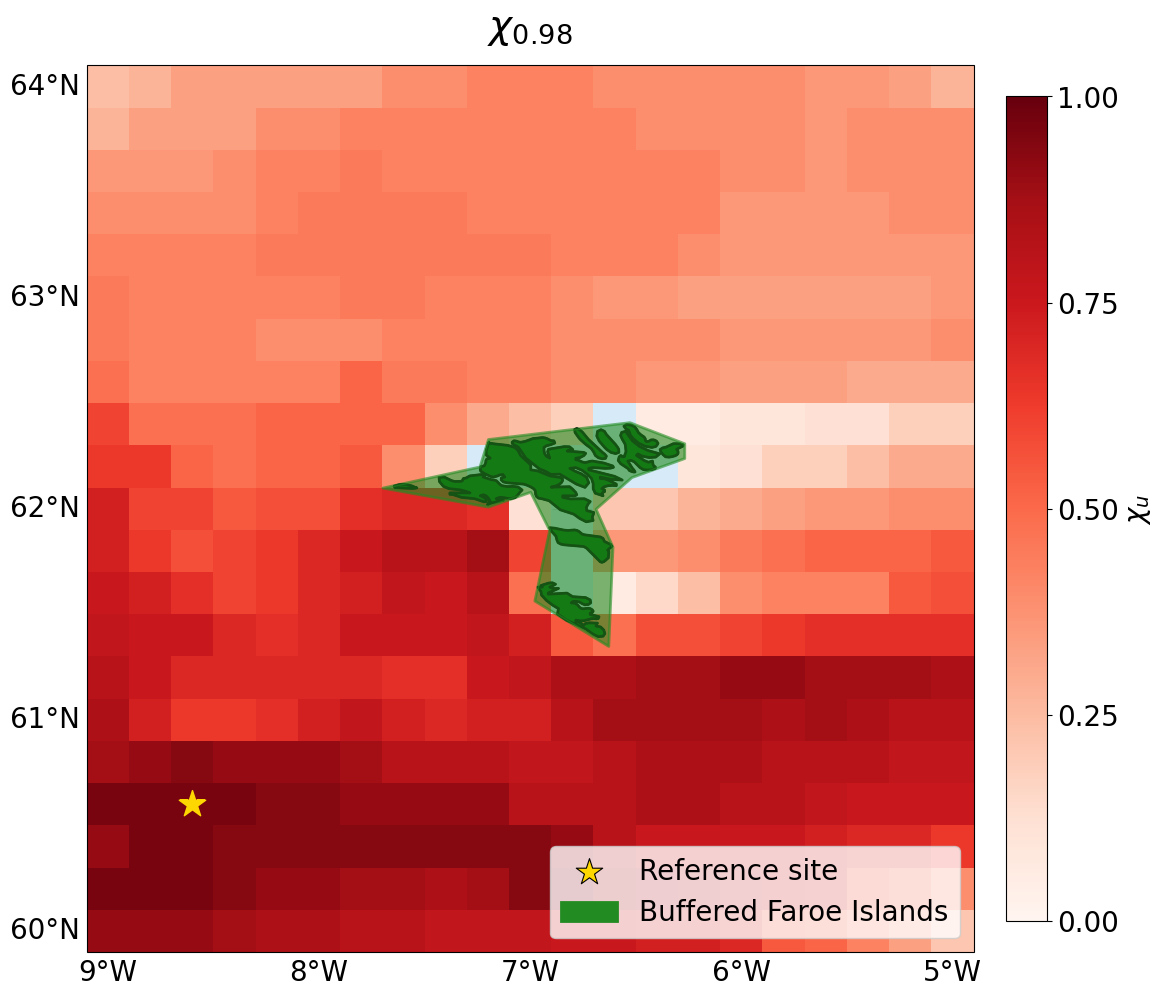} & 
			\includegraphics[width=0.2\textwidth, clip=true, trim=0 0 30 40]
			{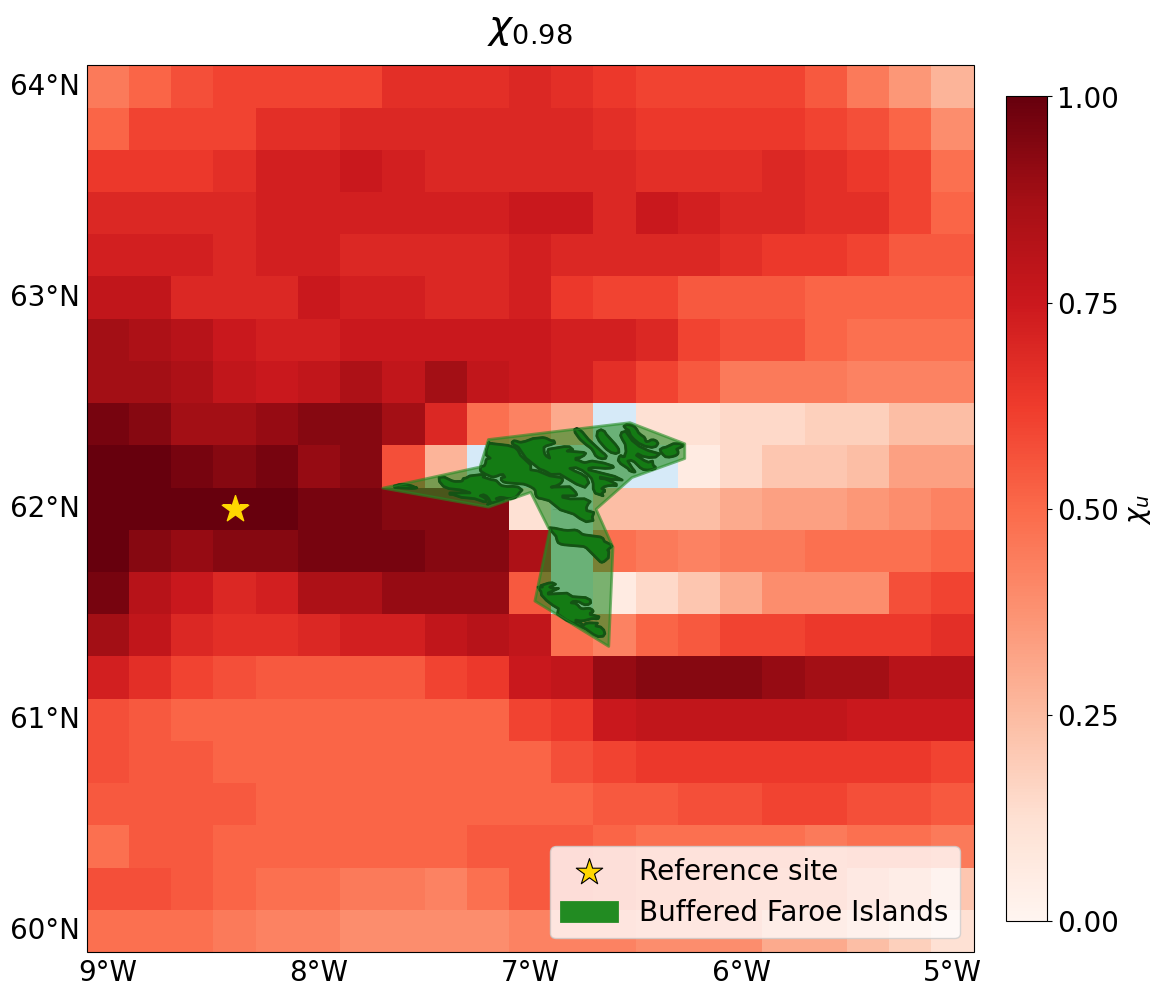} &
			\includegraphics[width=0.2\textwidth, clip=true, trim=0 0 30 40]
			{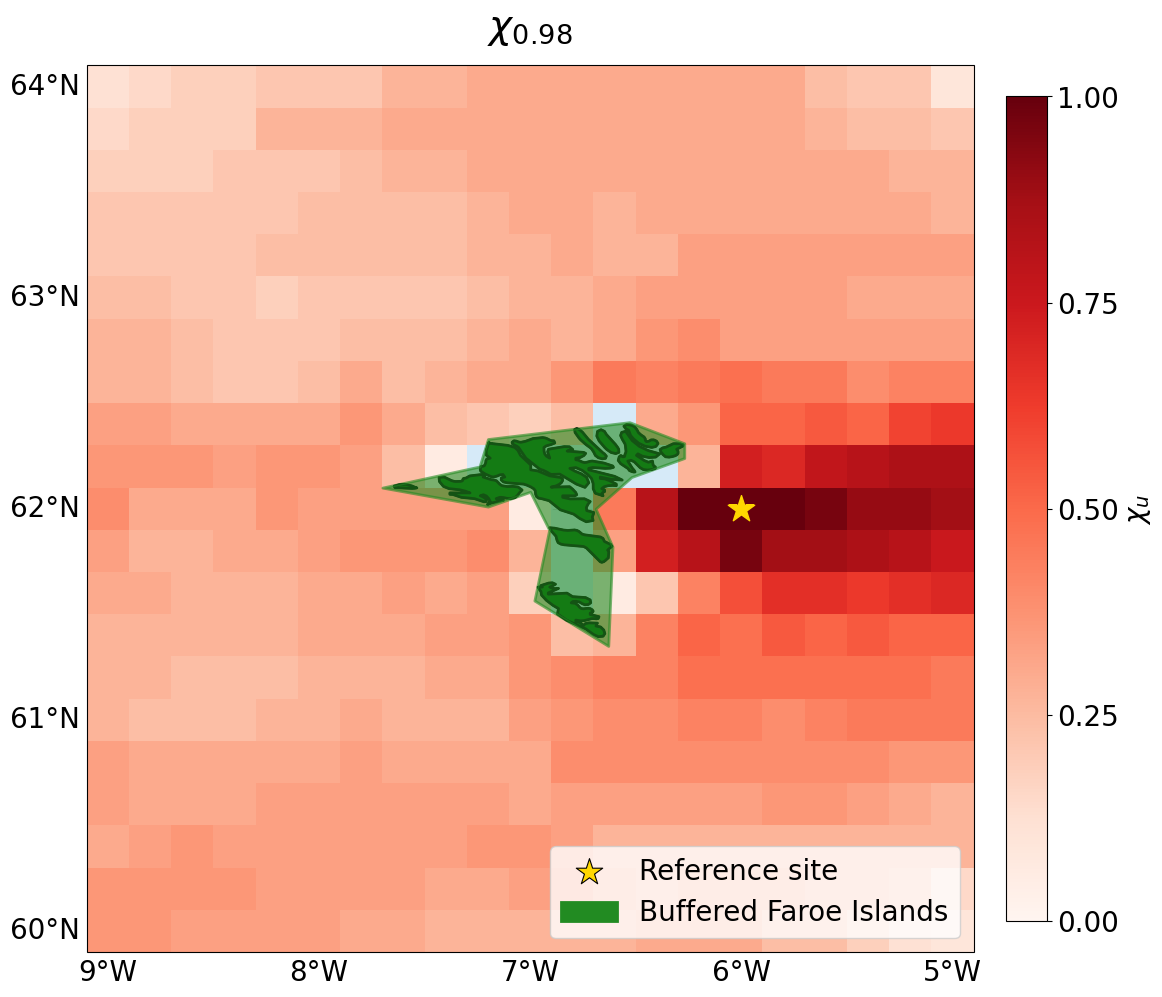} & 
			\includegraphics[width=0.2\textwidth, clip=true, trim=0 0 30 40]
			{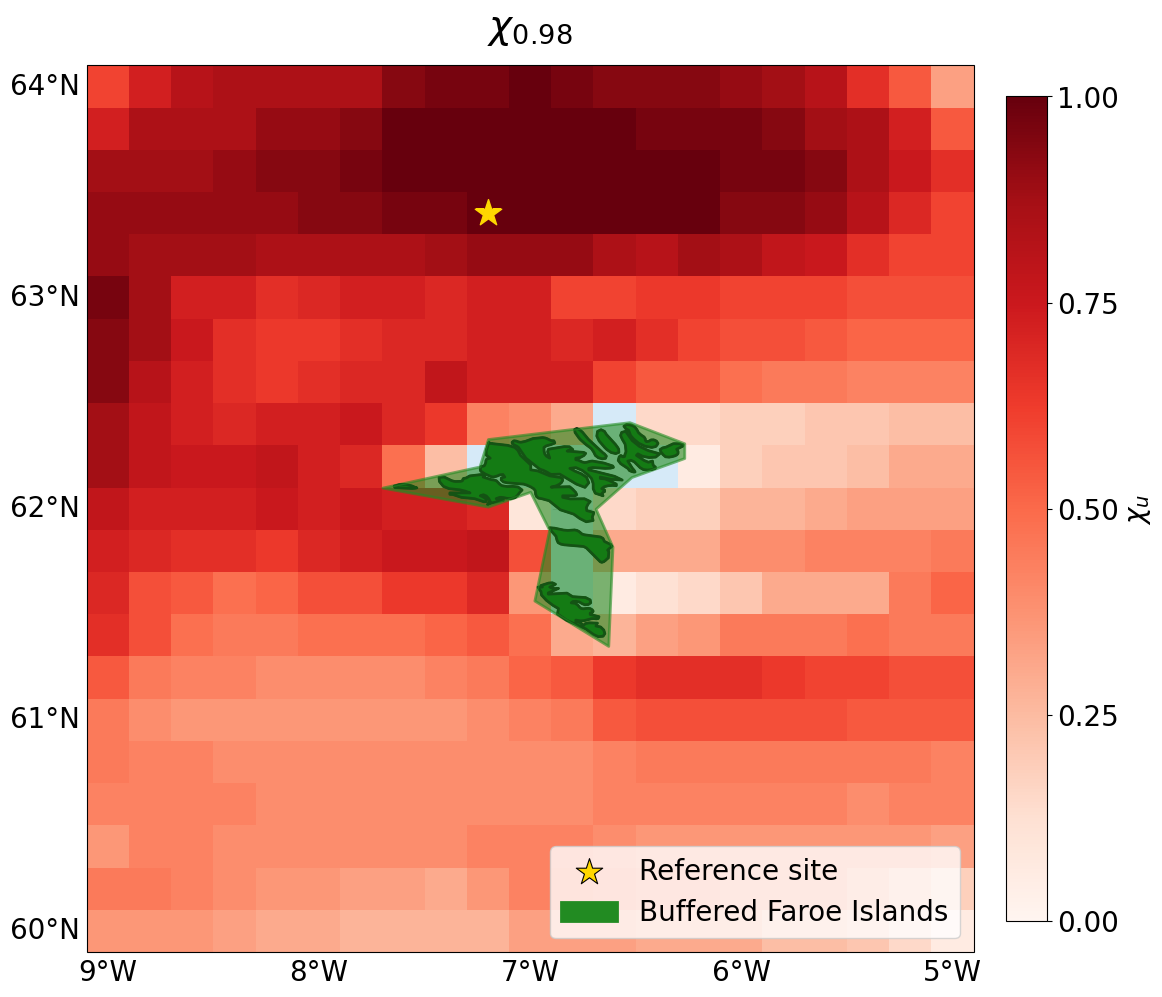} 
			\\
			\includegraphics[width=0.2\textwidth, clip=true, trim=0 0 30 40]
			{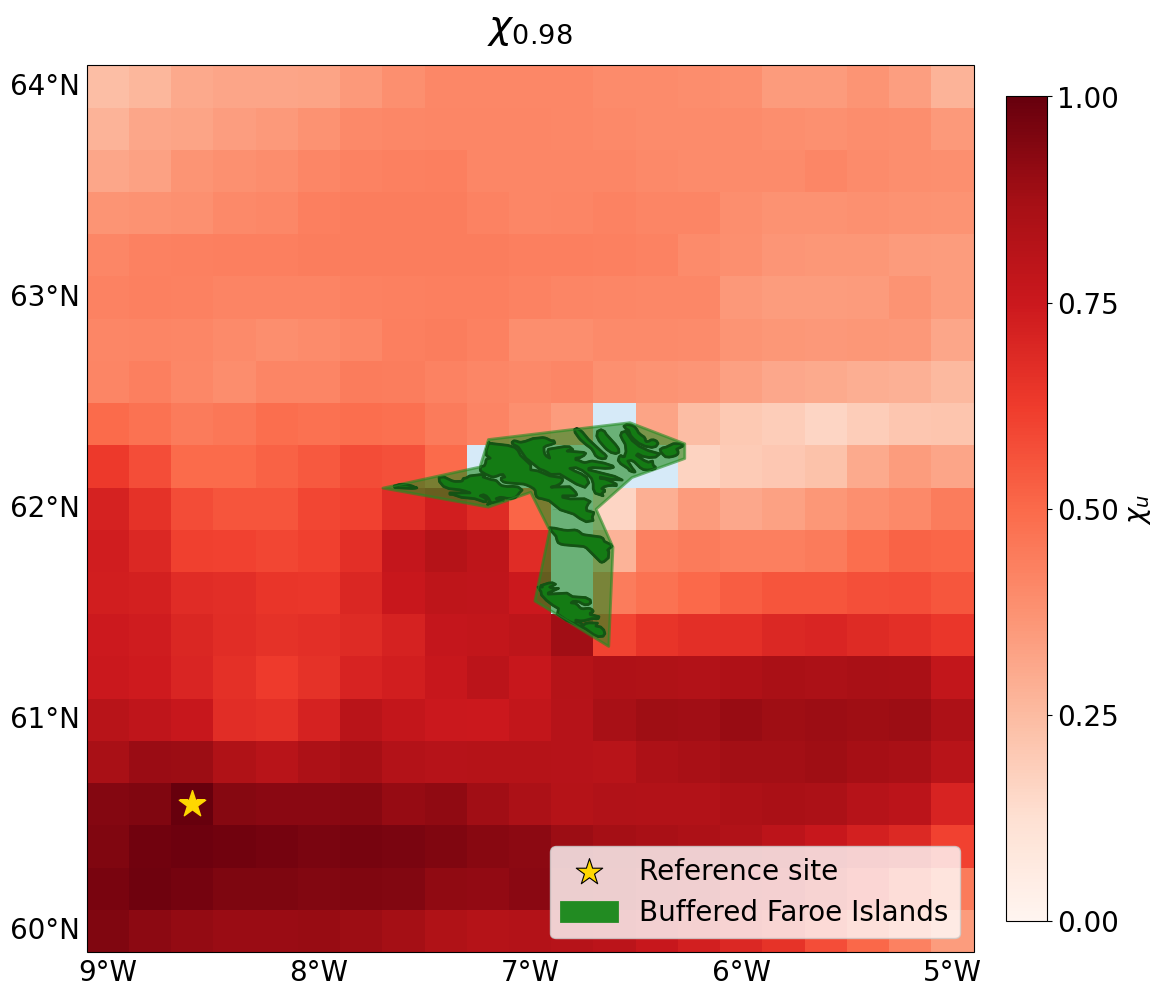} & 
			\includegraphics[width=0.2\textwidth, clip=true, trim=0 0 30 40]
			{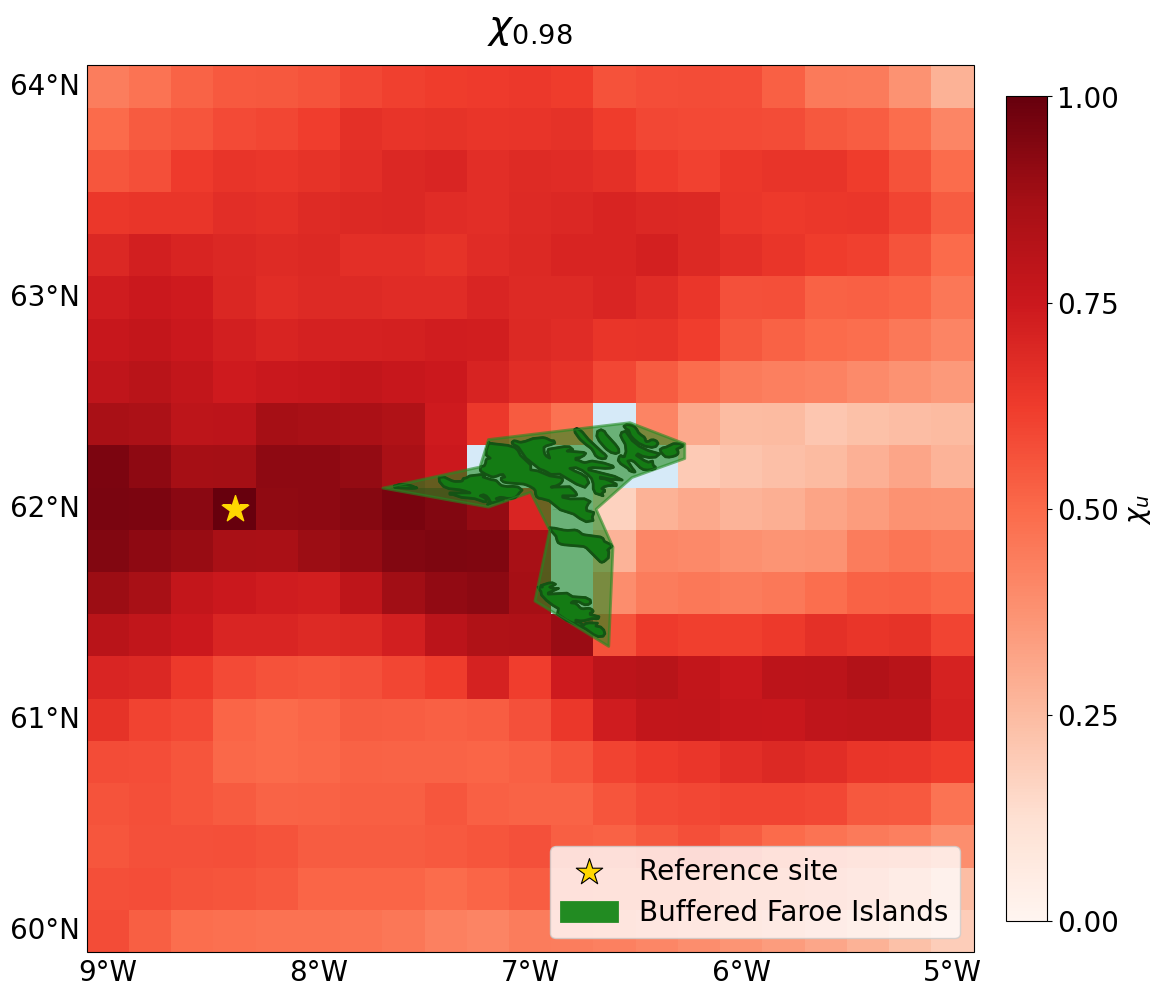} &
			\includegraphics[width=0.2\textwidth, clip=true, trim=0 0 30 40]
			{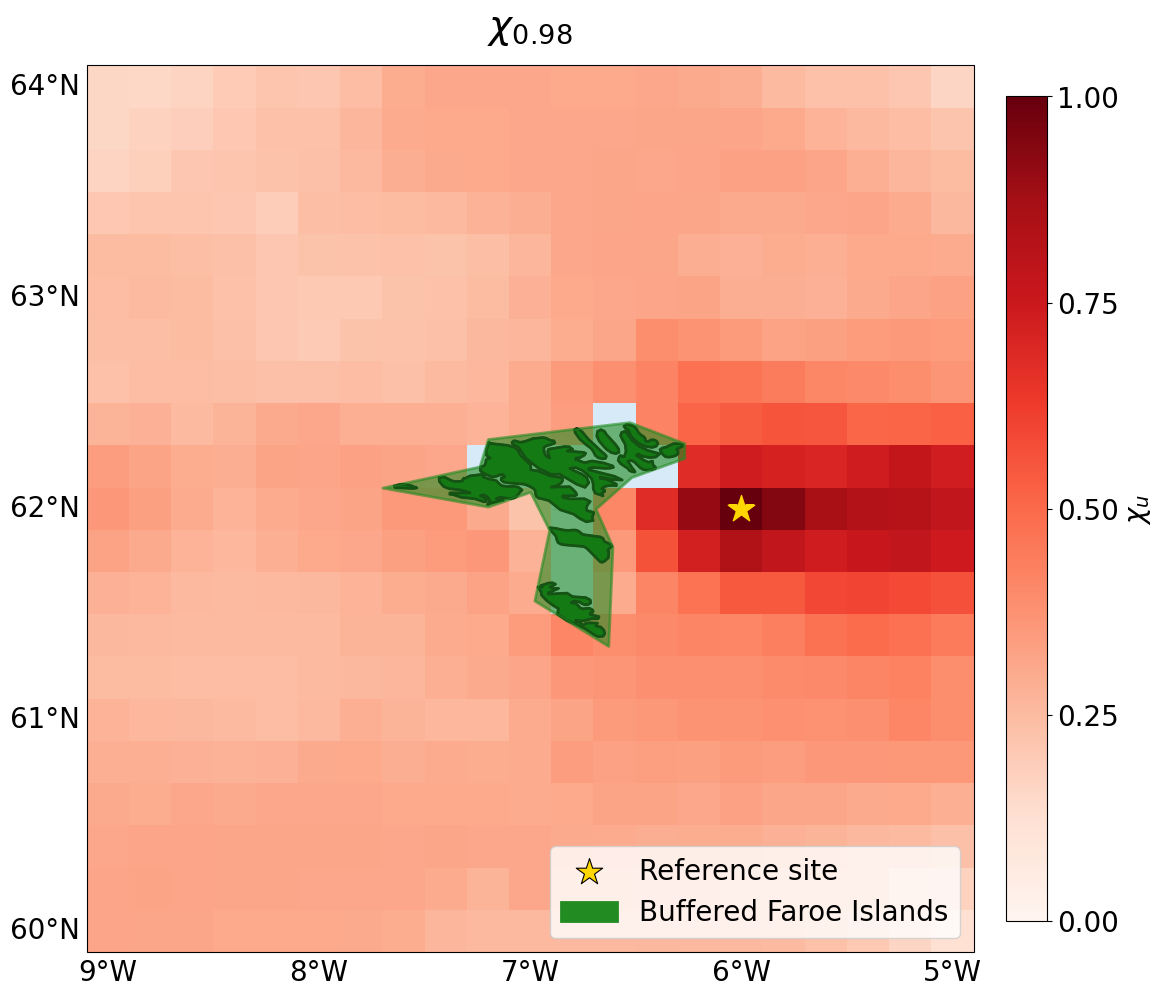} & 
			\includegraphics[width=0.2\textwidth, clip=true, trim=0 0 30 40]
			{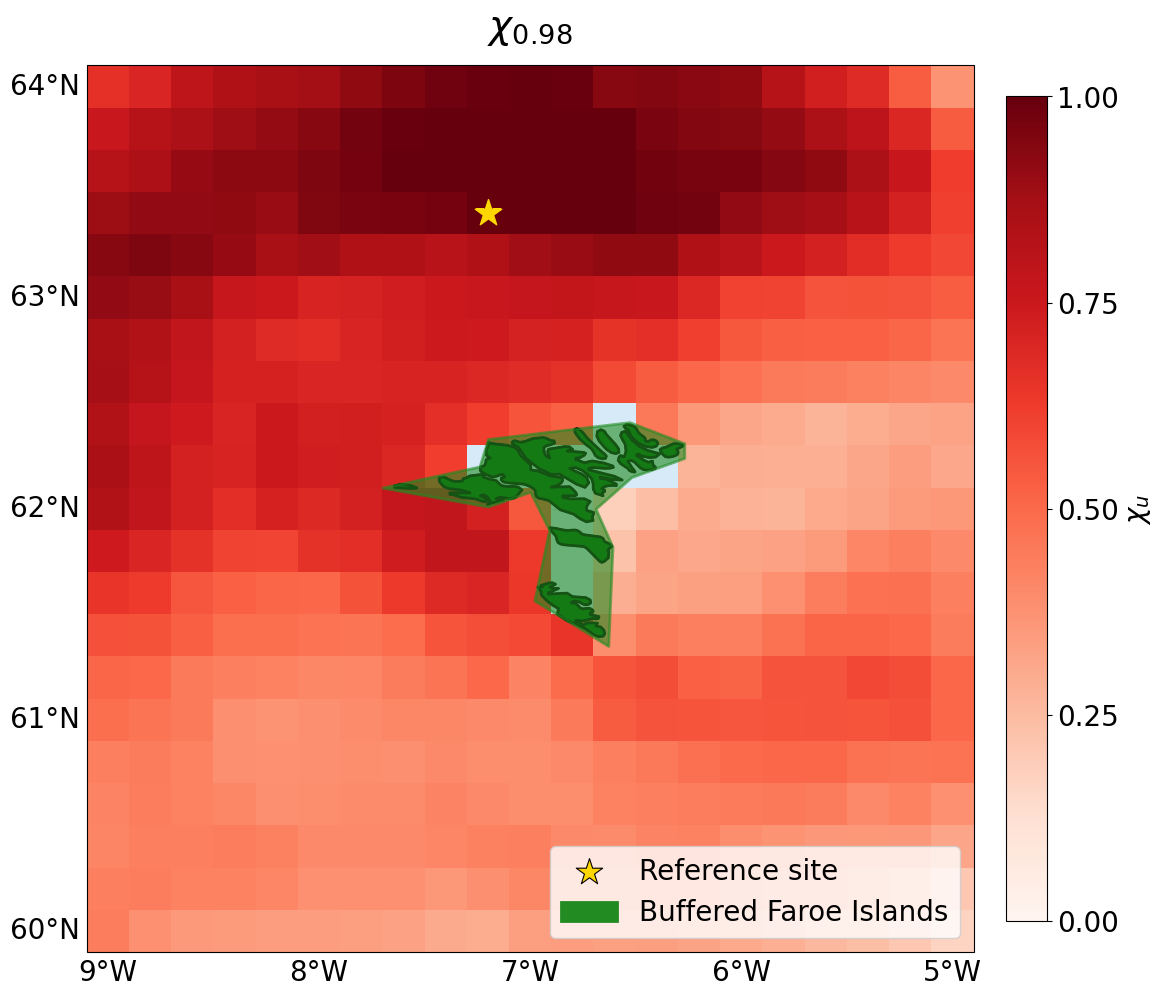} 
		\end{tabular}
		\caption{Empirical (top row within each threshold) and visibility-model-based (bottom row within each threshold) $\chi_{0.96}$ and $\chi_{0.98}$ surfaces for significant wave height around the Faroe Islands. Columns from left to right correspond to the four reference sites shown in the maps.}
		\label{fig:wave_chiu_emp_est_more}
	\end{figure}

	Figure~\ref{fig:wave_chiu_emp_est_more} compares the empirical and model-based conditional $\chi_u$ surfaces at thresholds $u = 0.96$ and $u = 0.98$ for four reference sites around the Faroe Islands. For each threshold, the top row displays the empirical estimates and the bottom row displays the corresponding visibility model estimates. The four columns represent the reference sites in the order shown in the figure. These results complement the $\chi_{0.99}$ surfaces present in the main text and provide a comparison of the estimated extremal dependnece patterns at less extreme thresholds. 
	
	The empirical surfaces exhibit substantial spatial variation across the domain, with the patterns depending on the location of the reference site. For the reference site in the western open ocean, elevated dependence extends across a broad region. For reference sites east of, north of, or closer to the islands, the areas of strongest dependence are more spatially localized. The visibility model estimates reproduce these broad reference-site-dependent patterns, including the spatial contrasts between open-ocean and island-adjacent regions, although some local differences remain between the empirical and fitted surfaces. 
	
	These comparisons illustrate how the visibility model represents spatially heterogeneous extremal dependence around the Faroe Islands. By incorporating the geometry of the islands into the dependence structure, the model can accommodate spatial patterns that vary with the reference site rather than relying solely on Euclidean separation. The results at $u = 0.96$ and $u = 0.98$ provide additional evidence for assessing the model's spatial fit beyond the highest threshold considered in the main text.
	
	\FloatBarrier
	\subsection{Comparison with Alternative Models} \label{supp:wave_model_comparison}
	
	\subsubsection{Posterior Summaries} \label{supp:wave_comparison_posterior}
	
	Tables~\ref{tab:wave_dependence_three_model_other} and \ref{tab:wave_dependence_two_model_knots} summarize the posterior means of the dependence parameters for the three models fitted to the training data. For the visibility and Shi et al.\ models, the tail dependence parameter varies spatially through a Gaussian kernel basis expansion, $\delta(\bm{s}) = \bm{G}(\bm{s})^\top \bm{\phi}$, with 41 knots and a shared bandwidth of $\sqrt{1/6}$. The posterior summaries of the knot coefficients are reported separately, while the resulting spatial patterns are illustrated in Figure~\ref{fig:wave_comparison_delta_posterior_mean}. In contrast, the Huser-Wadsworth model assumes a spatially constant tail dependence parameter, so its estimate is not directly comparable to individual coefficients in the spatially varying models. 
	
	\begin{table}[htbp]
		\centering
		\caption{Posterior mean and 95\% equal-tail credible interval, dependence hyperparameters, all three models. Visibility and Shi et al.\ both model tail dependence as spatially varying, $\delta(\bm{s}) = \bm{G}(\bm{s})^\top \bm{\phi}$, a 24-knot Gaussian-kernel expansion with shared bandwidth $\sqrt{1/6}$ (individual knot coefficients $\phi_1,\ldots,\phi_{41}$ shown in Figure~\ref{fig:wave_comparison_delta_posterior_mean} rather than tabulated); Huser--Wadsworth instead uses a single constant $\delta$, so its tail-dependence row is not comparable cell-by-cell with the other two.}
		\label{tab:wave_dependence_three_model_other}
		\begin{tabular}{ll ccc}
			\toprule
			\multicolumn{2}{c}{Parameter} & Visibility & Shi et al. & Huser--Wadsworth \\
			\midrule
			\multicolumn{2}{l}{Tail dependence, constant $\delta$} & -- & -- & $\underset{(0.173,\, 0.184)}{0.179}$ \\
			\midrule
			\multirow{3}{*}{Wendland} & $r$ & $\underset{(1.697,\, 1.716)}{1.707}$ & $\underset{(2.135,\, 2.156)}{2.147}$ & -- \\
			& $\lambda$ & $\underset{(0.625,\, 0.634)}{0.629}$ & $\underset{(0.569,\, 0.578)}{0.573}$ & -- \\
			& $\alpha$ & $\underset{(0.013,\, 0.040)}{0.025}$ & $\underset{(0.086,\, 0.095)}{0.090}$ & -- \\
			\midrule
			\multirow{2}{*}{Mat\'ern} & $\rho$ & $\underset{(3.139,\, 3.273)}{3.209}$ & -- & $\underset{(3.388,\, 3.532)}{3.455}$ \\
			& $\nu$ & $\underset{(1.002,\, 1.026)}{1.013}$ & $\underset{(1.270,\, 1.293)}{1.281}$ & $\underset{(0.939,\, 0.961)}{0.951}$ \\
			\midrule
			Nugget Effect & $\alpha_0$  & $\underset{(2.001,\, 2.002)}{2.001}$  & $\underset{(2.001,\, 2.003)}{2.002}$  &  $\underset{(2.001,\, 2.002)}{2.001}$ \\
			\bottomrule
		\end{tabular}
	\end{table}
	
	\begin{table}[htbp]
		\centering
		\caption{Posterior mean and 95\% credible interval of the tail-dependence knot coefficients $\phi_1,\ldots,\phi_{41}$, Visibility and Shi et al.\ models. Huser--Wadsworth is omitted here since it has no spatially varying tail dependence coefficients, using a single constant $\delta$ instead (Table~\ref{tab:wave_dependence_three_model_other}).}
		\label{tab:wave_dependence_two_model_knots}
			\begin{tabular}{c ccc @{\hskip 8mm} c cc}
				\toprule
				Model & Visibility & \multicolumn{2}{c}{Shi et al.} & Model & Visibility & Shi et al. \\ 
				\midrule
				$k$ & $\phi_k$ & $\phi_k$ & $\rho_k$ & Parameter & $\phi_k$ & $\phi_k$ \\
				\midrule
				1 & $\underset{(1.497,\, 1.588)}{1.543}$ & $\underset{(1.360,\, 1.439)}{1.399}$ & $\underset{(3.453,\, 3.631)}{3.542}$ & 22 & $\underset{(0.558,\, 0.641)}{0.600}$ & $\underset{(0.352,\, 0.425)}{0.388}$ \\
				2 & $\underset{(1.109,\, 1.189)}{1.148}$ & $\underset{(1.313,\, 1.403)}{1.362}$ & $\underset{(3.580,\, 3.762)}{3.670}$ & 23 & $\underset{(1.359,\, 1.548)}{1.454}$ & $\underset{(1.392,\, 1.483)}{1.441}$ \\
				3 & $\underset{(0.984,\, 1.052)}{1.017}$ & $\underset{(1.303,\, 1.379)}{1.340}$ & $\underset{(3.119,\, 3.306)}{3.214}$ & 24 & $\underset{(1.476,\, 1.542)}{1.510}$ & $\underset{(1.442,\, 1.484)}{1.463}$ \\
				4 & $\underset{(1.202,\, 1.276)}{1.239}$ & $\underset{(1.479,\, 1.547)}{1.514}$ & $\underset{(3.654,\, 3.843)}{3.751}$ & 25 & $\underset{(1.162,\, 1.213)}{1.189}$ & $\underset{(1.191,\, 1.233)}{1.213}$ \\
				5 & $\underset{(1.444,\, 1.527)}{1.484}$ & $\underset{(1.541,\, 1.593)}{1.567}$ & $\underset{(3.167,\, 3.304)}{3.236}$ & 26 & $\underset{(0.052,\, 0.224)}{0.131}$ & $\underset{(-0.449,\, -0.294)}{-0.372}$ \\
				6 & $\underset{(0.877,\, 0.925)}{0.902}$ & $\underset{(0.857,\, 0.895)}{0.876}$ & $\underset{(2.550,\, 2.874)}{2.711}$ & 27 & $\underset{(1.175,\, 1.270)}{1.225}$ & $\underset{(0.977,\, 1.062)}{1.023}$ \\
				7 & $\underset{(0.307,\, 0.390)}{0.345}$ & $\underset{(0.232,\, 0.296)}{0.265}$ & $\underset{(2.732,\, 2.914)}{2.822}$ & 28 & $\underset{(1.254,\, 1.343)}{1.298}$ & $\underset{(0.828,\, 0.919)}{0.876}$ \\
				8 & $\underset{(0.733,\, 0.805)}{0.770}$ & $\underset{(0.500,\, 0.552)}{0.526}$ & $\underset{(2.081,\, 2.385)}{2.235}$ & 29 & $\underset{(1.048,\, 1.130)}{1.087}$ & $\underset{(0.705,\, 0.797)}{0.754}$ \\
				9 & $\underset{(1.833,\, 1.922)}{1.874}$ & $\underset{(1.404,\, 1.466)}{1.436}$ & $\underset{(3.037,\, 3.205)}{3.119}$ & 30 & $\underset{(0.995,\, 1.268)}{1.120}$ & $\underset{(0.501,\, 0.694)}{0.594}$ \\
				10 & $\underset{(0.876,\, 0.957)}{0.915}$ & $\underset{(0.926,\, 0.971)}{0.949}$ & $\underset{(2.904,\, 3.060)}{2.979}$ & 31 & $\underset{(0.977,\, 1.066)}{1.021}$ & $\underset{(0.493,\, 0.575)}{0.533}$ \\
				11 & $\underset{(0.299,\, 0.350)}{0.323}$ & $\underset{(0.293,\, 0.337)}{0.314}$ & $\underset{(2.399,\, 2.714)}{2.557}$ & 32 & $\underset{(0.341,\, 0.427)}{0.382}$ & $\underset{(0.426,\, 0.511)}{0.470}$ \\
				12 & $\underset{(0.738,\, 0.810)}{0.773}$ & $\underset{(0.806,\, 0.872)}{0.840}$ & $\underset{(3.044,\, 3.217)}{3.127}$ & 33 & $\underset{(0.087,\, 0.169)}{0.129}$ & $\underset{(-0.129,\, -0.055)}{-0.094}$ \\
				13 & $\underset{(1.356,\, 1.435)}{1.397}$ & $\underset{(1.464,\, 1.533)}{1.498}$ & $\underset{(2.369,\, 2.655)}{2.510}$ & 34 & $\underset{(0.226,\, 0.326)}{0.275}$ & $\underset{(0.066,\, 0.141)}{0.101}$ \\
				14 & $\underset{(1.214,\, 1.273)}{1.245}$ & $\underset{(1.205,\, 1.260)}{1.232}$ & -- & 35 & $\underset{(0.780,\, 0.922)}{0.847}$ & $\underset{(0.477,\, 0.579)}{0.526}$ \\
				15 & $\underset{(1.007,\, 1.065)}{1.035}$ & $\underset{(1.026,\, 1.068)}{1.048}$ & -- & 36 & $\underset{(0.493,\, 0.581)}{0.537}$ & $\underset{(0.673,\, 0.736)}{0.703}$ \\
				16 & $\underset{(0.882,\, 0.953)}{0.918}$ & $\underset{(0.822,\, 0.875)}{0.848}$ & -- & 37 & $\underset{(1.277,\, 1.466)}{1.374}$ & $\underset{(1.040,\, 1.256)}{1.141}$ \\
				17 & $\underset{(1.457,\, 1.526)}{1.496}$ & $\underset{(1.532,\, 1.593)}{1.564}$ & -- & 38 & $\underset{(1.132,\, 1.216)}{1.173}$ & $\underset{(1.103,\, 1.182)}{1.141}$ \\ 
				18 & $\underset{(0.899,\, 1.021)}{0.961}$ & $\underset{(1.135,\, 1.260)}{1.196}$ & -- & 39 & $\underset{(1.483,\, 1.575)}{1.531}$ & $\underset{(0.991,\, 1.058)}{1.025}$ \\
				19 & $\underset{(0.922,\, 0.979)}{0.951}$ & $\underset{(1.243,\, 1.293)}{1.267}$ & -- & 40 & $\underset{(1.344,\, 1.439)}{1.394}$ & $\underset{(1.055,\, 1.125)}{1.093}$ \\ 
				20 & $\underset{(0.738,\, 0.793)}{0.766}$ & $\underset{(0.986,\, 1.026)}{1.006}$ & -- & 41 & $\underset{(1.704,\, 1.961)}{1.828}$ & $\underset{(1.241,\, 1.368)}{1.300}$ \\ 
				21 & $\underset{(0.316,\, 0.392)}{0.355}$ & $\underset{(0.060,\, 0.112)}{0.086}$ & -- &  &  & \\
				\bottomrule
			\end{tabular}
	\end{table}
	
	The Wendland kernel parameters govern the spatial extent and anisotropy of the dependence structure. The latent Gaussian process specifications differ across models. The visibility model incorporates the visibility-based Mat{\'e}rn covariance structure \citep{Gilbert2024}, the Shi et al.\ model uses a non-stationary, locally isotropic Mat{\'e}rn covariance \citep{paciorek2006, risser2015} constructed using 13 knots, and the Huser-Wadsworth model uses a stationary isotropic Mat{\'e}rn covariance. All three models include the log-Laplace nugget effect \citep{Shi2026log}. 
	
	For the wave height analysis, the second-step MCMC estimates the dependence parameters conditional on the marginal GPD parameter estimates obtained in the first step. Accordingly, the reported credible intervals quantify posterior uncertainty in the dependence parameters conditional on those fixed marginal estimates. These posterior summaries complement the holdout predictive scores reported in the main text by describing the fitted dependence structures for the three competing models. 
	
	\subsubsection{QQ Plots Diagnostics at Holdout Sites} \label{supp:wave_comparison_qq}
	
\begin{figure}[htbp]
		\centering
		\includegraphics[width=0.8\textwidth]{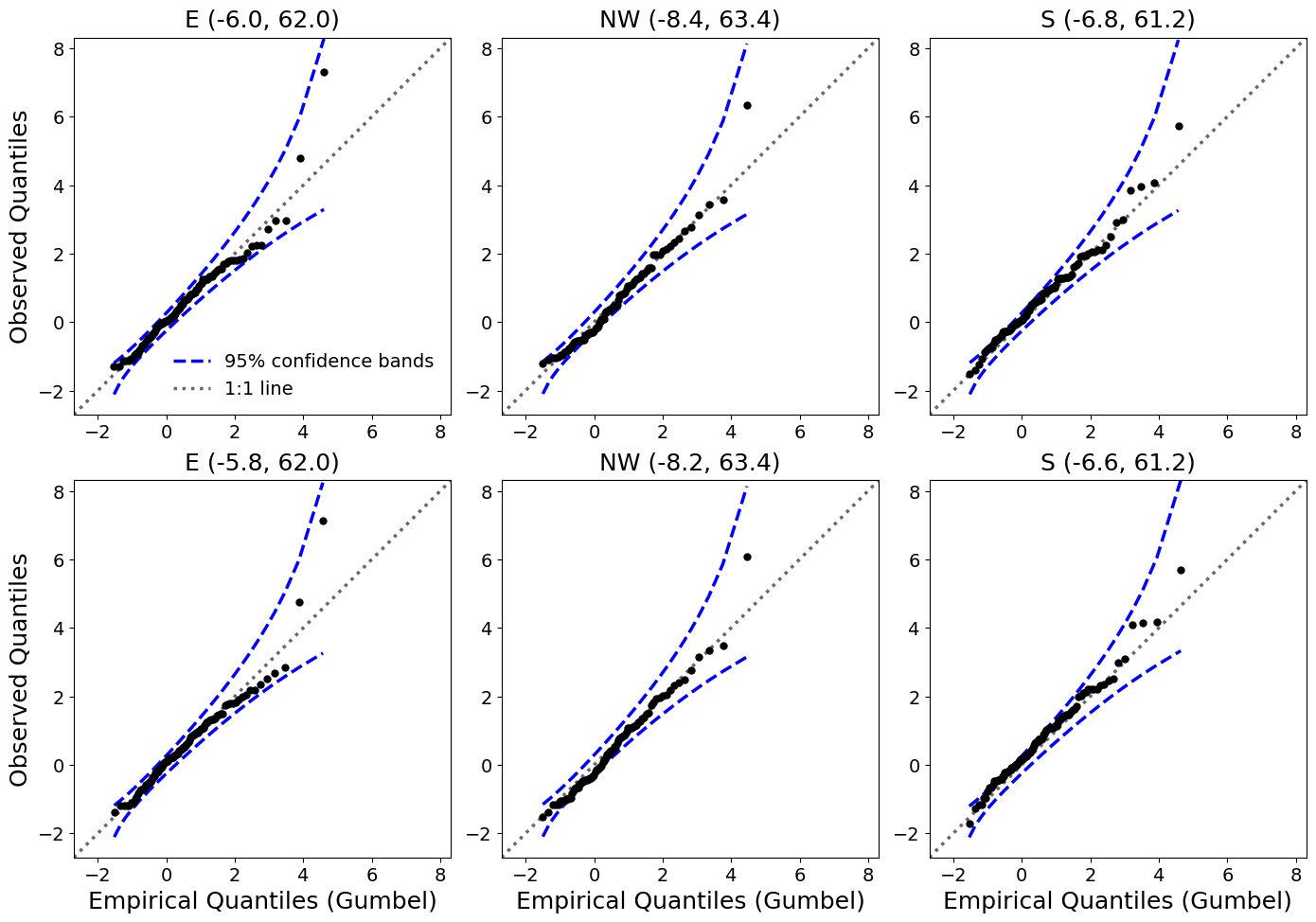}
		\caption{QQ plots for the first-step marginal GPD models at six selected holdout sites in the Faroe Islands wave height analysis. Observations are transformed to a standard Gumbel distribution, and the empirical quantiles are compared with theoretical Gumbel quantiles. The sites are drawn from the eastern (E), northwestern (NW), and southern (S) holdout blocks. The diagonal line indicates agreement with the theoretical distribution, and the 95\% confidence bands represent uncertainty in the estimated quantiles.}
		\label{fig:wave_qq_holdout}
	\end{figure}

	Figure~\ref{fig:wave_qq_holdout} presents the QQ plots for the first-step marginal GPD model at six selected holdout sites distributed across three spatial blocks, East (E), northwest (NW), and south (S) of the Faroe Islands. The marginal observations are transformed to a standard Gumbel distribution, and the resulting empirical quantiles are computed with the corresponding theoretical Gumbel quantiles. The plots also display the associated 95\% confidence bands. 
	
	Overall, the QQ plots show good agreement between the transformed observations and the theoretical Gumbel quantiles. The estimated quantiles and their 95\% confidence bands generally lie close to the diagonal reference line, indicating that the fitted GPD margins provide a reasonable representation of the exceedance distributions at these holdout sites. This agreement is observed across all three spatial blocks, supporting the adequacy of the first-step marginal fits for the subsequent dependence analysis. These diagnostics assess marginal fit at the selected sites and do not, by themselves, establish the adequacy of the dependence model. 
	
	\subsubsection{Wendland Kernel: Visibility and Shi et al. Model} \label{supp:wave_comparison_wendland}
	
\begin{figure}[htbp]
		\centering
		\includegraphics[width=0.7\textwidth]{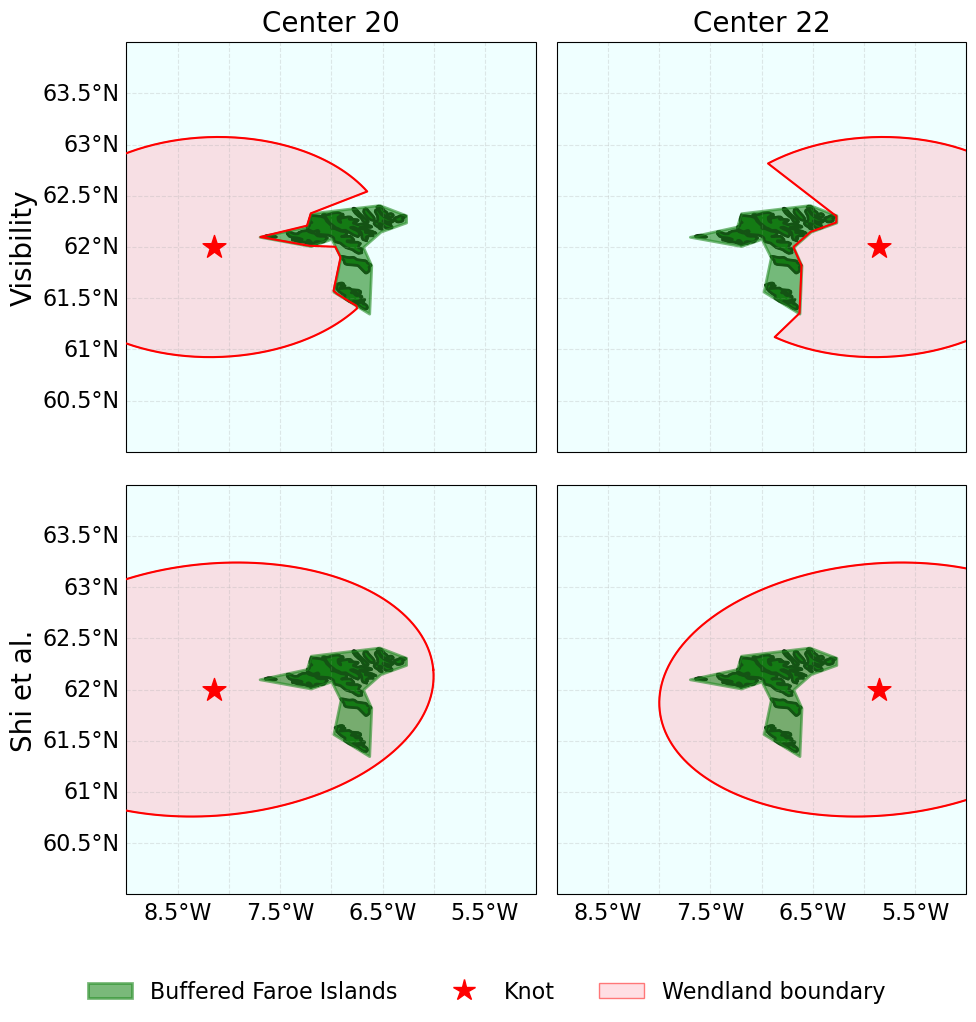}
		\caption{Anisotropic Wendland kernel supports for two representative knots west and east of the Faroe Islands (left and right columns, respectively) under the visibility model (top row) and the Shi et al.\ model (bottom row), both fitted using the $369$ training sites.}
		\label{fig:wave_compairson_wendland_examples}
	\end{figure}

	Figure~\ref{fig:wave_compairson_wendland_examples} compares the anisotropic Wendland kernel supports under the visibility model and the Shi et al.\ model for two representative knots, one west and one east of the Faroe Islands. Both models are fitted using only the $369$ training sites from the spatial holdout described in Section~\ref{subsec:wave_comparison} of the main text, with the same training-test split used throughout the model comparison. 
	
	The two model produce different support geometries despite using the same type of anisotropic Wendland kernel. Under the visibility model, the kernel support is restricted by the visibility graph, locations that cannot be connected to a knot by a line segment lying entirely within the ocean domain receive zero kernel weight. Consequently, the support is truncated where the islands obstruct visibility. In contrast, the Shi et al.\ model does not incorporate this visibility restriction. Its anisotropic Wendland kernel therefore retains support across the islands wherever locations fall within the kernel's elliptical boundary, even when land intervenes between a knot and a location. 
	
	These plots illustrate a structural distinction between the two dependence models rather than a difference arising solely from parameter estimation or the choice of training sites. Although both models accommodate anisotropy, only the visibility model explicitly modifies the kernel support to account for the physical barriers formed by the Faroe Islands. 
	
	\subsubsection{Tail Dependence Surface: Visibility and Shi et al. Model} \label{supp:wave_comparison_tail_surface}
	
\begin{figure}[htbp]
		\centering
		\includegraphics[width=0.7\textwidth]{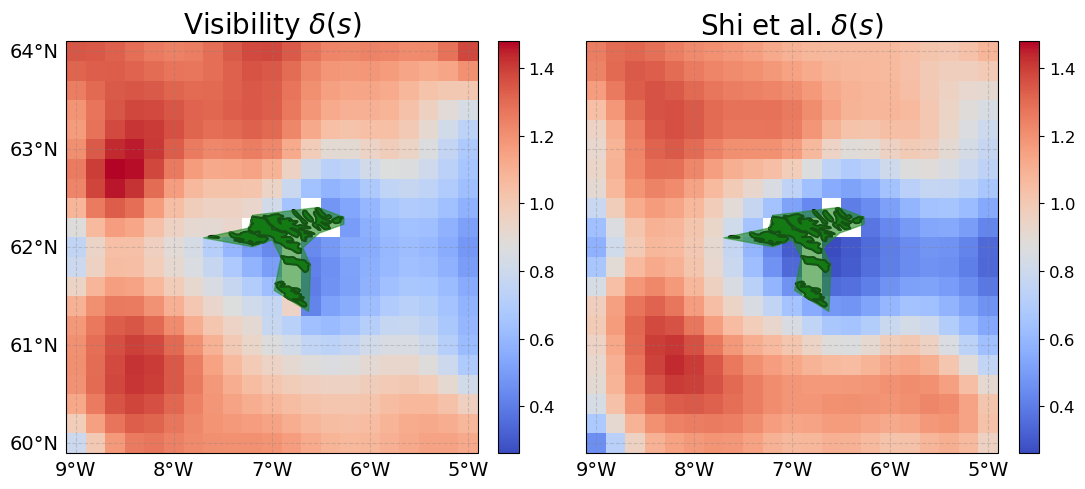}
		\caption{Posterior mean of the site-level tail dependence parameter $\delta(\bm{s})$ under the visibility model (left) and the Shi et al.\ model (right), both fitted using only the $369$ training sites.}
		\label{fig:wave_comparison_delta_posterior_mean}
	\end{figure}

	Figure~\ref{fig:wave_comparison_delta_posterior_mean} compares the posterior mean of the site-level tail dependence parameter $\delta(\bm{s})$ under the visibility model and Shi et al.\ model, both fitted using only the $369$ training sites. In both models, $\delta(\bm{s})$ is represented by a Gaussian kernel basis expansion with 41 knots and a shared bandwidth of $\sqrt{1/6}$. The key difference is that the visibility model incorporates the visibility graph into its dependence construction, whereas the Gaussian kernel expansion for the Shi et al.\ model is not restricted by the geometry of the islands. 
	
	The two posterior mean surfaces can be compared to across how the estimated spatial variation in tail dependence differs between the barrier-aware and barrier-blind specifications. In particular, the surfaces show whether the spatial contrasts between the open ocean and regions around or east of the islands are captured similarly by both models, despite their different constructions of tail dependence. Any differences between the estimated surfaces reflect the combined effects of the dependence specifications and the information available from the training data, rather than the visibility restriction alone. 
	
	Together with the Wendland support plots in Figure~\ref{fig:wave_compairson_wendland_examples}, this comparison distinguishes the geometric restriction imposed directly by the visibility model from the spatial variation in $\delta(\bm{s})$ estimated through the likelihood. The resulting surfaces provide a complementary view of how the two models represent extremal dependence around the Faroe Islands. 
	
	\subsubsection{Extremal Dependence: All Three Models} \label{supp:wave_comparison_chiu}
	
\begin{figure}[ht]
		\centering
		\begin{tabular}{cc cc}
			Empirical & Visibility & Shi et al. & Huser-Wadsworth \\
			\includegraphics[width=0.2\textwidth, clip=true, trim=0 0 30 40]
			{wave_emp_chiu_center_65_99.png} & 
			\includegraphics[width=0.2\textwidth, clip=true, trim=0 0 30 40]
			{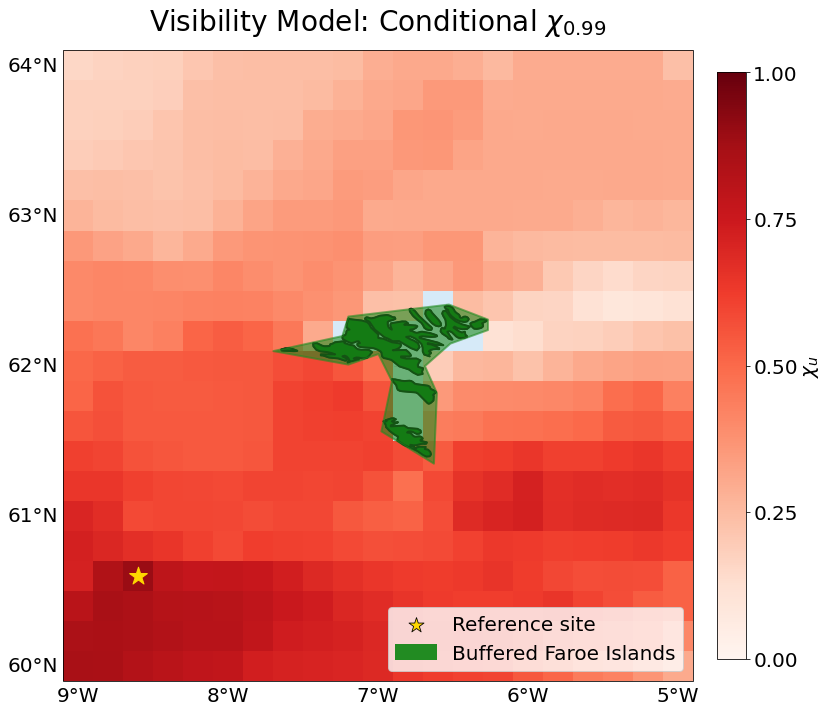} &
			\includegraphics[width=0.2\textwidth, clip=true, trim=0 0 30 40]
			{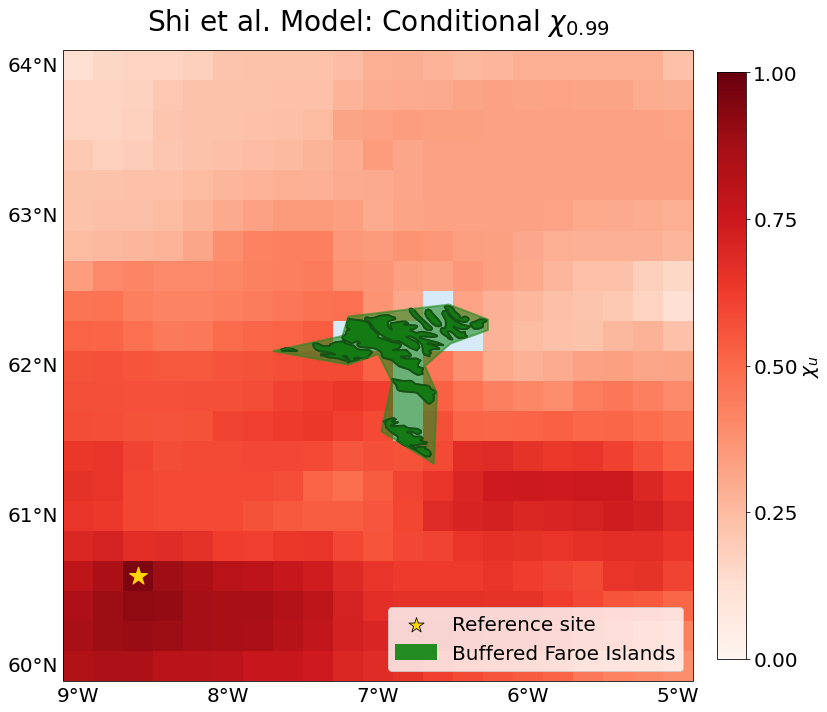} & 
			\includegraphics[width=0.2\textwidth, clip=true, trim=0 0 30 40]
			{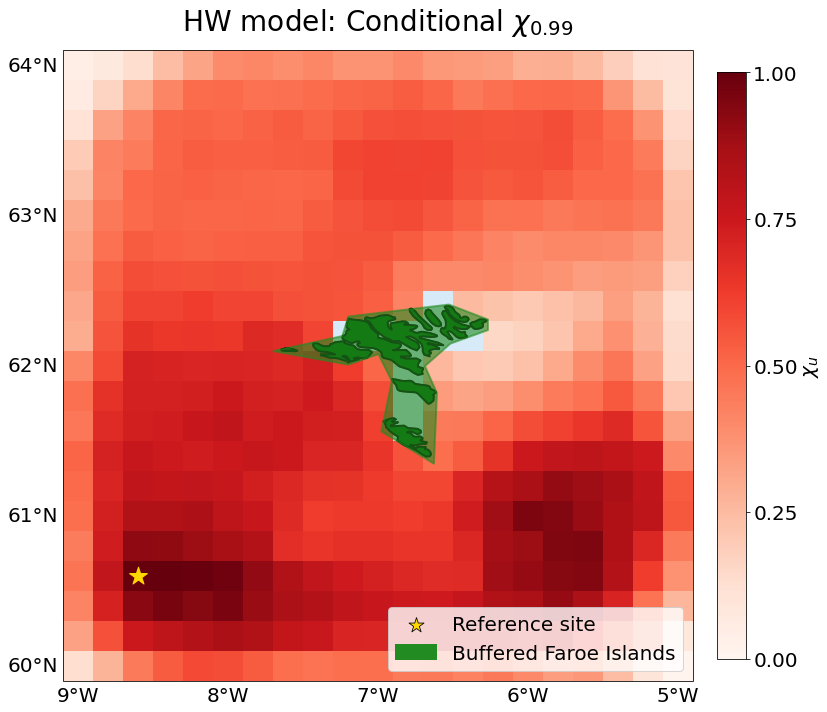} 
			\\
			\includegraphics[width=0.2\textwidth, clip=true, trim=0 0 30 40]
			{wave_emp_chiu_center_211_99.png} & 
			\includegraphics[width=0.2\textwidth, clip=true, trim=0 0 30 40]
			{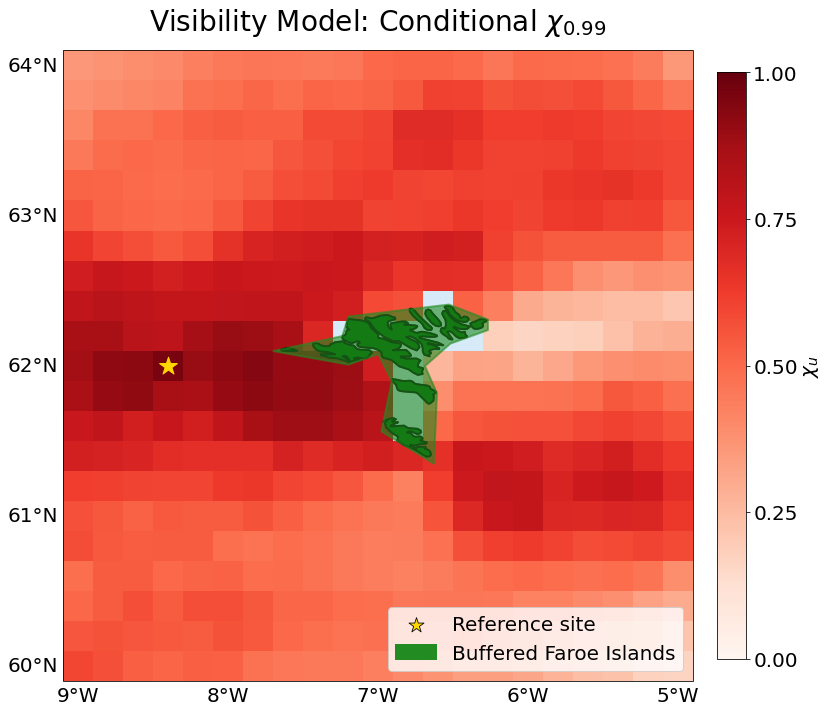} &
			\includegraphics[width=0.2\textwidth, clip=true, trim=0 0 30 40]
			{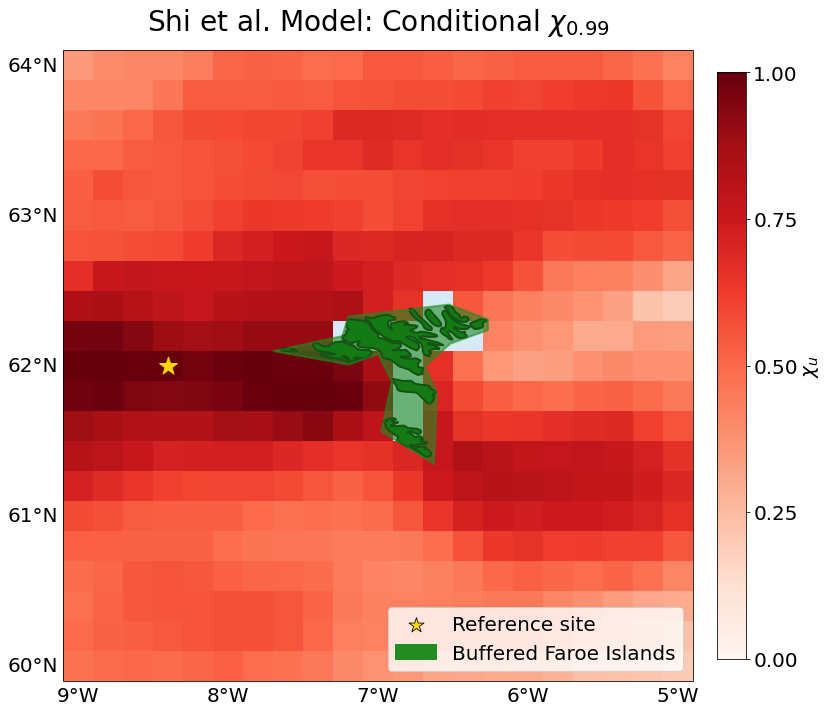} & 
			\includegraphics[width=0.2\textwidth, clip=true, trim=0 0 30 40]
			{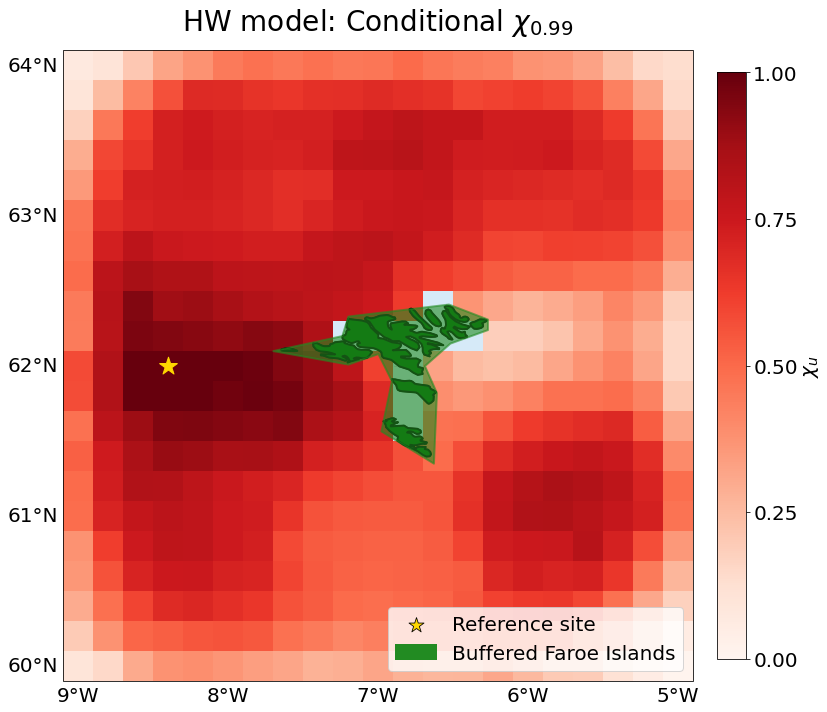} 
			\\
			\includegraphics[width=0.2\textwidth, clip=true, trim=0 0 30 40]
			{wave_emp_chiu_center_222_99.png} & 
			\includegraphics[width=0.2\textwidth, clip=true, trim=0 0 30 40]
			{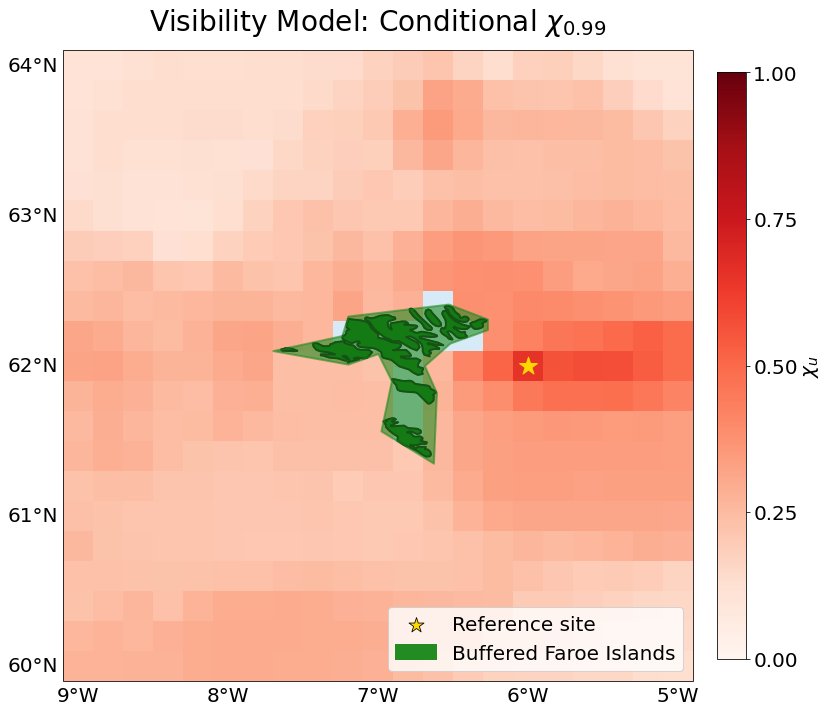} &
			\includegraphics[width=0.2\textwidth, clip=true, trim=0 0 30 40]
			{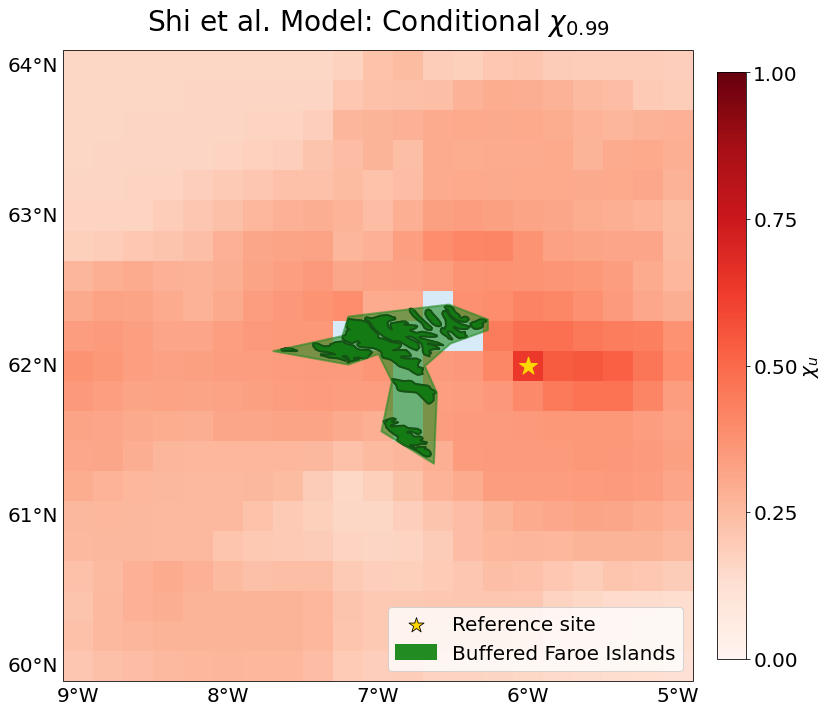} & 
			\includegraphics[width=0.2\textwidth, clip=true, trim=0 0 30 40]
			{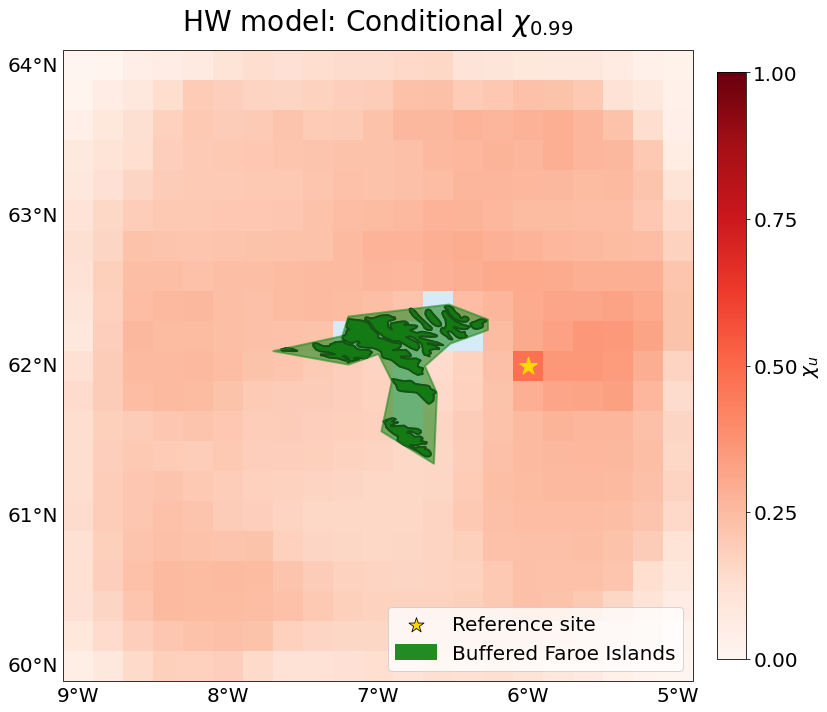} 
			\\
			\includegraphics[width=0.2\textwidth, clip=true, trim=0 0 30 40]
			{wave_emp_chiu_center_357_99.png} & 
			\includegraphics[width=0.2\textwidth, clip=true, trim=0 0 30 40]
			{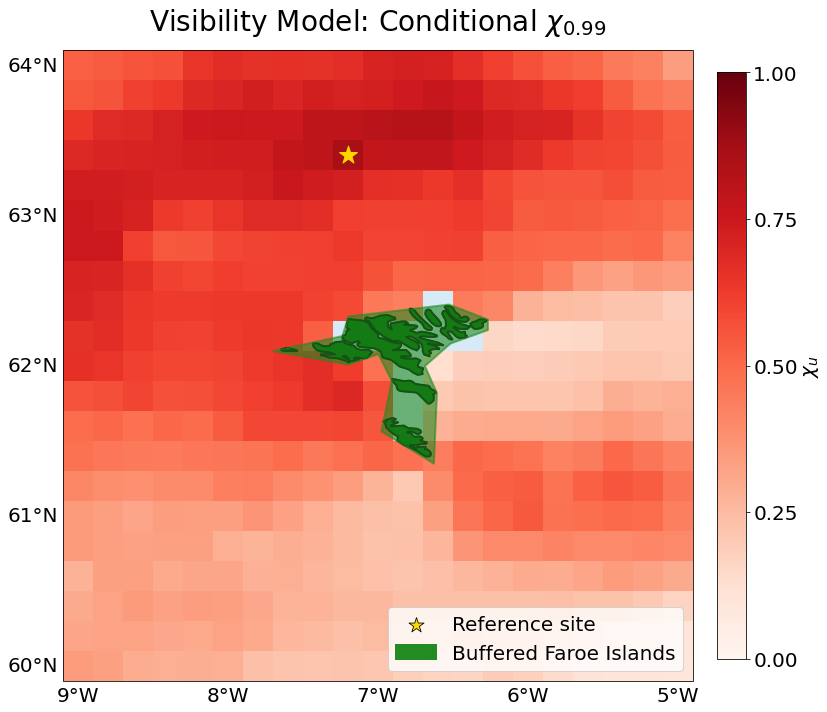} &
			\includegraphics[width=0.2\textwidth, clip=true, trim=0 0 30 40]
			{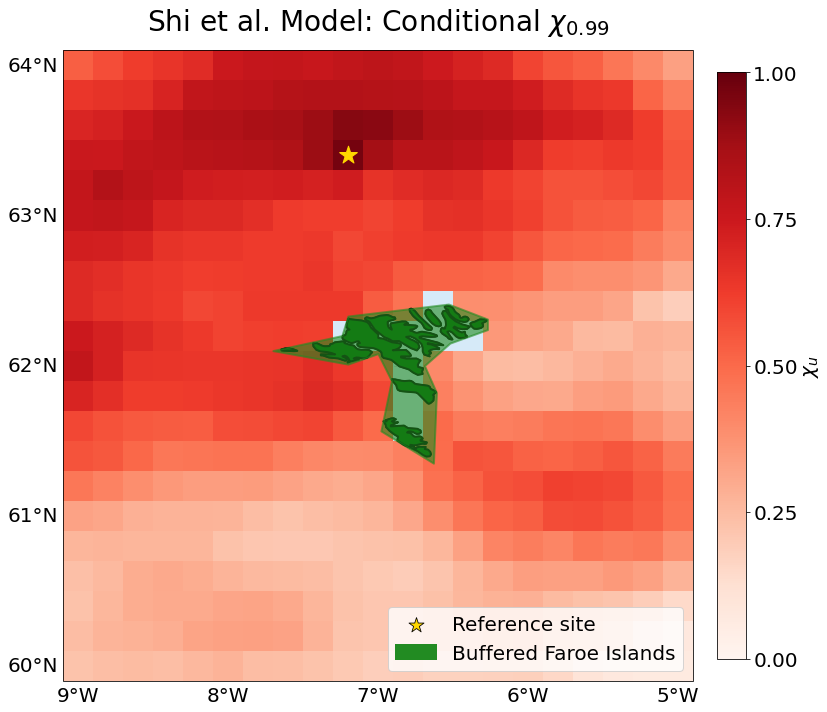} & 
			\includegraphics[width=0.2\textwidth, clip=true, trim=0 0 30 40]
			{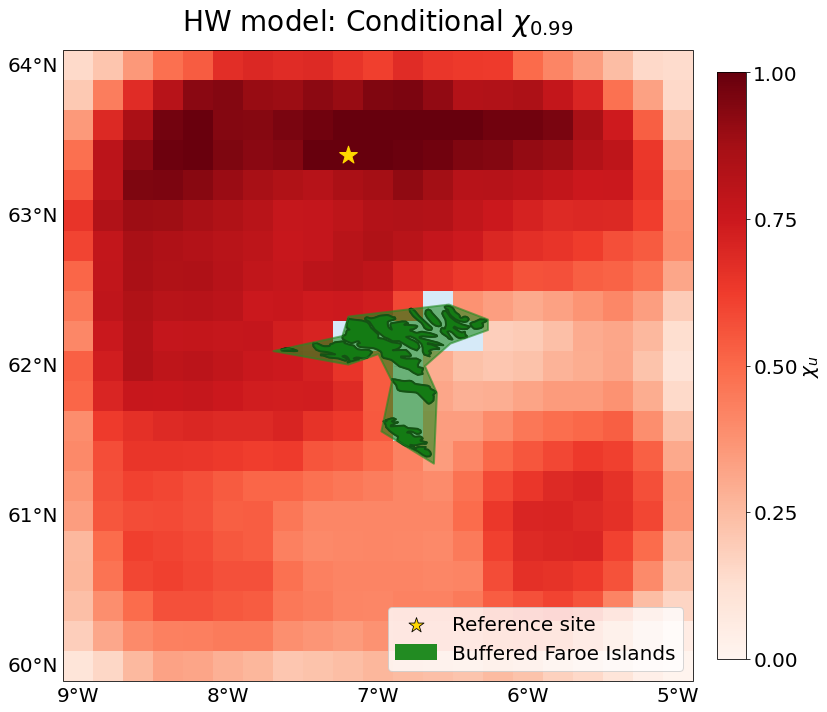} 
		\end{tabular}
		\caption{Empirical and model-based conditional $\chi_{0.99}$ surfaces around a reference site, for all three models.}
		\label{fig:wave_chiu_model_comparison_99}
	\end{figure}

\begin{figure}[htbp]
		\centering
		\begin{tabular}{cc cc}
			Empirical & Visibility & Shi et al. & Huser-Wadsworth \\
			\includegraphics[width=0.2\textwidth, clip=true, trim=0 0 30 40]
			{wave_emp_chiu_center_65_96.png} & 
			\includegraphics[width=0.2\textwidth, clip=true, trim=0 0 30 40]
			{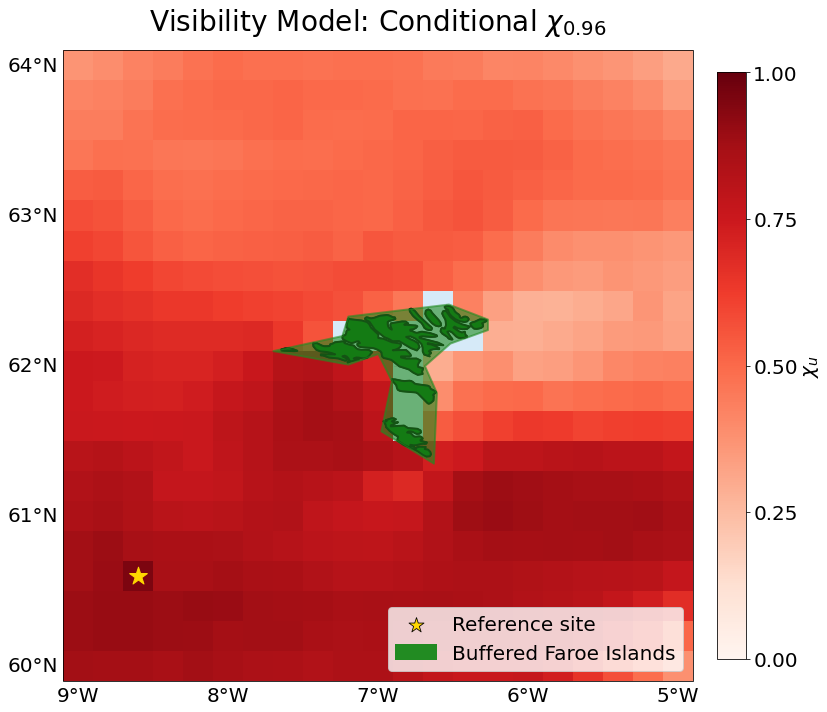} &
			\includegraphics[width=0.2\textwidth, clip=true, trim=0 0 30 40]
			{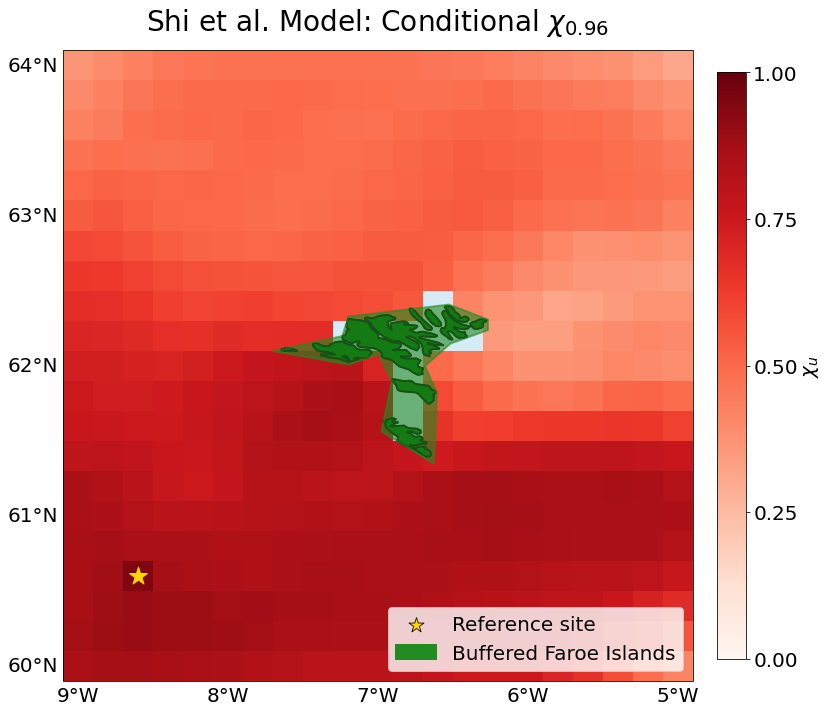} & 
			\includegraphics[width=0.2\textwidth, clip=true, trim=0 0 30 40]
			{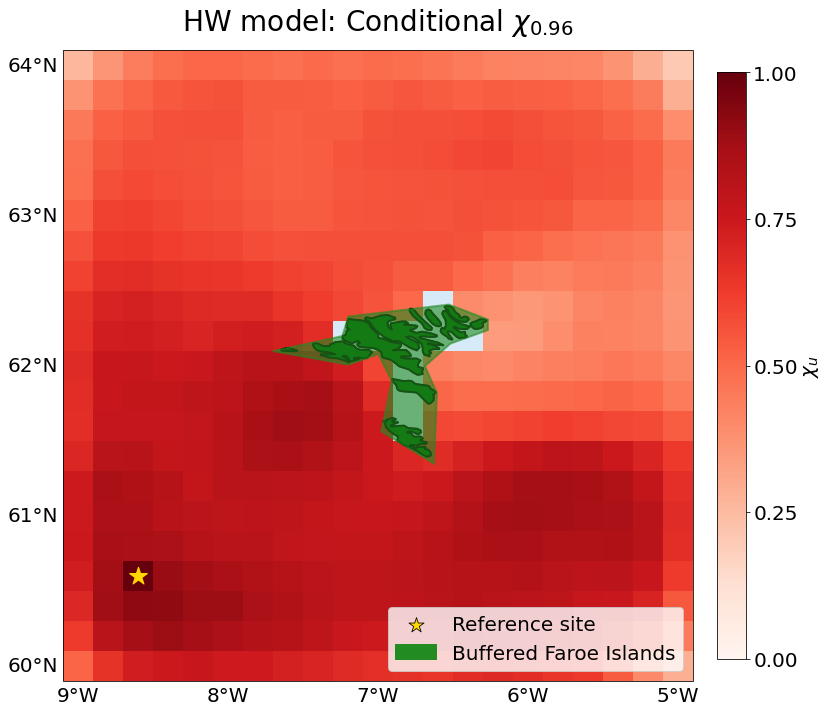} 
			\\
			\includegraphics[width=0.2\textwidth, clip=true, trim=0 0 30 40]
			{wave_emp_chiu_center_211_96.png} & 
			\includegraphics[width=0.2\textwidth, clip=true, trim=0 0 30 40]
			{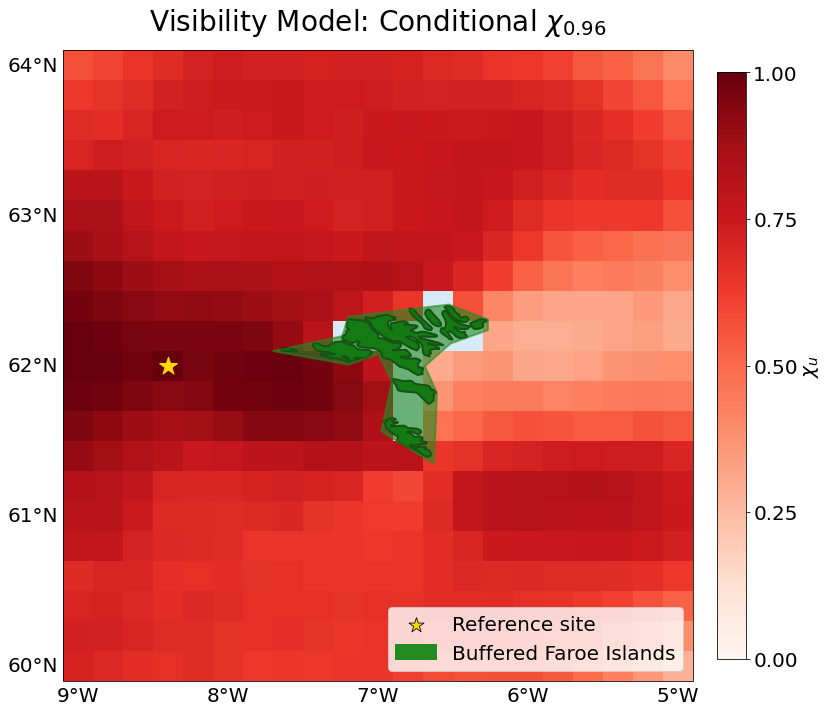} &
			\includegraphics[width=0.2\textwidth, clip=true, trim=0 0 30 40]
			{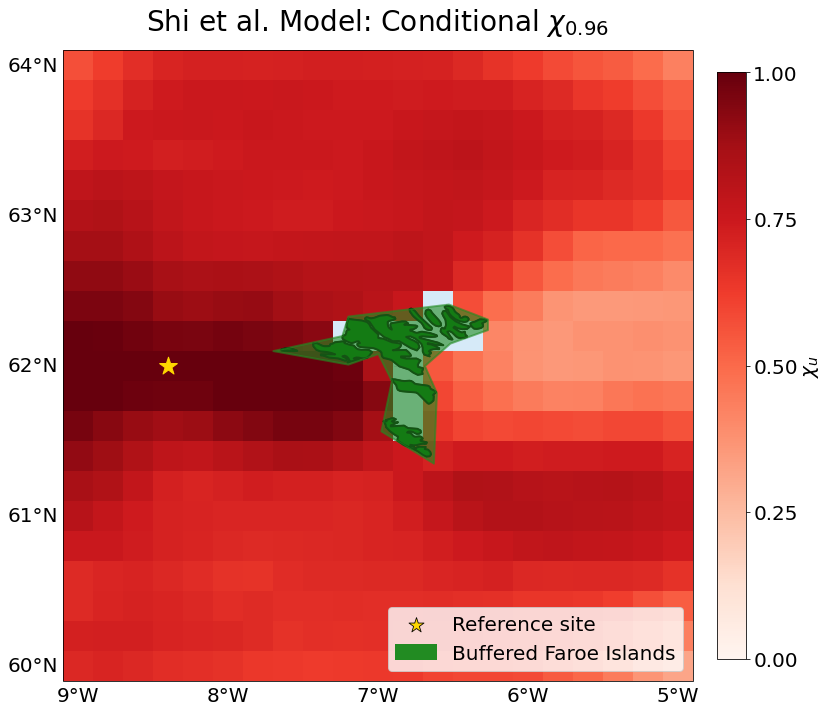} & 
			\includegraphics[width=0.2\textwidth, clip=true, trim=0 0 30 40]
			{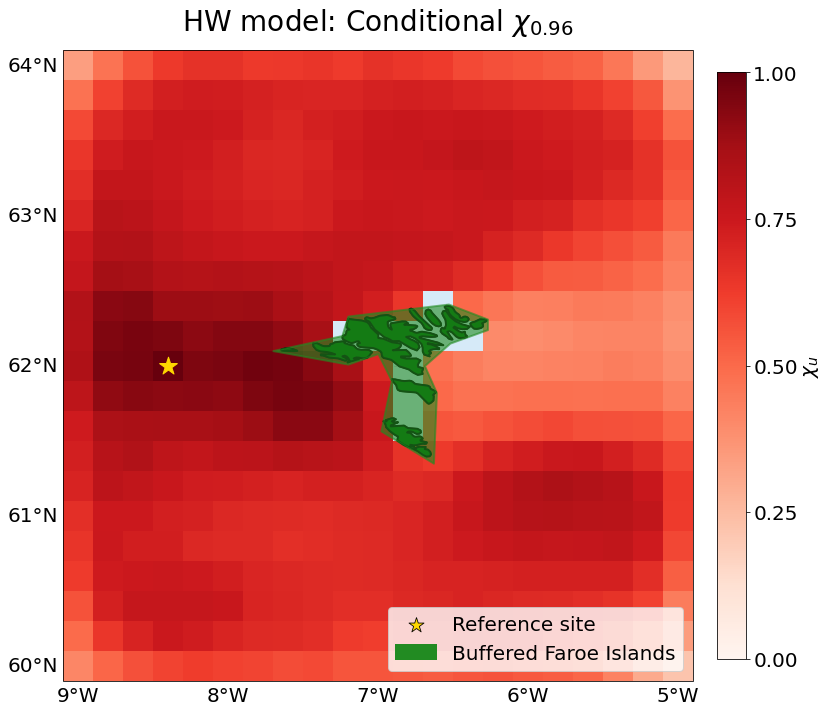} 
			\\
			\includegraphics[width=0.2\textwidth, clip=true, trim=0 0 30 40]
			{wave_emp_chiu_center_222_96.png} & 
			\includegraphics[width=0.2\textwidth, clip=true, trim=0 0 30 40]
			{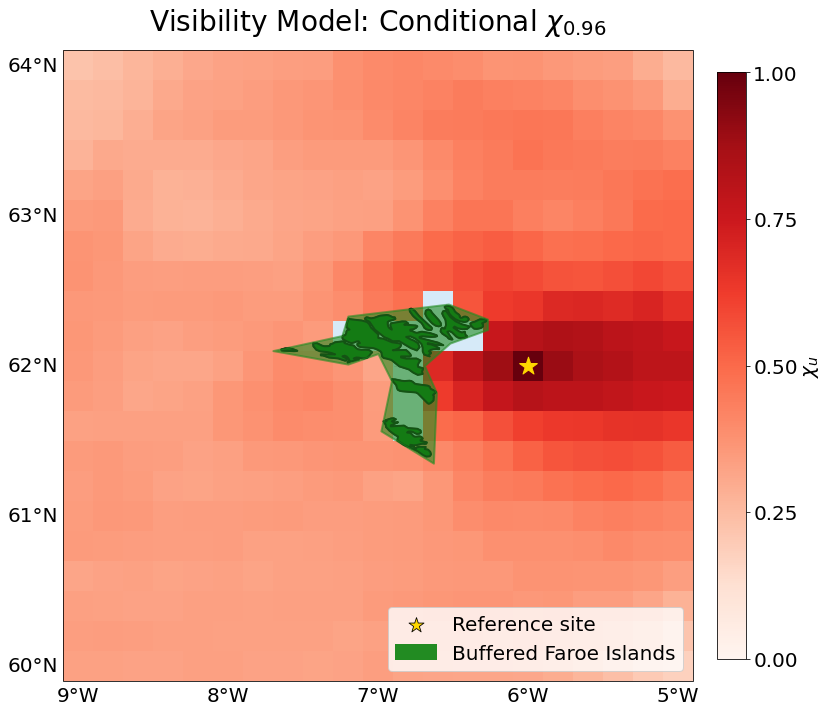} &
			\includegraphics[width=0.2\textwidth, clip=true, trim=0 0 30 40]
			{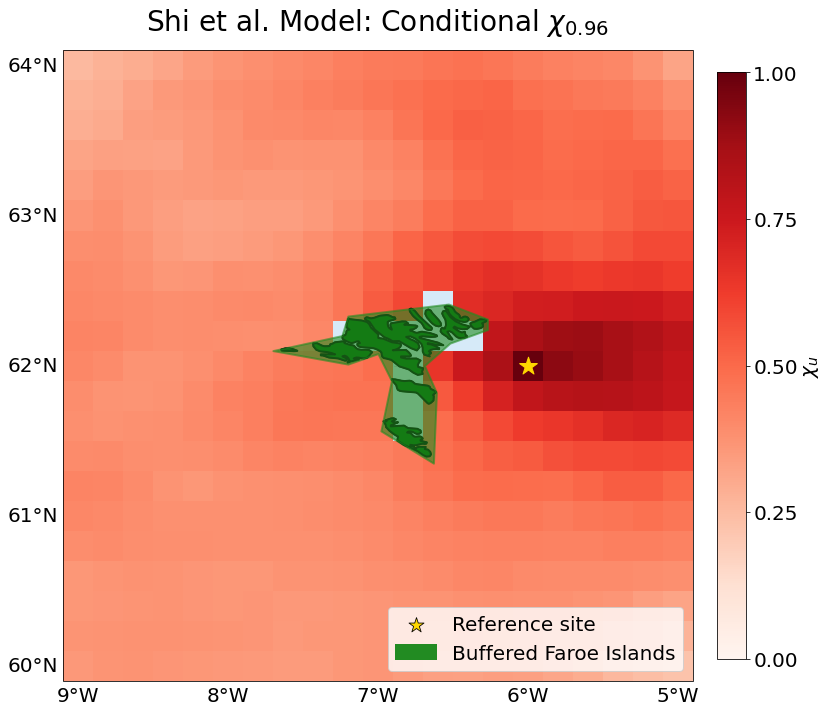} & 
			\includegraphics[width=0.2\textwidth, clip=true, trim=0 0 30 40]
			{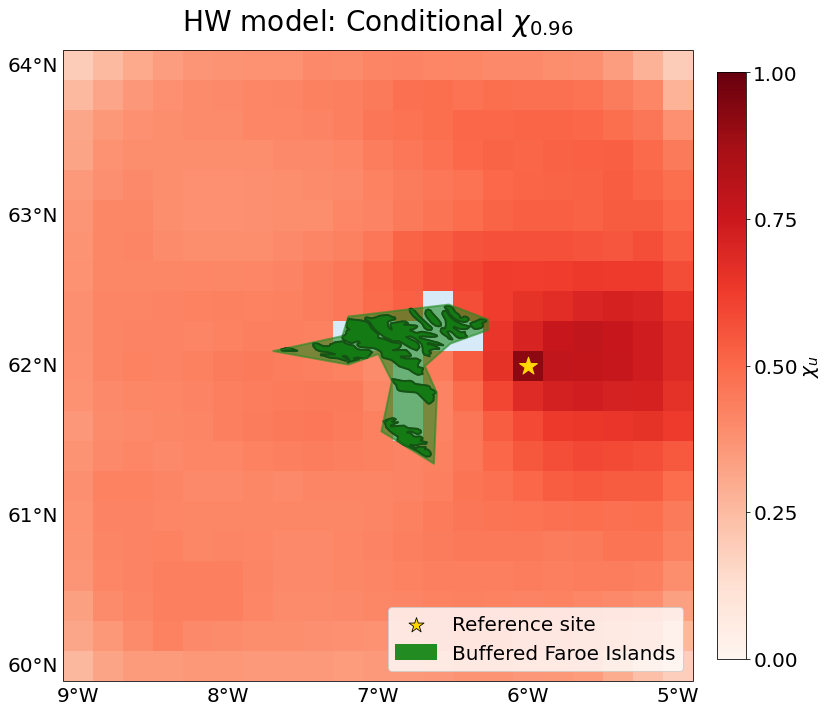} 
			\\
			\includegraphics[width=0.2\textwidth, clip=true, trim=0 0 30 40]
			{wave_emp_chiu_center_357_96.png} & 
			\includegraphics[width=0.2\textwidth, clip=true, trim=0 0 30 40]
			{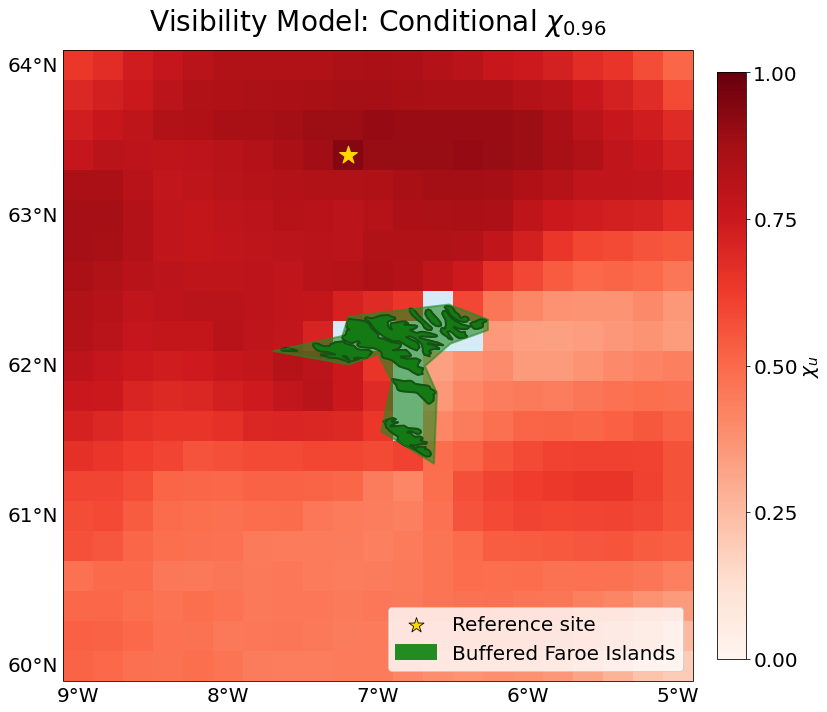} &
			\includegraphics[width=0.2\textwidth, clip=true, trim=0 0 30 40]
			{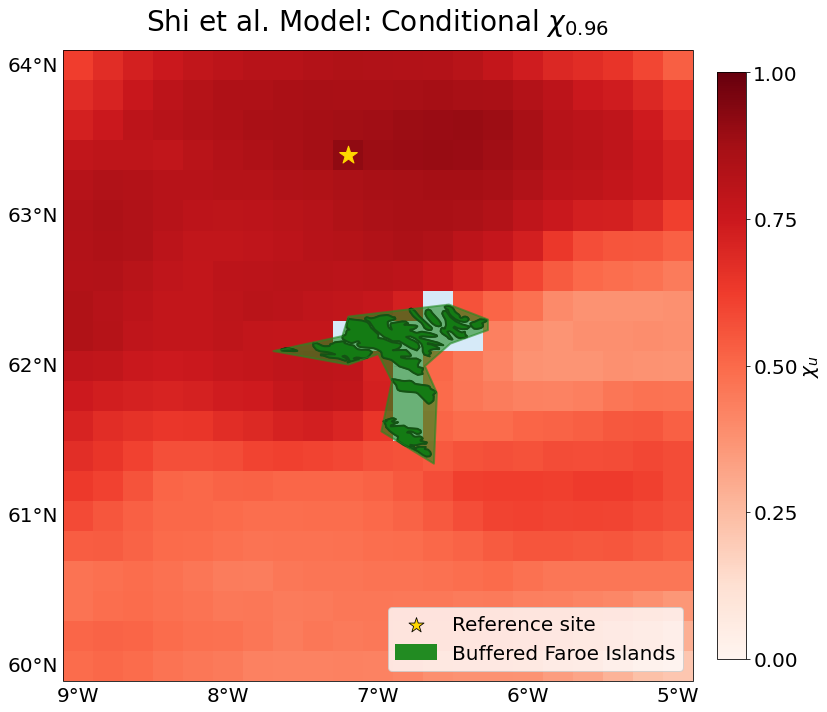} & 
			\includegraphics[width=0.2\textwidth, clip=true, trim=0 0 30 40]
			{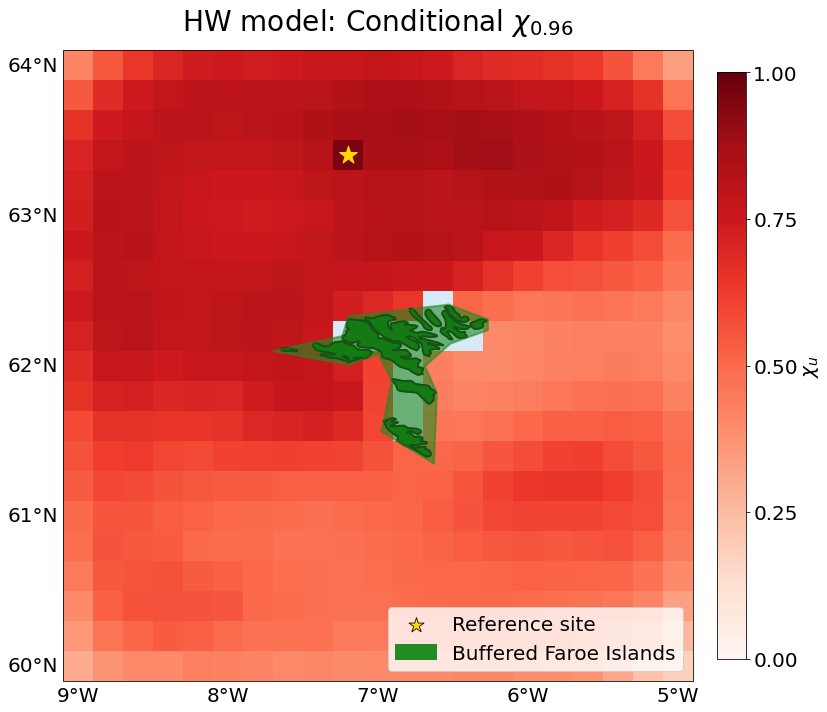} 
		\end{tabular}
		\caption{Empirical and model-based conditional $\chi_{0.96}$ surfaces around a reference site, for all three models.}
		\label{fig:wave_chiu_model_comparison_96}
	\end{figure}

\begin{figure}[htbp]
		\centering
		\begin{tabular}{cc cc}
			Empirical & Visibility & Shi et al. & Huser-Wadsworth \\
			\includegraphics[width=0.2\textwidth, clip=true, trim=0 0 30 40]
			{wave_emp_chiu_center_65_98.png} & 
			\includegraphics[width=0.2\textwidth, clip=true, trim=0 0 30 40]
			{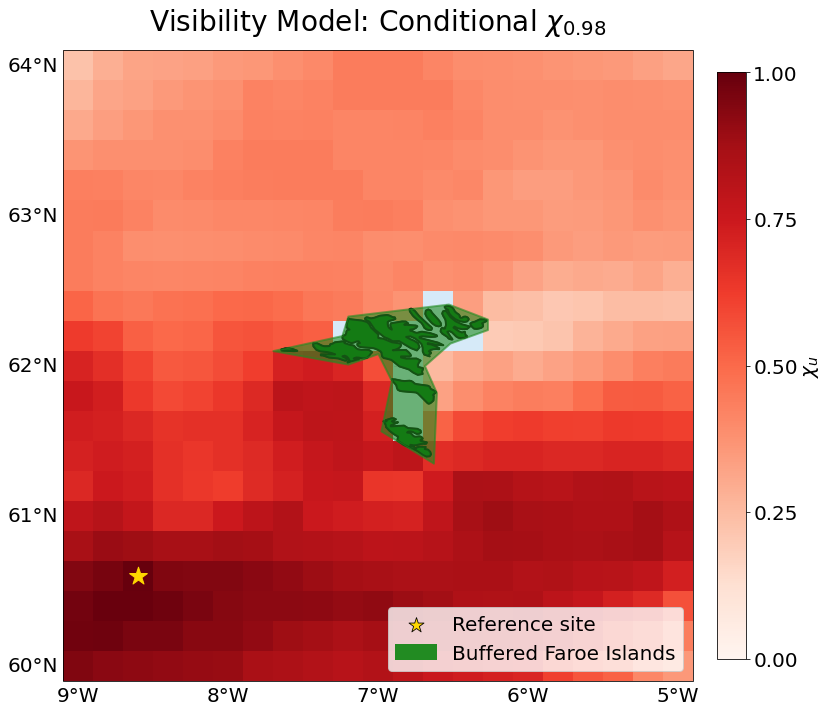} &
			\includegraphics[width=0.2\textwidth, clip=true, trim=0 0 30 40]
			{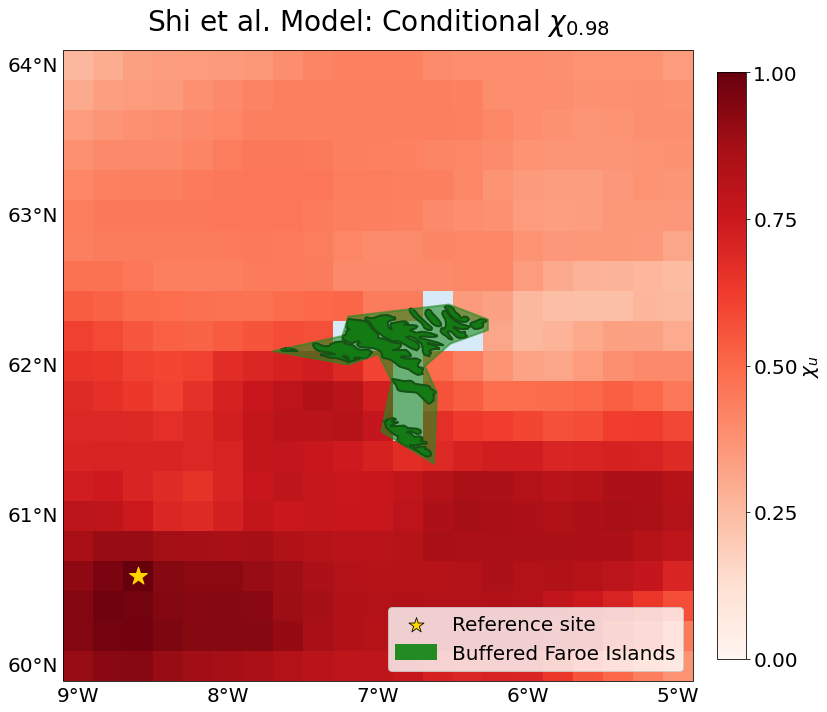} & 
			\includegraphics[width=0.2\textwidth, clip=true, trim=0 0 30 40]
			{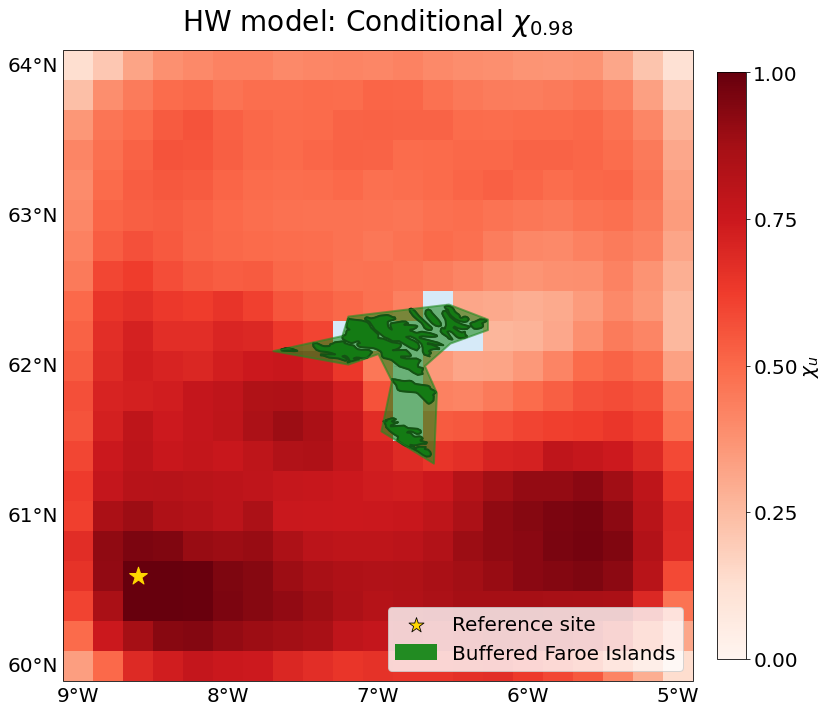} 
			\\
			\includegraphics[width=0.2\textwidth, clip=true, trim=0 0 30 40]
			{wave_emp_chiu_center_211_98.png} & 
			\includegraphics[width=0.2\textwidth, clip=true, trim=0 0 30 40]
			{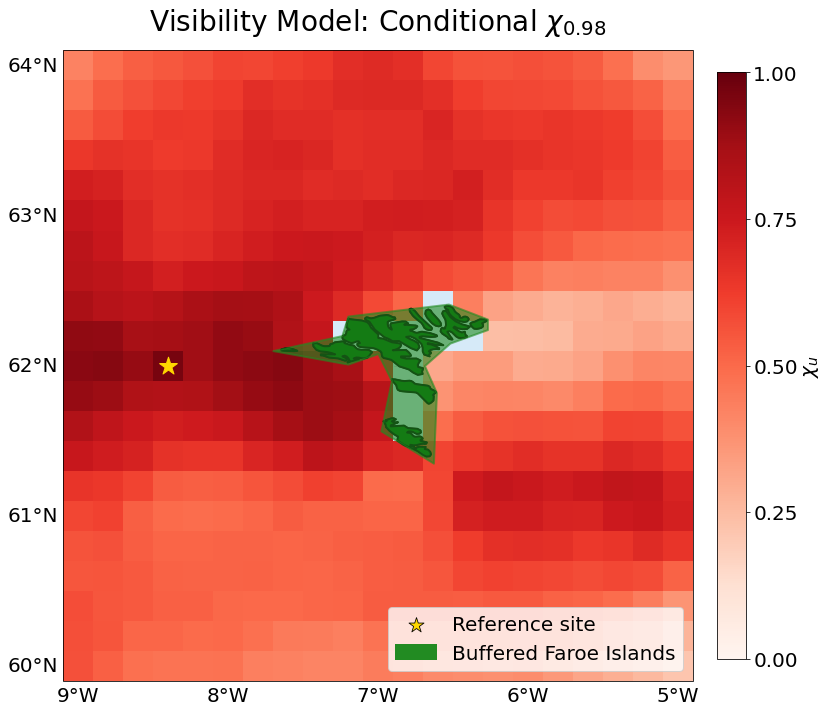} &
			\includegraphics[width=0.2\textwidth, clip=true, trim=0 0 30 40]
			{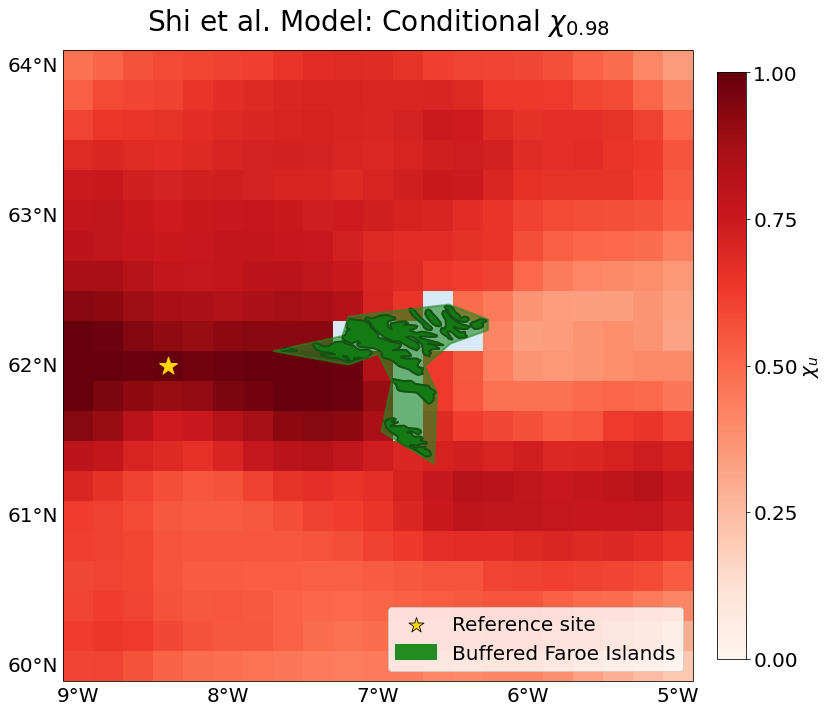} & 
			\includegraphics[width=0.2\textwidth, clip=true, trim=0 0 30 40]
			{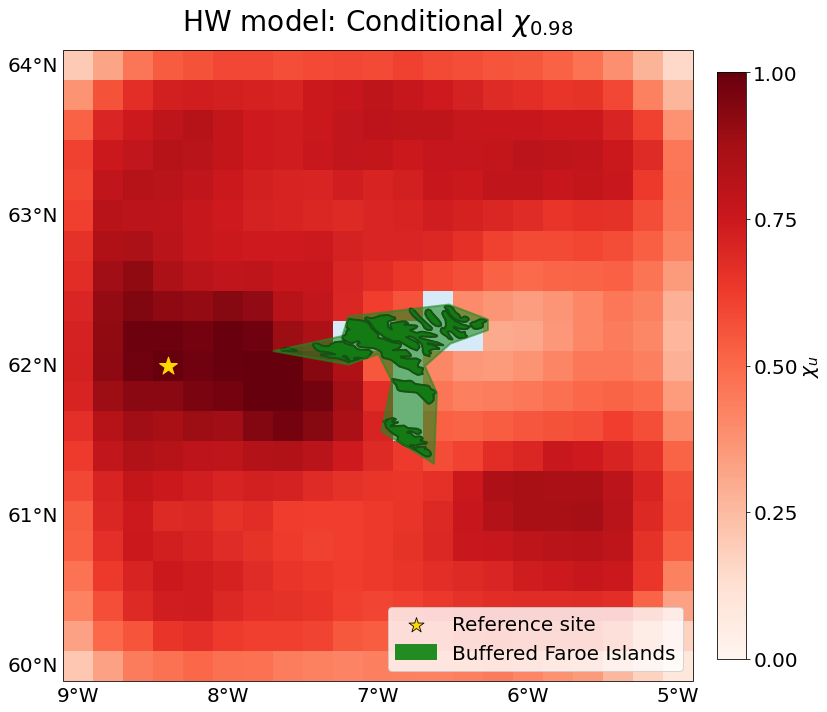} 
			\\
			\includegraphics[width=0.2\textwidth, clip=true, trim=0 0 30 40]
			{wave_emp_chiu_center_222_98.png} & 
			\includegraphics[width=0.2\textwidth, clip=true, trim=0 0 30 40]
			{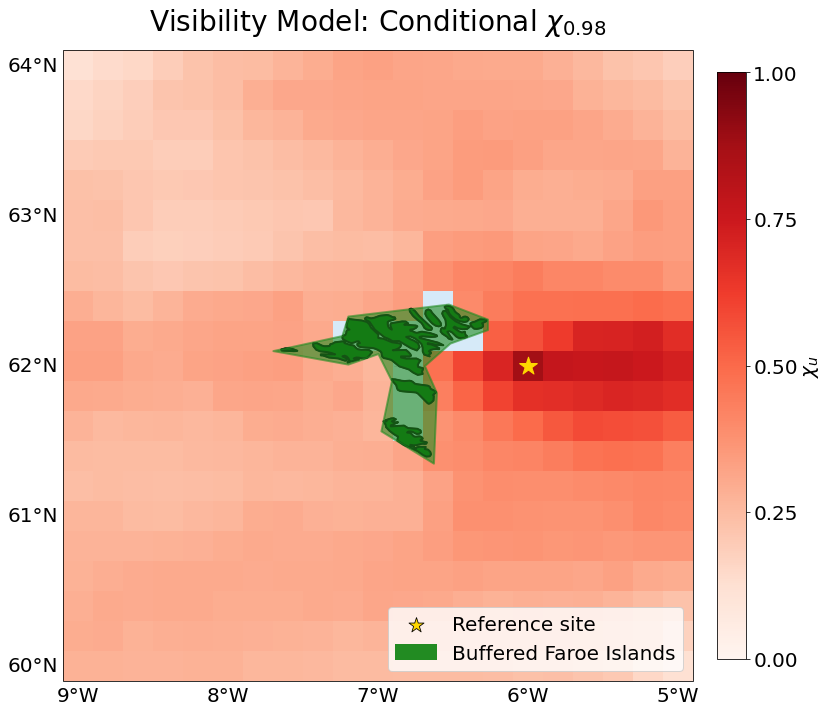} &
			\includegraphics[width=0.2\textwidth, clip=true, trim=0 0 30 40]
			{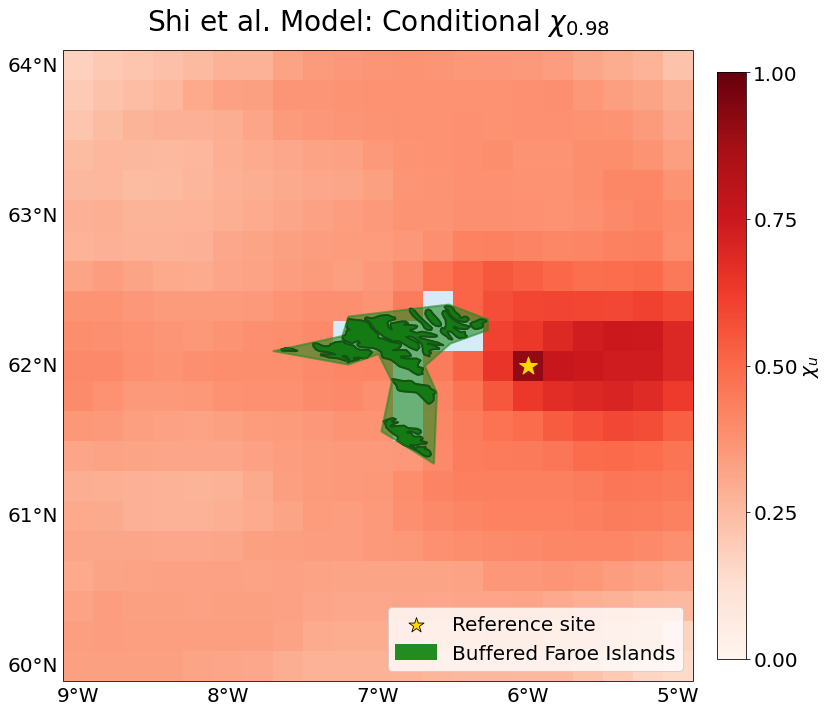} & 
			\includegraphics[width=0.2\textwidth, clip=true, trim=0 0 30 40]
			{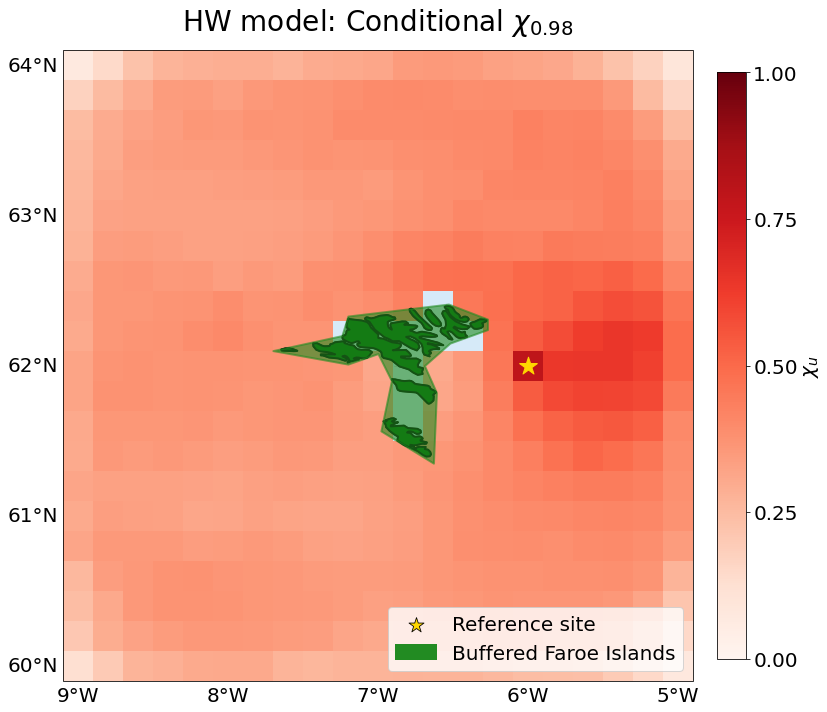} 
			\\
			\includegraphics[width=0.2\textwidth, clip=true, trim=0 0 30 40]
			{wave_emp_chiu_center_357_98.png} & 
			\includegraphics[width=0.2\textwidth, clip=true, trim=0 0 30 40]
			{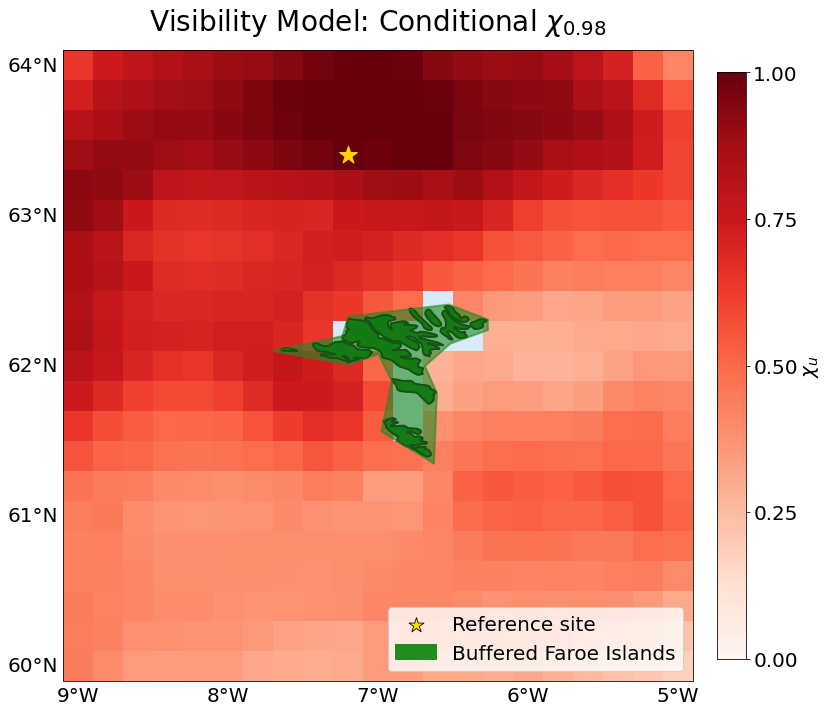} &
			\includegraphics[width=0.2\textwidth, clip=true, trim=0 0 30 40]
			{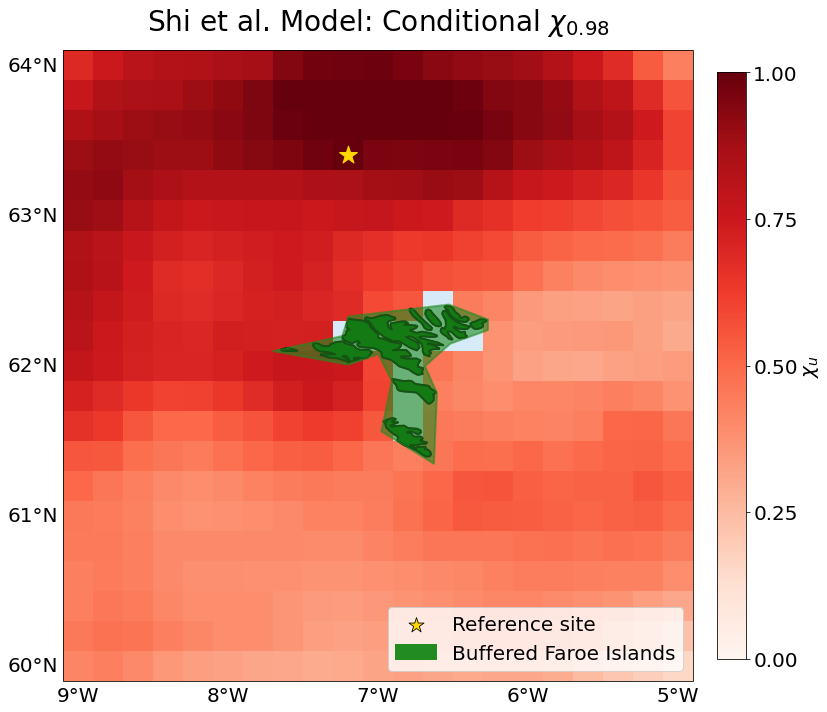} & 
			\includegraphics[width=0.2\textwidth, clip=true, trim=0 0 30 40]
			{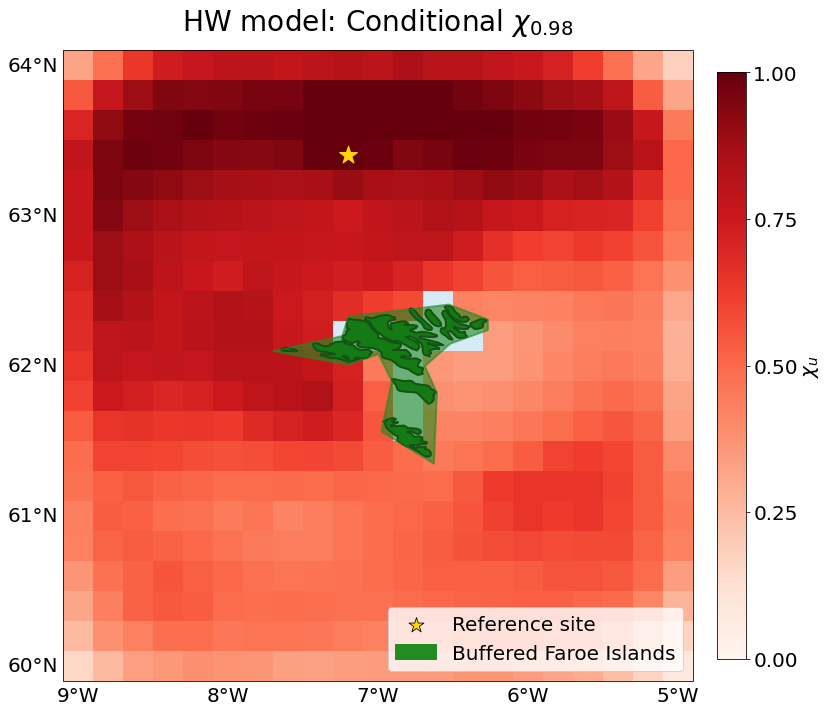} 
		\end{tabular}
		\caption{Empirical and model-based conditional $\chi_{0.98}$ surfaces around a reference site, for all three models.}
		\label{fig:wave_chiu_model_comparison_98}
	\end{figure}

	Figure~\ref{fig:wave_chiu_model_comparison_99}--\ref{fig:wave_chiu_model_comparison_98} compare empirical and model-based conditional extremal dependence estimates, $\chi_u$, at thresholds $u = 0.99$, $u = 0.96$, and $u = 0.98$, respectively, for the held-out wave-height locations. Each figure consider four reference sites, located in the southwestern open ocean and on the western, eastern, and northern sides of the Faroe Islands, from top to bottom. The columns show the empirical estimates and the corresponding estimates from the visibility, Shi et al., and Huser--Wadsworth models, from left to right.
	
	Across all thress thresholds, the empirical surfaces exhibit clear directional variation in extremal dependence around the islands. In particular, dependence is generally attenuated in directions obstructed by the islands, although the extent and spatial pattern of this attenuation vary with the reference site. The visibility model reproduces these broad spatial features, including the contrast between unobstructed ocean regions and areas shielded by the islands. This agreement is particularly apparent in the reference-site-specific patterns, which cannot be characterized by Euclidean distance alone. 
	
	The Shi et al. and Huser--Wadsworth models also capture some of the large-scale spatial variation, but their fitted surfaces tend to be smoother and less effective at representing the directional attenuation associated with the islands. The differences are particularly evident when the reference site is located to the east or north of the islands, where the empirical estimates exhibit distinct dependence patterns across the barrier. By incorporating line-of-sight visibility into the dependence structure, the proposed visibility model provides a more direct representation of these barrier-related features. 
	
	The overall directional patterns are broadly consistent across the three thresholds, suggesting that the main spatial features are not specific to a particular threshold choice. Nevertheless, the empirical surfaces remain spatially heterogeneous, and the agreement between the fitted and empirical estimates is not uniform across all reference sites. Taken together, these holdout results illustrate the value of incorporating physical barriers into the dependence structure when predicting extremal dependence at unobserved locations. 

\end{document}